\documentclass[12pt]{iopart}
\usepackage[utf8]{inputenc}
\usepackage{iopams}
\usepackage{overpic,bbm,bm,epsfig}
\usepackage{multirow}
\usepackage{color}
\usepackage{braket}

\begin{document}

\title[Zhi-zhong Xing]{The $\mu$-$\tau$ reflection symmetry of Majorana
neutrinos\footnote{Dedicated to the memory of Prof. Harald Fritzsch.}}

\author{Zhi-zhong Xing$^{1,2}$}
\address{$^1$Institute of High Energy Physics and School of Physical Sciences,
University of Chinese Academy of Sciences, Beijing 100049, China \\
$^2$Center of High Energy Physics, Peking University, Beijing 100871, China}
\vspace{10pt}
\begin{indented}
\item[] E-mail: xingzz@ihep.ac.cn
\end{indented}

\begin{abstract}
The observed pattern of lepton flavor mixing and CP violation strongly indicates
the possible existence of a simple flavor symmetry in the neutrino
sector --- the effective Majorana neutrino mass term keeps invariant when the
three left-handed neutrino fields transform as
$\nu^{}_{e \rm L} \to (\nu^{}_{e \rm L})^c$,
$\nu^{}_{\mu \rm L} \to (\nu^{}_{\tau \rm L})^c$ and
$\nu^{}_{\tau \rm L} \to (\nu^{}_{\mu \rm L})^c$. A direct application of
such a $\mu$-$\tau$ reflection symmetry to the canonical seesaw mechanism can help
a lot to constrain the flavor textures of active and sterile Majorana neutrinos.
The present article is intended to summarize the latest progress made in
exploring the properties of this minimal flavor symmetry, its translational
and rotational extensions, its soft breaking effects via radiative corrections
from a superhigh energy scale to the electroweak scale, and its various
phenomenological implications.
\end{abstract}


\tableofcontents

\def\thefootnote{\arabic{footnote}}
\setcounter{footnote}{0}


\setcounter{equation}{0}
\section{Introduction}
\label{section 1}

\subsection{Some key flavor issues of leptons and quarks}
\label{section 1.1}

\begin{figure}[t]
\begin{center}
\includegraphics[width=4.4in]{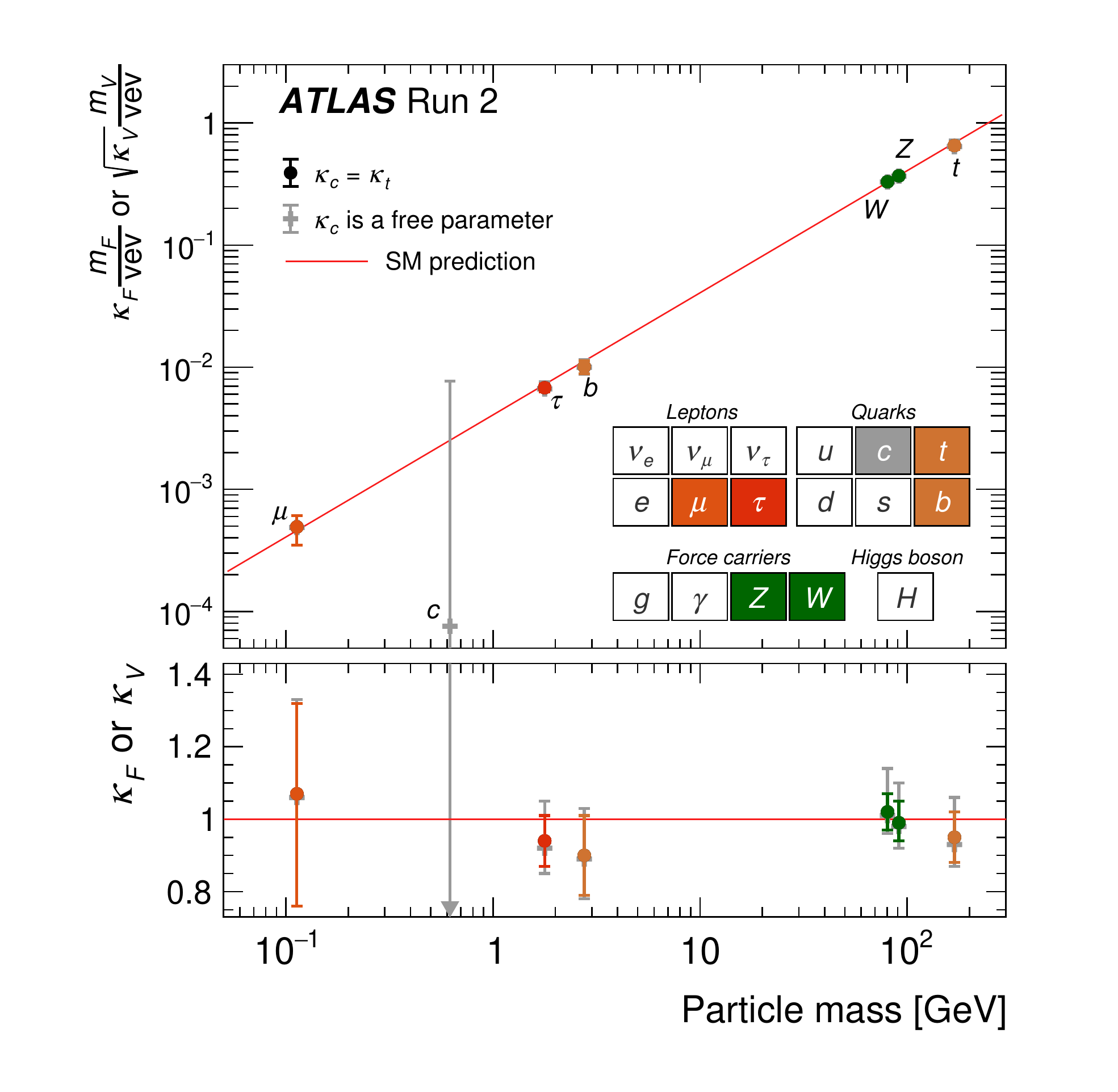}
\vspace{-0.65cm}
\caption{The reduced Higgs boson coupling strength modifiers defined as
$\kappa^{}_F m^{}_F/{\rm vev}$ for a charged fermion with mass $m^{}_F$ and
$\sqrt{\kappa^{}_V} m^{}_V/{\rm vev}$ for a vector boson with mass
$m^{}_V$~\cite{ATLAS:2022vkf}, where the
scale factors $\kappa^{}_F$ and $\kappa^{}_V$ satisfy $\kappa^{}_F =
\kappa^{}_V = 1$ in the SM and ``vev" denotes the vacuum expectation value
of the Higgs field. Two fit scenarios with $\kappa^{}_c = \kappa^{}_t$
(coloured circle markers) or $\kappa^{}_c$ left free-floating in the fit
(grey cross markers) are shown. The loop-induced processes are assumed to
have the SM structure, and the Higgs boson decays to the non-SM particles
are not allowed. The vertical bar on each point denotes the $68\%$ confidence
interval. The lower panel shows the values of the coupling strength modifiers.
The grey arrow points in the direction of the best-fit value and the
corresponding grey uncertainty bar extends beyond the lower panel range.
The relevant data are from the ATLAS Run II experiment~\cite{ATLAS:2022vkf}.}
\label{Fig:ATLAS}
\end{center}
\end{figure}

The standard model (SM) of modern particle physics is composed of a unified
theory of electromagnetic and weak interactions established in the
1960s~\cite{Glashow:1961tr,Weinberg:1967tq,Salam:1968rm} and
quantum chromodynamics (QCD) for strong interactions formulated in the
1970s~\cite{Fritzsch:1973pi,Gross:1973id,Politzer:1973fx}. This model has proved
to be a huge success in the past five decades, as its consistency and
predictive power have been tested extensively and accurately~\cite{Workman:2022ynf}.
In particular, recent ATLAS~\cite{ATLAS:2022vkf} and CMS~\cite{CMS:2022dwd}
measurements of the Higgs boson's couplings to the vector bosons (i.e., $W^\pm$
and $Z^0$) and the heavier charged fermions (e.g., $t$, $b$, $\tau$ and $\mu$)
have convincingly verified the mass generation mechanism of these fundamental
particles in the SM. Figure~\ref{Fig:ATLAS} illustrates the ATLAS result as an
example, and it is fully compatible with the CMS measurement. In this situation
one may naturally anticipate that the masses of those lighter charged fermions
should also originate from their corresponding Yukawa interactions with the
Higgs field.

But it is well known that the SM is actually incomplete and unsatisfactory in several
aspects, especially in its flavor sector~\cite{Xing:2019vks}. Let us highlight the
flavor puzzles from the following two perspectives. On the one hand, the SM does not
lay a foundation for nonzero neutrino masses and significant lepton flavor mixing
effects; on the other hand, all of the flavor parameters in the SM are theoretically
undetermined and their values can only be extracted from a variety of high-energy
and low-energy experiments~\cite{Workman:2022ynf}. Nevertheless, the SM has
provided a canonical pathway for qualitatively unravelling the mysteries of flavor
mixing and CP violation in the quark sector --- the Cabibbo-Kobayashi-Maskawa (CKM)
mechanism~\cite{Cabibbo:1963yz,Kobayashi:1973fv}. A quite similar picture, known as the
Pontecorvo-Maki-Nakagawa-Sakata (PMNS) mechanism~\cite{Pontecorvo:1957cp,Maki:1962mu},
is expected to work in the lepton sector once the neutrinos acquire their finite
but tiny masses in a natural and straightforward extension of the SM framework.

In this regard the key point is that the known fermions
of the same electric charge all have three families, and they interact not only
with the gauge fields but also with the scalar fields. As a consequence, there must
be an irremovable mismatch between the mass and flavor eigenstates of those fermion
fields. It should be noted that the mass eigenstates of quarks and leptons
are defined respectively by their participation in the strong interactions and by
their propagation in free space, and the flavor eigenstates of either leptons or
quarks are defined by their participation in the electroweak interactions.
Such a dynamical mismatch is highly nontrivial, and it is
described by the CKM quark flavor mixing matrix $V$ or the PMNS lepton flavor
mixing matrix $U$ in the standard weak charged-current interactions:
\begin{eqnarray}
&& -{\cal L}^{(q)}_{\rm cc} = \frac{g^{}_{\rm w}}{\sqrt{2}} \hspace{0.05cm}
\overline{\left(u \hspace{0.3cm} c \hspace{0.3cm} t\right)^{}_{\rm L}}
\hspace{0.05cm} \gamma^\mu \hspace{0.05cm} V \pmatrix{ \hspace{0.08cm} d
\hspace{0.08cm} \cr s \cr b \cr}^{}_{\hspace{-0.15cm} \rm L} W^+_\mu
+ {\rm h.c.} \; ,
\nonumber \\
&& -{\cal L}^{(l)}_{\rm cc} = \frac{g^{}_{\rm w}}{\sqrt{2}} \hspace{0.05cm}
\overline{\left(e ~~ \mu ~~ \tau\right)^{}_{\rm L}} \hspace{0.05cm} \gamma^\mu
\hspace{0.05cm} U \pmatrix{ \nu^{}_{1} \cr \nu^{}_{2} \cr
\nu^{}_{3} \cr}^{}_{\hspace{-0.15cm} \rm L} W^-_\mu + {\rm h.c.} \; ,
\label{eq:Lcc}
\end{eqnarray}
where $g^{}_{\rm w}$ denotes the coupling constant of weak interactions, and
all the relevant fermion fields are in their mass eigenstates and have the
left-handed chirality. The SM tells us that the $3\times 3$ matrix $V$ must be
unitary and hence can be fully described in terms of
four independent parameters, usually chosen as three Euler-like
flavor mixing angles (denoted by $\vartheta^{}_{12}$, $\vartheta^{}_{13}$ and
$\vartheta^{}_{23}$) and one CP-violating phase (denoted by $\delta^{}_q$) as
advocated by the Particle Data Group~\cite{Workman:2022ynf}:
\begin{eqnarray}
V = P^{}_{\rm u} \pmatrix{
c^{}_{12} c^{}_{13} & s^{}_{12} c^{}_{13} & \hat{s}^{*}_{13} \cr
-s^{}_{12} c^{}_{23} - c^{}_{12} \hat{s}^{}_{13} s^{}_{23} & c^{}_{12} c^{}_{23} -
s^{}_{12} \hat{s}^{}_{13} s^{}_{23} & c^{}_{13} s^{}_{23} \cr
s^{}_{12} s^{}_{23} - c^{}_{12} \hat{s}^{}_{13} c^{}_{23}
&- c^{}_{12} s^{}_{23} - s^{}_{12} \hat{s}^{}_{13}
c^{}_{23} &  c^{}_{13} c^{}_{23} \cr} P^{}_{\rm d} \; ,
\label{eq:V}
\end{eqnarray}
where $c^{}_{ij} \equiv \cos\vartheta^{}_{ij}$, $s^{}_{ij} \equiv
\sin\vartheta^{}_{ij}$ and $\hat{s}^{}_{ij} \equiv s^{}_{ij} e^{{\rm i} \delta^{}_q}$
(for $ij = 12, 13, 23$) with $\vartheta^{}_{ij}$ lying in the first quadrant,
$\delta^{}_q$ is the irreducible CP-violating phase,
$P^{}_{\rm u} = {\rm Diag}\{e^{{\rm i}\phi^{}_u} , e^{{\rm i}\phi^{}_c} ,
e^{{\rm i}\phi^{}_t}\}$ and $P^{}_{\rm d} = {\rm Diag}\{e^{{\rm i}\phi^{}_d} ,
e^{{\rm i}\phi^{}_s} , e^{{\rm i}\phi^{}_b}\}$ are two basis-dependent phase
matrices which have no physical significance. In comparison, whether the $3\times 3$
PMNS matrix $U$ is unitary or not depends on how the tiny
masses of three active neutrinos originate
\footnote{In a fermionic seesaw mechanism of either type-I~\cite{Fritzsch:1975sr,
Minkowski:1977sc,Yanagida:1979as,GellMann:1980vs,Glashow:1979nm,Mohapatra:1979ia}
or type-III~\cite{Foot:1988aq}, the PMNS matrix $U$ may slightly deviate from its exact
unitarity limit because of slight mixing between the active neutrino sector and the sterile
neutrino sector or the fermion triplet sector via the corresponding Yukawa interactions.
But $U$ is exactly unitary in the type-II
seesaw mechanism with a Higgs triplet being introduced into the SM~\cite{Konetschny:1977bn,Magg:1980ut,Schechter:1980gr,Cheng:1980qt,
Lazarides:1980nt,Mohapatra:1980yp}.}.
Fortunately, $U$ is expected to be essentially unitary at least at the ${\cal O}(1\%)$
accuracy level as constrained by both today's neutrino oscillation data and
precision measurements of various electroweak and lepton-flavor-violating processes  ~\cite{Antusch:2006vwa,Fernandez-Martinez:2016lgt,Blennow:2016jkn,Hu:2020oba,Wang:2021rsi}.
That is why we are going to treat $U$ as a unitary matrix unless otherwise specified.
In this case the $3\times 3$ unitary matrix $U$ can simply be parametrized in terms
of three lepton flavor mixing angles (denoted as $\theta^{}_{12}$, $\theta^{}_{13}$
and $\theta^{}_{23}$) and one or three nontrivial CP-violating phases (denoted as
$\delta^{}_\nu$ or as $\delta^{}_\nu$, $\rho$ and $\sigma$), corresponding to
the Dirac or Majorana nature of massive neutrinos. Similar to the CKM matrix $V$,
the standard parametrization of the PMNS matrix $U$ is
\begin{eqnarray}
U = P^{}_l \pmatrix{
c^{}_{12} c^{}_{13} & s^{}_{12} c^{}_{13} & \hat{s}^{*}_{13} \cr
-s^{}_{12} c^{}_{23} - c^{}_{12} \hat{s}^{}_{13} s^{}_{23} & c^{}_{12} c^{}_{23} -
s^{}_{12} \hat{s}^{}_{13} s^{}_{23} & c^{}_{13} s^{}_{23} \cr
s^{}_{12} s^{}_{23} - c^{}_{12} \hat{s}^{}_{13} c^{}_{23}
&- c^{}_{12} s^{}_{23} - s^{}_{12} \hat{s}^{}_{13}
c^{}_{23} &  c^{}_{13} c^{}_{23} \cr} P^{}_\nu \; ,
\label{eq:U}
\end{eqnarray}
where $c^{}_{ij} \equiv \cos\theta^{}_{ij}$, $s^{}_{ij} \equiv \sin\theta^{}_{ij}$
and $\hat{s}^{}_{ij} \equiv s^{}_{ij} e^{{\rm i} \delta^{}_\nu}$
(for $ij = 12, 13, 23$) with all $\theta^{}_{ij}$ lying in the first quadrant,
$\delta^{}_\nu$ is usually referred to as the Dirac CP-violating phase,
the diagonal Majorana phase matrix $P^{}_\nu$ is defined as
$P^{}_\nu \equiv {\rm Diag}\big\{e^{{\rm i}\rho}, e^{{\rm i}\sigma}, 1\big\}$, and
the arbitrary phase matrix $P^{}_l = {\rm Diag}\{e^{{\rm i}\phi^{}_e} ,
e^{{\rm i}\phi^{}_\mu} , e^{{\rm i}\phi^{}_\tau}\}$ has no physical significance.
Although $P^{}_{\rm u}$, $P^{}_{\rm d}$ and $P^{}_l$ are irrelevant to any
physical quantities, they should not be ignored when a specific flavor symmetry
is applied to constraining either the pattern of $V$ or that of $U$. This point
will become transparent later (see, e.g., section~\ref{section 2.1}).

\begin{figure}[t]
\begin{center}
\includegraphics[width=5.5in]{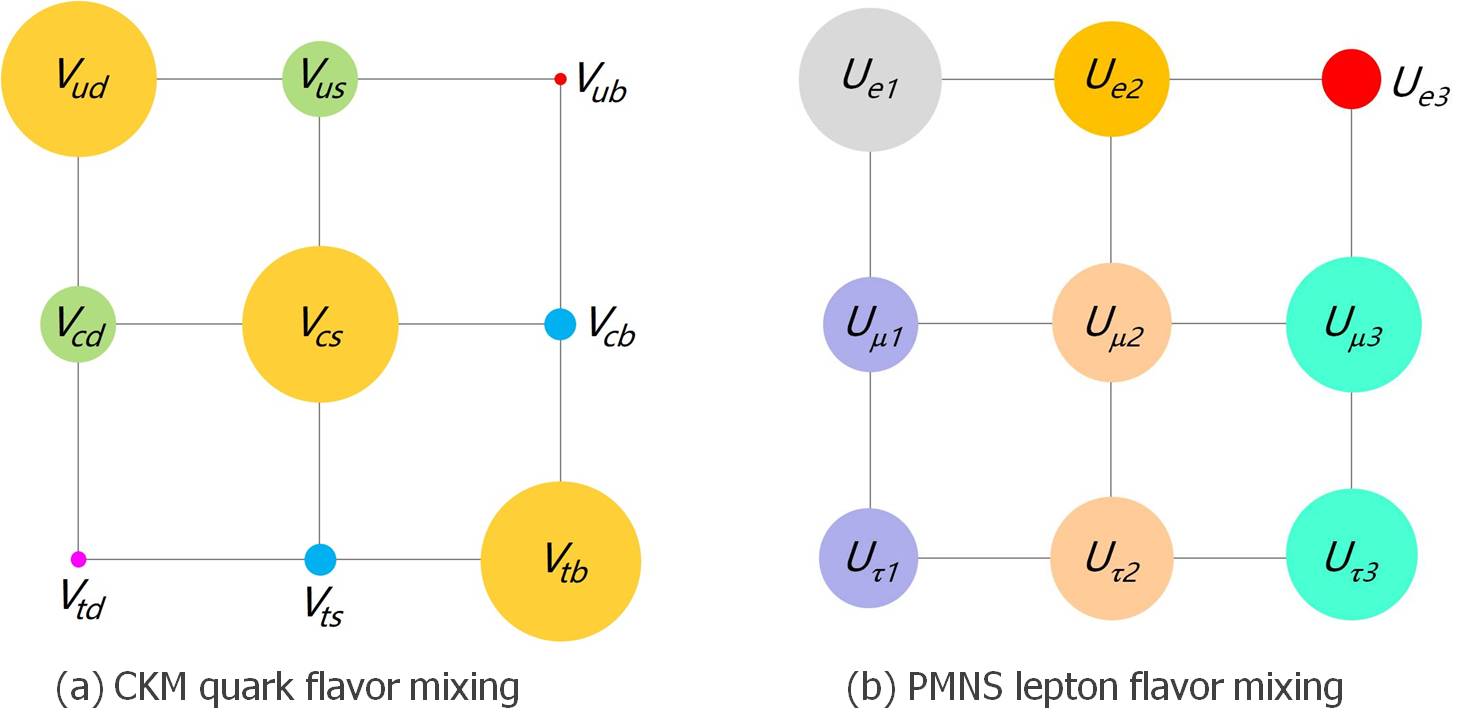}
\vspace{-0.1cm}
\caption{An illustration of the relative magnitudes of the CKM and PMNS matrix
elements, where the colored area of each circle is proportional to $|V^{}_{\alpha i}|$
(for $\alpha = u, c, t$ and $i = d, s, b$) or to $|U^{}_{\alpha i}|$ (for
$\alpha = e, \mu, \tau$ and $i = 1, 2, 3$) on the same scaling.}
\label{Fig:CKM-PMNS}
\end{center}
\end{figure}

A global analysis of current experimental data on quark flavor mixing and CP
violation leads us to the following results for nine elements of the CKM
matrix~\cite{Workman:2022ynf}:
\begin{eqnarray}
\big|V^{}_{ud}\big| = 0.97435 \pm 0.00016 \; , \quad &&
\big|V^{}_{us}\big| = 0.22500 \pm 0.00067 \; ,
\nonumber \\
\big|V^{}_{ub}\big| = 0.00369 \pm 0.00011 \; , \quad &&
\big|V^{}_{cd}\big| = 0.22486 \pm 0.00067 \; ,
\nonumber \\
\big|V^{}_{cs}\big| = 0.97349 \pm 0.00016 \; , \quad &&
\big|V^{}_{cb}\big| = 0.04182^{+0.00085}_{-0.00074} \; ,
\nonumber \\
\big|V^{}_{td}\big| = 0.00857^{+0.00020}_{-0.00018} \; , \quad &&
\big|V^{}_{ts}\big| = 0.04110^{+0.00083}_{-0.00072} \; ,
\nonumber \\
\big|V^{}_{tb}\big| = 0.999118^{+0.000031}_{-0.000036} \; .
\label{eq:CKM}
\end{eqnarray}
In comparison, the $3\sigma$ intervals of nine elements of the PMNS matrix obtained
from a global analysis of the latest atmospheric, solar, reactor and accelerator
neutrino (or antineutrino) oscillation data are~\cite{Esteban:2020cvm}
\begin{eqnarray}
\big|U^{}_{e 1}\big| = 0.801 \to 0.845 \; , \quad &&
\big|U^{}_{e 2}\big| = 0.513 \to 0.579 \; ,
\nonumber \\
\big|U^{}_{e 3}\big| = 0.143 \to 0.155 \; , \quad &&
\big|U^{}_{\mu 1}\big| = 0.234 \to 0.500 \; ,
\nonumber \\
\big|U^{}_{\mu 2}\big| = 0.471 \to 0.689 \; , \quad &&
\big|U^{}_{\mu 3}\big| = 0.637 \to 0.776 \; ,
\nonumber \\
\big|U^{}_{\tau 1}\big| = 0.271 \to 0.525 \; , \quad &&
\big|U^{}_{\tau 2}\big| = 0.477 \to 0.694 \; ,
\nonumber \\
\big|U^{}_{\tau 3}\big| = 0.613 \to 0.756 \; .
\label{eq:PMNS}
\end{eqnarray}
To show the most salient features of $V$ and $U$ in a more intuitive way, let us
illustrate the relative magnitudes of their eighteen elements in
Figure~\ref{Fig:CKM-PMNS}, where each circle represents a given matrix element
and its area is proportional to $|V^{}_{\alpha i}|$ (for $\alpha = u, c, t$ and
$i = d, s, b$) for the CKM quark flavor mixing or to $|U^{}_{\alpha i}|$ (for
$\alpha = e, \mu, \tau$ and $i = 1, 2, 3$) for the PMNS lepton flavor mixing on
the same scaling. Two observations are in order.
\begin{itemize}
\item     The CKM matrix $V$ can be expressed as $V = {\cal I} + \Delta V$,
where $\Delta V$ denotes a kind of perturbation to the identity matrix $\cal I$
and its magnitude is at most of ${\cal O}(\lambda)$ with
$\lambda \equiv |V^{}_{us}| \simeq 0.2$ being the
Wolfenstein expansion parameter~\cite{Wolfenstein:1983yz}. The strong structural
hierarchy of $V$ is characterized by three layers of its six off-diagonal elements,
ranging from ${\cal O}(\lambda)$ to ${\cal O}(\lambda^2)$ and from
${\cal O}(\lambda^2)$ to ${\cal O}(\lambda^3)$.

\item     The PMNS matrix $U$ does not have a clear structural hierarchy. Instead,
it exhibits an intriguing $\mu$-$\tau$ interchange symmetry~\cite{Xing:2015fdg}:
$|U^{}_{\mu i}| = |U^{}_{\tau i}|$ (for $i = 1, 2, 3$).
Some entries of $U$ are essentially equal or very close to a few simple but
typical Clebsh-Gordan coefficients of group theory (i.e., the square roots of
fractions formed from very small integers, such as $1/\sqrt{2}$, $1/\sqrt{3}$ and
$1/\sqrt{6}$), implying a kind of underlying
discrete flavor symmetry. It is therefore reasonable
to conjecture $U = U^{}_0 + \Delta U$, where $U^{}_0$ denotes a constant flavor
mixing pattern originating from an underlying flavor symmetry group, and
$\Delta U$ stands for small corrections or perturbations to $U^{}_0$ as a result
of some soft symmetry breaking effects.
\end{itemize}
Up to now many efforts have been made to try out various non-Abelian discrete
flavor symmetry groups which can help a lot to determine the explicit form of
$U^{}_0$ in a chosen flavor basis~\cite{Xing:2019vks,Xing:2015fdg,Altarelli:2010gt,
King:2013eh,Feruglio:2019ybq}. Most of those early attempts focused on $U = U^{}_0$,
simply because the available neutrino oscillation data were quite preliminary at
the time. To fit today's more precise experimental results, however, a viable
neutrino mass model usually needs the $\Delta U$ term which depends on either
spontaneous or explicit breaking of the flavor symmetry related to $U^{}_0$.
\begin{figure}[t]
\begin{center}
\includegraphics[width=3.95in]{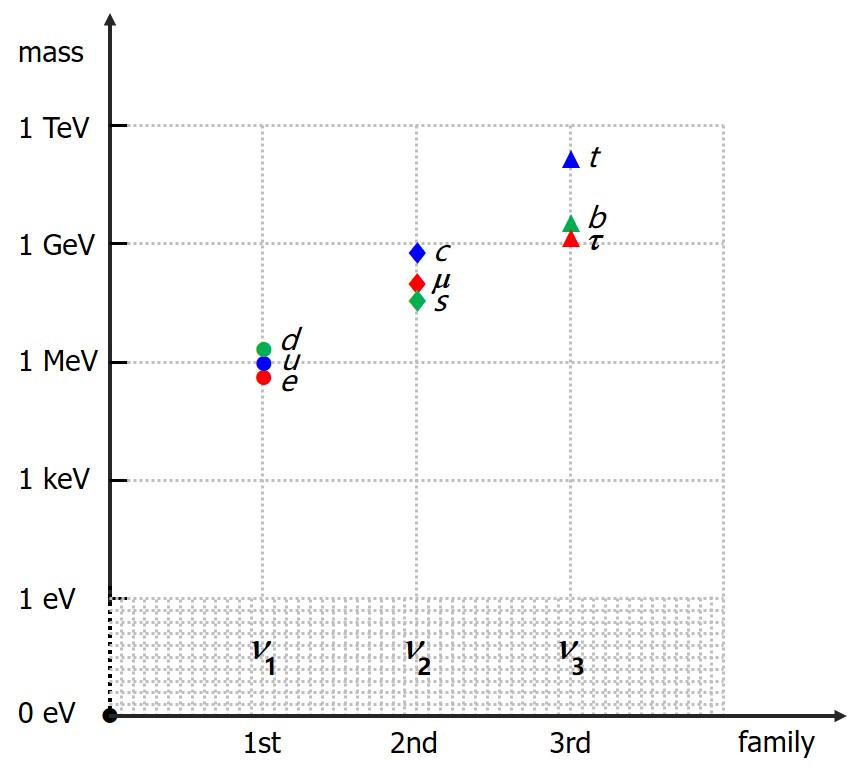}
\vspace{-0.25cm}
\caption{An illustration of the mass spectra of three charged leptons and six
quarks at the electroweak scale, where the vertical coordinate is logarithmic. The
currently allowed region of neutrino masses is also included.}
\label{Fig:fermion-masses}
\end{center}
\end{figure}

Another flavor puzzle is associated with the peculiar mass spectra of three
charged leptons and six quarks as shown in Figure~\ref{Fig:fermion-masses},
where the allowed region of three active neutrino masses is also included.
The quark mass values are extracted from a number of different
measurements~\cite{ParticleDataGroup:2020ssz} and can be renormalized to the
electroweak scale (i.e., $M^{}_Z = 91.1876 ~{\rm GeV}$) in the SM framework as follows~\cite{Xing:2007fb,Xing:2011aa,Huang:2020hdv}:
\begin{eqnarray}
m^{}_u = 1.23 \pm 0.21 ~{\rm MeV} \; , \quad &&
m^{}_d = 2.67 \pm 0.19 ~{\rm MeV} \; ,
\nonumber \\
m^{}_c = 0.620 \pm 0.017 ~{\rm GeV} \; , \quad &&
m^{}_s = 53.16 \pm 4.61 ~{\rm MeV} \; ,
\nonumber \\
m^{}_t = 168.26 \pm 0.75 ~{\rm GeV} \; , \quad &&
m^{}_b = 2.839 \pm 0.026 ~{\rm GeV} \; .
\label{eq:quark-mass}
\end{eqnarray}
The pole masses of three charged leptons have been precisely
measured~\cite{ParticleDataGroup:2020ssz}, and the values of their running masses
at the electroweak scale are given by~\cite{Xing:2007fb,Xing:2011aa,Huang:2020hdv}
\begin{eqnarray}
&& m^{}_e = 0.48307 \pm 0.00045 ~{\rm MeV} \; ,
\nonumber \\
&& m^{}_\mu = 101.766 \pm 0.023 ~{\rm MeV} \; ,
\nonumber \\
&& m^{}_\tau = 1728.56 \pm 0.28 ~{\rm MeV} \; .
\label{eq:lepton-mass}
\end{eqnarray}
Today's neutrino oscillation experiments have determined the two independent
neutrino mass-squared differences to a quite good degree of accuracy
(i.e., $\Delta m^2_{21} \equiv m^2_2 - m^2_1 = 7.42^{+0.21}_{-0.20} \times
10^{-5} ~{\rm eV}^2$ and $\Delta m^2_{31} \equiv m^2_3 - m^2_1 =
+2.517^{+0.026}_{-0.028} \times 10^{-3} ~{\rm eV}^2$ for the normal mass
ordering $m^{}_1 < m^{}_2 < m^{}_3$ or $\Delta m^2_{32} \equiv m^2_3 - m^2_2 =
-2.498^{+0.028}_{-0.028} \times 10^{-3} ~{\rm eV}^2$ for the inverted mass
ordering $m^{}_3 < m^{}_1 < m^{}_2$~\cite{Esteban:2020cvm}), but the absolute
neutrino mass scale remains unknown. The latest KATRIN experiment, which aims
to probe the electron energy spectrum's endpoint for the tritium beta decay
$^3{\rm H} \to \hspace{0.03cm}^3{\rm He} + e^- + \overline{\nu}^{}_e$, has set
an upper bound for the effective electron-neutrino mass
\begin{eqnarray}
\langle m\rangle^{}_e = \sqrt{ m^{2}_1 \left|U^{}_{e1}\right|^2 + m^{2}_2
\left|U^{}_{e2}\right|^2 + m^{2}_3 \left|U^{}_{e3}\right|^2} \; ;
\label{eq:beta-decay}
\end{eqnarray}
namely, $\langle m\rangle^{}_{e} < 0.8 ~{\rm eV}$ at the $90\%$ confidence
level~\cite{KATRIN:2021uub}. The future KATRIN experiment may ultimately reach
a sensitivity down to $0.2~{\rm eV}$~\cite{Osipowicz:2001sq}. Precision
measurements of both the cosmic microwave background anisotropies
and the large-scale structure formation of the Universe, especially the
one done by the Planck Collaboration, have put the most stringent bound on
the sum of three neutrino masses~\cite{Aghanim:2018eyx}:
$m^{}_1 + m^{}_2 + m^{}_3 < 0.12~{\rm eV}$ at the $95\%$ confidence level.
But one should keep in mind that such a stringent cosmological bound is dependent
more or less on some assumptions associated with the neutrino mass spectrum and
the standard $\Lambda$CDM model~\cite{Xing:2019vks,Loureiro:2018pdz},
where $\Lambda$ denotes the cosmological constant (or dark energy) and CDM
stands for cold dark matter. That is why we have simply set $m^{}_i < 1~{\rm eV}$
(for $i = 1, 2, 3$) in Figure~\ref{Fig:fermion-masses} as the most conservative
upper limit for the absolute neutrino mass scale.

If the three massive neutrinos are their own antiparticles (i.e.,
$\nu^{}_i = \nu^{c}_i$, where the superscript ``$c$" denotes the charge
conjugation of $\nu^{}_i$, and $i = 1, 2, 3$), as first conjectured
by Ettore Majorana in 1937~\cite{Majorana:1937vz}, they can mediate the
neutrinoless double-beta ($0\nu 2\beta$) decays of some even-even nuclei
(such as $^{76}_{32}{\rm Ge} \to \hspace{0.03cm}^{76}_{34}{\rm Se} + 2 e^-$,
$^{130}_{~52}{\rm Te} \to \hspace{0.03cm}^{130}_{~54}{\rm Xe} + 2 e^-$ and
$^{136}_{~54}{\rm Xe}
\to \hspace{0.03cm}^{136}_{~56}{\rm Ba} + 2 e^-$)~\cite{Furry:1939qr}.
The rates of such rare lepton-number-violating processes are governed by the
magnitude of the effective Majorana electron-neutrino mass
\begin{eqnarray}
\langle m\rangle^{}_{ee} =  m^{}_1 U^2_{e1} + m^{}_2 U^2_{e2} +
m^{}_3 U^2_{e3} \; .
\label{eq:2beta-decay}
\end{eqnarray}
Today's stringent upper bound on
$|\langle m\rangle^{}_{ee}|$ is $0.06$ eV to $0.2$ eV at the $95\%$
confidence level~\cite{KamLAND-Zen:2016pfg,EXO-200:2019rkq,GERDA:2020xhi},
where the large uncertainty originates mainly from the calculations of relevant
nuclear matrix elements. The next-generation $0\nu 2\beta$ experiments are
expected to bring the sensitivity down to the ${\cal O}(10)$ meV level, and
the ultimate sensitivity of such experiments is likely to reach the
${\cal O}(1)$ meV limit~\cite{Cao:2019hli}.

It is commonly believed that the smallness of all the three quark flavor
mixing angles should be closely related to the strong quark mass
hierarchies~\cite{Fritzsch:1999ee}.
The reason is simply that in an arbitrary flavor basis the CKM matrix $V$
is actually a product of two unitary matrices used to diagonalize the up-
and down-type quark mass matrices $M^{}_{\rm u,d}$
from their left-hand sides; i.e., $V \equiv O^\dagger_{\rm u} O^{}_{\rm d}$
with $O^\dagger_{\rm u} M^{}_{\rm u} O^\prime_{\rm u} = {\rm Diag}\big\{
m^{}_u, m^{}_c, m^{}_t\big\}$ and $O^\dagger_{\rm d} M^{}_{\rm d} O^\prime_{\rm d}
= {\rm Diag}\big\{m^{}_d, m^{}_s, m^{}_b\big\}$, where $O^{}_{\rm u,d}$ and
$O^\prime_{\rm u,d}$ are all unitary. So the elements of $V$ are naturally
expected to rely on $m^{}_u/m^{}_c \sim m^{}_c/m^{}_t \sim \lambda^4$ and
$m^{}_d/m^{}_s \sim m^{}_s/m^{}_b \sim \lambda^2$,
although their specific relations depend on specific textures of quark
mass matrices. For example, the simple but phenomenologically successful
relations $\big|V^{}_{us}\big| \simeq \big|V^{}_{cd}\big| \simeq
\sqrt{m^{}_d/m^{}_s}$ and $\big|V^{}_{td}\big|/\big|V^{}_{ts}\big| \simeq
\sqrt{m^{}_d/m^{}_s}$ can be easily obtained from the well-known Fritzsch
texture of Hermitian $M^{}_{\rm u}$ and
$M^{}_{\rm d}$~\cite{Fritzsch:1977vd,Fritzsch:1979zq}.

In the lepton sector it seems less convincing to ascribe the largeness of two of
the three lepton flavor mixing angles to a very weak mass hierarchy of three
neutrinos, given the fact that the charged leptons have a rather strong mass
hierarchy and hence should not contribute much to the PMNS matrix $U$.
Especially in the $m^{}_1 \to 0$ (or $m^{}_3 \to 0$) limit, in which the
neutrino mass spectrum is expected to be most hierarchical (or to have a near
degeneracy between $m^{}_1$ and $m^{}_2$), one will be left
with $m^{}_1/m^{}_2 \to 0$ (or $m^{}_1/m^{}_2 \to 1$) and
$m^{}_2/m^{}_3 \sim \lambda$ (or $m^{}_3/m^{}_2 \to 0$). But it has been found
that some viable flavor textures of up- and down-type quarks can also be
used to successfully describe the flavor textures of charged leptons and massive
neutrinos as constrained by current neutrino oscillation data~\cite{Xing:2019vks}.
Note that small $\Delta m^2_{21} \simeq 7.4 \times 10^{-5}~{\rm eV}^2$ and
$|\Delta m^2_{31}| \simeq 2.5 \times 10^{-3}~{\rm eV}^2$
{\it do} allow for a nearly degenerate neutrino mass spectrum, but
this special possibility does not mean that the large flavor mixing angles
$\theta^{}_{12}$ and $\theta^{}_{23}$ should be directly associated with
the mass ratios $m^{}_1/m^{}_2$ and $m^{}_2/m^{}_3$. Instead, a kind of
non-Abelian discrete flavor symmetry is more likely to lie behind such a
neutrino mass spectrum and two large flavor mixing
angles~\cite{Xing:2019vks,Xing:2015fdg,Altarelli:2010gt,
King:2013eh,Feruglio:2019ybq}.

All in all, the observed fermion mass spectra and flavor mixing patterns
strongly suggest that charged leptons,
massive neutrinos and quarks should have their respective
flavor structures instead of a random nature. A real challenge is therefore to
explore how the flavors are structured, especially how the active and sterile
neutrino flavors are structured. In the absence of a fundamental guiding principle
regarding the origin of tiny neutrino masses, it is realistic and greatly helpful
to follow a phenomenological approach to the flavor issues of massive neutrinos.
In this regard a consensus is that no matter how large or complicated the true
flavor symmetry group can be, its residual symmetry at low energies should serve
as a {\it minimal} flavor symmetry in the neutrino sector. The physical meaning of
“minimal” is essentially two-fold: (1) this flavor symmetry can be described by
one of the simplest discrete or continuous symmetry groups; (2) the pattern of
lepton flavor mixing and CP violation determined or constrained by such a simple
flavor symmetry should be as close as possible to the one extracted from current
and future experimental data on neutrino oscillations.

So the strategy of a successful study along this line of thought is also two-fold.
On the one hand, one may make every effort to identify the {\it minimal} flavor
symmetry behind the observed neutrino mass spectrum and flavor mixing pattern.
On the other hand, it makes a lot of sense to investigate possible flavor
structures of the active and sterile neutrinos as constrained by such an
instructive flavor symmetry. Up to now a $\rm Z^{}_2$ flavor symmetry, the
$\mu$-$\tau$ interchange symmetry, has been commonly accepted to be
the desired minimal flavor symmetry associated with massive neutrinos. We are
going into details step by step in the following sections.

\subsection{A minimal symmetry of the neutrino flavors}
\label{section 1.2}

As shown in Figure~\ref{Fig:CKM-PMNS},
current neutrino oscillation data apparently favor
$|U^{}_{\mu i}| = |U^{}_{\tau i}|$ (for $i = 1, 2, 3$), implying an intriguing
$\mu$-$\tau$ interchange symmetry for three pairs of the $(\mu,\tau)$-associated
PMNS matrix elements. This observation is a strong reflection of the most
salient feature of lepton flavor mixing, and it should imply a kind of underlying
flavor symmetry in the lepton sector. Note that a lepton flavor symmetry of this
kind must be basis-dependent, so let us choose a convenient basis in which the
flavor eigenstates of three charged leptons $(l^{}_e , l^{}_\mu , l^{}_\tau)$ are
identified with their mass eigenstates $(e , \mu , \tau)$. Given the charged lepton
mass term
\begin{eqnarray}
-{\cal L}^{}_{l} = \overline{\pmatrix{l^{}_e & \hspace{-0.25cm} l^{}_\mu &
\hspace{-0.25cm} l^{}_\tau}^{}_{\rm L}}
\hspace{0.07cm} M^{}_l \pmatrix{l^{}_e \cr l^{}_\mu \cr
l^{}_\tau \cr}^{}_{\hspace{-0.15cm} \rm R} + {\rm h.c.} \; ,
\label{eq:lepton-mass}
\end{eqnarray}
where $M^{}_l$ is in general neither Hermitian nor symmetric,
a bi-unitary transformation $O^\dagger_l M^{}_l O^\prime_l = D^{}_l \equiv
{\rm Diag}\big\{m^{}_e, m^{}_\mu, m^{}_\tau\big\}$ allows us to change the
flavor basis $(l^{}_e , l^{}_\mu , l^{}_\tau)$ to the mass basis
$(e , \mu , \tau)$. Given the strong mass hierarchy of three charged leptons,
the flavor structure of $M^{}_l$ itself is expected to have no way to respect
a $\mu$-$\tau$ interchange (either permutation or reflection) symmetry.
Without loss of generality, we simply take $M^{}_l = D^{}_l$
or equivalently $O^{}_l = O^\prime_l = {\cal I}$ with $\cal I$ being the identity
matrix, such that $l^{}_\alpha = \alpha$ (for $\alpha = e, \mu, \tau$) holds. In
this flavor basis the PMNS matrix $U$ linearly links the mass eigenstates
of three active neutrinos to their flavor eigenstates as follows:
\begin{eqnarray}
\pmatrix{ \nu^{}_e \cr \nu^{}_\mu \cr \nu^{}_\tau \cr}^{}_{\hspace{-0.15cm} \rm L}
= \pmatrix{ U^{}_{e 1} & U^{}_{e 2} & U^{}_{e 3} \cr
U^{}_{\mu 1} & U^{}_{\mu 2} & U^{}_{\mu 3} \cr
U^{}_{\tau 1} & U^{}_{\tau 2} & U^{}_{\tau 3} \cr} \hspace{-0.1cm}
\pmatrix{ \nu^{}_{1} \cr \nu^{}_{2} \cr
\nu^{}_{3} \cr}^{}_{\hspace{-0.15cm} \rm L} \; .
\label{eq:UPMNS}
\end{eqnarray}
Therefore, a special pattern of $U$ is strongly suggestive of an underlying
flavor symmetry associated with the flavor structure of massive neutrinos.
Now that the $\mu$-$\tau$ interchange symmetry is a sort of $\rm Z^{}_2$
transformation symmetry with respect to the $\nu^{}_\mu$ and $\nu^{}_\tau$
flavors, it should be the simplest {\it residual} symmetry of any promising
discrete flavor symmetry groups that can be used to build viable neutrino mass
models~\cite{Xing:2019vks,Xing:2015fdg,Altarelli:2010gt,King:2013eh,Feruglio:2019ybq}.
That is why we are well motivated to explore what can be maximally learnt from
this minimal flavor symmetry in the neutrino sector.

For the sake of illustration, we consider the effective Majorana mass term of
three active neutrinos in the chosen flavor basis:
\begin{eqnarray}
-{\cal L}^{}_{\nu} = \frac{1}{2} \hspace{0.05cm}
\overline{\pmatrix{\nu^{}_e & \hspace{-0.25cm} \nu^{}_\mu &
\hspace{-0.25cm} \nu^{}_\tau}^{}_{\rm L}}
\hspace{0.05cm} M^{}_\nu \pmatrix{\nu^{c}_e \cr \nu^{c}_\mu \cr
\nu^{c}_\tau \cr}^{}_{\hspace{-0.15cm} \rm R} + {\rm h.c.} \; ,
\label{eq:Majorana-mass}
\end{eqnarray}
where the Majorana neutrino mass matrix $M^{}_\nu$ is symmetric,
$\nu^c_\alpha \equiv {\cal C}\overline{\nu^{}_\alpha}^T$ and
$\big(\nu^c_\alpha\big)^{}_{\rm R} = \big(\nu^{}_{\alpha \rm L}\big)^c$
stand respectively for the charge conjugates of $\nu^{}_\alpha$ and
$\nu^{}_{\alpha \rm L}$ (for $\alpha = e, \mu ,\tau$), and $\cal C$ is the charge
conjugation operator satisfying ${\cal C}\gamma^{T}_\mu {\cal C}^{-1} =
-\gamma^{}_\mu$, ${\cal C}\gamma^{T}_5 {\cal C}^{-1} = \gamma^{}_5$ and
${\cal C}^{-1} = {\cal C}^\dagger = {\cal C}^{T} = -{\cal C}$~\cite{Xing:2011zza}.
Then the $\mu$-$\tau$ interchange symmetry for ${\cal L}^{}_{\nu}$ can be
classified into the following two subsidiary sets~\cite{Xing:2015fdg}.

(1) {\it The $\mu$-$\tau$ permutation symmetry}, which requires ${\cal L}^{}_{\nu}$
to keep unchanged under the partially flavor-changing transformations
\begin{eqnarray}
\nu^{}_{e \rm L} \to \nu^{}_{e \rm L} \; , \quad
\nu^{}_{\mu \rm L} \to \nu^{}_{\tau \rm L} \; , \quad
\nu^{}_{\tau \rm L} \to \nu^{}_{\mu \rm L} \; .
\label{eq:permutation}
\end{eqnarray}
In this case the $3\times 3$ symmetric Majorana neutrino mass matrix $M^{}_\nu$
is constrained to have the following texture~\cite{Grimus:2012hu}:
\begin{eqnarray}
M^{}_\nu = \pmatrix{\langle m\rangle^{}_{ee} & \langle m\rangle^{}_{e\mu} &
\langle m\rangle^{}_{e\mu} \cr \langle m\rangle^{}_{e\mu} &
\langle m\rangle^{}_{\mu\mu} & \langle m\rangle^{}_{\mu\tau} \cr
\langle m\rangle^{}_{e\mu} & \langle m\rangle^{}_{\mu\tau} &
\langle m\rangle^{}_{\mu\mu} \cr} \; ,
\label{eq:M-permutation}
\end{eqnarray}
where all the four independent matrix elements are in general complex. To diagonalize
$M^{}_\nu$ with the Autonne-Takagi transformation $U^\dagger M^{}_\nu U^* = D^{}_\nu
\equiv {\rm Diag}\big\{m^{}_1, m^{}_2, m^{}_3\big\}$~\cite{Dreiner:2008tw},
one immediately finds that the unitary PMNS matrix $U$ contains two special
flavor mixing angles in its standard parametrization given by Eq.~(\ref{eq:U}):
\begin{eqnarray}
\theta^{}_{13} = 0 \; , \quad \theta^{}_{23} = \frac{\pi}{4} \; ,
\label{eq:U-permutation}
\end{eqnarray}
a result which assures the equalities $|U^{}_{\mu i}| = |U^{}_{\tau i}|$
(for $i = 1, 2, 3$) to hold. To generate nonzero $\theta^{}_{13}$ and hence
nontrivial $\delta^{}_\nu$, one may break the above $\mu$-$\tau$ permutation
symmetry by simply introducing a single phase parameter such that
$\langle m\rangle^{}_{e\tau} = \langle m\rangle^*_{e\mu}$,
$\langle m\rangle^{}_{\tau\tau} = \langle m\rangle^*_{\mu\mu}$ and
$\arg\big(\langle m\rangle^{}_{e\mu}\big) =
\arg\big(\langle m\rangle^{}_{\mu\mu}\big)$ hold~\cite{Chamoun:2019pbh} --- a
special texture of $M^{}_\nu$ to be discussed below.

(2) {\it The $\mu$-$\tau$ reflection symmetry}, which requires ${\cal L}^{}_{\nu}$
to be invariant under the partially flavor-changing CP transformations
\begin{eqnarray}
\nu^{}_{e \rm L} \to \big(\nu^{}_{e \rm L}\big)^c \; , \quad
\nu^{}_{\mu \rm L} \to \big(\nu^{}_{\tau \rm L}\big)^c \; , \quad
\nu^{}_{\tau \rm L} \to \big(\nu^{}_{\mu \rm L}\big)^c \; .
\label{eq:reflection}
\end{eqnarray}
In this case the $3\times 3$ Majorana neutrino mass matrix $M^{}_\nu$ takes the
special form
\begin{eqnarray}
M^{}_\nu = \pmatrix{\langle m\rangle^{}_{ee} & \langle m\rangle^{}_{e\mu} &
\langle m\rangle^{*}_{e\mu} \cr \langle m\rangle^{}_{e\mu} &
\langle m\rangle^{}_{\mu\mu} & \langle m\rangle^{}_{\mu\tau} \cr
\langle m\rangle^{*}_{e\mu} & \langle m\rangle^{}_{\mu\tau} &
\langle m\rangle^{*}_{\mu\mu} \cr} \; ,
\label{eq:M-reflection}
\end{eqnarray}
where the matrix elements $\langle m\rangle^{}_{ee}$ and
$\langle m\rangle^{}_{\mu\tau}$ are real. After $M^{}_\nu$ is diagonalized by
means of the transformation $U^\dagger M^{}_\nu U^* = D^{}_\nu$, we are left with
\begin{eqnarray}
\theta^{}_{23} = \frac{\pi}{4} \; , \quad \delta^{}_\nu = \pm\frac{\pi}{2} \; ,
\label{eq:U-reflection}
\end{eqnarray}
which give rise to the rephasing-independent equalities
$|U^{}_{\mu i}| = |U^{}_{\tau i}|$ (for $i = 1, 2, 3$) too. Of course,
the Majorana CP-violating phases of $U$ can also be
determined or constrained by use of the $\mu$-$\tau$ reflection symmetry~\cite{Xing:2019vks,Xing:2015fdg}, as will be discussed in
section~\ref{section 2.2}.

At this point it is worth mentioning that the so-called {\it trimaximal} neutrino
mixing matrix proposed by Cabibbo~\cite{Cabibbo:1977nk} should be the first constant
flavor mixing pattern respecting the $\mu$-$\tau$ reflection symmetry:
\begin{eqnarray}
U^{}_{\rm C} = \pmatrix{
\frac{1}{\sqrt 3} & \frac{1}{\sqrt 3} & \frac{1}{\sqrt 3} \cr
\frac{1}{\sqrt 3} & \frac{\omega}{\sqrt 3}
& \frac{\omega^*}{\sqrt 3} \cr
\frac{1}{\sqrt 3} & \frac{\omega^*}{\sqrt 3}
& \frac{\omega}{\sqrt 3}} \; ,
\label{eq:U-Cabibbo}
\end{eqnarray}
where $\omega = \exp\left({\rm i} 2\pi/3\right) = -1/2 + {\rm i} \sqrt{3}/2$.
But $U^{}_{\rm C}$ leads us to a too large value of $\theta^{}_{13}$ (i.e.,
$\theta^{}_{13} = \arcsin\left(1/\sqrt{3}\right) \simeq 35.3^\circ$) and hence
has been ruled out by the experimental data on neutrino oscillations. A more
realistic constant neutrino mixing matrix, which exactly respects the
$\mu$-$\tau$ reflection symmetry, is the {\it tetramaximal} flavor mixing
pattern~\cite{Xing:2008ie}
\begin{eqnarray}
U^{}_{\rm 4M} = \pmatrix{
\frac{1}{2} + \frac{1}{2\sqrt{2}}
& \frac{1}{2}
& \frac{1}{2} - \frac{1}{2\sqrt{2}} \cr
-\frac{1}{2\sqrt{2}} + \frac{\rm i}{4\left(\sqrt{2} + 1\right)}
& \frac{1}{2} - \frac{\rm i}{2\sqrt{2}}
& \frac{1}{2\sqrt{2}} + \frac{\rm i}{4\left(\sqrt{2} - 1\right)} \cr
-\frac{1}{2\sqrt{2}} - \frac{\rm i}{4\left(\sqrt{2} + 1\right)}
& \frac{1}{2} + \frac{\rm i}{2\sqrt{2}}
& \frac{1}{2\sqrt{2}} - \frac{\rm i}{4\left(\sqrt{2} - 1\right)}} \; .
\label{eq:U-tetramaximal}
\end{eqnarray}
It predicts $\theta^{}_{12} = \arctan\big(2 - \sqrt{2}\big) \simeq 30.4^\circ$ and
$\theta^{}_{13} = \arcsin\big[\left(\sqrt{2} - 1\right)/\left(2\sqrt{2}\right)\big]
\simeq 8.4^\circ$ besides Eq.~(\ref{eq:U-reflection}).
We find that $U^{}_{\rm 4M}$ can fit current neutrino oscillation data after slight
$\mu$-$\tau$ reflection symmetry breaking effects are taken into
account~\cite{Zhang:2011aw,Nath:2019asv}. A number of interesting
constant neutrino mixing patterns respecting the $\mu$-$\tau$ permutation symmetry
have also been proposed and discussed in the literature~\cite{Xing:2019vks}.

Here the emphasis is on the fact that the $\mu$-$\tau$ interchange symmetry has
proved to be powerful enough to constrain the texture of $M^{}_\nu$ and thus
interpret the most salient feature of $U$ as revealed by the present experimental
data. It can be regarded as a very simple {\it generalized} CP
symmetry~\cite{Neufeld:1987wa,Grimus:1995zi,Feruglio:2012cw,Rodejohann:2017lre}.
Note, however, that in this regard one usually focuses on a possible $\mu$-$\tau$
interchange symmetry of the effective neutrino mass term {\it after} spontaneous gauge
symmetry breaking. Otherwise, the transformations made in Eq.~(\ref{eq:permutation})
or Eq.~(\ref{eq:reflection}) would affect some other parts of the Lagrangian of
electroweak interactions. To construct a consistent neutrino mass model which
respect both the electroweak gauge symmetry and the $\mu$-$\tau$ permutation or
reflection symmetry, it is in most cases unavoidable to introduce some additional
scalar fields and require them to couple to the charged-lepton and neutrino
sectors in quite different ways (see, e.g.,
Refs.~\cite{Xing:2015fdg,Grimus:2012hu,Mohapatra:2015gwa,Dey:2022qpu} for details).
In some viable models of this kind, such new hypothetical
degrees of freedom may lead to rich phenomenology in particle physics and
thus enhance their experimental testability.

That is why a less ambitious strategy has been commonly adopted: one concentrates
on how the flavor textures of three known active neutrinos and possible sterile
neutrinos can be constrained by the respective $\mu$-$\tau$ interchange symmetry
and an analogous symmetry associated with the sterile flavors, and how the
constrained textures are further corrected by the relevant spontaneous or
explicit symmetry breaking effects. A very comprehensive review of such
phenomenological attempts was first made in 2016~\cite{Xing:2015fdg}, and since
then a lot of new and in-depth research works have been done along this line of
thought. In particular, the $\mu$-$\tau$ reflection symmetry has attracted much
more attention after the smallest lepton flavor mixing angle $\theta^{}_{13}$ was
measured by the Daya Bay Collaboration~\cite{DayaBay:2012fng} and a nearly
$3\sigma$ evidence for the nonzero CP-violating phase $\delta^{}_\nu$ was achieved
by the T2K Collaboration~\cite{T2K:2019bcf}.

The present paper is therefore intended to summarize both the latest progress
in studying possible flavor textures of the active neutrinos constrained by the
$\mu$-$\tau$ reflection symmetry and its $\rm S^{}_3$ extension, which can also
be referred to as the flavor-changing CP symmetry as a whole, and the recent
developments in exploring possible flavor textures of the sterile neutrinos with
the help of a similar symmetry and the seesaw mechanism. Let us stress that a
flavor-changing CP transformation of the neutrino field $\nu^{}_{\alpha \rm L}$,
namely $\nu^{}_{\alpha \rm L} \to (\nu^{}_{\beta \rm L})^c$ (for
$\alpha, \beta = e, \mu, \tau$), is actually a combined operation of the standard CP
transformation of $\nu^{}_{\alpha \rm L}$ itself and the subsequent flavor-changing
$(\nu^{}_{\alpha \rm L})^c \to (\nu^{}_{\beta \rm L})^c$
transformation, and thus it makes CP violation possible in the lepton sector.

The remaining parts of this article are organized as follows. In section 2 we
are going to describe the flavor-changing CP transformation for the three active
neutrino fields and identify the $\mu$-$\tau$ reflection symmetry as a minimal
flavor symmetry associated with either the effective Majorana neutrino mass
term or the canonical seesaw mechanism. An $\rm S^{}_3$ extension of the
$\mu$-$\tau$ reflection symmetry is also discussed. Sections 3 and 4 are devoted
to the {\it translational} and {\it rotational} $\mu$-$\tau$ reflection symmetries,
respectively. Both of them can give rise to the phenomenologically favored
$\rm TM^{}_1$ flavor mixing pattern, and they can also be combined with the
canonical seesaw mechanism. After a brief introduction to the
possibilities of explicit breaking of the $\mu$-$\tau$ reflection symmetry in the
beginning of section 5, we pay particular attention to its soft breaking
effects induced by the renormalization-group equations (RGEs) from a
superhigh-energy scale to the electroweak scale. The relevant RGEs are
simplified in the case of either $m^{}_1 = 0$ or $m^{}_3 = 0$. Some remarkable
consequences of the $\mu$-$\tau$ reflection symmetry in neutrino phenomenology,
including its implications for lepton flavor violation and lepton number
violation, its constraint on the flavor distribution of ultrahigh energy cosmic
neutrinos at a neutrino telescope, its application to the minimal seesaw model
and its impact on leptogenesis, are explored in section 6. In section 7 we
give a few more specific examples for how to extend or utilize the $\mu$-$\tau$
reflection symmetry, such as the littlest seesaw model, the zero textures of
fermion mass matrices with a $\rm Z^{}_2$ reflection symmetry, the
$\mu$-$\tau$ reflection symmetry in a seesaw-induced $(3+1)$ active-sterile
flavor mixing scenario, and matter effects and nonstandard interactions
associated with neutrino oscillations. Section 8 is devoted to a brief summary
of what we have discussed, together with an outlook for the prospects of
applying Occam's razor in neutrino physics.

\setcounter{equation}{0}
\section{On the flavor-changing CP symmetry}
\label{section 2}

\subsection{On the CP transformation and CP violation}
\label{section 2.1}

Within the framework of the SM there is no leptonic CP violation in weak
interactions because a renormalizable gauge- and Lorentz-invariant neutrino
mass term is absent. To take a close look at how CP violation may naturally occur
in the lepton sector as in the quark sector, let us simply consider the
effective Majorana neutrino mass term ${\cal L}^{}_{\nu}$ given in
Eq.~(\ref{eq:Majorana-mass}) and rewrite it in the following way:
\begin{eqnarray}
-{\cal L}^{}_{\nu} = \frac{1}{2} \left[\overline{\nu^{}_{\rm
L}} \hspace{0.1cm} M^{}_\nu \hspace{0.05cm} (\nu^{}_{\rm L})^c +
\overline{(\nu^{}_{\rm L})^c} \hspace{0.1cm}
M^{*}_\nu \hspace{0.05cm} \nu^{}_{\rm L} \right] \; ,
\label{eq:M-mass}
\end{eqnarray}
in which $\nu^{}_{\rm L} = (\nu^{}_{e \rm L}, \nu^{}_{\mu \rm L},
\nu^{}_{\tau \rm L})^T$ denotes the column vector of three left-handed
neutrino fields. By definition, a left-handed neutrino field
$\nu^{}_{\alpha \rm L}(t, {\bf x})$ (for $\alpha = e$, $\mu$ or $\tau$)
transforms under charge conjugation (C) and parity (P) as~\cite{Xing:2011zza}
\begin{eqnarray}
\nu^{}_{\alpha \rm L} (t, {\bf x}) \to
\big[\nu^{}_{\alpha \rm L} (t, -{\bf x}) \big]^c = {\cal C} \hspace{0.05cm}
\overline{\nu^{}_{\alpha \rm L} (t, -{\bf x})}^T ,
\label{eq:CP}
\end{eqnarray}
where the properties of $\cal C$ have been elucidated below
Eq.~(\ref{eq:Majorana-mass}). Substituting Eq.~(\ref{eq:CP}) into
Eq.~(\ref{eq:M-mass}), we find that ${\cal L}^{}_\nu$ becomes
\begin{eqnarray}
-{\cal L}^{\prime}_{\nu} = \frac{1}{2} \left[\overline{\nu^{}_{\rm
L}} \hspace{0.1cm} M^{*}_\nu \hspace{0.05cm} (\nu^{}_{\rm L})^c +
\overline{(\nu^{}_{\rm L})^c} \hspace{0.1cm}
M^{}_\nu \hspace{0.05cm} \nu^{}_{\rm L} \right] \; ,
\label{eq:M-mass-CP}
\end{eqnarray}
where ${\bf x} \to -{\bf x}$ is implied. One can immediate see that
${\cal L}^\prime_{\nu} = {\cal L}^{}_{\nu}$ will hold (to be more exact,
their corresponding actions will be equal) if $M^{}_\nu$ is real. That is
to say, CP will be a good symmetry if $M^{}_\nu$ does not contain any
nontrivial CP-violating phase.

In general, $M^{}_\nu$ is a symmetric complex matrix which certainly contains
a number of nontrivial phases. It can be diagonalized in an arbitrary basis
by the transformation
\begin{eqnarray}
O^\dagger_\nu M^{}_\nu O^*_\nu = D^{}_\nu \equiv
\pmatrix{m^{}_1 & 0 & 0 \cr 0 & m^{}_2 & 0 \cr 0 & 0 & m^{}_3 \cr} \; ,
\label{eq:M-mass-diagonalization}
\end{eqnarray}
where $O^{}_\nu$ is a $3\times 3$ unitary matrix, and $m^{}_i$ (for $i = 1, 2, 3$)
denote the neutrino masses. Eq.~(\ref{eq:M-mass-diagonalization}) is actually
equivalent to the basis transformation
$\nu^{}_{\rm L} \to \tilde{\nu}^{}_{\rm L} \equiv O^\dagger_\nu \nu^{}_{\rm L}$
for ${\cal L}^{}_\nu$, where
$\tilde{\nu}^{}_{\rm L} = (\nu^{}_{1 \rm L}, \nu^{}_{2 \rm L}, \nu^{}_{3 \rm L})^T$
stands for the column vector of three left-handed neutrino mass eigenstates.
So the PMNS matrix $U$ appearing in ${\cal L}^{}_{\rm cc}$ is given by
$U = O^\dagger_l O^{}_\nu$, where the unitary matrix $O^{}_l$ has been used
to diagonalize the charged lepton mass matrix $M^{}_l$ below Eq.~(\ref{eq:lepton-mass}).
In the flavor basis of $M^{}_l = D^{}_l$, we have $O^{}_l = {\cal I}$ and thus
$U = O^{}_\nu$. It is then possible to reconstruct the elements
$\langle m\rangle^{}_{\alpha\beta}$ of $M^{}_\nu$
in terms of $U$ and $m^{}_i$ as follows:
\begin{eqnarray}
\langle m\rangle^{}_{\alpha\beta} \equiv \left(U D^{}_\nu U^T\right)^{}_{\alpha\beta}
= \sum^3_{i=1} m^{}_i U^{}_{\alpha i} U^{}_{\beta i} \; .
\label{eq:M-reconstruction}
\end{eqnarray}
It is obvious that a real $M^{}_\nu$ will lead to a real $U$ in the chosen
flavor basis, or vice versa. The standard source of leptonic CP violation is
therefore the complex $M^{}_l$ and (or) $M^{}_\nu$, or equivalently the
nontrivial CP-violating phases of $U$ in the weak charged-current
interactions~\cite{Branco:2011zb}.
In other words, fermion mass generation, flavor mixing and CP violation are closely
related to one another and constitute the three key issues of flavor physics.
But in almost all the cases a discrete or continuous flavor
symmetry is not powerful enough to simultaneously predict the neutrino mass
spectrum, the flavor mixing pattern and the CP-violating phases.

Note that what can be really measured in the lepton-number-violating processes are
the rephasing invariants $|\langle m\rangle^{}_{\alpha\beta}|$ instead of
$\langle m\rangle^{}_{\alpha\beta}$ themselves (for $\alpha, \beta = e, \mu, \tau$).
So one may define the observable effective neutrino masses
$\langle m\rangle^{}_\alpha$ (for $\alpha = e, \mu, \tau$) as follows:
\begin{eqnarray}
\langle m\rangle^{2}_{\alpha} \equiv \sum_\beta \left|\langle m\rangle^{}_{\alpha\beta}
\right|^2 = \sum_i m^2_i \left|U^{}_{\alpha i}\right|^2 \; ,
\label{eq:new-mass-squared}
\end{eqnarray}
where the unitarity of $U$ and Eq.~(\ref{eq:M-reconstruction}) have been used. It
is clear that $\langle m\rangle^{}_{e}$ is just the effective electron-neutrino
mass in Eq.~(\ref{eq:beta-decay}), which can be probed in the beta decays. In
comparison, the effective neutrino masses $\langle m\rangle^{}_\mu$ and
$\langle m\rangle^{}_\tau$ are associated respectively with the decay modes
$\pi^+ \to \mu^+ + \nu^{}_\mu$ and $\tau \to 5\pi + \nu^{}_\tau$ (or
$\tau \to 3\pi + \nu^{}_\tau$)~\cite{Xing:2003tp}. We easily see that the three
effective mass terms satisfy the sum rule
\begin{eqnarray}
\langle m\rangle^{2}_{e} + \langle m\rangle^{2}_{\mu} +
\langle m\rangle^{2}_{\tau} = m^2_1 + m^2_2 + m^2_3 \; .
\label{eq:sum-rule}
\end{eqnarray}
It is worth pointing out that each $\big|\langle m\rangle^{}_{\alpha\beta}\big|$
contains at least two of the three independent CP-violating phases of $U$, but each
$\langle m\rangle^{}_\alpha$ contains at most a single phase parameter --- the
so-called Dirac phase $\delta^{}_\nu$
which can be determined from the lepton-flavor-violating but
lepton-number-conserving neutrino oscillations.
If the massive neutrinos are the Majorana particles, $\delta^{}_\nu$ may also
affect those lepton-number-violating processes such as the $0\nu 2\beta$ decays.
This point can be seen from Eq.~(\ref{eq:2beta-decay}), where $\delta^{}_\nu$
is hidden in $U^{}_{e 3}$; and it will become more transparent in
section~\ref{section 6.2}.

\subsection{How to identify the $\mu$-$\tau$ reflection symmetry}
\label{section 2.2}

Now let us show that behind the observed equalities
$\big|U^{}_{\mu i}\big| = \big|U^{}_{\tau i}\big|$
(for $i = 1, 2, 3$) and the T2K hint of CP violation in neutrino
oscillations~\cite{T2K:2019bcf} should lie a minimal flavor symmetry --- the
$\mu$-$\tau$ reflection symmetry for the effective Majorana neutrino mass term.
The same proof is applicable to the effective Dirac neutrino mass term.

Given $\big|U^{}_{\mu i}\big| = \big|U^{}_{\tau i}\big|$ for the unitary PMNS
matrix $U$, the question is how to transform $U^{}_{\mu i}$ to $U^{}_{\tau i}$ or
$U^*_{\tau i}$ in a given flavor basis, in which $U^{}_{e i}$ either stays unchanged
or becomes its complex conjugate such that $\big|U^{}_{e i}\big|$ keeps invariant.
We find that there are two typical possibilities of this kind~\cite{Xing:2022oob}.

(1) One possibility is $U = {\cal P} U \zeta$, where $\cal P$ is a real
orthogonal $(\mu, \tau)$-associated permutation matrix of the form
\begin{eqnarray}
{\cal P} = {\cal P}^T = {\cal P}^\dagger =
\pmatrix{ 1 & 0 & 0 \cr
0 & 0 & 1 \cr 0 & 1 & 0 \cr} \; ,
\label{eq:S-tranformation}
\end{eqnarray}
and $\zeta = {\rm Diag}\{\eta^{}_1 , \eta^{}_2 , \eta^{}_3\}$ with $\eta^{}_i = \pm 1$.
In this case the $(\mu, \tau)$-associated orthogonality condition of $U$ becomes
\begin{eqnarray}
\sum^3_{i=1} U^{}_{\mu i} U^*_{\tau i}
= \sum^3_{i=1} \eta^{}_i \big|U^{}_{\mu i}\big|^2 = 0 \; ,
\label{eq:mu-tau-orthogonality}
\end{eqnarray}
implying the corresponding unitarity triangle in the complex plane collapses
into lines. So the very possibility of $\eta^{}_1 = \eta^{}_2 = \eta^{}_3$ is utterly
incompatible with the normalization condition of $U^{}_{\mu i}$ (for $i = 1, 2, 3$)
and hence has to be abandoned. Considering that $|U^{}_{\mu 1}|$ is definitely smaller
than $|U^{}_{\mu 3}|$, and $|U^{}_{\mu 2}|$ is most likely smaller than $|U^{}_{\mu 3}|$
as one can clearly see in Eq.~(\ref{eq:PMNS}), we find that the possibility of
$\eta^{}_1 = \eta^{}_2 = -\eta^{}_3$ is phenomenologically interesting because it
leads us to
\begin{eqnarray}
\big|U^{}_{\mu 1}\big|^2 + \big|U^{}_{\mu 2}\big|^2 =
\big|U^{}_{\mu 3}\big|^2 = \frac{1}{2} \; ,
\label{eq:mu-tau-permutation}
\end{eqnarray}
and a similar relation for $|U^{}_{\tau i}|^2$ (for $i = 1, 2, 3$). As a consequence,
$|U^{}_{e 3}| = 0$ must hold. Combining these results with the parametrization of $U$
in Eq.~(\ref{eq:U}), we immediately arrive at $\theta^{}_{13} = 0$
and $\theta^{}_{23} = \pi/4$ as given by Eq.~(\ref{eq:U-permutation}).

This observation implies that $U = {\cal P} U \zeta$ should give rise to the
$\mu$-$\tau$ permutation symmetry; namely, the effective neutrino mass term
${\cal L}^{}_\nu$ in Eq.~(\ref{eq:Majorana-mass}) or Eq.~(\ref{eq:M-mass}) should
keep unchanged under the transformations made in Eq.~(\ref{eq:permutation}). To
see this point more clearly, we substitute $U = {\cal P} U \zeta$ into
Eq.~(\ref{eq:M-mass-diagonalization}) by taking $O^{}_\nu = U$ in the chosen
$M^{}_l = D^{}_l$ basis. Then we obtain
\begin{eqnarray}
M^{}_\nu = U D^{}_\nu U^T = {\cal P} U \zeta D^{}_\nu \zeta U^T {\cal P}
= {\cal P} U D^{}_\nu U^T {\cal P} = {\cal P} M^{}_\nu {\cal P} \; .
\label{eq:M-permutation2}
\end{eqnarray}
So the texture of $M^{}_\nu$ can be constrained to the form given
in Eq.~(\ref{eq:M-permutation}). If Eq.~(\ref{eq:M-permutation2}) is inserted
into Eq.~(\ref{eq:M-mass}), one will have
\begin{eqnarray}
-{\cal L}^{}_{\nu} = \frac{1}{2} \hspace{0.05cm} \overline{{\cal P} \nu^{}_{\rm
L}} \hspace{0.1cm} M^{}_\nu \hspace{0.05cm} {\cal P} (\nu^{}_{\rm L})^c +
{\rm h.c.} \; ,
\label{eq:M-mass2}
\end{eqnarray}
implying that ${\cal L}^{}_\nu$ is invariant under the transformation
$\nu^{}_{\rm L} \to {\cal P} \nu^{}_{\rm L}$. The latter is simply equivalent
to the $\mu$-$\tau$ permutation transformation given in Eq.~(\ref{eq:permutation}).
In this way we have identified that behind a special pattern of $U$ as constrained
by $U = {\cal P} U \zeta$ is the invariance of ${\cal L}^{}_\nu$ under the
$\mu$-$\tau$ permutation transformation, or vice versa.

(2) The other typical and more intriguing possibility is $U = {\cal P} U^* \zeta$,
where $\cal P$ and $\zeta$ have been given in and below
Eq.~(\ref{eq:S-tranformation}). In this case the $(\mu,\tau)$-associated
orthogonality condition of $U$ turns out to be
\begin{eqnarray}
\sum^3_{i=1} U^{}_{\mu i} U^*_{\tau i}
= \sum^3_{i=1} \eta^{}_i U^{2}_{\mu i} = 0 \; ,
\label{eq:mu-tau-orthogonality2}
\end{eqnarray}
which defines a normal unitarity triangle in the complex plane. Taking
$\eta^{}_1 = \eta^{}_2 = \eta^{}_3$ for example, one may combine
Eq.~(\ref{eq:mu-tau-orthogonality2}) with the $(\mu,\tau)$-associated
normalization condition of $U$ and then obtain the interesting results
\begin{eqnarray}
&& \sum^3_{i=1} \left({\rm Re} U^{}_{\mu i}\right)^2 = \sum^3_{i=1}
\left({\rm Im} U^{}_{\mu i}\right)^2 = \frac{1}{2} \; , \hspace{1.2cm}
\nonumber \\
&& \sum^3_{i=1} \left({\rm Re} U^{}_{\mu i} \hspace{0.05cm}
{\rm Im} U^{}_{\mu i}\right) = 0 \; .
\label{eq:mu-tau-orthogonality3}
\end{eqnarray}
A combination of Eqs.~(\ref{eq:U}) and (\ref{eq:mu-tau-orthogonality3}) will
allow us to arrive at $\theta^{}_{23} =\pi/4$ and $\delta^{}_\nu = \pm \pi/2$
in the $\mu$-$\tau$ reflection symmetry, as already given
in Eq.~(\ref{eq:U-reflection}).
\begin{table}[t]
\caption{A comparison between the $\mu$-$\tau$-interchanging CP transformation
and the traditional CP transformation for the left-handed neutrino
fields~\cite{Xing:2022hst}.} \vspace{0.2cm}
\label{table1}
\begin{indented}
\item[]\begin{tabular}{cc} \br $\bigstar$ Traditional CP transformation
~&~ $\bigstar$ $\mu$-$\tau$-interchanging CP transformation \\
$\left(t, {\bf x}\right) \to \left(t, -{\bf x}\right)$
& $\left(t, {\bf x}\right) \to \left(t, -{\bf x}\right)$ \\ \mr
$\left\{ \hspace{-0.15cm} \begin{array}{rcl}
\nu^{}_{e \rm L} \hspace{-0.2cm} & \to & \hspace{-0.2cm}
\big(\nu^{}_{e \rm L}\big)^c \\
\nu^{}_{\mu \rm L} \hspace{-0.2cm} & \to & \hspace{-0.2cm}
\big(\nu^{}_{\mu \rm L}\big)^c \\
\nu^{}_{\tau \rm L} \hspace{-0.2cm} & \to & \hspace{-0.2cm}
\big(\nu^{}_{\tau \rm L}\big)^c
\end{array}
\right.$
& $\left\{ \hspace{-0.15cm} \begin{array}{rcl}
\nu^{}_{e \rm L} \hspace{-0.2cm} & \to & \hspace{-0.2cm}
\big(\nu^{}_{e \rm L}\big)^c \\
\nu^{}_{\mu \rm L} \hspace{-0.2cm} & \to & \hspace{-0.2cm}
\big(\nu^{}_{\tau \rm L}\big)^c \\
\nu^{}_{\tau \rm L} \hspace{-0.2cm} & \to & \hspace{-0.2cm}
\big(\nu^{}_{\mu \rm L}\big)^c
\end{array}
\right.$ \\ \br
\end{tabular}
\end{indented}
\end{table}

Now let us prove that behind $U = {\cal P} U^* \zeta$ is the $\mu$-$\tau$
reflection symmetry of ${\cal L}^{}_\nu$ in Eq.~(\ref{eq:M-mass}). Thanks
to $U = {\cal P} U^* \zeta$, we simply have
\begin{eqnarray}
M^{}_\nu = U D^{}_\nu U^T = {\cal P} U^* \zeta D^{}_\nu \zeta U^\dagger {\cal P}
= {\cal P} \left(U D^{}_\nu U^T\right)^* {\cal P} = {\cal P} M^{*}_\nu {\cal P} \; .
\label{eq:M-reflection2}
\end{eqnarray}
That is why the texture of $M^{}_\nu$ can be constrained to the
form given in Eq.~(\ref{eq:M-reflection}). After Eq.~(\ref{eq:M-reflection2})
is substituted into Eq.~(\ref{eq:M-mass}), one will have
\begin{eqnarray}
-{\cal L}^{}_{\nu} = \frac{1}{2} \hspace{0.05cm} \overline{{\cal P} (\nu^{}_{\rm
L})^c} \hspace{0.1cm} M^{}_\nu \hspace{0.05cm} {\cal P} \nu^{}_{\rm L} +
{\rm h.c.} \; ,
\label{eq:M-mass3}
\end{eqnarray}
implying that ${\cal L}^{}_\nu$ keeps invariant under the transformation
$\nu^{}_{\rm L} \to {\cal P} (\nu^{}_{\rm L})^c$. Such a very simple
$\mu$-$\tau$-interchanging CP transformation is just equivalent to the
$\mu$-$\tau$ reflection transformation given in Eq.~(\ref{eq:reflection}),
and its similarity to or difference from the traditional flavor-conserving
CP transformation is illustrated in Table~\ref{table1}.
In this manner we have identified that behind a
special pattern of $U$ as constrained by $U = {\cal P} U^* \zeta$ is actually
the invariance of ${\cal L}^{}_\nu$ under the $\mu$-$\tau$ reflection
transformation, or vice versa. Since the $\mu$-$\tau$ reflection symmetry
of ${\cal L}^{}_\nu$ can naturally accommodate CP violation, it is expected
to be more helpful than the $\mu$-$\tau$ permutation symmetry of ${\cal L}^{}_\nu$
in revealing the salient features of current neutrino oscillation data.

Concentrating on the $\mu$-$\tau$ reflection symmetry of ${\cal L}^{}_\nu$,
we take a close look at the constraints of $U = {\cal P} U^* \zeta$ on the
angles and phases of the standard parametrization of $U$ in Eq.~(\ref{eq:U}).
It is easy to find that the basis-dependent equality $U = {\cal P} U^* \zeta$
leads us to the following constraint conditions:
\begin{eqnarray}
e^{{\rm i} 2\left(\phi^{}_e + \rho\right)} = \eta^{}_1 \; , \quad
e^{{\rm i} 2\left(\phi^{}_e + \sigma\right)} = \eta^{}_2 \; , \quad
e^{{\rm i} 2\left(\phi^{}_e - \delta^{}_\nu\right)} = \eta^{}_3 \;
\label{eq:phase1}
\end{eqnarray}
arising from $U^{}_{e i} = \eta^{}_i U^*_{e i}$ (for $i = 1, 2, 3$), and
\begin{eqnarray}
&& \left(s^{}_{12} c^{}_{23} + c^{}_{12} s^{}_{13} s^{}_{23}
e^{{\rm i} \delta^{}_\nu}\right)
e^{{\rm i}\left(\phi^{}_\mu + \phi^{}_\tau + 2\rho\right)} =
- \eta^{}_1 \left(s^{}_{12} s^{}_{23} - c^{}_{12} s^{}_{13} c^{}_{23}
e^{-{\rm i}\delta^{}_\nu}\right) \; ,
\nonumber \\
&& \left(c^{}_{12} c^{}_{23} - s^{}_{12} s^{}_{13} s^{}_{23}
e^{{\rm i} \delta^{}_\nu}\right)
e^{{\rm i}\left(\phi^{}_\mu + \phi^{}_\tau + 2\sigma\right)} =
- \eta^{}_2 \left(c^{}_{12} s^{}_{23} + s^{}_{12} s^{}_{13} c^{}_{23}
e^{-{\rm i}\delta^{}_\nu}\right) \; ,
\hspace{1cm}
\nonumber \\
&& s^{}_{23} e^{{\rm i}\left(\phi^{}_\mu + \phi^{}_\tau\right)} =
\eta^{}_3 c^{}_{23} \;
\label{eq:phase2}
\end{eqnarray}
coming from $U^{}_{\mu i } = \eta^{}_i U^*_{\tau i}$ (for $i = 1, 2, 3$) with
$\eta^{}_i = \pm 1$. We are therefore left with
\begin{eqnarray}
\theta^{}_{23} = \frac{\pi}{4} \; , \quad \delta^{}_\nu = \pm\frac{\pi}{2} \; ,
\quad \rho = 0 \hspace{0.2cm} {\rm or} \hspace{0.2cm} \frac{\pi}{2} \; , \quad
\sigma = 0 \hspace{0.2cm} {\rm or} \hspace{0.2cm} \frac{\pi}{2} \; ,
\label{eq:reflection-phases}
\end{eqnarray}
together with $\phi^{}_e = 0$ or $\pi/2$ and $\phi^{}_\mu + \phi^{}_\tau =
\left(2 n + 1\right)\pi - 2\phi^{}_e$ with $n$ being an integer, in which the
ambiguity of $\eta^{}_i$ has been taken into account. Three immediate comments
are in order. First, the same constraints on $\theta^{}_{23}$, $\delta^{}_\nu$,
$\rho$, $\sigma$ and $\phi^{}_\alpha$ (for $\alpha = e, \mu, \tau$) can actually be
achieved from the texture of $M^{}_\nu$ constrained by Eq.~(\ref{eq:M-reflection2})
without any ambiguity of $\eta^{}_i$~\cite{Zhou:2014sya,Xing:2019edp}. Second, it
is very easy to check that Eq.~(\ref{eq:reflection-phases}) allows us to
confirm the parametrization-independent constraint conditions obtained in
Eq.~(\ref{eq:mu-tau-orthogonality3}) with $\eta^{}_1 = \eta^{}_2 = \eta^{}_3$.
Third, the recent T2K measurement of leptonic CP violation in neutrino
oscillations hints that the possibility of $\delta^{}_\nu = -\pi/2$
is somewhat preferable to that of $\delta^{}_\nu = \pi/2$~\cite{T2K:2019bcf}.
Moreover, one may figure out that the area of the $(\mu, \tau)$-associated
unitarity triangle defined by Eq.~(\ref{eq:mu-tau-orthogonality2}) is given by
\begin{eqnarray}
S^{}_\triangle = \frac{1}{4} \big|U^{}_{e1}\big| \big|U^{}_{e2}\big|
\big|U^{}_{e3}\big| \sim 1.5\% \;
\label{eq:unitarity-triangle}
\end{eqnarray}
in the $\mu$-$\tau$ reflection symmetry limit,
where Eq.~(\ref{eq:PMNS}) has been considered to make the numerical estimate.
So a significant CP-violating effect in normal neutrino-neutrino or
antineutrino-antineutrino oscillations, which is essentially
measured by the magnitude of $S^{}_\triangle$, is naturally expected (see, e.g.,
Ref.~\cite{Zhu:2018dvj} for an analysis of all the six leptonic unitarity
triangles which have the same area $S^{}_\triangle$).

\subsection{The Jarlskog-like invariants of CP violation}
\label{section 2.3}

Note that the geometric quantity $S^{}_\triangle$ is exactly equivalent
to the well-known Jarlskog invariant of CP violation defined by the
rephasing-independent relationship~\cite{Jarlskog:1985ht}
\begin{eqnarray}
{\rm Im}\left(U^{}_{\alpha i} U^{}_{\beta j} U^*_{\alpha j} U^*_{\beta i}\right)
= {\cal J}^{}_\nu \sum_\gamma \sum_k \epsilon^{}_{\alpha\beta\gamma}
\epsilon^{}_{ijk} \; ,
\label{eq:J}
\end{eqnarray}
where $\epsilon^{}_{\alpha\beta\gamma}$ and $\epsilon^{}_{ijk}$ are the
three-dimensional Levi-Civita symbols with the Greek and Latin subscripts
running respectively over $(e, \mu, \tau)$ and $(1, 2, 3)$. It is very easy
to prove $S^{}_\triangle = \big|{\cal J}^{}_\nu\big|/2$ and
obtain ${\cal J}^{}_\nu = c^{}_{12} s^{}_{12} c^2_{13} s^{}_{13} c^{}_{23} s^{}_{23}
\sin\delta^{}_\nu$ in the standard parametrization of $U$, and therefore
$\big|{\cal J}^{}_\nu\big| = c^{}_{12} s^{}_{12} c^2_{13} s^{}_{13}/2 \sim 3\%$ holds
in the $\mu$-$\tau$ reflection symmetry limit. In comparison, the Jarlskog invariant
of quark flavor mixing and CP violation has been well determined by current
experimental data: ${\cal J}^{}_q = \left(3.08^{+0.15}_{-0.13}\right) \times
10^{-5}$~\cite{Workman:2022ynf}, which is about three orders of magnitude smaller
than its leptonic analogue ${\cal J}^{}_\nu$ given above. Whether ${\cal J}^{}_\nu$
is related to the observed matter-antimatter asymmetry of the Universe has been
a burning question in today's neutrino cosmology~\cite{Xing:2011zza}.

If the three active neutrinos have the Majorana nature, then the
lepton-number-violating neutrino-antineutrino oscillations can in
principle happen~\cite{Pontecorvo:1957cp,Bahcall:1978jn,Chang:1980qw,
Schechter:1980gk,Li:1981um,Bernabeu:1982vi,Vergados:1985pq,Langacker:1998pv,
deGouvea:2002gf,Delepine:2009qg,Xing:2013ty,Xing:2013woa,Kimura:2021juc},
although such rare processes are strongly suppressed and almost
unobservable in practice. Nevertheless, we find that it is conceptually interesting
to examine how the Jarlskog-like invariants of CP violation, which govern the
strengths of CP-violating effects in possible neutrino-antineutrino oscillations,
are simplified by the $\mu$-$\tau$ reflection symmetry.

The probabilities of lepton-number-violating $\nu^{}_\alpha \to \overline{\nu}^{}_\beta$
and $\overline{\nu}^{}_\alpha \to \nu^{}_\beta$ oscillations (for $\alpha, \beta
= e, \mu, \tau$) can be expressed as~\cite{Xing:2013ty,Xing:2013woa}
\begin{eqnarray}
P\left(\nu^{}_\alpha \to \overline{\nu}^{}_\beta\right) & = &
\frac{\big| \langle m \rangle^{}_{\alpha \beta}\big|^2}{E^2}
- 4 \sum_{i<j} \frac{m^{}_i m^{}_j}{E^2} {\rm Re}\left(U^{}_{\alpha i} U^{}_{\beta i}
U^*_{\alpha j} U^*_{\beta j} \right) \sin^2 \frac{\Delta m^2_{ji} L}{4 E}
\nonumber \\
&& + 2 \sum_{i<j} \frac{m^{}_i m^{}_j}{E^2} {\rm Im}\left(U^{}_{\alpha i} U^{}_{\beta i}
U^*_{\alpha j} U^*_{\beta j} \right) \sin \frac{\Delta m^2_{ji} L}{2 E} \; ,
\nonumber \\
P\left(\overline{\nu}^{}_\alpha \to \nu^{}_\beta\right) & = &
\frac{\big| \langle m \rangle^{}_{\alpha \beta}\big|^2}{E^2}
- 4 \sum_{i<j} \frac{m^{}_i m^{}_j}{E^2} {\rm Re}\left(U^{}_{\alpha i} U^{}_{\beta i}
U^*_{\alpha j} U^*_{\beta j} \right) \sin^2 \frac{\Delta m^2_{ji} L}{4 E}
\nonumber \\
&& - 2 \sum_{i<j} \frac{m^{}_i m^{}_j}{E^2} {\rm Im}\left(U^{}_{\alpha i} U^{}_{\beta i}
U^*_{\alpha j} U^*_{\beta j} \right) \sin \frac{\Delta m^2_{ji} L}{2 E} \; ,
\label{eq:neutrino-antineutrino}
\end{eqnarray}
where a universal kinematical factor has been neglected for
these two probabilities, $E$ is the average beam energy of neutrinos or antineutrinos,
$\Delta m^2_{ji} \equiv m^2_j - m^2_i$ denote the neutrino mass-squared differences,
$L$ stands for the baseline length between the neutrino (or antineutrino)
source and the detector, and the effective Majorana neutrino mass term
$\langle m\rangle^{}_{\alpha\beta}$ has been defined in
Eq.~(\ref{eq:M-reconstruction}). We see that CP violation in neutrino-antineutrino
oscillations is measured by the Jarlskog-like invariants
\begin{eqnarray}
{\cal J}^{ij}_{\alpha\beta} \equiv
{\rm Im}\left(U^{}_{\alpha i} U^{}_{\beta i}
U^*_{\alpha j} U^*_{\beta j} \right) \; ,
\label{eq:Jarlskog-like}
\end{eqnarray}
where $i, j = 1, 2, 3$ and $\alpha, \beta = e, \mu, \tau$.
Different from the canonical Jarlskog invariant ${\cal J}^{}_\nu$ defined
in Eq.~(\ref{eq:J}), which is only sensitive to the Dirac CP-violating phase
$\delta^{}_\nu$, ${\cal J}^{ij}_{\alpha\beta}$ are in general sensitive to all
the three CP-violating phases of $U$~\cite{Xing:2013woa}.
Taking account of Eq.~(\ref{eq:reflection-phases})
in the $\mu$-$\tau$ reflection symmetry limit, however, we can obtain the
following simplified results of ${\cal J}^{ij}_{\alpha\beta}$:
\begin{eqnarray}
{\cal J}_{ee}^{12} = {\cal J}_{ee}^{13} = {\cal J}_{ee}^{23} =
{\cal J}_{\mu\tau}^{12} = {\cal J}_{\mu\tau}^{13} = {\cal J}_{\mu\tau}^{23}
= 0 \; ,
\label{eq:Jarlskog-like1}
\end{eqnarray}
and
\begin{eqnarray}
{\cal J}_{\mu\mu}^{12} & = & -{\cal J}_{\tau\tau}^{12} =
-{\cal J}_{e\mu}^{12} = {\cal J}_{e\tau}^{12} = {\cal J}^{}_\nu \eta^{}_\rho
\eta^{}_\sigma \; ,
\nonumber \\
{\cal J}_{\mu\mu}^{13} & = & -{\cal J}_{\tau\tau}^{13} =
-{\cal J}_{e\mu}^{13} = {\cal J}_{e\tau}^{13} = +{\cal J}^{}_\nu \eta^{}_\rho \; ,
\nonumber \\
{\cal J}_{\mu\mu}^{23} & = & -{\cal J}_{\tau\tau}^{23} =
-{\cal J}_{e\mu}^{23} = {\cal J}_{e\tau}^{23} = -{\cal J}^{}_\nu \eta^{}_\sigma \; ,
\label{eq:Jarlskog-like2}
\end{eqnarray}
where ${\cal J}^{}_\nu = \pm c^{}_{12} s^{}_{12} c^2_{13} s^{}_{13}/2$ with
``$\pm$" corresponding to $\delta^{}_\nu = \pm\pi/2$, and
$\eta^{}_\rho \equiv \cos 2\rho$ and $\eta^{}_\sigma \equiv \cos 2\sigma$
are defined. As a result, the $\mu$-$\tau$ reflection symmetry guarantees
\begin{eqnarray}
P\big(\nu^{}_e \to \overline{\nu}^{}_e\big) =
P\big(\overline{\nu}^{}_e \to \nu^{}_e\big) \; , \quad &&
P\big(\nu^{}_\mu \to \overline{\nu}^{}_\tau\big) =
P\big(\overline{\nu}^{}_\mu \to \nu^{}_\tau\big) \; ,
\label{eq:neutrino-antineutrino2}
\end{eqnarray}
which are both T-conserving and CP-conserving; and
\begin{eqnarray}
P\big(\nu^{}_e \to \overline{\nu}^{}_\mu\big) =
P\big(\nu^{}_e \to \overline{\nu}^{}_\tau\big) \; , \quad &&
P\big(\nu^{}_\mu \to \overline{\nu}^{}_\mu\big) =
P\big(\nu^{}_\tau \to \overline{\nu}^{}_\tau\big) \; ,
\label{eq:neutrino-antineutrino3}
\end{eqnarray}
which contain the CP-violating term. Of course,
Eqs.~(\ref{eq:Jarlskog-like1})---(\ref{eq:neutrino-antineutrino3}) do not
hold when the effects of $\mu$-$\tau$ reflection symmetry breaking are taken
into account~\cite{Xing:2013ty,Xing:2013woa}. If the unitarity of $U$ is
slightly broken, the above results will accordingly be
modified~\cite{Wang:2021rsi}.

It is worth pointing out that how to experimentally
determine or constrain the Majorana phases of $U$ will become a burning question
once the Majorana nature of massive neutrinos is verified in the CP-conserving
$0\nu 2\beta$-decay experiments someday. Although it is extremely difficult (if
not impossible) to observe neutrino-antineutrino oscillations even in the far
future, the above discussions should conceptually make sense as they can at least
give one a ball-park feeling of some interesting constraints on the Jarlskog-like
invariants of CP violation in such extremely rare processes as a natural consequence
of the $\mu$-$\tau$ reflection symmetry.

\subsection{A generalized case in the seesaw mechanism}
\label{section 2.4}

Given the fact that the canonical seesaw mechanism~\cite{Fritzsch:1975sr,
Minkowski:1977sc,Yanagida:1979as,GellMann:1980vs,Glashow:1979nm,Mohapatra:1979ia}
is currently the most convincing and popular mechanism for interpreting the origin
of tiny neutrino masses, we are going to combine it with the $\mu$-$\tau$ reflection
symmetry in a rather generic approach that has recently been developed in
Ref.~\cite{Xing:2022oob}. First of all, let us consider the SM extended
by adding three right-handed neutrino fields $N^{}_{\alpha \rm R}$
(for $\alpha = e, \mu, \tau$) and allowing for lepton number violation. In this case
the gauge- and Lorentz-invariant neutrino mass terms can be written as
\begin{eqnarray}
-{\cal L}^{}_{\rm SS} = \overline{\ell^{}_{\rm L}} \hspace{0.05cm} Y^{}_\nu
\widetilde{H} N^{}_{\rm R} + \frac{1}{2} \hspace{0.05cm} \overline{(N^{}_{\rm R})^c}
\hspace{0.05cm} M^{}_{\rm R} N^{}_{\rm R} + {\rm h.c.} \; ,
\label{eq:seesaw1}
\end{eqnarray}
where $\ell^{}_{\rm L}$ denotes the $\rm SU(2)^{}_{\rm L}$ doublet of the left-handed
lepton fields, $\widetilde{H} \equiv {\rm i} \sigma^{}_2 H^*$ with
$H$ being the Higgs doublet and $\sigma^{}_2$ being the second Pauli matrix,
$N^{}_{\rm R} = (N^{}_{e \rm R} , N^{}_{\mu \rm R} , N^{}_{\tau \rm R})^T$ is the column
vector of three right-handed neutrino fields, $(N^{}_{\rm R})^c \equiv {\cal C}
\overline{N^{}_{\rm R}}^T$ with $\cal C$ having been described below
Eq.~({\ref{eq:Majorana-mass}), $Y^{}_\nu$ represents an arbitrary $3\times 3$
Yukawa coupling matrix, and $M^{}_{\rm R}$ stands for a symmetric $3\times 3$
Majorana mass matrix. After spontaneous electroweak symmetry breaking,
Eq.~(\ref{eq:seesaw1}) becomes
\begin{eqnarray}
-{\cal L}^\prime_{\rm SS} = \frac{1}{2} \hspace{0.05cm} \overline{\pmatrix{
\nu^{}_{\rm L} \hspace{-0.2cm} & (N^{}_{\rm R})^c}}
\pmatrix{ {\bf 0} & M^{}_{\rm D} \cr M^T_{\rm D} & M^{}_{\rm R} \cr}
\pmatrix{ (\nu^{}_{\rm L})^c \cr N^{}_{\rm R} \cr} + {\rm h.c.} \; ,
\label{eq:seesaw-mass-matrix}
\end{eqnarray}
where $M^{}_{\rm D} \equiv Y^{}_\nu \langle H\rangle$ with $\langle H\rangle$ being the
vacuum expectation value of the Higgs field, and $\bf 0$ denotes
the $3\times 3$ zero matrix. Note that $M^{}_{\rm D}$
is in general neither Hermitian nor symmetric. The overall $6\times 6$
neutrino mass matrix in Eq.~(\ref{eq:seesaw-mass-matrix}) can be diagonalized by the
following unitary transformation:
\begin{eqnarray}
\pmatrix{ U & R \cr S & Q \cr}^{\hspace{-0.05cm} \dagger} \pmatrix{ {\bf 0}
& M^{}_{\rm D} \cr M^{T}_{\rm D} & M^{}_{\rm R} \cr}
\pmatrix{ U & R \cr S & Q \cr}^{\hspace{-0.05cm} *}
= \pmatrix{ D^{}_\nu & {\bf 0} \cr {\bf 0} & D^{}_N \cr} \; ,
\label{eq:seesaw-diagonalization}
\end{eqnarray}
where $D^{}_\nu$ has already been defined in Eq.~(\ref{eq:M-mass-diagonalization}), and
$D^{}_N \equiv {\rm Diag}\{M^{}_1, M^{}_2, M^{}_3 \}$ with $M^{}_i$ (for $i = 1, 2, 3$)
being the masses of three heavy Majorana neutrinos. The four $3\times 3$ flavor mixing
submatrices in Eq.~(\ref{eq:seesaw-diagonalization}) satisfy the unitarity conditions
\begin{eqnarray}
&& U U^\dagger + RR^\dagger = SS^\dagger + Q Q^\dagger = {\cal I} \; ,
\nonumber \\
&& U^\dagger U + S^\dagger S = R^\dagger R + Q^\dagger Q = {\cal I} \; ,
\nonumber \\
&& U S^\dagger + R Q^\dagger = U^\dagger R + S^\dagger Q = {\bf 0} \; ;
\label{eq:seesaw-unitarity}
\end{eqnarray}
and among them, $U$ and $R$ also satisfy the {\it exact} seesaw
relation~\cite{Xing:2007zj,Xing:2011ur}
\begin{eqnarray}
U D^{}_\nu U^T + R D^{}_N R^T = {\bf 0} \; .
\label{eq:exact-seesaw}
\end{eqnarray}
To be more specific, Eq.~(\ref{eq:exact-seesaw}) leads us to the {\it exact} mass
relations between light (active) and heavy (sterile) neutrinos:
\begin{eqnarray}
\sum^3_{i=1} \left(m^{}_i U^{}_{\alpha i} U^{}_{\beta i}\right) =
- \sum^3_{i=1} \left(M^{}_i R^{}_{\alpha i} R^{}_{\beta i}\right) \; ,
\label{eq:exact-seesaw-mass}
\end{eqnarray}
where $\alpha$ and $\beta$ run over $e$, $\mu$ and $\tau$.
In this case $\nu^{}_{\alpha}$
(for $\alpha = e, \mu, \tau$) can be expressed as a linear combination of
the mass eigenstates of three active neutrinos $\nu^{}_{i} = \nu^c_i$
and three sterile neutrinos $N^{}_{i} = N^c_i$ (for $i = 1, 2, 3$).
The standard weak charged-current interactions of these six Majorana neutrinos
are therefore expressed as
\begin{eqnarray}
-{\cal L}^{\prime}_{\rm cc} = \frac{g^{}_{\rm w}}{\sqrt{2}} \hspace{0.1cm}
\overline{\pmatrix{e & \hspace{-0.25cm} \mu & \hspace{-0.25cm} \tau}^{}_{\rm L}}
\hspace{0.1cm} \gamma^\mu
\left[ U \pmatrix{ \nu^{}_{1} \cr \nu^{}_{2} \cr
\nu^{}_{3} \cr}^{}_{\hspace{-0.15cm} \rm L} + R
\pmatrix{ N^{}_{1} \cr N^{}_{2} \cr N^{}_{3}
\cr}^{}_{\hspace{-0.15cm} \rm L} \right] W^-_\mu + {\rm h.c.} \; ,
\label{eq:cc-seesaw}
\end{eqnarray}
in which the PMNS matrix $U$ is responsible for describing the phenomenology
of active neutrino oscillations, and the matrix $R$ characterizes the strengths of
weak charged-current interactions of those heavy neutrinos. As a consequence,
the $(\mu, \tau)$-associated orthogonality condition of $U$ and $R$ defines a
unitarity hexagon in the complex plane as illustrated by
Figure~\ref{Fig:unitarity-hexagon},
but the sides $R^{}_{\mu i} R^*_{\tau i}$ (for $i = 1, 2, 3$) are expected to be
very short and thus can be treated as an effective vertex of the {\it deformed}
unitarity triangle formed by the sides $U^{}_{\mu i} U^*_{\tau i}$ (for $i = 1, 2, 3$).
\begin{figure}[t]
\begin{center}
\includegraphics[width=3.4in]{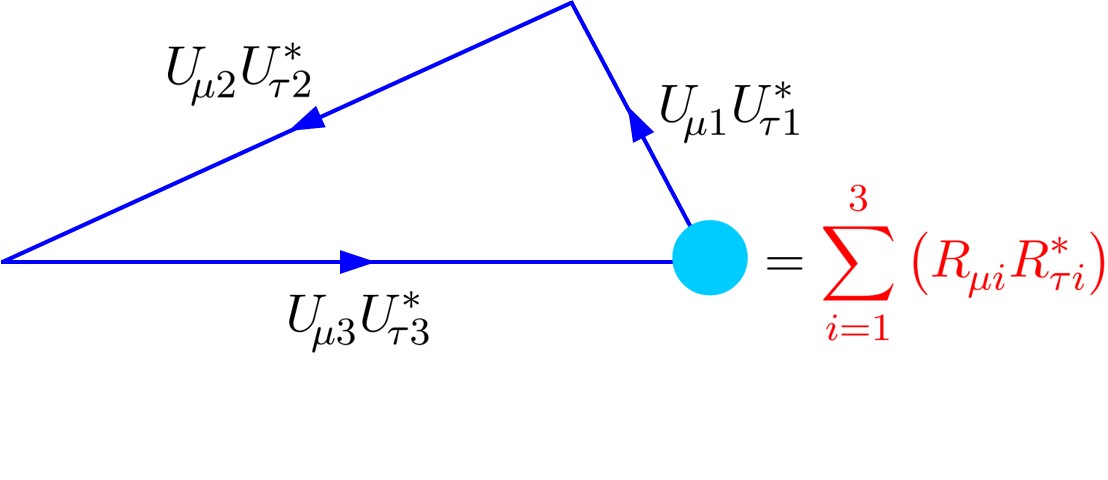}
\vspace{-1.1cm}
\caption{An illustration of the unitarity hexagon defined by the
$(\mu,\tau)$-associated orthogonality condition in the complex plane,
where the {\it effective} vertex in sky blue denotes small corrections of
the active-sterile flavor mixing effects to the three-flavor unitarity triangle
in the seesaw framework~\cite{Xing:2022oob}.}
\label{Fig:unitarity-hexagon}
\end{center}
\end{figure}

Although $U$ is not exactly unitary in the canonical seesaw framework, a careful
analysis of current electroweak precision measurements and neutrino oscillation
data has constrained $U$ to be essentially unitary at least at the
${\cal O}(10^{-2})$ level~\cite{Antusch:2006vwa,Fernandez-Martinez:2016lgt,
Blennow:2016jkn,Hu:2020oba,Wang:2021rsi}. So possible non-unitarity effects hiding
in $U$ must be below ${\cal O}(10^{-2})$ and hence cannot be identified up to the
accuracy level of today's neutrino oscillation experiments. In other words, the
possibility of $U = {\cal P} U^* \zeta$ as revealed by the numerical results listed
in Eq.~(\ref{eq:PMNS}) is still expected even though $U$ is slightly non-unitary.
This observation is the starting point of view for our subsequent discussions
about the $\mu$-$\tau$ reflection symmetry associated with the seesaw mechanism.

Now let us begin with $U = {\cal P} U^* \zeta$. Inserting this $\mu$-$\tau$
reflection symmetry relation into the exact seesaw formula in
Eq.~(\ref{eq:exact-seesaw}) and taking the complex conjugate for the whole
equation, we simply arrive at
\begin{eqnarray}
U D^{}_\nu U^T + {\cal P} R^* D^{}_N ({\cal P} R^*)^T = {\bf 0} \; .
\label{eq:exact-seesaw2}
\end{eqnarray}
A comparison between Eqs.~(\ref{eq:exact-seesaw}) and (\ref{eq:exact-seesaw2})
immediately leads us to $R = {\cal P} R^* \zeta^\prime$~\cite{Xing:2022oob},
where $\zeta^\prime = {\rm Diag}\{\eta^\prime_1 , \eta^\prime_2 , \eta^\prime_3\}$
with $\eta^\prime_i = \pm 1$ (for $i = 1, 2, 3$). This result implies that the
novel prediction $\big|R^{}_{\mu i}\big| = \big|R^{}_{\tau i}\big|$ is actually a
natural consequence of $\big|U^{}_{\mu i}\big| = \big|U^{}_{\tau i}\big|$ in the
canonical seesaw mechanism. Given a full Euler-like parametrization of the
$3\times 3$ matrices $U$, $R$, $S$ and $Q$
in Eq.~(\ref{eq:seesaw-diagonalization})~\cite{Xing:2007zj,Xing:2011ur}, the
$\mu$-$\tau$ reflection symmetry allows us to obtain some quite strong
constraints on the relevant flavor mixing angles and CP-violating phases appearing
in $U$ and $R$ (see~\ref{Appendix A} for a detailed discussion~\cite{Xing:2022oob}).

After substituting $U = {\cal P} U^* \zeta$ and
$R = {\cal P} R^* \zeta^\prime$ into Eq.~(\ref{eq:seesaw-unitarity}), we find
$S = {\cal T} S^* \zeta$ and $Q = {\cal T} Q^* \zeta^\prime$ with $\cal T$
being an arbitrary unitary matrix. We proceed to insert these four relations
into Eq.~(\ref{eq:seesaw-diagonalization}) and take the complex conjugate for
the whole equation. In this case we are left with
\begin{eqnarray}
\pmatrix{ U & R \cr S & Q \cr}^{\hspace{-0.05cm} \dagger}
\pmatrix{ {\bf 0} & {\cal P} M^{*}_{\rm D} {\cal T}
\cr {\cal T}^T M^{\dagger}_{\rm D} {\cal P}
& {\cal T}^T M^{*}_{\rm R} {\cal T} \cr}
\pmatrix{ U & R \cr S & Q \cr}^{\hspace{-0.05cm} *}
= \pmatrix{ D^{}_\nu & {\bf 0} \cr {\bf 0} & D^{}_N \cr} \; ,
\label{eq:seesaw-diagonalization2}
\end{eqnarray}
which is apparently independent of the ambiguities of $\zeta$ and $\zeta^\prime$.
Then a direct comparison between Eqs.~(\ref{eq:seesaw-diagonalization}) and
(\ref{eq:seesaw-diagonalization2}) yields
\begin{eqnarray}
M^{}_{\rm D} = {\cal P} M^{*}_{\rm D} {\cal T} \; , \quad
M^{}_{\rm R} = {\cal T}^T M^{*}_{\rm R} {\cal T} \; .
\label{eq:seesaw-constraint}
\end{eqnarray}
Note that the arbitrary unitary matrix $\cal T$ can be fixed
in a given seesaw model of massive neutrinos. In this case it will be possible to
constrain the textures of $M^{}_{\rm D}$ and $M^{}_{\rm R}$. Taking ${\cal T} = {\cal P}$
for example, just as considered in Refs.~\cite{Mohapatra:2015gwa,Xing:2019edp},
we have the flavor textures of $M^{}_{\rm D}$ and $M^{}_{\rm R}$ as follows:
\begin{eqnarray}
M^{}_{\rm D} = \pmatrix{ X^{}_{11} & X^{}_{12} & X^*_{12} \cr
X^{}_{21} & X^{}_{22} & X^{}_{23} \cr
X^*_{21} & X^*_{23} & X^*_{22} \cr} \; , \quad
M^{}_{\rm R} = \pmatrix{ Z^{}_{11} & \hspace{0.08cm} Z^{}_{12}
\hspace{0.08cm} & Z^*_{12} \cr Z^{}_{12} & Z^{}_{22} & Z^{}_{23} \cr
Z^*_{12} & Z^{}_{23} & Z^*_{22} \cr} \; ,
\label{eq:seesaw-constraint2}
\end{eqnarray}
where $X^{}_{11}$, $Z^{}_{11}$ and $Z^{}_{23}$ are real. In such a seesaw
scenario the number of real free parameters of $M^{}_{\rm D}$ (or $M^{}_{\rm R}$)
is reduced by half, from eighteen (or twelve) to nine (or six).

After Eq.~(\ref{eq:seesaw-constraint}) is substituted into
Eq.~(\ref{eq:seesaw-mass-matrix}), the overall neutrino mass
term ${\cal L}^\prime_{\rm SS}$ turns out to be
\begin{eqnarray}
-{\cal L}^{\prime}_{\rm SS} & = & \frac{1}{2} \hspace{0.05cm} \overline{\pmatrix{
{\cal P} \nu^{}_{\rm L} \hspace{-0.2cm} & {\cal T}^* (N^{}_{\rm R})^c}}
\pmatrix{ {\bf 0} & M^{*}_{\rm D} \cr
M^\dagger_{\rm D} & M^{*}_{\rm R} \cr}
\pmatrix{ {\cal P} (\nu^{}_{\rm L})^c \cr
{\cal T} N^{}_{\rm R} \cr} + {\rm h.c.}
\nonumber \\
& = & \frac{1}{2} \hspace{0.05cm} \overline{\pmatrix{
{\cal P} (\nu^{}_{\rm L})^c \hspace{-0.2cm} & {\cal T} N^{}_{\rm R}}}
\pmatrix{ {\bf 0} & M^{}_{\rm D} \cr
M^T_{\rm D} & M^{}_{\rm R} \cr}
\pmatrix{ {\cal P} \nu^{}_{\rm L} \cr
{\cal T}^* (N^{}_{\rm R})^c \cr} + {\rm h.c.} \; ,
\label{eq:seesaw-mass-matrix2}
\end{eqnarray}
in which the apparent mass term of the second equation comes from the Hermitian
conjugation term of the first equation. Comparing between
Eq.~(\ref{eq:seesaw-mass-matrix}) and Eq.~(\ref{eq:seesaw-mass-matrix2}),
one may easily find that ${\cal L}^\prime_{\rm SS}$ keeps unchanged under
the transformations
\begin{eqnarray}
\nu^{}_{\rm L} \to {\cal P} (\nu^{}_{\rm L})^c \; , \quad
N^{}_{\rm R} \to {\cal T}^* (N^{}_{\rm R})^c \; .
\label{eq:seesaw-reflection}
\end{eqnarray}
It is well known that CP would be a good symmetry in the canonical seesaw
mechanism if ${\cal L}^\prime_{\rm SS}$  were invariant under the {\it charge
conjugation} and {\it parity} transformations $\nu^{}_{\alpha \rm L} (t, {\bf x})
\to \big[\nu^{}_{\alpha \rm L} (t, -{\bf x}) \big]^c$ and $N^{}_{\alpha \rm R}
(t, {\bf x}) \to \big[N^{}_{\alpha \rm R} (t, -{\bf x}) \big]^c$
(for $\alpha = e, \mu, \tau$)~\cite{Xing:2011zza}. But the $\mu$-$\tau$
reflection transformation of the left-handed neutrino fields, together with an
arbitrary {\it unitary} CP transformation of the right-handed
neutrino fields in Eq.~(\ref{eq:seesaw-reflection}), makes CP violation
possible even though ${\cal L}^\prime_{\rm SS}$ itself keeps invariant
in this case. The issue will certainly be simplified if a specific
form of ${\cal T}$ is assumed, such as ${\cal T} = {\cal P}$ as taken above.

Integrating out those heavy degrees of freedom in the canonical seesaw mechanism,
one may get the effective Majorana mass term ${\cal L}^{}_\nu$ for three active
neutrinos in Eq.~(\ref{eq:M-mass}) with the approximate but well-known seesaw
formula
\begin{eqnarray}
M^{}_\nu \simeq - M^{}_{\rm D} M^{-1}_{\rm R} M^T_{\rm D} \; .
\label{eq:seesaw-formula}
\end{eqnarray}
Substituting Eq.~(\ref{eq:seesaw-constraint}) into Eq.~(\ref{eq:seesaw-formula}),
we immediately arrive at $M^{}_\nu = {\cal P} M^*_\nu {\cal P}$. This result
is fully compatible with Eq.~(\ref{eq:M-reflection2}), in which the PMNS matrix
$U$ is unitary and satisfies $U = {\cal P} U^* \zeta$. As a direct consequence,
${\cal L}^{}_\nu$ is invariant under the $\mu$-$\tau$ reflection transformations
$\nu^{}_{e \rm L} \to (\nu^{}_{e \rm L})^c$,
$\nu^{}_{\mu \rm L} \to (\nu^{}_{\tau \rm L})^c$ and
$\nu^{}_{\tau \rm L} \to (\nu^{}_{\mu \rm L})^c$.

\subsection{An $\rm S^{}_3$ extension of the $\mu$-$\tau$ flavor symmetry}
\label{section 2.5}

It is well known that the real orthogonal $(\mu, \tau)$-associated permutation
matrix $\cal P$ in Eq.~(\ref{eq:S-tranformation}) is just one of the six elements
of the non-Abelian $\rm S^{}_3$ group --- a permutation group of three
objects~\cite{Pakvasa:1977in,Harari:1978yi}:
\begin{eqnarray}
&& {\cal S}^{(123)} = \pmatrix{
1 & 0 & 0 \cr 0 & 1 & 0 \cr 0 & 0 & 1} \; , \quad
{\cal S}^{(231)} = \pmatrix{
0 & 1 & 0 \cr 0 & 0 & 1 \cr 1 & 0 & 0} \; ,
\nonumber \\
&& {\cal S}^{(312)} = \pmatrix{
0 & 0 & 1 \cr 1 & 0 & 0 \cr 0 & 1 & 0} \; , \quad
{\cal S}^{(213)} = \pmatrix{
0 & 1 & 0 \cr 1 & 0 & 0 \cr 0 & 0 & 1} \; ,
\nonumber \\
&& {\cal S}^{(132)} = \pmatrix{
1 & 0 & 0 \cr 0 & 0 & 1 \cr 0 & 1 & 0} \; , \quad
{\cal S}^{(321)} = \pmatrix{
0 & 0 & 1 \cr 0 & 1 & 0 \cr 1 & 0 & 0} \; .
\label{eq:S3}
\end{eqnarray}
Namely, ${\cal S}^{(123)} = {\cal I}$ and ${\cal S}^{(132)} = {\cal P}$ hold.
Note that ${\cal S}^{(231)}$ and ${\cal S}^{(312)}$ are non-symmetric
but can be expressed as the products of two symmetric $\rm S^{}_3$
transformations:
\begin{eqnarray}
&& {\cal S}^{(231)} = {\cal S}^{(213)} {\cal S}^{(321)} =
{\cal S}^{(321)} {\cal S}^{(132)} = {\cal S}^{(132)} {\cal S}^{(213)} \; ,
\nonumber \\
&& {\cal S}^{(312)} = {\cal S}^{(321)} {\cal S}^{(213)} =
{\cal S}^{(213)} {\cal S}^{(132)} = {\cal S}^{(132)} {\cal S}^{(321)} \; .
\label{eq:S3relation}
\end{eqnarray}
Note also that the real orthogonal matrices ${\cal S}^{(213)}$ and ${\cal S}^{(321)}$
describe the corresponding $e$-$\mu$ and $e$-$\tau$ flavor permutations, but neither
of them is phenomenologically favored to serve as a simple working flavor symmetry
in describing the observed pattern of lepton flavor mixing. We have two immediate
observations as follows.

(1) If the effective neutrino mass matrix $M^{}_\nu$ satisfies
\begin{eqnarray}
{\cal S}^{(ijk)} M^{}_\nu = M^{}_\nu {\cal S}^{(ijk)} \; ,
\label{eq:S3Mnu}
\end{eqnarray}
where ${\cal S}^{(ijk)}$ stands for any of the six matrices in Eq.~(\ref{eq:S3}),
then it will be constrained to have a very particular texture:
\begin{eqnarray}
M^{}_\nu = a {\cal I} + b {\cal D} \; ,
\label{eq:S3Mnu2}
\end{eqnarray}
where the free parameters $a$ and $b$ are in general
complex~\cite{Jora:2006dh,Jora:2009gz}, and $\cal D$ is a rank-one
constant matrix of the form
\begin{eqnarray}
{\cal D} \equiv {\cal S}^{(132)} + {\cal S}^{(213)} + {\cal S}^{(321)}
= \pmatrix{ 1 & 1 & 1 \cr 1 & 1 & 1 \cr 1 & 1 & 1} \; .
\label{eq:Democracy}
\end{eqnarray}
The pattern of $\cal D$ is usually referred to as the {\it flavor democracy}
texture~\cite{Fritzsch:1995dj}, which has often been used to define a special
flavor basis of the charged leptons, up-type quarks or down-type quarks by taking
account of their respective strong mass hierarchies~\cite{Fritzsch:1999ee,Xing:2010iu}.
In this case $M^{}_\nu$ in Eq.~(\ref{eq:S3Mnu2}) can be diagonalized by either
the so-called ``democratic" flavor mixing pattern~\cite{Fritzsch:1995dj,Fritzsch:1998xs}
\begin{eqnarray}
U^{}_{\rm DEM} = \pmatrix{ \frac{1}{\sqrt 2} & \frac{-1}{\sqrt 2} & 0 \cr
\vspace{-0.45cm} \cr
\frac{1}{\sqrt 6} & \frac{1}{\sqrt 6} & \frac{-2}{\sqrt 6} \cr
\vspace{-0.45cm} \cr
\frac{1}{\sqrt 3} & \frac{1}{\sqrt 3} & \frac{1}{\sqrt 3}} \;
\label{eq:DEM}
\end{eqnarray}
via the real orthogonal transformation
\begin{eqnarray}
U^{}_{\rm DEM} M^{}_\nu U^T_{\rm DEM} =
\pmatrix{a & 0 & 0 \cr 0 & a & 0 \cr 0 & 0 & a + 3b}  \; ;
\label{eq:DEM2}
\end{eqnarray}
or the so-called ``tribimaximal" flavor mixing
pattern~\cite{Harrison:2002er,Xing:2002sw,He:2003rm}
\begin{eqnarray}
U^{}_{\rm TBM} = \pmatrix{ \frac{2}{\sqrt 6} & \frac{1}{\sqrt 3} & 0 \cr
\vspace{-0.45cm} \cr
\frac{-1}{\sqrt 6} & \frac{1}{\sqrt 3} & \frac{-1}{\sqrt 2} \cr
\vspace{-0.45cm} \cr
\frac{-1}{\sqrt 6} & \frac{1}{\sqrt 3} & \frac{1}{\sqrt 2}} \;
\label{eq:TBM}
\end{eqnarray}
via the real orthogonal transformation
\begin{eqnarray}
U^T_{\rm TBM} M^{}_\nu U^{}_{\rm TBM} =
\pmatrix{a & 0 & 0 \cr 0 & a + 3b & 0 \cr 0 & 0 & a}  \; .
\label{eq:TBM2}
\end{eqnarray}
It is obvious that both $U^{}_{\rm DEM}$ and $U^{}_{\rm TBM}$ are far away
from the pattern of $U$ extracted from current neutrino oscillation data
as shown in Eq.~(\ref{eq:PMNS}), and hence proper perturbations to
the $\rm S^{}_3$-constrained texture of $M^{}_\nu$ in Eq.~(\ref{eq:S3Mnu2})
have to be introduced
\footnote{Note that there exists a mass eigenvalue degeneracy
in either Eq.~(\ref{eq:DEM2}) or Eq.~(\ref{eq:TBM2}), and hence the corresponding
flavor mixing pattern cannot be uniquely fixed. In other words, the $(1,2)$
mixing part of $U^{}_{\rm DEM}$ in Eq.~(\ref{eq:DEM}) or the $(1,3)$ mixing
part of $U^{}_{\rm TBM}$ in Eq.~(\ref{eq:TBM}) involves a degree of
arbitrariness. But such uncertainties can be removed by introducing
appropriate perturbations to $M^{}_\nu$ in Eq.~(\ref{eq:S3Mnu2}) from a
point of view of model building, as done
in Refs.~\cite{Xing:2010iu,Tanimoto:2000fz,Branco:2001hn,Fritzsch:2004xc}
for example.}

(2) If the effective neutrino mass matrix $M^{}_\nu$ satisfies
\begin{eqnarray}
{\cal S}^{(ijk)} M^{}_\nu = M^{*}_\nu {\cal S}^{(ijk)} \; ,
\label{eq:S3MnuCP}
\end{eqnarray}
where ${\cal S}^{(ijk)}$ can be any of the six matrices listed in
Eq.~(\ref{eq:S3}), it will be constrained to the following specific form:
\begin{eqnarray}
M^{}_\nu = a {\cal I} + b {\cal S}^{(231)} + b^* {\cal S}^{(312)}
= \pmatrix{a & b & b^* \cr b^* & a & b \cr b & b^* & a} \; ,
\label{eq:S3Mnu3}
\end{eqnarray}
where $a$ is real and $b$ is in general complex. Now that $M^{}_\nu$ is in
general Hermitian, it cannot coincide with the Majorana nature of massive neutrinos
unless $b$ is real. But in this case $M^{}_\nu$ becomes real and thus needs
certain complex and symmetric perturbations so as to accommodate CP violation
for the Majorana neutrinos. Of course, a Hermitian texture of $M^{}_\nu$ is
suitable for describing the Dirac neutrinos.

Note that the six $\rm S^{}_3$ group elements can be categorized into three
conjugacy classes ${\cal C}^{}_{0} = \left\{ {\cal S}^{(123)} \right\}$,
${\cal C}^{}_{1} = \left\{ {\cal S}^{(231)} , {\cal S}^{(312)} \right\}$ and
${\cal C}^{}_{2} = \left\{ {\cal S}^{(213)} , {\cal S}^{(132)},
{\cal S}^{(321)} \right\}$. It is straightforward to show that the invariant
subgroup of $\rm S^{}_3$ is the cyclic group of order three
\begin{eqnarray}
{\cal Z}^{}_{3} = \left\{{\cal S}^{(123)} , {\cal S}^{(231)} , {\cal S}^{(312)}
\right\} = \left\{ {\cal I} , {\cal Z} , {\cal Z}^2\right\} \; ,
\label{eq:Z3}
\end{eqnarray}
where the generator of the $\rm Z^{}_3$ group has been defined as
${\cal Z} \equiv {\cal S}^{(231)}$. One may easily verify ${\cal Z}^3 = {\cal I}$
with the help of Eq.~(\ref{eq:S3}). Moreover, it is worth pointing out that
the $\rm S^{}_3$ group has three $\rm Z^{}_2$ subgroups of order two,
\begin{eqnarray}
{\cal Z}^{(12)}_{2} = \left\{{\cal S}^{(123)} , {\cal S}^{(213)}\right\} \; ,
\nonumber \\
{\cal Z}^{(23)}_{2} = \left\{{\cal S}^{(123)} , {\cal S}^{(132)}\right\} \; ,
\nonumber \\
{\cal Z}^{(31)}_{2} = \left\{{\cal S}^{(123)} , {\cal S}^{(321)}\right\} \; ,
\label{eq:Z3}
\end{eqnarray}
which can be identified as the $e$-$\mu$, $\mu$-$\tau$ and $e$-$\tau$
permutation symmetries, respectively. The symmetry breaking chains
$\rm S^{}_3 \to Z^{}_3 \to 0$ and $\rm S^{}_3 \to Z^{}_2 \to 0$ may help constrain
the flavor textures of charged leptons and massive neutrinos so as to explain
current neutrino oscillation data~\cite{Dicus:2010iq,Zhou:2011nu}.

One may also apply the $\rm S^{}_3$ flavor symmetry to the canonical seesaw
framework in order to constrain the flavor textures of $M^{}_{\rm D}$ and
$M^{}_{\rm R}$ (see, e.g., Ref.~\cite{Xing:2019edp} for a systematic analysis).
In this case it is found that the most favored $\rm S^{}_3$ transformation
for left- and right-handed neutrino fields, which can naturally lead to a
phenomenologically favored pattern of the PMNS matrix $U$, is just the
${\cal S}^{(132)}$ transformation --- or equivalently the $(\mu, \tau)$-associated
reflection transformation. As shown in section~\ref{section 2.3}, such
an observation is also true even if the seesaw relation between light and heavy
neutrino masses is exact and the unitarity of $U$ is slightly violated. That is
why we are going to focus on the $\mu$-$\tau$ reflection symmetry and its possible
breaking effects for both active and sterile neutrinos in the subsequent sections.

\setcounter{equation}{0}
\section{Translational $\mu$-$\tau$ reflection symmetry}
\label{section 3}

\subsection{The translation of a massless neutrino field}
\label{section 3.1}

It is well known that current experimental data, either from the neutrino oscillation
experiments or from those non-oscillation experiments, {\it do} allow the existence of a
massless neutrino species --- either $\nu^{}_1$ with $m^{}_1 =0$ or $\nu^{}_3$ with
$m^{}_3 =0$. That is why the minimal seesaw model and some other neutrino mass models,
which can naturally predict $m^{}_1 = 0$ or $m^{}_3 = 0$ at least at the tree level,
have aroused quite a lot of interest in the past two decades (see, e.g.,
Ref.~\cite{Xing:2020ald} for a recent review). One may try to see the issue from a
different perspective: is there any simple flavor symmetry that can dictate one of
the three active neutrinos to be massless?

The answer is actually affirmative. It was found that the Dirac equation of a free
neutrino would be invariant under a translation of the neutrino field in the spinor
space if the neutrino were a massless Goldstone-like
fermion~\cite{Volkov:1973ix,deWit:1975xci}.  Taking account of the very fact that
there are three neutrino species and their mass eigenstates (i.e., $\nu^{}_i$ for
$i = 1, 2, 3$) do not match with their flavor eigenstates (i.e., $\nu^{}_\alpha$ for
$\alpha = e, \mu, \tau$), Friedberg and Lee put forward a novel idea that a Dirac
neutrino mass term of the form $\overline{\nu^{}_{\rm L}} \hspace{0.05cm} M^{}_{\rm D}
N^{}_{\rm R} + {\rm h.c.}$ should keep unchanged under the transformations
$\nu^{}_{\alpha \rm L} \to \nu^{}_{\alpha \rm L} + z^{}_{\rm L}$ and
$N^{}_{\alpha \rm R} \to N^{}_{\alpha \rm R} + z^{}_{\rm R}$
in the flavor space, where $z^{}_{\rm L}$ and $z^{}_{\rm R}$ are the
constant spinor fields which anticommute with the corresponding neutrino fields
$\nu^{}_{\alpha \rm L}$ and $N^{}_{\alpha \rm R}$~\cite{Friedberg:2006it,Lee:2008zzh,
Friedberg:2007ba,Friedberg:2007uk,Friedberg:2009fb,Friedberg:2010zt}.
Such a {\it translational} flavor symmetry, which is also referred to as the
Friedberg-Lee symmetry, can not only assure one of the neutrino masses $m^{}_i$ to
be vanishing as a consequence of $\det\big(M^{}_{\rm D}\big) = 0$ but also predict an
instructive neutrino mixing matrix which is equivalent to a modified form of the
tribimaximal flavor mixing pattern $U^{}_{\rm TBM}$ in Eq.~(\ref{eq:TBM}). In
particular, the original Friedberg-Lee symmetry can be regarded as a simple extension
of the $\mu$-$\tau$ reflection symmetry~\cite{Xing:2006xa}, and some generalized
scenarios of this kind have been proposed and discussed (see, e.g.,
Refs.~\cite{Luo:2008yc,Huang:2008ri,Jarlskog:2007qy,He:2009pt,Araki:2009grl,Araki:2009kp,
Razzaghi:2012rr,Sinha:2018uop,Xing:2021zxf,Bao:2022kon}).

It is worth pointing out that 't Hooft once considered an analogous translational
displacement $\phi(x) \to \phi(x) + \Phi$ for the Lagrangian ${\cal L}^{}_\phi$
in a renormalizable scalar field theory, where $\Phi$ is spacetime-independent
and commutes with $\phi$~\cite{tHooft:1979rat}. In this case both the mass of
$\phi$ and its self-coupling parameter have to vanish so as to assure the
invariance of ${\cal L}^{}_\phi$, which can be referred to as a Goldstone-type
symmetry~\cite{Goldstone:1961eq,Goldstone:1962es}, under the above translational
transformation. So it is generally expected that the translational symmetry
of an effective Lagrangian may provide an easy way to understand why the mass
of a fermion or boson in this physical system is vanishing or vanishingly small
\footnote{If there were no flavor mixing in the lepton or quark sector, the
flavor and mass eigenstates of a fermion would be identical with each other.
In this case the kinetic energy and mass terms of a fermion field, which
involves no self-interaction, would be rather analogous to those of a scalar
field~\cite{Xing:2021zxf}.},
although its deep meaning remains a big puzzle.

We find that there is also a close correlation between the zero mass of a particle
and the translational symmetry of its equation of motion~\cite{Xing:2021zxf}.
For example, the massless photon travels at the speed of light in free
space and its electromagnetic field obeys the equation of motion
$\square {\bf A} = 0$. This equation is invariant under the translational
transformation ${\bf A} \to {\bf A} + {\bf A}^{}_0$, where ${\bf A}^{}_0$ stands
for a constant vector field commuting with ${\bf A}$. Analogously, a massless
neutrino field $\nu^{}_i$ (for $i = 1, 2$ or $3$) also travels at the speed of
light in free space and it satisfies the Dirac equation
${\rm i}\gamma^\mu \partial^{}_\mu \nu^{}_i = 0$. The latter is invariant
under the translational transformation
\begin{eqnarray}
\nu^{}_i \to \nu^{}_i + z^{}_\nu \; ,
\label{eq:mass-translation}
\end{eqnarray}
where $z^{}_\nu$ is a constant spinor field anticommuting with $\nu^{}_i$.
One may argue that such a translation of the massless neutrino field is
fully compatible with Einstein's principle of constancy of light velocity
for all the inertial reference systems in vacuum, just like the massless
photon itself. But different from the photon, the massless neutrino field
$\nu^{}_{1 \rm L}$ (or $\nu^{}_{3 \rm L}$) and its two massive counterparts
$\nu^{}_{2 \rm L}$ and $\nu^{}_{3 \rm L}$ (or $\nu^{}_{1 \rm L}$) form three
quantum superposition states --- the neutrino flavor eigenstates
$\nu^{}_{e \rm L}$, $\nu^{}_{\mu \rm L}$ and $\nu^{}_{\tau \rm L}$.
The mismatch between flavor and mass eigenstates of the
three active neutrinos is described by the $3\times 3$ PMNS matrix $U$,
as can be seen from Eq.~(\ref{eq:UPMNS}). In this case the translational
transformation made for the massless neutrino field $\nu^{}_i$ (i.e.,
$m^{}_i =0$) in Eq.~(\ref{eq:mass-translation}) demands that the neutrino
flavor eigenstates transform in the following way~\cite{Xing:2021zxf}:
\begin{eqnarray}
\nu^{}_{\alpha \rm L} \to \nu^{}_{\alpha \rm L} + U^{}_{\alpha i} z^{}_\nu \; ,
\label{eq:flavor-translation}
\end{eqnarray}
where $U^{}_{\alpha i}$ denote the elements in the $i$th column of $U$.
We must emphasize that the coefficient $U^{}_{\alpha i}$ of $z^{}_\nu$ in
Eq.~(\ref{eq:flavor-translation}) is a natural consequence of
Eq.~(\ref{eq:mass-translation}) and lepton flavor mixing instead of a
phenomenological assumption.
So the aforementioned Friedberg-Lee transformation is equivalent to
assuming $U^{}_{e 2} = U^{}_{\mu 2} = U^{}_{\tau 2} = 1/\sqrt{3}$,
corresponding to $m^{}_2 =0$. That is why the Friedberg-Lee symmetry can give
rise to a special neutrino mixing scenario which is very close to the
tribimaximal flavor mixing pattern $U^{}_{\rm TBM}$, but a diagonal and
flavor-blind symmetry breaking term has to be introduced to acquire
$m^{}_2 \neq 0$ and keep the flavor mixing matrix
unchanged~\cite{Friedberg:2006it}.

Now let us start from Eq.~(\ref{eq:flavor-translation}) to prove that the
invariance of the effective Majorana neutrino mass term ${\cal L}^{}_\nu$
in Eq.~(\ref{eq:Majorana-mass}) under this translational transformation must
give rise to $m^{}_i = 0$. To do so, we rewrite ${\cal L}^{}_\nu$ in the
following form:
\begin{eqnarray}
-{\cal L}^{}_\nu = \frac{1}{2} \sum_\alpha\sum_\beta \overline{\nu^{}_{\alpha \rm L}}
\hspace{0.1cm} \langle m\rangle^{}_{\alpha\beta} \left(\nu^{}_{\beta \rm L}\right)^c
+ {\rm h.c.} \; ,
\label{eq:Majorana-mass-elements}
\end{eqnarray}
where $\alpha$ and $\beta$ run over the flavor indices $e$, $\mu$ and $\tau$,
and $\langle m\rangle^{}_{\alpha\beta}$ has been given in
Eq.~(\ref{eq:M-reconstruction}). The unitarity of $U$ leads us to
\begin{eqnarray}
\sum_\alpha U^{*}_{\alpha i} \langle m\rangle^{}_{\alpha\beta}
= m^{}_i U^{}_{\beta i} \; ,
\nonumber \\
\sum_\beta \langle m\rangle^{}_{\alpha\beta} U^{*}_{\beta i}
= m^{}_i U^{}_{\alpha i} \; ,
\nonumber \\
\sum_\alpha \sum_\beta U^{*}_{\alpha i} \langle m\rangle^{}_{\alpha\beta}
U^{*}_{\beta i} = m^{}_i \; ,
\label{eq:unitarity-relations}
\end{eqnarray}
which are basically equivalent to one another and all proportional to the
neutrino mass $m^{}_i$ (for $i=1$, $2$ or $3$). Under the translational
transformation made in Eq.~(\ref{eq:flavor-translation}), the effective
neutrino mass term ${\cal L}^{}_\nu$ turns out to be
\begin{eqnarray}
-\widehat{\cal L}^{}_\nu = -{\cal L}^{}_\nu + \frac{1}{2} m^{}_i \left[
\overline{z^{}_\nu} \hspace{0.05cm} z^c_\nu +
\sum_\alpha U^{}_{\alpha i} \hspace{0.05cm}\overline{\nu^{}_{\alpha \rm L}}
z^c_\nu + \overline{z^{}_\nu} \sum_\beta U^{}_{\beta i}
\left(\nu^{}_{\beta \rm L}\right)^c\right] \; .
\label{eq:Majorana-mass-elements2}
\end{eqnarray}
It is therefore obvious that $\widehat{\cal L}^{}_\nu = {\cal L}^{}_\nu$ will
hold if and only if $m^{}_i = 0$ holds. Namely, one of the three active
neutrinos must be massless if the effective Majorana neutrino mass term
${\cal L}^{}_\nu$ keeps invariant under the discrete shifts of
$\nu^{}_{\alpha \rm L}$ made above. Such a translational transformation
helps provide a novel link between the two sides of one coin --- the mass
and flavor mixing issues of the Majorana neutrinos. The point is that
the three flavor-dependent coefficients $U^{}_{\alpha i}$ of $z^{}_\nu$
constitute the $i$th column of the PMNS matrix $U$ which corresponds to
$m^{}_i =0$~\cite{Xing:2021zxf}. In comparison, most of the popular global
discrete flavor symmetries can predict very specific neutrino mixing
patterns but leave the neutrino mass spectrum
unconstrained \cite{Altarelli:2010gt,King:2013eh,Ishimori:2010au}.
Given the fact of $m^{}_2 > m^{}_1$, we are mainly interested in either
$m^{}_1 =0$ (normal mass ordering) or $m^{}_3 =0$ (inverted mass ordering)
for the neutrino mass spectrum.

If the translational symmetry of a Majorana neutrino mass term is realized
at a superhigh energy scale $\Lambda$, it is in general expected to be broken
at the electroweak scale $\Lambda^{}_{\rm EW}$ due to the RGE-induced quantum
corrections RGEs~\cite{Ohlsson:2013xva}. If the evolution of
neutrino flavor parameters starts from the heaviest Majorana neutrino mass scale
$M^{}_{\rm max}$ to the lightest Majorana neutrino mass scale $M^{}_{\rm min}$
and then to $\Lambda^{}_{\rm EW}$, it will in general be possible to generate a
finite value of $m^{}_1$ (or $m^{}_3$) at $\Lambda^{}_{\rm EW}$ from $m^{}_1 = 0$
(or $m^{}_3 = 0$) at $M^{}_{\rm max}$ via the one-loop
RGEs~\cite{Antusch:2002rr,Antusch:2003kp,Antusch:2005gp,Mei:2005qp}.
In this case the seesaw threshold effects play a critical role, but the numerical
output of $m^{}_1$ (or $m^{}_3$) strongly depends on the numerical inputs of
those unknown Yukawa coupling matrix elements. If the evolution of neutrino flavor
parameters simply starts from $M^{}_{\rm min}$, the so-called seesaw scale where
the heavy degrees of freedom have been integrated out and the unique dimension-five
Weinberg operator is responsible for the origin of three active neutrino
masses~\cite{Weinberg:1979sa}, then a nonzero value of $m^{}_1$ (or $m^{}_3$) at
$\Lambda^{}_{\rm EW}$ cannot be generated from $m^{}_1 =0$ (or $m^{}_3 =0$) at
$M^{}_{\rm min}$ unless the two-loop RGE running behaviors are taken into account~\cite{Davidson:2006tg,Xing:2020ezi,Zhou:2021bqs,Benoit:2022ohv}. In
the latter case the resulting mass of the lightest Majorana neutrino $\nu^{}_1$
or $\nu^{}_3$ is typically of ${\cal O}(10^{-13}) ~{\rm eV}$ or smaller --- such
a vanishingly small neutrino mass and the associated Majorana CP-violating phase
do not cause any seeable effects in neutrino physics.

\subsection{A $\rm TM^{}_1$ flavor mixing scenario with $m^{}_1 = 0$}
\label{section 3.2}

Now that the normal neutrino mass ordering is somewhat favored over the
inverted one by current neutrino oscillation data~\cite{Workman:2022ynf,
Esteban:2020cvm,Gonzalez-Garcia:2021dve}, let us focus on the translational
flavor symmetry of ${\cal L}^{}_\nu$ in Eq.~(\ref{eq:Majorana-mass-elements})
by assuming~\cite{Xing:2014zka,Zhao:2015bza}
\begin{eqnarray}
U^{}_{e 1} = -2 U^{}_{\mu 1} = -2 U^{}_{\tau 1} = \frac{2}{\sqrt 6} \; ,
\label{eq:Zhao}
\end{eqnarray}
which corresponds to $m^{}_1 = 0$.
We find that the standard parametrization of $U$ given in
Eq.~(\ref{eq:U}) is not very suitable for this case, because its first column
involves all the three flavor mixing angles. Instead, it is more suitable to
adopt the Kobayashi-Maskawa parametrization of $U$~\cite{Kobayashi:1973fv}
to accommodate the condition in Eq.~(\ref{eq:Zhao}), simply because its first
column is much simpler:
\begin{eqnarray}
U = P^{}_l \pmatrix{ c^{}_1 & s^{}_1 c^{}_3 & s^{}_1 \hat{s}^{*}_3 \cr
-s^{}_1 c^{}_2 & c^{}_1 c^{}_2 c^{}_3 + s^{}_2 \hat{s}^{}_3 &
c^{}_1 c^{}_2 \hat{s}^{*}_3 - s^{}_2 c^{}_3 \cr
-s^{}_1 s^{}_2 & c^{}_1 s^{}_2 c^{}_3 - c^{}_2 \hat{s}^{}_3 &
c^{}_1 s^{}_2 \hat{s}^{*}_3 + c^{}_2 c^{}_3 \cr} P^{}_\nu \; ,
\label{eq:U-KM}
\end{eqnarray}
where $c^{}_i \equiv \cos\theta^{}_i$, $s^{}_i \equiv \sin\theta^{}_i$,
$\hat{s}^{}_3 \equiv s^{}_3 e^{{\rm i}\phi}$ and
$P^{}_\nu \equiv {\rm diag}\{1 , e^{{\rm i}\sigma} , 1\}$ with $\sigma$
being the single Majorana CP-violating phase. A combination of
Eqs.~(\ref{eq:Zhao}) and (\ref{eq:U-KM}) leads us to
$\theta^{}_1 = \arcsin\left(1/\sqrt{3}\right) \simeq 35.3^\circ$ and
$\theta^{}_2 = 45^\circ$. Then the PMNS matrix $U$ is simplified to
\begin{eqnarray}
U & = & P^{}_l \hspace{0.05cm} U^{}_{\rm TBM} \pmatrix{ 1 & 0 & 0 \cr
0 & c^{}_3 & \hat{s}^{*}_3 \cr
0 & - \hat{s}^{}_3 & c^{}_3 \cr} P^{}_\nu
\nonumber \\
& = & P^{}_l \pmatrix{ \frac{2}{\sqrt 6} & \frac{1}{\sqrt 3} c^{}_3
& \frac{1}{\sqrt 3} \hat{s}^{*}_3 \cr
\vspace{-0.45cm} \cr
\frac{-1}{\sqrt 6} & \frac{1}{\sqrt 3} c^{}_3 + \frac{1}{\sqrt 2}
\hat{s}^{}_3 & \frac{-1}{\sqrt 2} c^{}_3 + \frac{1}{\sqrt 3} \hat{s}^{*}_3 \cr
\vspace{-0.45cm} \cr
\frac{-1}{\sqrt 6} & \frac{1}{\sqrt 3} c^{}_3 - \frac{1}{\sqrt 2} \hat{s}^{}_3
& \frac{1}{\sqrt 2} c^{}_3 + \frac{1}{\sqrt 3} \hat{s}^{*}_3} P^{}_\nu \; ,
\label{eq:TM1}
\end{eqnarray}
the so-called $\rm TM^{}_1$ flavor mixing pattern as its ``first" column is
the same as that of the tribimaximal flavor mixing pattern
$U^{}_{\rm TBM}$~\cite{Xing:2006ms,Lam:2006wm,Albright:2010ap}.
The $3\times 3$ Majorana neutrino mass matrix $M^{}_\nu$ can therefore be
reconstructed from $M^{}_\nu = U D^{}_\nu U^T$ with the help of the $\rm TM^{}_1$
form of $U$ in Eq.~(\ref{eq:TM1}). To be explicit,
\begin{eqnarray}
M^{}_\nu = P^{}_l \left[ a {\cal D} +
\pmatrix{ 0 & c & -c \cr
c & b + 2c & -b \cr
-c & -b & b - 2c \cr} \right] P^T_l \; ,
\label{eq:TM1-mass-matrix}
\end{eqnarray}
where the flavor democracy texture $\cal D$ has been defined in Eq.~(\ref{eq:Democracy}),
and
\begin{eqnarray}
a & = & \frac{1}{3} \left( \overline{m}^{}_2 c^2_3 + m^{}_3 \hat{s}^{*2}_3
\right) \; ,
\nonumber \\
b & = & \frac{1}{2} \left( \overline{m}^{}_2 \hat{s}^{2}_3 + m^{}_3 c^2_3
\right) \; ,
\nonumber \\
c & = & \frac{1}{\sqrt 6} \left( \overline{m}^{}_2 \hat{s}^{}_3 - m^{}_3
\hat{s}^{*}_3 \right) c^{}_3 \;
\label{eq:TM1-mass-matrix2}
\end{eqnarray}
with $\overline{m}^{}_2 \equiv m^{}_2 e^{2{\rm i}\sigma}$. As a consequence, the two
nonzero neutrino masses and the Majorana CP-violating phase are given by
\begin{eqnarray}
m^{}_2 = \frac{1}{2} \left| 3 a + 2 b \hspace{0.03cm} e^{-2{\rm i}\phi} + \frac{\sqrt{6}
\hspace{0.05cm} c}{c^{}_3 s^{}_3} e^{-{\rm i}\phi}\right| \; ,
\nonumber \\
m^{}_3 = \frac{1}{2} \left| 3 a \hspace{0.03cm} e^{+2{\rm i}\phi} + 2 b - \frac{\sqrt{6}
\hspace{0.05cm} c}{c^{}_3 s^{}_3} e^{+{\rm i}\phi}\right| \; ,
\nonumber \\
\tan 2\sigma = \frac{ c^{}_3 s^{}_3 {\rm Im} \left(3 a + 2 b \hspace{0.03cm} e^{-2{\rm i}
\phi}\right) + \sqrt{6} \hspace{0.05cm} {\rm Im}\left(c \hspace{0.03cm}
e^{-{\rm i}\phi}\right)}{ c^{}_3 s^{}_3 {\rm Re} \left(3 a + 2 b \hspace{0.05cm}
e^{-2{\rm i} \phi}\right) + \sqrt{6} \hspace{0.05cm} {\rm Re}\left(c \hspace{0.05cm}
e^{-{\rm i}\phi}\right)} \; .
\label{eq:TM1-mass-matrix3}
\end{eqnarray}
One can see that the simple texture of $M^{}_\nu$ depends on a very simple choice of
$U^{}_{\alpha 1}$ and is strongly suggestive of certain simple discrete flavor
symmetries which can be used for a specific model building exercise~\cite{Lam:2006wm}.

Note that the $\rm TM^{}_1$ flavor mixing pattern obtained in Eq.~(\ref{eq:TM1})
only possesses a partial $\mu$-$\tau$ reflection symmetry characterized by
$\big|U^{}_{\mu 1}\big| = \big|U^{}_{\tau 1}\big|$~\cite{Xing:2014zka}, if
$\phi \neq \pm \pi/2$ holds. But when $\phi = \pm \pi/2$ is taken, the
$\rm TM^{}_1$ scenario will assure $\big|U^{}_{\mu i}\big| = \big|U^{}_{\tau i}\big|$
(for $i = 1, 2, 3$) to hold and thus fully respect the $\mu$-$\tau$ reflection
symmetry~\cite{Xing:2006ms}. Comparing the standard parametrization
of $U$ in Eq.~(\ref{eq:U}) with the $\rm TM^{}_1$ pattern in Eq.~(\ref{eq:TM1}),
we arrive at
\begin{eqnarray}
s^{2}_{13} = \frac{1}{3} s^{2}_3 \; ,
\quad
s^2_{12} = \frac{1}{3} \left(1 - t^2_{13}\right) \; ,
\quad
s^2_{23} = \frac{1}{2} - \sqrt{6} \hspace{0.05cm} s^{}_{12} t^{}_{13}
\cos\phi \;
\label{eq:TM1-correlation1}
\end{eqnarray}
for the three flavor mixing angles, and
\begin{eqnarray}
\sin\delta^{}_\nu = \frac{\sin\phi}{\sqrt{\left(1 - 4 t^2_{13}\right)^2 +
8 t^2_{13} \left(1 - 2 t^2_{13}\right) \sin^2\phi}} \;
\label{eq:TM1-correlation2}
\end{eqnarray}
for the Dirac CP-violating phase, where $t^{}_{13} \equiv \tan\theta^{}_{13}$ has
been defined~\cite{Zhao:2015bza}. It is obvious that
$\phi = \pm\pi/2$ leads us to $\theta^{}_{23} = \pi/4$ and $\delta^{}_\nu =
\pm \pi/2$ as in the $\mu$-$\tau$ reflection symmetry limit. Furthermore, a combination
of Eqs.~(\ref{eq:TM1-correlation1}) and (\ref{eq:TM1-correlation2}) predicts an
instructive correlation between $\delta^{}_\nu$ and $\theta^{}_{23}$:
\begin{eqnarray}
\sin^2\delta^{}_\nu = \frac{8 s^2_{13} \cos 2\theta^{}_{13} - c^4_{13}
\cos^2 2\theta^{}_{23}}{8 s^2_{13} \left( \cos 2\theta^{}_{13}
\sin^2 2\theta^{}_{23} + s^2_{13} \cos^2 2\theta^{}_{23}\right)} \; .
\label{eq:TM1-correlation3}
\end{eqnarray}
This interesting relation, together with $s^2_{12} = \left( 1 - t^2_{13}\right)$
in Eq.~(\ref{eq:TM1-correlation1}), can be used to test the $\rm TM^{}_1$
flavor mixing scenario in the future precision neutrino experiments. Inputting
the best-fit and $3\sigma$ interval of $\theta^{}_{13}$~\cite{Esteban:2020cvm,
Gonzalez-Garcia:2021dve} for the normal neutrino mass ordering
which is consistent with $m^{}_1 = 0$, we
illustrate the numerical correlation between $\theta^{}_{23}$ and $\delta^{}_\nu$
in Figure~\ref{Fig:TM1}, where the grey bands of $\theta^{}_{23}$
and $\delta^{}_\nu$ cover their respective $3\sigma$ intervals. One can see that
the best-fit point of $\theta^{}_{23}$ and $\delta^{}_\nu$ obtained today
(the red star) is not far away from the $\mu$-$\tau$ reflection symmetry point
(the blue star), and it is not far away from the correlation curve and band (the
red curve and orange band) either.
\begin{figure}[t]
\begin{center}
\includegraphics[width=3.3in]{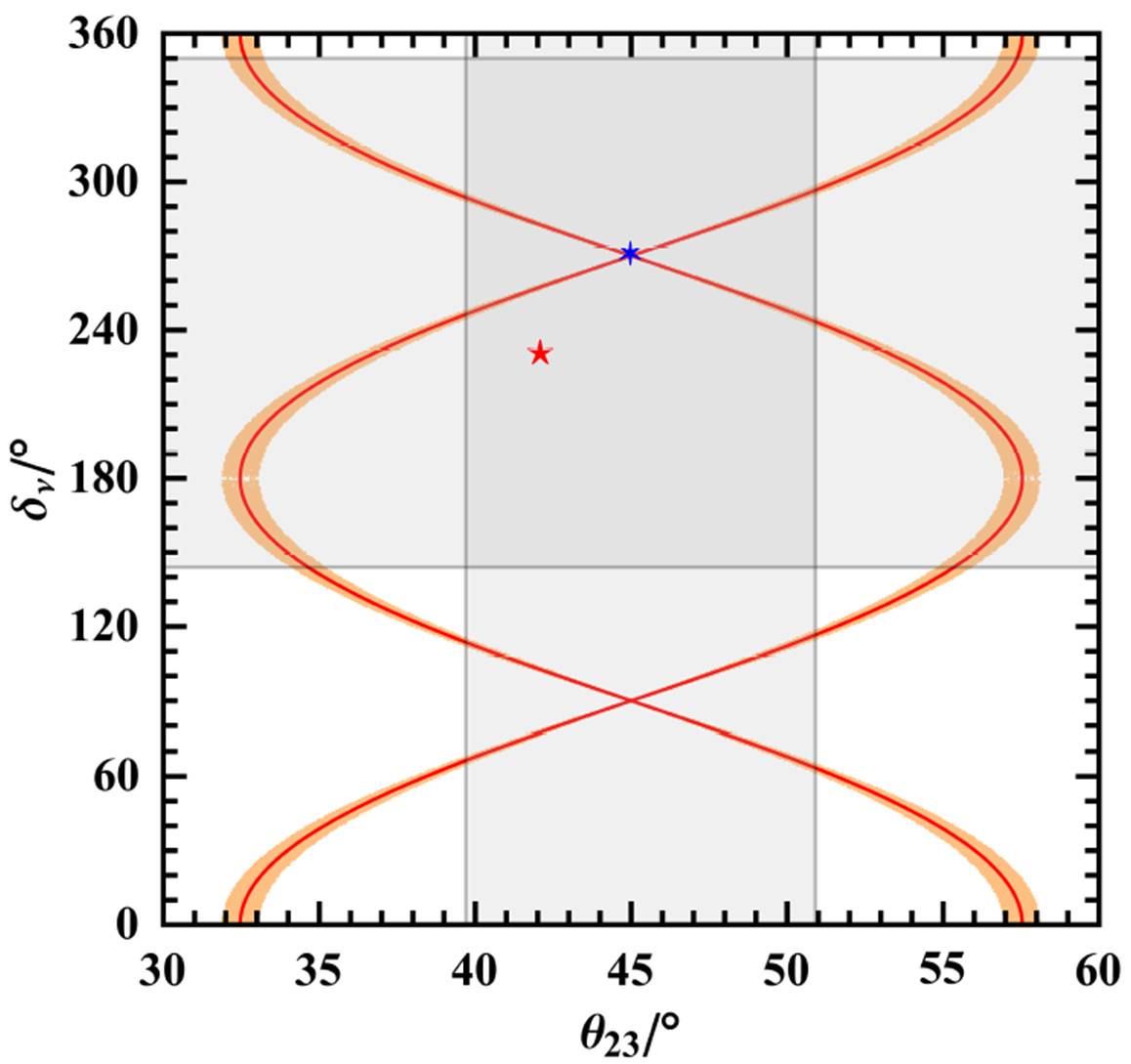}
\vspace{-0.2cm}
\caption{A numerical illustration of the correlation between $\theta^{}_{23}$
and $\delta^{}_\nu$ as predicted by the $\rm TM^{}_1$ flavor mixing pattern
with $m^{}_1 = 0$, where the best-fit (red curve) and $3\sigma$ interval (orange
band) of $\theta^{}_{13}$~\cite{Esteban:2020cvm,Gonzalez-Garcia:2021dve} have been
input. The grey bands of $\theta^{}_{23}$ and $\delta^{}_\nu$ cover their respective
$3\sigma$ intervals, and the red star ``$\star$" denotes their best-fit values. The
blue star ``$\ast$" stands for the $\mu$-$\tau$ reflection symmetry point.}
\label{Fig:TM1}
\end{center}
\end{figure}

It is certainly the assumption made in Eq.~(\ref{eq:Zhao}) for the invariance of
${\cal L}^{}_\nu$ under the translational transformation in
Eq.~(\ref{eq:flavor-translation}) that allows us to simultaneously
obtain $m^{}_1 = 0$ and the $\rm TM^{}_1$ flavor mixing pattern. To avoid such
an empirical assumption, a possible way out is simply to combine the translational
flavor symmetry under discussion and the $\mu$-$\tau$ reflection symmetry for the
effective Majorana neutrino mass term. Some interesting attempts have been made
along this line of thought~\cite{Xing:2022hst,Araki:2009grl,Araki:2009kp,
Sinha:2018uop,Bao:2022kon,Yang:2022yqw}. Let us proceed to go into details about
this idea.

\subsection{The translational $\mu$-$\tau$ reflection symmetry}
\label{section 3.3}

Given the effective Majorana neutrino mass term ${\cal L}^{}_\nu$ in
Eq.~(\ref{eq:Majorana-mass}) or Eq.~(\ref{eq:M-mass}), one may consider
the translational $\mu$-$\tau$ reflection transformation
\begin{eqnarray}
\nu^{}_{\rm L} \to {\cal P} \left[(\nu^{}_{\rm L})^c + n^*_i z^c_\nu
\right] \; ,
\label{eq:T-mu-tau}
\end{eqnarray}
and its charge-conjugated counterpart
$(\nu^{}_{\rm L})^c \to {\cal P} \left(\nu^{}_{\rm L} + n^{}_i z^{}_\nu\right)$,
where $\cal P$ has been given in Eq.~(\ref{eq:S-tranformation}),
and $n^{}_i \equiv (U^{}_{e i} , U^{}_{\mu i} , U^{}_{\tau i})^T$ is a
column vector of the $3\times 3$ PMNS matrix $U$ corresponding to $\nu^{}_i$
(for $i = 1, 2, 3$). Under these transformations ${\cal L}^{}_\nu$ becomes
\begin{eqnarray}
-\widehat{\cal L}^{}_\nu = \frac{1}{2} \Big[\overline{\nu^{}_{\rm L}}
\left({\cal P} M^{*}_\nu {\cal P}\right) (\nu^{}_{\rm L})^c +
\overline{\nu^{}_{\rm L}} \left({\cal P} M^{*}_\nu {\cal P}\right) n^*_i z^c_\nu
+ \overline{z^{}_\nu} \hspace{0.05cm} n^\dagger_i
\left({\cal P} M^{*}_\nu {\cal P}\right) (\nu^{}_{\rm L})^c
\nonumber \\
\hspace{1.42cm} + \hspace{0.05cm}
\overline{z^{}_\nu} \hspace{0.05cm} n^\dagger_i
\left({\cal P} M^{*}_\nu {\cal P}\right) n^*_i z^c_\nu\Big] + {\rm h.c.} \; .
\label{eq:T-mu-tau2}
\end{eqnarray}
The conditions for
$\widehat{\cal L}^{}_\nu = {\cal L}^{}_\nu$ turn out to be
$M^{}_\nu = {\cal P} M^{*}_\nu {\cal P}$ obtained
in Eq.~(\ref{eq:M-reflection2}) and
\begin{eqnarray}
M^{}_\nu n^*_i = {\bf 0} \; , \quad
n^\dagger_i M^{}_\nu = {\bf 0}^T \; , \quad
n^\dagger_i M^{}_\nu n^*_i = 0 \; ,
\label{eq:T-mu-tau3}
\end{eqnarray}
where $\bf 0$ denotes the zero column vector. Note that the three dependent
conditions in Eq.~(\ref{eq:T-mu-tau3}) are exactly equivalent to setting those
three relations in Eq.~(\ref{eq:unitarity-relations}) to be zero, leading to
$m^{}_i = 0$. Moreover, the texture of $M^{}_\nu$ takes the $\mu$-$\tau$
reflection symmetry form shown in Eq.~(\ref{eq:M-reflection}). So the
translational $\mu$-$\tau$ reflection symmetry of ${\cal L}^{}_\nu$ can
predict not only $m^{}_1 = 0$ or $m^{}_3 = 0$ for the neutrino mass spectrum
but also $\theta^{}_{23} = \pi/4$, $\delta^{}_\nu = \pm \pi/2$ and
$\sigma = 0$ or $\pi/2$ for the PMNS flavor mixing parameters. The experimental
testability of such a simple flavor symmetry is therefore impressive.

The above observation reminds us to consider a  combination of the
translational $\mu$-$\tau$ reflection symmetry with the canonical seesaw
mechanism at a superhigh energy scale, such that we can naturally arrive
at a minimal seesaw scenario constrained by the $\mu$-$\tau$ reflection
symmetry. To do so, let us rewrite ${\cal L}^\prime_{\rm SS}$ in
Eq.~(\ref{eq:seesaw-mass-matrix}) as
\begin{eqnarray}
-{\cal L}^{\prime}_{\rm SS} = \overline{\nu^{}_{\rm L}} \hspace{0.05cm}
M^{}_{\rm D} N^{}_{\rm R} + \frac{1}{2} \hspace{0.05cm}
\overline{(N^{}_{\rm R})^c} \hspace{0.05cm} M^{}_{\rm R} N^{}_{\rm R}
+ {\rm h.c.} \; ,
\label{eq:T-mu-tau-SS}
\end{eqnarray}
and require it to be invariant under the transformations
\footnote{One may also consider the possibility of
allowing $\nu^{}_{\rm L}$ to have an independent translational $\mu$-$\tau$
reflection symmetry instead of just the $\mu$-$\tau$ reflection symmetry.
In this case we find that the resulting constraint on the texture of
$M^{}_{\rm D}$ are so stringent that all the elements of $M^{}_{\rm D}$ have
to vanish, making the seesaw mechanism completely collapse.}
\begin{eqnarray}
\nu^{}_{\rm L} \to {\cal P} (\nu^{}_{\rm L})^c \; , \quad
N^{}_{\rm R} \to {\cal P} \big[(N^{}_{\rm R})^c + n^{\prime *}_i z^c_N
\big] \; ,
\label{eq:T-mu-tau-SS2}
\end{eqnarray}
together with $(\nu^{}_{\rm L})^c \to {\cal P} \nu^{}_{\rm L}$ and
$(N^{}_{\rm R})^c \to {\cal P} \left(N^{}_{\rm R} + n^{\prime}_i z^{}_N\right)$,
where $n^{\prime}_i \equiv
(U^{\prime}_{e i} , U^{\prime}_{\mu i} , U^{\prime}_{\tau i})^T$ is defined to
be a column vector of the $3\times 3$ unitary flavor mixing matrix $U^\prime$
used to diagonalize $M^{}_{\rm R}$ in correspondence to the heavy neutrino
mass eigenstate $N^{}_i$ with the mass $M^{}_i$ (for $i = 1, 2, 3$) in the
canonical seesaw mechanism, and $z^{}_N$ is another constant spinor field similar
to $z^{}_\nu$ in Eq.~(\ref{eq:flavor-translation}). Then the invariance of
${\cal L}^{\prime}_{\rm SS}$ with respect to the above transformations
leads us to~\cite{Xing:2022hst}
\begin{eqnarray}
M^{}_{\rm D} = {\cal P} M^{*}_{\rm D} {\cal P} \; , \quad
M^{}_{\rm R} = {\cal P} M^{*}_{\rm R} {\cal P} \; ;
\nonumber \\
M^{}_{\rm D} \hspace{0.05cm} n^{\prime}_i = {\bf 0} \; , \quad
M^{}_{\rm R} \hspace{0.05cm} n^{\prime}_i = {\bf 0} \; , \quad
n^{\prime T}_i M^{}_{\rm R} = {\bf 0}^T \; , \quad
n^{\prime T}_i M^{}_{\rm R} \hspace{0.05cm} n^{\prime}_i = 0 \; ,
\label{eq:T-mu-tau-SS3}
\end{eqnarray}
The textures of $M^{}_{\rm D}$ and $M^{}_{\rm R}$ turn out to
have the familiar $\mu$-$\tau$ reflection symmetry form
\begin{eqnarray}
M^{}_{\rm D} = \pmatrix{
a & b & b^* \cr
e & c & d \cr
e^* & d^* & c^*} \; ,
\quad
M^{}_{\rm R} = \pmatrix{
A & B & B^* \cr
B & C & D \cr
B^* & D & C^*} \; ,
\label{eq:T-mu-tau-SS4}
\end{eqnarray}
where the matrix elements $a$, $A$ and $D$ are real, and one of the mass
eigenvalues of $M^{}_{\rm R}$ is vanishing (i.e., $M^{}_i = 0$ corresponding to
$n^{\prime}_i$). Without loss of generality, we simply take $M^{}_1 = 0$
for the sake of illustration~\cite{Jarlskog:2007qy}.
Now that there is no way to measure $U^\prime$ in any realistic
experiments, let us just choose a flavor basis defined by
$n^{\prime}_1 \equiv (1, 0, 0)^T$, which is consistent with $M^{}_1 = 0$.
In this case Eq.~(\ref{eq:T-mu-tau-SS4}) becomes
\begin{eqnarray}
M^{}_{\rm D} = \pmatrix{
0 & b & b^* \cr
0 & c & d \cr
0 & \hspace{0.06cm} d^* \hspace{0.06cm} & c^*} \; ,
\quad
M^{}_{\rm R} = \pmatrix{
0 & 0 & 0 \cr
0 & C & D \cr
\hspace{0.04cm} 0 \hspace{0.04cm} & \hspace{0.1cm} D \hspace{0.1cm} & C^*} \; .
\label{eq:T-mu-tau-SS5}
\end{eqnarray}
As argued by Jarlskog~\cite{Jarlskog:2007qy}, the flavor
eigenstate of the massless sterile neutrino $N^{}_1$ should completely
decouple from other particles (both the SM particles and those massive
sterile neutrinos) and have only gravitational interactions due to its kinetic
energy term~\cite{Jarlskog:2007qy}. In other words, such a massless sterile 
neutrino state should have no impact on the effective number of neutrino 
species, the Big Bang nucleosynthesis and some other physical processes 
in the early Universe, but it might more or less affect the formation of 
the large scale structure of the Universe. Here let us simply focus on the
point that the translational $\mu$-$\tau$ reflection symmetry of
${\cal L}^{\prime}_{\rm SS}$ can be treated as a phenomenologically viable
way to reduce the degrees of freedom associated with the canonical seesaw
mechanism and constrain its flavor textures. In this way
${\cal L}^{\prime}_{\rm SS}$ in Eq.~(\ref{eq:T-mu-tau-SS}) is just
simplified to
\begin{eqnarray}
-{\cal L}^\prime_{\rm MSS} = \overline{\nu^{}_{\rm L}} \hspace{0.05cm}
\widehat{M}^{}_{\rm D} N^{\prime}_{\rm R} + \frac{1}{2}
\overline{(N^{\prime}_{\rm R})^c} \hspace{0.05cm} \widehat{M}^{}_{\rm R}
N^{\prime}_{\rm R} + {\rm h.c.} \; ,
\label{eq:minimal-seesaw-mass-matrix}
\end{eqnarray}
where $N^{\prime}_{\rm R} = (N^{}_{\mu \rm R} , N^{}_{\tau \rm R})^T$ is
the reduced column vector,
$\widehat{M}^{}_{\rm D}$ and $\widehat{M}^{}_{\rm R}$ are the $3\times 2$
and $2\times 2$ mass matrices of the forms
\begin{eqnarray}
\widehat{M}^{}_{\rm D} = \pmatrix{
b & b^* \cr
c & d \cr
d^* & c^*} \; , \quad
\widehat{M}^{}_{\rm R} = \pmatrix{
C & D \cr
D & C^*} \; ,
\label{eq:minimal-seesaw-textures}
\end{eqnarray}
with $D$ being real. It is obvious that $\widehat{M}^{}_{\rm D}$ only contains
six real free parameters, and $\widehat{M}^{}_{\rm R}$ consists of only three
real free parameters. In this case the approximate seesaw formula for the
effective Majorana mass matrix of three active neutrinos is
\begin{eqnarray}
\widehat{M}^{}_\nu \simeq -\widehat{M}^{}_{\rm D} \widehat{M}^{-1}_{\rm R}
\widehat{M}^{T}_{\rm D} =
\pmatrix{\langle m\rangle^{}_{ee} & \langle m\rangle^{}_{e\mu} &
\langle m\rangle^{*}_{e\mu} \cr \langle m\rangle^{}_{e\mu} &
\langle m\rangle^{}_{\mu\mu} & \langle m\rangle^{}_{\mu\tau} \cr
\langle m\rangle^{*}_{e\mu} & \langle m\rangle^{}_{\mu\tau} &
\langle m\rangle^{*}_{\mu\mu} \cr} \; ,
\label{eq:minimal-seesaw-formula}
\end{eqnarray}
where
\begin{eqnarray}
\langle m\rangle^{}_{ee} & = & \frac{1}{\big|C\big|^2 - D^2}
\Big[2 \big| b\big|^2 D - 2 {\rm Re}\left( b^2 C^*\right) \Big] \; ,
\nonumber \\
\langle m\rangle^{}_{e\mu} & = & \frac{1}{\big|C\big|^2 - D^2}
\Big[b \left(d D - c C^*\right) + b^* \left(c D - d C\right)\Big] \; ,
\nonumber \\
\langle m\rangle^{}_{\mu\mu} & = & \frac{1}{\big|C\big|^2 - D^2}
\Big[2 cd D -c^2 C^* - d^2 C\Big] \; ,
\nonumber \\
\langle m\rangle^{}_{\mu\tau} & = & \frac{1}{\big|C\big|^2 - D^2}
\Big[\left(\big| c\big|^2 + \big| d\big|^2 \right) D -2 {\rm Re}
\left( c d^* C^*\right)\Big] \; .
\label{eq:minimal-seesaw-elements}
\end{eqnarray}
So the texture of $\widehat{M}^{}_\nu$ also exhibits the $\mu$-$\tau$ reflection
symmetry and contains only five real free parameters. In particular, the rank of
$\widehat{M}^{}_\nu$ is two instead of three, implying that one of its three
mass eigenvalues must be vanishing (i.e., $m^{}_1 = 0$ or $m^{}_3 = 0$).
Note that the most salient feature of the minimal seesaw mechanism is just the
existence of a massless neutrino species, and this observation is actually
independent of the approximate seesaw formula~\cite{Xing:2007uq}.
A more generic analysis of the minimal seesaw model constrained by the
$\mu$-$\tau$ reflection symmetry will be made in section~\ref{section 6.4}.

\setcounter{equation}{0}
\section{Rotational $\mu$-$\tau$ reflection symmetry}
\label{section 4}

\subsection{A rotational $\mu$-$\tau$ reflection transformation}
\label{section 4.1}

Given the effective Majorana neutrino mass term ${\cal L}^{}_\nu$ in
Eq.~(\ref{eq:M-mass}), let us now conjecture that it keeps invariant
under a proper real and orthogonal transformation
$\nu^{}_{\rm L} \to {\cal R} \nu^{}_{\rm L}$ with
${\cal R} = {\cal R}^T = {\cal R}^\dagger$. In this case it is
easy to verify that the condition $M^{}_\nu = {\cal R} M^{}_\nu {\cal R}$
must hold for the symmetric mass matrix $M^{}_\nu$. We proceed to assume that
${\cal L}^{}_\nu$ is also invariant under the $\mu$-$\tau$ reflection
transformation $\nu^{}_{\rm L} \to {\cal P} (\nu^{}_{\rm L})^c$ as
discussed in section~\ref{section 2.2}, which requires $M^{}_\nu$ to
satisfy $M^{}_\nu = {\cal P} M^*_\nu {\cal P}$. If ${\cal P}$ and
${\cal R}$ commute with each other such that
\begin{eqnarray}
M^{}_\nu = {\cal R} {\cal P} M^{*}_\nu {\cal P} {\cal R} =
{\cal P} {\cal R} M^*_\nu {\cal R} {\cal P}
\label{eq:R-mu-tau-matrix}
\end{eqnarray}
holds together with $M^{}_\nu = {\cal P} M^*_\nu {\cal P}$ and
$M^{}_\nu = {\cal R} M^{}_\nu {\cal R}$, then ${\cal L}^{}_\nu$ will be
invariant under the separate but mutually exchangeable transformations
\begin{eqnarray}
\nu^{}_{\rm L} \to {\cal R} \nu^{}_{\rm L} \; , \quad
\nu^{}_{\rm L} \to {\cal P} (\nu^{}_{\rm L})^c \; .
\label{eq:R-mu-tau-transformation}
\end{eqnarray}
In this case ${\cal L}^{}_\nu$ is of course invariant under the joint
transformation $\nu^{}_{\rm L} \to {\cal R} {\cal P} (\nu^{}_{\rm L})^c$
or $\nu^{}_{\rm L} \to {\cal P} {\cal R} (\nu^{}_{\rm L})^c$.
Such a simple flavor symmetry of ${\cal L}^{}_\nu$ can therefore be
referred to as the {\it rotational} $\mu$-$\tau$ reflection symmetry,
in the sense that an extra $\rm Z^{}_2$ symmetry is imposed on the
$\mu$-$\tau$ reflection symmetry.
Note that the $\mu$-$\tau$ reflection symmetry itself
can essentially be regarded as a kind of generalized CP
symmetry~\cite{Neufeld:1987wa,Grimus:1995zi}, and hence the rotational
$\mu$-$\tau$ reflection symmetry is actually a combination of the
$\rm Z^{}_2$ flavor symmetry and the generalized CP symmetry which
satisfy the consistency conditions (see, e.g., Refs.~\cite{Lam:2008sh,
Ge:2010js,Ge:2011qn,Ge:2011ih,Holthausen:2012dk}).
The resulting texture of $M^{}_\nu$
should thus be more strongly constrained, unless ${\cal R}$ has
a rather trivial form (e.g., ${\cal R} = {\cal I}$ or ${\cal P}$).

Up to an overall sign ambiguity, the real orthogonal matrix ${\cal R}$
that represents another residual $\rm Z^{}_2$ symmetry and satisfies the
commutativity with ${\cal P}$ (i.e., ${\cal R} {\cal P} = {\cal P} {\cal R}$)
can in general be expressed as
\begin{eqnarray}
{\cal R} = \pmatrix{\cos 2\theta & \frac{1}{\sqrt 2} \sin 2\theta &
\frac{1}{\sqrt 2} \sin 2\theta \cr
\frac{1}{\sqrt 2} \sin 2\theta & -\cos^2\theta & \sin^2\theta \cr
\frac{1}{\sqrt 2} \sin 2\theta & \sin^2\theta & -\cos^2\theta} \; ,
\label{eq:R-matrix}
\end{eqnarray}
where $\theta$ is real. In Refs.~\cite{Ge:2010js,Ge:2011qn,Ge:2011ih} another
parametrization of ${\cal R}$, which is equivalent to taking
$\tan\theta = k/\sqrt{2}$ with $k$ being real, has been proposed. A special
but extremely interesting case is $\tan\theta = 1/\sqrt{2}$ or $k = 1$, as
first pointed out in Ref.~\cite{Lam:2008sh}. Substituting ${\cal P}$ in
Eq.~(\ref{eq:S-tranformation}) and ${\cal R}$ in Eq.~(\ref{eq:R-matrix})
into $M^{}_\nu = {\cal P} M^*_\nu {\cal P}$ and
$M^{}_\nu = {\cal R} M^{}_\nu {\cal R}$, we obtain the constraint
conditions for the elements $\langle m\rangle^{}_{\alpha\beta}$
of $M^{}_\nu$:
\begin{eqnarray}
\langle m\rangle^{}_{ee} = \langle m\rangle^{*}_{ee} \; , \quad &&
\langle m\rangle^{}_{\mu\mu} = \langle m\rangle^{*}_{\tau\tau} \; ,
\nonumber \\
\langle m\rangle^{}_{e\mu} = \langle m\rangle^{*}_{e\tau} \; , \quad &&
\langle m\rangle^{}_{\mu\tau} = \langle m\rangle^{*}_{\mu\tau} \; ,
\label{eq:R-mu-tau-condition1}
\end{eqnarray}
which are equivalent to the usual $\mu$-$\tau$ reflection symmetry
conditions; and
\begin{eqnarray}
{\rm Im}\langle m\rangle^{}_{\mu\mu} = \sqrt{2} \hspace{0.03cm}
\cot\theta \hspace{0.03cm} {\rm Im}\langle m\rangle^{}_{e\tau} \; ,
\nonumber \\
\langle m\rangle^{}_{ee} = {\rm Re}\langle m\rangle^{}_{\mu\mu}
+ \langle m\rangle^{}_{\mu\tau} + 2\sqrt{2} \hspace{0.03cm}
\cot 2\theta \hspace{0.03cm} {\rm Re}\langle m\rangle^{}_{e\mu} \; ,
\label{eq:R-mu-tau-condition2}
\end{eqnarray}
which arise from $M^{}_\nu = {\cal R} M^{}_\nu {\cal R}$. It is
clear that the rotational $\mu$-$\tau$ reflection symmetry may
provide two additional constraints on $M^{}_\nu$ as compared with
the normal $\mu$-$\tau$ reflection symmetry, if the free angle
$\theta$ is fixed by an explicit flavor symmetry group. Taking
the notations $\langle m\rangle^{}_{e\mu} = a + {\rm i} a^\prime$,
${\rm Re}\langle m\rangle^{}_{\mu\mu} = b$ and
$\langle m\rangle^{}_{\mu\tau} = c$, let us parametrize the flavor
texture of $M^{}_\nu$ as follows:
\begin{eqnarray}
M^{}_\nu = \pmatrix{2\sqrt{2} \hspace{0.05cm} a \cot 2\theta
+ b + c & a + {\rm i} a^\prime & a - {\rm i} a^\prime \cr
a + {\rm i} a^\prime & b - {\rm i} \sqrt{2} \hspace{0.05cm}
a^\prime \cot \theta & c \cr
a - {\rm i} a^\prime & c & b + {\rm i} \sqrt{2} \hspace{0.05cm}
a^\prime \cot \theta} \;
\label{eq:Mnu-matrix}
\end{eqnarray}
with $a$, $a^\prime$, $b$ and $c$ being all real.

In the basis where the flavor eigenstates of three charged leptons
are identical with their mass eigenstates (i.e., $M^{}_l = D^{}_l$),
the effective Majorana neutrino mass matrix $M^{}_\nu$ can be
diagonalized by the PMNS matrix $U$ via the transformation
$U^\dagger M^{}_\nu U^* = D^{}_\nu$. After the rotational $\mu$-$\tau$
reflection symmetry is taken into account, we have not only
\begin{eqnarray}
U D^{}_\nu U^T = \left({\cal P} U^*\right) D^{}_\nu
\left({\cal P} U^*\right)^T \; , \quad
U D^{}_\nu U^T = \left({\cal R} U\right) D^{}_\nu
\left({\cal R} U\right)^T \; ,
\label{eq:R-mu-tau-U1}
\end{eqnarray}
but also the joint constraint relationship
\begin{eqnarray}
U D^{}_\nu U^T = \left({\cal P} {\cal R} U^*\right) D^{}_\nu
\left({\cal P} {\cal R} U^*\right)^T \; .
\label{eq:R-mu-tau-U2}
\end{eqnarray}
As a result, we obtain
\begin{eqnarray}
U = {\cal P} U^* \zeta^{}_{\cal P} \; , \quad
U = {\cal R} U \zeta^{}_{\cal R} \; , \quad
U = {\cal P} {\cal R} U^* \zeta \; ,
\label{eq:R-mu-tau-U}
\end{eqnarray}
where $\zeta^{}_{\cal P} =
{\rm Diag}\{\eta^{\prime}_1 , \eta^{\prime}_2 , \eta^{\prime}_3\}$,
$\zeta^{}_{\cal R} = {\rm Diag}\{\eta^{\prime\prime}_1 ,
\eta^{\prime\prime}_2 , \eta^{\prime\prime}_3\}$ and
$\zeta = {\rm Diag}\{\eta^{}_1 , \eta^{}_2 , \eta^{}_3\}$ with the
elements $\eta^{\prime}_i = \pm 1$, $\eta^{\prime\prime}_i = \pm 1$
and $\eta^{}_i = \pm 1$ (for $i = 1, 2, 3$). So the pattern of $U$ is
not only constrained by the ${\cal P}$ transformation but also
restricted by the ${\cal R}$ transformation. Nevertheless, the concrete
form of ${\cal R}$ should be determined by the residual symmetry of an
underlying flavor symmetry group in a given neutrino mass model.

\subsection{A constrained $\rm TM^{}_1$ flavor mixing scenario}
\label{section 4.2}

There have been some interesting model-building exercises
based either on the rotational $\mu$-$\tau$ reflection symmetry or on its
analogs~\cite{Xing:2015fdg,Altarelli:2010gt,King:2013eh,Feruglio:2019ybq,
Rodejohann:2017lre,Lam:2008sh,Ge:2010js,Ge:2011qn,Ge:2011ih,
Holthausen:2012dk}, and among them a particularly successful example is
the derivation of a constrained $\rm TM^{}_1$ lepton flavor mixing scenario
--- a phenomenologically favored pattern of $U$ which can be expressed as
a product of the tribimaximal flavor mixing matrix $U^{}_{\rm TBM}$ in
Eq.~(\ref{eq:TBM}) and a special rotation matrix in the $(2,3)$ complex
plane like that shown in Eq.~(\ref{eq:TM1})~\cite{Xing:2006ms,Lam:2006wm,
Albright:2010ap}. Instead of repeating a specific flavor symmetry model of
this kind, here let us start from the flavor texture of $M^{}_\nu$ obtained
in section~\ref{section 4.1} and explore its consequence on flavor mixing
and CP violation. For this purpose we first make a transformation of $M^{}_\nu$
in Eq.~(\ref{eq:Mnu-matrix}) as follows:
\begin{eqnarray}
\left(P^{\prime}_l U^{}_{\rm TBM}\right)^\dagger M^{}_\nu
\left(P^{\prime}_l U^{}_{\rm TBM}\right)^*
= M^\prime_\nu \; ,
\label{eq:R-mu-tau-M-transformation}
\end{eqnarray}
where $P^{\prime}_l = {\rm Diag}\{1, -1, -1\}$, and the matrix elements
$\langle m\rangle^\prime_{\alpha\beta}$ of $M^\prime_\nu$ are given by
\begin{eqnarray}
\langle m\rangle^\prime_{ee} & = & \frac{4}{3}\left(1 + \sqrt{2}
\hspace{0.03cm} \cot 2\theta\right) a + b + c \; ,
\nonumber \\
\langle m\rangle^\prime_{e\mu} & = & \frac{\sqrt 2}{3}\left(2\sqrt{2}
\hspace{0.03cm} \cot 2\theta - 1\right) a \; ,
\nonumber \\
\langle m\rangle^\prime_{e\tau} & = & {\rm i} \frac{1}{\sqrt 3}
\left(2 - \sqrt{2} \hspace{0.03cm} \cot\theta\right) a^\prime \; ,
\nonumber \\
\langle m\rangle^\prime_{\mu\mu} & = & \frac{2}{3}\left(\sqrt{2}
\hspace{0.03cm} \cot 2\theta - 2\right) a + b + c \; ,
\nonumber \\
\langle m\rangle^\prime_{\mu\tau} & = & {\rm i} \frac{2}{\sqrt 6}
\left(1 + \sqrt{2} \hspace{0.03cm} \cot\theta\right) a^\prime \; ,
\nonumber \\
\langle m\rangle^\prime_{\tau\tau} & = & b - c \; .
\label{eq:R-mu-tau-M-transformation2}
\end{eqnarray}
It is obvious that $\langle m\rangle^\prime_{e\mu} =
\langle m\rangle^\prime_{e\tau} = 0$, or equivalently
$\cot\theta = \sqrt{2}$ and $\cot 2\theta = 1/\left(2\sqrt{2}\right)$,
is the unique condition for the PMNS matrix
$U$ to have the $\rm TM^{}_1$ flavor mixing pattern. In this case
the real orthogonal matrix ${\cal R}$ in Eq.~(\ref{eq:R-matrix})
is fully determined as~\cite{Lam:2008sh}
\begin{eqnarray}
{\cal R} = \frac{1}{3} \pmatrix{1 & 2 & 2 \cr
2 & -2 & 1 \cr
2 & 1 & -2} \; ,
\label{eq:R-matrix2}
\end{eqnarray}
and thus the effective Majorana neutrino mass matrix $M^{}_\nu$ in
Eq.~(\ref{eq:Mnu-matrix}) and its reduced version $M^\prime_\nu$
in Eq.~(\ref{eq:R-mu-tau-M-transformation}) are of the more
specific textures~\cite{Rodejohann:2017lre}
\begin{eqnarray}
M^{}_\nu = \pmatrix{a + b + c & \hspace{0.19cm} a + {\rm i} a^\prime
\hspace{0.19cm} & a - {\rm i} a^\prime \cr
a + {\rm i} a^\prime & b - {\rm i} 2 a^\prime & c \cr
a - {\rm i} a^\prime & c & b + {\rm i} 2 a^\prime} \; ,
\nonumber \\
M^{\prime}_\nu = \pmatrix{2 a + b + c & 0 & 0 \cr
0 & b + c - a & {\rm i} \sqrt{6} \hspace{0.05cm} a^\prime \cr
0 & {\rm i} \sqrt{6} \hspace{0.05cm} a^\prime & b -c} \; .
\label{eq:Mnu-matrix2}
\end{eqnarray}
The latter can be diagonalized via the transformation
$O^\dagger_{23} M^\prime_\nu O^*_{23} =
{\rm Diag}\{m^\prime_1 , m^\prime_2 , m^\prime_3\}$, where the unitary
matrix $O^{}_{23}$ is given by
\begin{eqnarray}
O^{}_{23} = \pmatrix{1 & 0 & 0 \cr
0 & c^{}_* & -{\rm i} s^{}_* \cr
0 & -{\rm i} s^{}_* & c^{}_*} \;
\label{eq:Mnu-matrix2-diagonalization}
\end{eqnarray}
with $c^{}_* \equiv \cos\theta^{}_*$, $s^{}_* \equiv \sin\theta^{}_*$ and
$\tan 2\theta^{}_* = 2\sqrt{6} \hspace{0.05cm} a^\prime/
\left(a - 2 b\right)$, and the eigenvalues $m^\prime_i$ are
\begin{eqnarray}
m^\prime_1 = 2 a + b + c \; ,
\nonumber \\
m^\prime_2 = \pm \frac{1}{2} \sqrt{\left(a - 2 b\right)^2 + 24 a^{\prime 2}}
- \frac{1}{2} a + c \; ,
\nonumber \\
m^\prime_3 = \pm \frac{1}{2} \sqrt{\left(a - 2 b\right)^2 + 24 a^{\prime 2}}
+ \frac{1}{2} a - c \; .
\label{eq:Mnu-matrix2-eigenvalues}
\end{eqnarray}
Note that the physical neutrino masses are $m^{}_i = \left|m^\prime_i\right|$
(for $i = 1, 2, 3$). Then one may easily figure out the PMNS matrix $U$ from
$U^\dagger M^{}_\nu U^* = D^{}_\nu =
{\rm Diag}\{m^{}_1 , m^{}_2, m^{}_3\}$~\cite{Xing:2006ms}:
\begin{eqnarray}
U = P^{\prime}_l U^{}_{\rm TBM} O^{}_{23} =
P^{\prime}_l \pmatrix{ \frac{2}{\sqrt 6} & \frac{1}{\sqrt 3} c^{}_*
& \frac{\rm -i}{\sqrt 3} s^{}_* \cr
\vspace{-0.45cm} \cr
\frac{-1}{\sqrt 6} & \frac{1}{\sqrt 3} c^{}_* + \frac{\rm i}{\sqrt 2}
s^{}_* & \frac{-1}{\sqrt 2} c^{}_* - \frac{\rm i}{\sqrt 3} s^{}_* \cr
\vspace{-0.45cm} \cr
\frac{-1}{\sqrt 6} & \frac{1}{\sqrt 3} c^{}_* - \frac{\rm i}{\sqrt 2}
s^{}_* & \frac{1}{\sqrt 2} c^{}_* - \frac{\rm i}{\sqrt 3} s^{}_*}
P^\prime_\nu\; ,
\label{eq:TM1R}
\end{eqnarray}
where $P^\prime_\nu$ contains the trivial Majorana CP phases arising
from possible negative signs of $m^\prime_i$. Comparing this constrained
$\rm TM^{}_1$ flavor mixing pattern with the more general one in
Eq.~(\ref{eq:TM1}), we see that the rotational $\mu$-$\tau$ reflection
symmetry can further enhance the predictive power of the $\mu$-$\tau$
reflection symmetry by fixing the phase parameter of $U$.
It is easy to verify $U = {\cal P} {\cal R} U^* \zeta$ with
$\zeta = {\rm Diag}\{1, -1, 1\}$, as originally expected.

It is worth mentioning that the specific rotational $\mu$-$\tau$ reflection
symmetry discussed above has been referred to as the ``trimaximal
$\mu$-$\tau$ reflection symmetry" in Ref.~\cite{Rodejohann:2017lre},
where a detailed analysis of its phenomenological implications and a careful
study of the soft symmetry breaking effects induced by radiative corrections
in the canonical seesaw framework have been made. One may certainly consider
some other possible forms of ${\cal R}$ to obtain different but viable
patterns of the PMNS matrix $U$ (see, e.g., Ref.~\cite{Zhao:2018vxy}),
especially when the future precision neutrino oscillation
data become available.

\subsection{An extension to active-sterile flavor mixing}
\label{section 4.3}

A combination of the rotational $\mu$-$\tau$ reflection symmetry with
the canonical seesaw mechanism is expected to help constrain the flavor
structures of both active and sterile neutrinos. Following the discussions
in section~\ref{section 2.4}, let us simply assume that the overall neutrino
mass term ${\cal L}^\prime_{\rm SS}$ in Eq.~(\ref{eq:seesaw-mass-matrix})
is invariant not only under the transformations
\begin{eqnarray}
\nu^{}_{\rm L} \to {\cal P} (\nu^{}_{\rm L})^c \; , \quad
N^{}_{\rm R} \to {\cal T}^* (N^{}_{\rm R})^c \; ,
\label{eq:R-mu-tau-transformation1}
\end{eqnarray}
but also under the transformations
\begin{eqnarray}
\nu^{}_{\rm L} \to {\cal R} (\nu^{}_{\rm L})^c \; , \quad
N^{}_{\rm R} \to {\cal T}^* (N^{}_{\rm R})^c \; ,
\label{eq:R-mu-tau-transformation2}
\end{eqnarray}
where ${\cal P}$ and ${\cal R}$ are mutually exchangeable and their
expressions are given in Eqs.~(\ref{eq:S-tranformation})
and (\ref{eq:R-matrix}) respectively, and $\cal T$ is an arbitrary unitary
matrix. Then we find that the invariance of ${\cal L}^\prime_{\rm SS}$ under the
transformations made in Eqs.~(\ref{eq:R-mu-tau-transformation1}) and
(\ref{eq:R-mu-tau-transformation2}) requires
\begin{eqnarray}
\pmatrix{{\cal P} & {\bf 0} \cr
{\bf 0} & {\cal T}^{\dagger} \cr}
\pmatrix{ {\bf 0} & M^{}_{\rm D} \cr
M^T_{\rm D} & M^{}_{\rm R} \cr}
\pmatrix{ {\cal R} & 0 \cr
{\bf 0} & {\cal T}^* \cr}
= \pmatrix{ {\bf 0} & M^{*}_{\rm D} \cr
M^\dagger_{\rm D} & M^{*}_{\rm R} \cr} \; ,
\label{eq:R-mu-tau-matrix-overall}
\end{eqnarray}
where the symmetry of $M^{}_{\rm R}$ has been used. As a result,
we immediately arrive at
\begin{eqnarray}
M^{}_{\rm D} = {\cal P} M^{*}_{\rm D} {\cal T} \; , \quad
M^{}_{\rm D} = {\cal R} M^{*}_{\rm D} {\cal T} \; , \quad
M^{}_{\rm R} = {\cal T}^T M^{*}_{\rm R} {\cal T} \; .
\label{eq:R-mu-tau-seesaw-constraint}
\end{eqnarray}
Substituting these relations into the approximate seesaw formula
for the active Majorana neutrino mass matrix $M^{}_\nu$ in
Eq.~(\ref{eq:seesaw-formula}), for example, we have
\begin{eqnarray}
M^{}_\nu & \simeq & - {\cal P} {\cal R} M^{*}_{\rm D} {\cal T}
\left({\cal T}^T M^{*}_{\rm R} {\cal T}\right)^{-1}
{\cal T}^T M^\dagger_{\rm D} {\cal R} {\cal P}
\nonumber \\
& = & {\cal P} {\cal R} \left(- M^{}_{\rm D} M^{-1}_{\rm R}
M^T_{\rm D}\right)^* {\cal P} {\cal R}
\nonumber \\
& \simeq & {\cal P} {\cal R} M^*_\nu {\cal P} {\cal R} \; .
\label{eq:R-mu-tau-seesaw-formula}
\end{eqnarray}
So $M^{}_\nu$ respects the rotational $\mu$-$\tau$ reflection symmetry.
As discussed in section~\ref{section 2.4}, the textures of $M^{}_{\rm D}$
and $M^{}_{\rm R}$ can be explicitly constrained by the rotational
$\mu$-$\tau$ reflection symmetry if both the patterns of $\cal R$ and
$\cal T$ are fixed by an underlying flavor symmetry group.
As for the arbitrariness of $\cal T$, one may simply
assume ${\cal T} = {\cal I}$, ${\cal P}$, ${\cal R}$ or ${\cal P} {\cal R}$
as a phenomenological example.
Section~\ref{section 4.2} has given a
typical pattern of $\cal R$ to produce a constrained $\rm TM^{}_1$
flavor mixing pattern of $U$. An extension of that interesting example
to the seesaw mechanism is certainly straightforward.

Now let us insert the rotational $\mu$-$\tau$ reflection symmetry
constraints on $U$ obtained in Eq.~(\ref{eq:R-mu-tau-U}) into the
exact seesaw formula in Eq.~(\ref{eq:exact-seesaw}) and take the
complex conjugate for the whole equation. Then we are left with
\begin{eqnarray}
U D^{}_\nu U^T + {\cal P} R^* D^{}_N ({\cal P} R^*)^T = {\bf 0} \; ,
\nonumber \\
U D^{}_\nu U^T + {\cal R} R^* D^{}_N ({\cal R} R^*)^T = {\bf 0} \; ,
\nonumber \\
U D^{}_\nu U^T + {\cal P} {\cal R} R^* D^{}_N ({\cal P}
{\cal R} R^*)^T = {\bf 0} \; .
\label{eq:R-mu-tau-exact-seesaw}
\end{eqnarray}
A comparison between Eqs.~(\ref{eq:exact-seesaw}) and
(\ref{eq:R-mu-tau-exact-seesaw}) can therefore lead us to
\begin{eqnarray}
R = {\cal P} R^* \zeta^{}_{\cal P} \; , \quad
R = {\cal R} R \zeta^{}_{\cal R} \; , \quad
R = {\cal P} {\cal R} R^* \zeta \; ,
\label{eq:R-mu-tau-R}
\end{eqnarray}
where $\zeta^{}_{\cal P}$, $\zeta^{}_{\cal R}$ and $\zeta$ have been
given below Eq.~(\ref{eq:R-mu-tau-U}). Such simple but instructive
results mean that the active-sterile flavor mixing matrix $R$ is
constrained by the same rotational $\mu$-$\tau$ reflection symmetry
as the PMNS matrix $U$. Note that both $U$ and $R$ appear in the standard
weak charged-current interactions of massive Majorana neutrinos as
described by Eq.~(\ref{eq:cc-seesaw}), and they are correlated with
each other via the unitarity condition in Eq.~(\ref{eq:seesaw-unitarity})
and the exact seesaw relation in Eq.~(\ref{eq:exact-seesaw}). So it is
actually a natural result that both of them are constrained by the
same flavor symmetry.

In \ref{Appendix A}, a full Euler-like parametrization of
the active-sterile flavor mixing matrix $R$ has been presented. It
is therefore straightforward to constrain the relevant flavor mixing
angles and CP-violating phases of $U$ and $R$ in a model-independent
way after a specific pattern of ${\cal R}$ is assumed.

\setcounter{equation}{0}
\section{Soft $\mu$-$\tau$ reflection symmetry breaking}
\label{section 5}

\subsection{General remarks on $\mu$-$\tau$ symmetry breaking}
\label{section 5.1}

In the $\mu$-$\tau$ reflection symmetry limit the unitary PMNS matrix $U$
satisfies $U = {\cal P} U^* \zeta$, and therefore its flavor mixing angle
$\theta^{}_{23}$ and three CP-violating phases $\delta^{}_\nu$, $\rho$ and
$\sigma$ are well constrained, as can be seen from Eq.~(\ref{eq:reflection-phases}).
To make sure of how far away such constraints are from the corresponding
experimental results, let us quote the best-fit values of six neutrino
oscillation parameters extracted from a three-flavor global analysis of the
latest experimental data \cite{Esteban:2020cvm,Gonzalez-Garcia:2021dve},
together with their 1$\sigma$ and 3$\sigma$ ranges, as shown in
Table~\ref{table2}. It is then trivial to see
\begin{eqnarray}
&& \Delta \theta^{}_{23} \equiv \theta^{}_{23} - 45^\circ
= \left\{ \begin{array}{l}
\hspace{-0.15cm} -2.9^{+1.1^\circ}_{-0.9^\circ} ~({\rm NMO}) \\
\vspace{-0.4cm} \\
\hspace{-0.15cm} +4.0^{+0.9^\circ}_{-1.3^\circ} ~({\rm IMO}) \end{array} \right. \;
\nonumber \\
&& \Delta \delta^{}_\nu \equiv \delta^{}_\nu - 270^\circ
= \left\{ \begin{array}{l}
\hspace{-0.15cm} -40^{+36^\circ}_{-25^\circ} ~({\rm NMO}) \\
\vspace{-0.4cm} \\
\hspace{-0.15cm} +8^{+22^\circ}_{-30^\circ} ~({\rm IMO}) \end{array} \right. \;
\label{eq:octant}
\end{eqnarray}
at the $\pm 1\sigma$ level, where ``NMO" (or ``IMO") denotes the normal (or
inverted) neutrino mass ordering, and the possibility of $\delta^{}_\nu = 90^\circ$
in the $\mu$-$\tau$ reflection symmetry limit has tentatively been discarded
as it seems less compatible with current neutrino oscillation data.
We find that $|\Delta \theta^{}_{23}| / \theta^{}_{23} \lesssim 10\%$ holds up
to the $\pm 1\sigma$ confidence level for the global-fit result of $\theta^{}_{23}$
as given in Table~\ref{table2}, and the ratio $|\Delta \delta^{}_\nu| / \delta^{}_\nu$
involves much larger uncertainties which are attributed to the large experimental
uncertainties of $\delta^{}_\nu$. This observation implies that it is rather safe
to treat $\Delta \theta^{}_{23}$ as a small perturbation to the $\mu$-$\tau$
reflection symmetry value $\theta^{}_{23} = 45^\circ$, but whether
$\Delta \delta^{}_\nu$ can be treated as a small perturbation to the $\mu$-$\tau$
reflection symmetry value $\delta^{}_\nu = 270^\circ$ depends on which neutrino mass
ordering is assumed and what value of $\delta^{}_\nu$ is adopted. For instance,
$\Delta \delta^{}_\nu / \delta^{}_\nu \lesssim 10\%$ is expected to hold in the
IMO case at the $\pm 1\sigma$ confidence level for the present global-fit result of
$\delta^{}_\nu$, but the situation will be worse in the NMO case as reflected by
Eq.~(\ref{eq:octant}).
\begin{table}[t]
\caption{The best-fit values and the 1$\sigma$ and 3$\sigma$ ranges of six
neutrino oscillation parameters in the standard parametrization of $U$,
obtained from a three-flavor global analysis of the latest
experimental data~\cite{Esteban:2020cvm,Gonzalez-Garcia:2021dve}.} \vspace{0.2cm}
\label{table2}
\begin{indented}
\item[]\begin{tabular}{lll} \br Parameter & Best fit $\pm 1\sigma$
& 3$\sigma$ range \\ \mr \multicolumn{3}{c}{Normal
mass ordering $(m^{}_1 < m^{}_2 < m^{}_3$)} \\ \mr \vspace{0.1cm}
$\Delta m^2_{21}/10^{-5} ~{\rm eV}^2$ \hspace{1cm}
& $7.42^{+0.21}_{-0.20}$
& $6.82 \longrightarrow 8.04$ \\ \vspace{0.1cm}
$\Delta m^2_{31}/10^{-3} ~ {\rm eV}^2$
& $+2.510^{+0.027}_{-0.027}$ \hspace{1cm}
& $+2.430 \longrightarrow +2.593$ \\ \vspace{0.1cm}
$\theta^{}_{12}/^\circ$ & $33.45^{+0.77}_{-0.75}$
& $31.27 \longrightarrow 35.87$ \\ \vspace{0.1cm}
$\theta^{}_{13}/^\circ$ & $8.62^{+0.12}_{-0.12}$
& $8.25 \longrightarrow 8.98$ \\ \vspace{0.1cm}
$\theta^{}_{23}/^\circ$ & $42.1^{+1.1}_{-0.9}$
& $39.7 \longrightarrow 50.9$ \\ \vspace{0.1cm}
$\delta^{}_\nu/^\circ$ &  $230^{+36}_{-25}$
& $144 \longrightarrow 350$ \\ \br
\multicolumn{3}{c}{Inverted mass ordering $(m^{}_3 < m^{}_1 <
m^{}_2$)} \\ \mr \vspace{0.1cm}
$\Delta m^2_{21}/10^{-5} ~{\rm eV}^2$ \hspace{1cm}
& $7.42^{+0.21}_{-0.20}$
& $6.82 \longrightarrow 8.04$ \\ \vspace{0.1cm}
$\Delta m^2_{32}/10^{-3} ~ {\rm eV}^2$
& $-2.490^{+0.026}_{-0.028}$
& $-2.574 \longrightarrow -2.410$ \\ \vspace{0.1cm}
$\theta^{}_{12}/^\circ$ & $33.45^{+0.78}_{-0.75}$ \hspace{1cm}
& $31.27 \longrightarrow 35.87$ \\ \vspace{0.1cm}
$\theta^{}_{13}/^\circ$ & $8.61^{+0.14}_{-0.12}$
& $8.24 \longrightarrow 9.02$ \\ \vspace{0.1cm}
$\theta^{}_{23}/^\circ$ & $49.0^{+0.9}_{-1.3}$
& $39.8 \longrightarrow 51.6$ \\ \vspace{0.1cm}
$\delta^{}_\nu/^\circ$ & $278^{+22}_{-30}$
& $194 \longrightarrow 345$ \\ \br
\end{tabular}
\end{indented}
\end{table}

On the other hand, there has been no convincing experimental information on the
Majorana nature of massive neutrinos, although a lot of efforts have been made in
searching for the $0\nu 2\beta$ decays and some other lepton-number-violating
processes~\cite{ParticleDataGroup:2020ssz}. So the two Majorana CP-violating phases
$\rho$ and $\sigma$ are completely unconstrained on the experimental side.
Now that we have no idea about what values of $\rho$ and $\sigma$ can be at
low energies, it certainly makes no sense for us to discuss to what extent
they may deviate from their corresponding $\mu$-$\tau$ reflection symmetry
values $\rho = 0^\circ$ or $90^\circ$ and $\sigma = 0^\circ$ or $90^\circ$ in an
{\it explicit} way. The key point is that an explicit symmetry breaking scenario is
usually guided by the available experimental data or a kind of phenomenological
judgement~\cite{Fritzsch:1999ee}. In view of the insurmountable uncertainties
associated with the three phase parameters of Majorana neutrinos (especially,
$\rho$ and $\sigma$), we conclude that it is very hard to describe how the
$\mu$-$\tau$ reflection symmetry can be explicitly broken for the whole
PMNS flavor mixing matrix $U$ at the present time.

But one may introduce the following three rephasing invariants to describe the
breaking of $\mu$-$\tau$ reflection symmetry for lepton flavor mixing and
CP violation no matter whether it is spontaneously or explicitly
realized~\cite{Luo:2014upa}:
\begin{eqnarray}
\Delta^{}_1 & \equiv & |U^{}_{\tau 1}|^2 - |U^{}_{\mu 1}|^2 =
\left(\cos^2\theta^{}_{12} \sin^2\theta^{}_{13} - \sin^2\theta^{}_{12}\right)
\cos 2\theta^{}_{23}
\nonumber \\
&& \hspace{3.07cm} - \sin 2\theta^{}_{12} \sin\theta^{}_{13}
\sin 2\theta^{}_{23} \cos\delta^{}_\nu \; ,
\nonumber \\
\Delta^{}_2 & \equiv & |U^{}_{\tau 2}|^2 - |U^{}_{\mu 2}|^2 =
\left(\sin^2\theta^{}_{12} \sin^2\theta^{}_{13} - \cos^2\theta^{}_{12}\right)
\cos 2\theta^{}_{23}
\nonumber \\
&& \hspace{3.07cm} + \sin 2\theta^{}_{12} \sin\theta^{}_{13}
\sin 2\theta^{}_{23} \cos\delta^{}_\nu \; ,
\nonumber \\
\Delta^{}_3 & \equiv & |U^{}_{\tau 3}|^2 - |U^{}_{\mu 3}|^2 =
\cos^{2}\theta^{}_{13} \cos 2\theta^{}_{23} \; ,
\label{eq:mu-tau-U-breaking}
\end{eqnarray}
which satisfy the sum rule $\Delta^{}_1 + \Delta^{}_2 + \Delta^{}_3 = 0$. At
the energy scale $\Lambda^{}_{\mu\tau}$ where the $\mu$-$\tau$ reflection
symmetry exactly holds, we immediately have
\begin{eqnarray}
\Delta^{}_1(\Lambda^{}_{\mu\tau}) = \Delta^{}_2(\Lambda^{}_{\mu\tau}) =
\Delta^{}_3(\Lambda^{}_{\mu\tau}) = 0 \;
\label{eq:mu-tau-U-symmetry}
\end{eqnarray}
thanks to $\theta^{}_{23} (\Lambda^{}_{\mu\tau}) = \pi/4$ and
$\delta^{}_\nu (\Lambda^{}_{\mu\tau}) = \pm\pi/2$. Provided $\Delta\theta^{}_{23}$
and $\Delta\delta^{}_\nu$ defined in Eq.~(\ref{eq:octant}) are small enough
and thus can be regarded as small perturbations to the respective values of
$\theta^{}_{23}$ and $\delta^{}_\nu$ at $\Lambda^{}_{\mu\tau}$, the
corresponding effects of $\mu$-$\tau$ reflection symmetry breaking will
be referred to as ``soft".
Substituting small $\Delta\theta^{}_{23}$ and $\Delta\delta^{}_\nu$
into Eq.~(\ref{eq:mu-tau-U-breaking}), we obtain the approximate $\mu$-$\tau$
asymmetries at the electroweak scale $\Lambda^{}_{\rm EW}$:
\begin{eqnarray}
\Delta^{}_1 (\Lambda^{}_{\rm EW}) & \simeq &
-2\left(c^2_{12} s^2_{13} - s^2_{12}\right) \Delta \theta^{}_{23}
- 2 c^{}_{12} s^{}_{12} s^{}_{13} \Delta\delta^{}_\nu \; ,
\nonumber \\
\Delta^{}_2 (\Lambda^{}_{\rm EW}) & \simeq &
-2\left(s^2_{12} s^2_{13} - c^2_{12}\right) \Delta\theta^{}_{23}
+ 2 c^{}_{12} s^{}_{12} s^{}_{13} \Delta\delta^{}_\nu \; ,
\nonumber \\
\Delta^{}_3 (\Lambda^{}_{\rm EW}) & \simeq & -2 c^{2}_{13} \Delta\theta^{}_{23} \; ,
\label{eq:mu-tau-U-breaking2}
\end{eqnarray}
where the three flavor mixing angles $\theta^{}_{ij}$ (for $ij = 12, 13, 23$)
take their values at $\Lambda^{}_{\rm EW}$. This result implies that the
finite effects of $\mu$-$\tau$ reflection symmetry breaking are sensitive to
the octant of $\theta^{}_{23}$ and the quadrant of $\delta^{}_\nu$, or vice versa.
The so-called {\it octant} and {\it quadrant} issues of lepton flavor mixing,
namely the signs of $\Delta\theta^{}_{23}$ and $\Delta\delta^{}_\nu$ as illustrated
by Figure~\ref{Fig:octant}, can therefore be resolved after the strengths of
$\mu$-$\tau$ reflection symmetry breaking are finally determined from some more
precise experimental data on neutrino oscillations in the near future. On
the other hand, the accurate results of $\Delta\theta^{}_{23}$ and
$\Delta\delta^{}_\nu$ may serve as a very sensitive model discriminator.
\begin{figure}[t]
\begin{center}
\includegraphics[width=2.7in]{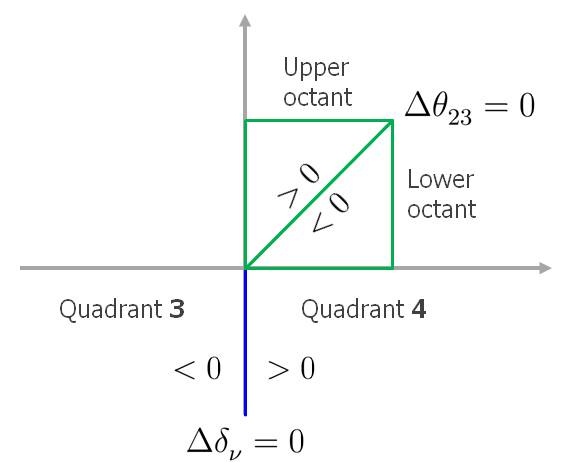}
\vspace{-0.1cm}
\caption{An illustration of the so-called octant problem of $\theta^{}_{23}$
(i.e., whether it is in the upper or lower octant) and the quadrant problem
of $\delta^{}_\nu$ (i.e., whether it is in the third or fourth quadrant) as
compared with their respective values in the $\mu$-$\tau$ reflection symmetry
limit (i.e., $\theta^{}_{23} = \pi/4$ and $\delta^{}_\nu = -\pi/2$).}
\label{Fig:octant}
\end{center}
\end{figure}

As for the effective Majorana neutrino mass matrix $M^{}_\nu$, its flavor texture
in the $\mu$-$\tau$ reflection symmetry limit has been given in
Eq.~(\ref{eq:M-reflection}). Slight symmetry breaking effects on this special
form of $M^{}_\nu$ should in general include not only small imaginary corrections
to real $\langle m\rangle^{}_{ee}$ and $\langle m\rangle^{}_{\mu\tau}$ but also
small complex corrections to the equalities
$\langle m\rangle^{}_{e\tau} = \langle m\rangle^*_{e\mu}$ and
$\langle m\rangle^{}_{\tau\tau} = \langle m\rangle^{*}_{\mu\mu}$~\cite{Xing:2015fdg}.
In this case the modified texture of $M^{}_\nu$ can be generally parametrized
as follows:
\begin{eqnarray}
M^{}_\nu = \pmatrix{a \left(1 + {\rm i} \varepsilon^{}_a\right) &
b \left(1 + \varepsilon^{}_b\right) & b^* \left(1 - \varepsilon^{*}_b\right) \cr
b \left(1 + \varepsilon^{}_b\right) & c \left(1 + \varepsilon^{}_c\right) &
d \left(1 + {\rm i} \varepsilon^{}_d\right) \cr
b^* \left(1 - \varepsilon^{*}_b\right) & d \left(1 + {\rm i} \varepsilon^{}_d\right)
& c^* \left(1 - \varepsilon^{*}_c\right)} \; ,
\label{eq:M-symmetry-breaking}
\end{eqnarray}
in which $a$ and $d$ are real, and
\begin{eqnarray}
\varepsilon^{}_a = \frac{{\rm Im}\langle m\rangle^{}_{ee}}
{{\rm Re}\langle m\rangle^{}_{ee}} \; , \quad
\varepsilon^{}_d = \frac{{\rm Im}\langle m\rangle^{}_{\mu\tau}}
{{\rm Re}\langle m\rangle^{}_{\mu\tau}}
\label{eq:SB-parameter1}
\end{eqnarray}
are real perturbations which should be very small in magnitude (typically,
$\lesssim 0.2$), and
\begin{eqnarray}
\varepsilon^{}_b = \frac{\langle m\rangle^{}_{e\mu} - \langle m\rangle^*_{e\tau}}
{\langle m\rangle^{}_{e\mu} + \langle m\rangle^*_{e\tau}} \; , \quad
\varepsilon^{}_c = \frac{\langle m\rangle^{}_{\mu\mu} - \langle m\rangle^*_{\tau\tau}}
{\langle m\rangle^{}_{\mu\mu} + \langle m\rangle^*_{\tau\tau}}
\label{eq:SB-parameter2}
\end{eqnarray}
are complex perturbations which are also required to be very small in magnitude
(typically, $\lesssim 0.2$). It should be noted
that a specific texture of $M^{}_\nu$ is always dependent upon a chosen flavor basis.
That is why the four perturbation parameters in Eqs.~(\ref{eq:SB-parameter1}) and
(\ref{eq:SB-parameter2}) are rephasing-dependent, unlike $\Delta^{}_i$ (for
$i = 1, 2, 3$). Of course, one may or may not switch off any of $\varepsilon^{}_x$ (for
$x = a, b , c, d$) in Eq.~(\ref{eq:M-symmetry-breaking}) to partly or completely
break the $\mu$-$\tau$ reflection symmetry, and the corresponding PMNS flavor mixing
matrix $U$ can be obtained from the unitary transformation matrix used to diagonalize
$M^{}_\nu$ in such cases~\cite{Xing:2015fdg,Zhao:2017yvw}.

The {\it linear} perturbations introduced in Eq.~(\ref{eq:M-symmetry-breaking})
actually constitute a kind of {\it explicit} breaking of the $\mu$-$\tau$ reflection
symmetry of $M^{}_\nu$. The latter can certainly be broken in a {\it spontaneous}
way, but such a symmetry breaking mechanism is usually strongly model-dependent
and hence are not elaborated here (see Refs.~\cite{Mohapatra:2006un}
and~\cite{Joshipura:2009tg} for examples of this kind in the $\rm SU(5)$ and
$\rm SO(10)$ grand unified models, respectively).

A kind of {\it nonlinear} correction to the texture of $M^{}_\nu$ that has been
constrained by the $\mu$-$\tau$ reflection symmetry at a given superhigh energy scale
$\Lambda^{}_{\mu\tau}$, which can be regarded as an effective {\it spontaneous}
$\mu$-$\tau$ symmetry breaking mechanism to naturally achieve a realistic structure
of $M^{}_\nu$ at the electroweak scale $\Lambda^{}_{\rm EW}$, is the well-known
RGE-induced quantum correction. Such nontrivial $\mu$-$\tau$ reflection symmetry
breaking effects may help resolve the octant issue of $\theta^{}_{23}$ and the quadrant
issue of $\delta^{}_\nu$~\cite{Luo:2014upa,Xing:2017mkx,Huang:2018wqp,Huang:2020kgt}.
More details in this respect will be discussed later on.

\subsection{RGE-induced effects on Majorana neutrinos}
\label{section 5.2}

In quantum field theories, which certainly include the SM and its natural
extensions, the RGE approach was originally invented and developed in the
1950s~\cite{Stueckelberg:1953dz,GellMann:1954fq,Bogolyubov:1956gh}. This
powerful theoretical tool was successfully applied to the study of critical
phenomena in condensed matter physics in 1971~\cite{Wilson:1971bg,Wilson:1971dh}
and to the discovery of asymptotic freedom of strong interactions in particle
physics in 1973~\cite{Gross:1973id,Politzer:1973fx} --- both of these two outstanding
applications have been recognized by the Nobel Prize in physics. The key point is
that the physical parameters at one renormalization point or energy scale can be
related to their counterparts at another renormalization point or energy scale with
the help of the relevant RGEs, such that the theory keeps its form invariance or
self similarity between the two points or energy scales. As far as the flavor issues
of charged fermions and massive neutrinos are concerned, we are especially interested
in the RGE-induced bridge between a superhigh energy scale characterized by a
fundamental flavor theory or flavor symmetry and the electroweak scale where the
physical flavor parameters can be experimentally determined or
constrained~\cite{Xing:2019vks,Ohlsson:2013xva}.

It is well known that the dimension-five Weinberg operator is unique to describe
the lowest-order origin of neutrino masses in the effective field theory (EFT)
of the SM itself by allowing for lepton number violation or in the EFT of the
minimal supersymmetric SM (MSSM)~\cite{Weinberg:1979sa}. The corresponding Wilson
coefficients are equivalent to the effective Majorana neutrino coupling matrix
elements $\kappa^{}_{\alpha\beta}$ (for $\alpha, \beta = e, \mu, \tau$):
\begin{eqnarray}
{\cal O}^{(\rm SM)}_{\rm Weinberg} = \frac{1}{2} \kappa^{}_{\alpha\beta}
\Big[\overline{\ell^{}_{\alpha \rm L}} \hspace{0.05cm} \widetilde{H} \widetilde{H}^T
\left(\ell^{}_{\beta \rm L} \right)^c\Big] + {\rm h.c.}
\label{eq:Weinberg-SM}
\end{eqnarray}
in the SM, or
\begin{eqnarray}
{\cal O}^{(\rm MSSM)}_{\rm Weinberg} = \frac{1}{2} \kappa^{}_{\alpha\beta}
\Big[\overline{\ell^{}_{\alpha \rm L}} \hspace{0.05cm} H^{}_2 H^T_2
\left(\ell^{}_{\beta \rm L} \right)^c\Big] + {\rm h.c.}
\label{eq:Weinberg-SM}
\end{eqnarray}
in the MSSM, where $\ell^{}_{\rm L}$ denotes the $\rm SU(2)^{}_{\rm L}$ doublet
of lepton fields, $\widetilde{H} \equiv {\rm i} \sigma^{}_2 H^*$ with
$H$ standing for the Higgs doublet
with a hypercharge $+1/2$, and $H^{}_2$ denotes the second Higgs doublet with a
hypercharge $-1/2$ in the MSSM. After spontaneous electroweak symmetry breaking,
the effective Majorana neutrino mass matrix reads as $M^{}_\nu = \kappa v^2/2$ in the
SM or $M^{}_\nu = \kappa (v\sin\beta)^2/2$ in the MSSM, where $v \simeq 246~{\rm GeV}$
is the overall vacuum expectation value of the Higgs fields and $\sin\beta$ originates
from $\tan\beta$ which describes the ratio of the vacuum expectation value of
$H^{}_2$ to that of $H^{}_1$ (the first Higgs doublet with a hypercharge
$+1/2$) in the MSSM. We accordingly have the charged lepton
mass matrix $M^{}_l = Y^{}_l v /\sqrt{2}$ in the SM or
$M^{}_l = Y^{}_l v \cos\beta/\sqrt{2}$ in the MSSM.

The tiny masses of three active neutrinos imply that it is enough to consider the
RGE-induced quantum corrections to their flavor texture at the one-loop level in
most cases. To be explicit, the one-loop RGE for the effective Majorana neutrino
coupling matrix $\kappa$ is given by~\cite{Chankowski:1993tx,Babu:1993qv,Antusch:2001ck}
\begin{eqnarray}
16\pi^2 \frac{{\rm d}\kappa}{{\rm d}t} =
\alpha^{}_\kappa \kappa + C^{}_l \left[ \left(Y^{}_l Y^\dagger_l\right) \kappa
+ \kappa \left(Y^{}_l Y^\dagger_l\right)^T \right] \; ,
\label{eq:one-loop-RGE}
\end{eqnarray}
where $t\equiv \ln (\mu/\Lambda^{}_{\rm EW})$ with $\mu$ being a given
renormalization scale between $\Lambda^{}_{\rm EW}$ and $\Lambda^{}_{\mu\tau}$,
$C^{}_l = -3/2$ and $\alpha^{}_\kappa \simeq -3 g^2_2 + 6 y^2_t + \lambda$
in the SM case with $g^{}_2$ being the gauge coupling constant associated with
the weak interactions, $y^{}_t$ being the top-quark Yukawa coupling eigenvalue and
$\lambda$ being the Higgs self-coupling parameter; or $C^{}_l = 1$ and
$\alpha^{}_\kappa \simeq -6 g^2_1/5 - 6 g^2_2 + 6 y^2_t$
in the MSSM case with $g^{}_1$ being the gauge coupling constant associated
with the electromagnetic interactions in the grand-unified-theory charge
normalization. Working in the flavor basis where $Y^{}_l$
is diagonal, one may obtain the integral solution to
Eq.~(\ref{eq:one-loop-RGE}) at the electroweak scale $\Lambda^{}_{\rm EW}$
as follows~\cite{Xing:2019vks,Fritzsch:1999ee,Ellis:1999my}:
\begin{eqnarray}
M^{}_\nu (\Lambda^{}_{\rm EW}) =
I^{}_\kappa (\Lambda^{}_{\rm EW}) \left[T^{}_l (\Lambda^{}_{\rm EW})
\cdot M^{}_\nu (\Lambda^{}_{\mu\tau})
\cdot T^{}_l (\Lambda^{}_{\rm EW}) \right] \; ,
\label{eq:one-loop-inregral}
\end{eqnarray}
where $T^{}_l = {\rm Diag}\{I^{}_e , I^{}_\mu , I^{}_\tau\}$, and the
scale-dependent one-loop RGE evolution functions $I^{}_\kappa$ and
$I^{}_\alpha$ (for $\alpha = e, \mu, \tau$) are defined as~\cite{Zhang:2020lsd}
\begin{eqnarray}
I^{}_\kappa (\mu) & \equiv & \exp\left[-\frac{1}{16 \pi^2}
\int^{\ln(\Lambda^{}_{\mu\tau}/\Lambda^{}_{\rm EW})}_{\ln(\mu/\Lambda^{}_{\rm EW})}
\alpha^{}_\kappa(t) \hspace{0.05cm} {\rm d} t\right] \; ,
\nonumber \\
I^{}_\alpha (\mu) & \equiv & \exp\left[-\frac{C^{}_l}{16 \pi^2}
\int^{\ln(\Lambda^{}_{\mu\tau}/\Lambda^{}_{\rm EW})}_{\ln(\mu/\Lambda^{}_{\rm EW})}
y^2_\alpha(t) \hspace{0.05cm} {\rm d} t\right] \; ,
\label{eq:evolution-function}
\end{eqnarray}
where $\Lambda^{}_{\rm EW} \leq \mu \leq \Lambda^{}_{\mu\tau}$.
The smallness of $y^{}_e$ and $y^{}_\mu$ as a consequence of the smallness of
$m^{}_e$ and $m^{}_\mu$ with respect to the value of $v$, together with the
smallness of the loop factor $1/16\pi^2$, leads us to
$I^{}_e \simeq I^{}_\mu \simeq 1$ as an excellent approximation in the SM
or in the MSSM with $\tan\beta \lesssim 30$. So it is $y^{}_\tau$ that dominates
a slight deviation of the texture of $M^{}_\nu (\Lambda^{}_{\rm EW})$ from that
of $M^{}_\nu (\Lambda^{}_{\mu\tau})$ via a relatively larger departure of
the loop function $I^{}_\tau$ from $1$. In this case let us simply express
$I^{}_\tau$ as $I^{}_\tau = 1 + \Delta^{}_\tau$, where
\begin{eqnarray}
\Delta^{}_\tau (\mu) \simeq -\frac{C^{}_l}{16 \pi^2}
\int^{\ln(\Lambda^{}_{\mu\tau}/\Lambda^{}_{\rm EW})}_{\ln(\mu/\Lambda^{}_{\rm EW})}
y^2_\tau(t) \hspace{0.05cm} {\rm d} t \; ,
\label{eq:Delta-tau}
\end{eqnarray}
which is expected to be a very small quantity and vanish at
$\mu = \Lambda^{}_{\mu\tau}$. Figure~\ref{Fig:loop-functions}
provides a numerical illustration of
$I^{}_\kappa$ and $\Delta^{}_\tau$ evolving with the energy scale $\mu$ in the SM
or in the MSSM with $\tan\beta = 10$ or $30$ as two typical input values
\footnote{If $\tan\beta$ is very small (e.g., $\tan\beta <3$), the RGE-induced effects
will be insignificant as in the SM case. On the other hand, the possibility of
$\tan\beta >50$ is strongly disfavored as the heavy-quark Yukawa coupling eigenvalues
may fall into the non-perturbation region in this case.}.
\begin{figure}[t]
\begin{center}
\includegraphics[width=5.7in]{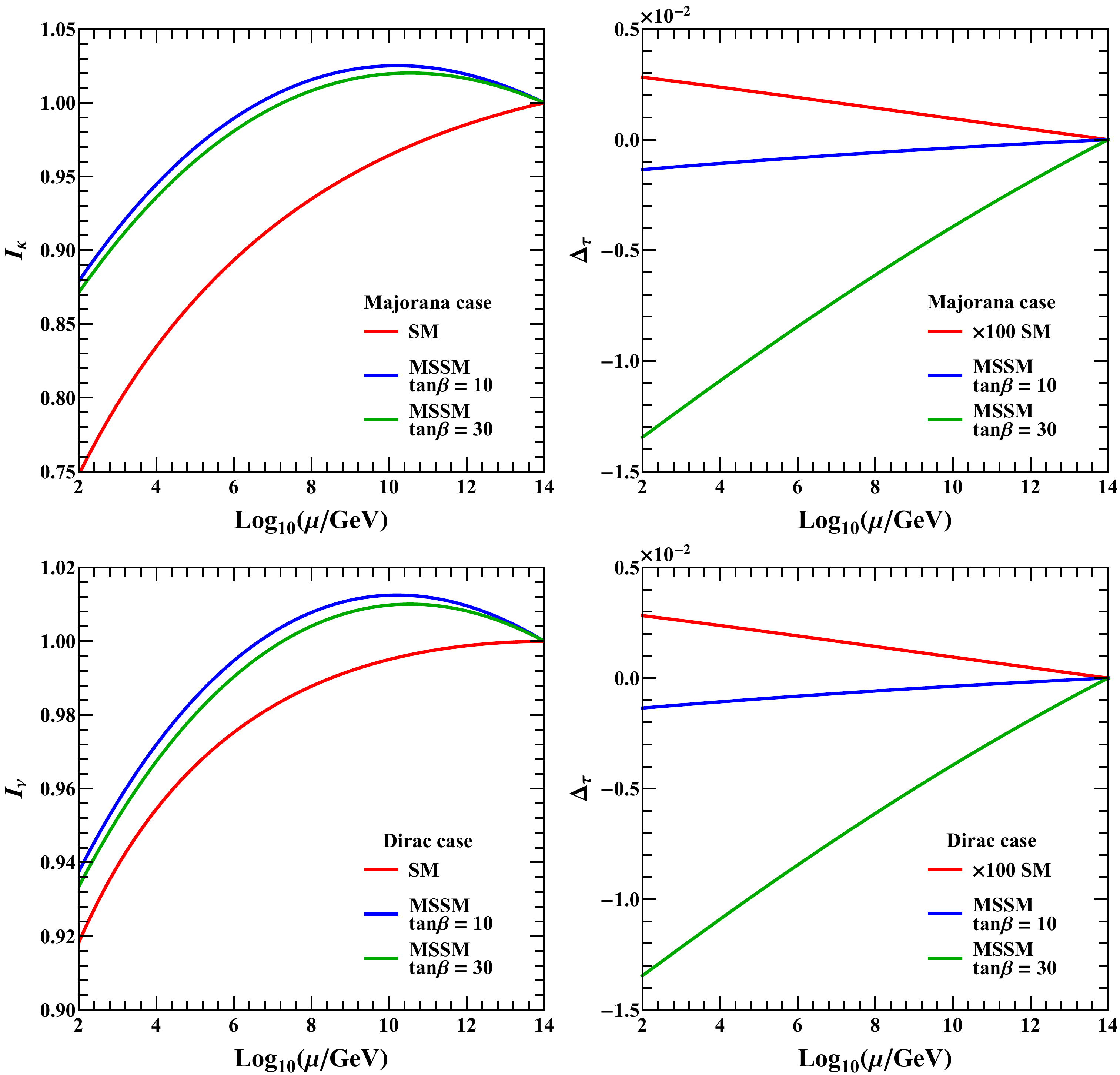}
\vspace{-0.1cm}
\caption{A numerical illustration of $I^{}_\kappa$ and $\Delta^{}_\tau$ versus
the energy scale $\mu$ in the Majorana case (upper panels), or $I^{}_\nu$ and
$\Delta^{}_\tau$ versus $\mu$ in the Dirac case (lower panels) in the SM or
MSSM with $\tan\beta = 10$ or $30$, where $\Lambda^{}_{\mu\tau} = 10^{14}$
GeV has typically been taken, and the notation ``$\times 100$" in the right
panels means that the value of $\Delta^{}_\tau$ in the SM is artificially
enhanced by a factor $100$~\cite{Zhang:2020lsd}.}
\label{Fig:loop-functions}
\end{center}
\end{figure}

To characterize possible deviations of the observed values of the PMNS flavor
mixing and CP-violating parameters at
$\Lambda^{}_{\rm EW} \sim {\cal O}(10^2)~{\rm GeV}$ from their values at
$\Lambda^{}_{\mu\tau}$ where the $\mu$-$\tau$ reflection symmetry is exact,
let us define
\begin{eqnarray}
&& \Delta \theta^{}_{ij} \equiv \theta^{}_{ij}(\Lambda_{\rm EW}^{})
- \theta^{}_{ij}(\Lambda_{\mu\tau}^{}) \; ,
\nonumber \\
&& \Delta \delta^{}_\nu \equiv \delta^{}_\nu(\Lambda_{\rm EW}^{})
- \delta^{}_\nu(\Lambda_{\mu\tau}^{}) \; ,
\nonumber \\
&& \Delta \rho \equiv \rho (\Lambda_{\rm EW}^{})
-\rho (\Lambda_{\mu\tau}^{}) \; ,
\nonumber \\
&& \Delta \sigma \equiv \sigma (\Lambda_{\rm EW}^{})
-\sigma (\Lambda_{\mu\tau}^{}) \; .
\label{eq:Deviation}
\end{eqnarray}
With the help of the generic analytical expressions of $m^{}_i$ (for $i = 1, 2, 3$),
$\Delta \theta^{}_{ij}$ (for $ij = 12, 13, 23$), $\Delta\delta^{}_\nu$, $\Delta\rho$
and $\Delta\sigma$ derived in the leading-order approximation~\cite{Zhang:2020lsd},
one may take $\theta^{}_{23} (\Lambda^{}_{\mu\tau}) = \pi/4$,
$\delta^{}_\nu (\Lambda^{}_{\mu\tau}) = \pm\pi/2$,
$\rho (\Lambda^{}_{\mu\tau}) = 0$ or $\pi/2$ and
$\sigma (\Lambda^{}_{\mu\tau}) = 0$ or $\pi/2$
as given by Eq.~(\ref{eq:reflection-phases}) in the $\mu$-$\tau$ reflection symmetry
limit to obtain the explicit results of neutrino masses and flavor mixing parameters
at the electroweak scale. The relations between $m^{}_i (\Lambda^{}_{\rm EW})$ and
$m^{}_i (\Lambda^{}_{\mu\tau})$ are found to be
\begin{eqnarray}
m^{}_1 (\Lambda^{}_{\rm EW}) & \simeq & I^{}_\kappa
\Big[ 1 + \Delta^{}_\tau \left( 1 - c^2_{12} c^2_{13} \right)
\Big] m^{}_1 (\Lambda^{}_{\mu\tau}) \; ,
\nonumber \\
m^{}_2 (\Lambda^{}_{\rm EW}) & \simeq & I^{}_\kappa
\Big[ 1 + \Delta^{}_\tau \left( 1 - s^2_{12} c^2_{13} \right)
\Big] m^{}_2 (\Lambda^{}_{\mu\tau}) \; ,
\nonumber \\
m^{}_3 (\Lambda^{}_{\rm EW}) & \simeq & I^{}_\kappa
\Big[ 1 + \Delta^{}_\tau \hspace{0.05cm} c^2_{13} \Big]
m^{}_3 (\Lambda^{}_{\mu\tau}) \; ,
\label{eq:mass-evolution}
\end{eqnarray}
where $I^{}_\kappa$, $\Delta^{}_\tau$, $\theta^{}_{12}$ and $\theta^{}_{13}$ take their
values at $\Lambda^{}_{\rm EW}$. The RGE-induced corrections to three flavor
mixing angles can be expressed as
\begin{eqnarray}
\Delta \theta^{}_{12} & \simeq & -\frac{\Delta^{}_\tau}{2} c^{}_{12} s^{}_{12}
\Big[ c^2_{13} \hspace{0.05cm} \zeta^{-\eta^{}_\rho \eta^{}_\sigma}_{21}
+ s^2_{13} \left(\zeta^{\eta^{}_\rho}_{31} - \zeta^{\eta^{}_\sigma}_{32}
\right)\Big] \; ,
\nonumber \\
\Delta \theta^{}_{13} & \simeq & -\frac{\Delta^{}_\tau}{2} c^{}_{13} s^{}_{13}
\Big[ c^2_{12} \hspace{0.05cm} \zeta^{\eta^{}_\rho}_{31} + s^2_{12} \hspace{0.05cm}
\zeta^{\eta^{}_\sigma}_{32}\Big] \; ,
\nonumber \\
\Delta \theta^{}_{23} & \simeq & -\frac{\Delta^{}_\tau}{2}
\Big[ s^2_{12} \hspace{0.05cm} \zeta^{-\eta^{}_\rho}_{31} + c^2_{12} \hspace{0.05cm}
\zeta^{-\eta^{}_\sigma}_{32}\Big] \; ,
\label{eq:angle-evolution}
\end{eqnarray}
where $\zeta^{}_{ij} \equiv \left(m^{}_i - m^{}_j\right)/\left(m^{}_i + m^{}_j\right)
= \Delta m^2_{ij}/\left(m^{}_i + m^{}_j\right)^2$ at the electroweak scale (for
$ij = 21, 31, 32$), $\eta^{}_\rho \equiv \cos 2\rho = \pm 1$
and $\eta^{}_\sigma \equiv \cos 2\sigma = \pm 1$ denote two
possible options of $\rho (\Lambda^{}_{\mu\tau})$ and $\sigma (\Lambda^{}_{\mu\tau})$
in the $\mu$-$\tau$ reflection symmetry limit. In the same approximations the deviations
of three CP-violating phases between $\Lambda^{}_{\rm EW}$ and $\Lambda^{}_{\mu\tau}$
are found to be
\begin{eqnarray}
&& \Delta \delta^{}_\nu \simeq \Delta^{}_\tau \frac{c^{}_{12} s^{}_{12}
\eta^{}_\nu}{2 s^{}_{13}}
\Big[\zeta^{-\eta^{}_\sigma}_{32} - \zeta^{-\eta^{}_\rho}_{31}
- \frac{s^{2}_{13}}{c^{2}_{12} s^{2}_{12}}
\left(c^4_{12} \hspace{0.05cm} \zeta^{-\eta^{}_\sigma}_{32}
- s^4_{12} \hspace{0.05cm} \zeta^{-\eta^{}_\rho}_{31}
+ \zeta^{\eta^{}_\rho \eta^{}_\sigma}_{21}\right)\Big] \; ,
\nonumber \\
&& \Delta \rho \simeq \Delta^{}_\tau \frac{c^{}_{12} s^{}_{13} \eta^{}_\nu}{s^{}_{12}}
\Big[s^2_{12} \left(\zeta^{-\eta^{}_\rho}_{31} - \zeta^{-\eta^{}_\sigma}_{32}\right)
+ \frac{1}{2} \left(\zeta^{-\eta^{}_\sigma}_{32} + \zeta^{\eta^{}_\rho
\eta^{}_\sigma}_{21}\right) \Big] \; ,
\nonumber \\
&& \Delta \sigma \simeq \Delta^{}_\tau \frac{s^{}_{12} s^{}_{13} \eta^{}_\nu}{2 c^{}_{12}}
\Big[c^2_{12} \left(\zeta^{-\eta^{}_\rho}_{31} - 2\zeta^{-\eta^{}_\sigma}_{32}\right)
- s^2_{12} \hspace{0.05cm} \zeta^{-\eta^{}_\rho}_{31}
+ \zeta^{\eta^{}_\rho \eta^{}_\sigma}_{21} \Big] \; ,
\label{eq:phase-evolution}
\end{eqnarray}
where $\eta^{}_\nu \equiv \sin\delta^{}_\nu = \pm 1$ stands for two possible options
of $\delta^{}_\nu (\Lambda^{}_{\mu\tau})$ in the $\mu$-$\tau$ reflection symmetry limit.
Taking $\delta^{}_\nu = -\pi/2$ and $C^{}_l = 1$ in the MSSM, one may use
Eqs.~(\ref{eq:mass-evolution}), (\ref{eq:angle-evolution}) and (\ref{eq:phase-evolution})
to reproduce the corresponding results given in Refs.~\cite{Huang:2018wqp,Huang:2020kgt}
\footnote{Note that the sign of $\Delta^{}_\tau$ defined in
Refs.~\cite{Huang:2018wqp,Huang:2020kgt} is opposite to the one defined here in
Eq.~(\ref{eq:Delta-tau}).}.

Some brief comments on what we have obtained in Eqs.~(\ref{eq:mass-evolution}),
(\ref{eq:angle-evolution}) and (\ref{eq:phase-evolution}) are in order.
(1) The smallness of $\Delta^{}_\tau$ assures that the three neutrino masses
change from $\Lambda^{}_{\mu\tau}$ down to $\Lambda^{}_{\rm EW}$ almost to the
same extent --- a generic one-loop observation which is actually independent of any
specific patterns of the PMNS matrix (see, e.g., Refs.~\cite{Casas:1999tg,Antusch:2003kp,
Antusch:2005gp,Mei:2005qp}). (2) Among the three flavor mixing angles, the RGE running
effect of $\theta^{}_{12}$ will be most significant provided
$\eta^{}_\rho \eta^{}_\sigma = 1$ is satisfied because in this case
$\Delta\theta^{}_{12} \propto \zeta^{-1}_{21} = \left(m^{}_1 + m^{}_2\right)^2/
\Delta m^2_{21}$ can be enhanced by the smallness of $\Delta m^2_{21}$. In comparison,
the evolution of $\theta^{}_{13}$ is most insignificant as $\Delta\theta^{}_{13}$ is
suppressed by both the smallness of $\Delta^{}_\tau$ and the smallness of $s^{}_{13}$.
(3) Among the three CP-violating phases, $\delta^{}_\nu$ may have a relatively
appreciable running effect because of $\Delta\delta^{}_\nu \propto s^{-1}_{13}$.
Although $\Delta\rho$ and $\Delta\sigma$ are both suppressed by $s^{}_{13}$, they
both contain a term proportional to $\zeta^{\eta^{}_\rho \eta^{}_\sigma}_{21}$
which may offer an enhancement in the $\eta^{}_\rho \eta^{}_\sigma = -1$ case.

To numerically illustrate how appreciable the nine physical quantities in
Eqs.~(\ref{eq:mass-evolution}), (\ref{eq:angle-evolution}) and (\ref{eq:phase-evolution})
can be, however, one has to assume specific values for both the absolute neutrino mass
scale (e.g., the value of $m^{}_1$) and the energy scale $\Lambda^{}_{\mu\tau}$. In
the MSSM case a specific value of $\tan\beta$ should also be assumed. Note that
$\Delta^{}_\tau \propto C^{}_l$ with $C^{}_l = -3/2$ (SM) or $1$ (MSSM),
and thus $\Delta\theta^{}_{ij}$ (for $ij = 12, 13, 23$), $\Delta\delta^{}_\nu$,
$\Delta\rho$ and $\Delta\sigma$ have the opposite signs in the SM and MSSM frameworks.
This observation implies that the octant and quadrant issues of $\theta^{}_{23}$ and
$\delta^{}_\nu$ are likely to be resolved with the help of the RGE-induced $\mu$-$\tau$
reflection symmetry breaking effects in either the SM framework or the MSSM
framework (see, e.g., Refs.~\cite{Luo:2014upa,Huang:2018wqp,Huang:2020kgt}),
depending upon the signs and sizes of $\Delta\theta^{}_{23}$ and $\Delta\delta^{}_\nu$
as shown in Figure~\ref{Fig:octant}. Here let us quote an instructive numerical example
presented in Ref.~\cite{Huan:2018lzd} for the {\it normal} neutrino mass ordering, namely
Figures~\ref{Fig:theta23} and \ref{Fig:delta}, as this mass ordering is currently
favored over the {\it inverted} one almost at the $3\sigma$ level.
These two figures illustrate the allowed regions
of $\theta^{}_{23}$ and $\delta^{}_\nu$ at $\Lambda^{}_{\rm EW}$ as functions of
$m^{}_1 \in [0, 0.1]$ eV and $\tan\beta \in [10, 50]$ in the MSSM framework with
$m^{}_1 < m^{}_2 < m^{}_3$, where
$\Lambda^{}_{\mu\tau} \sim 10^{14}~{\rm GeV}$ has been taken and the best-fit values
plus $1\sigma$ ranges of six neutrino oscillation parameters obtained in
Ref.~\cite{Capozzi:2018ubv} have been typically input.
\begin{figure}[t!]
\begin{center}
\includegraphics[width=15.4cm]{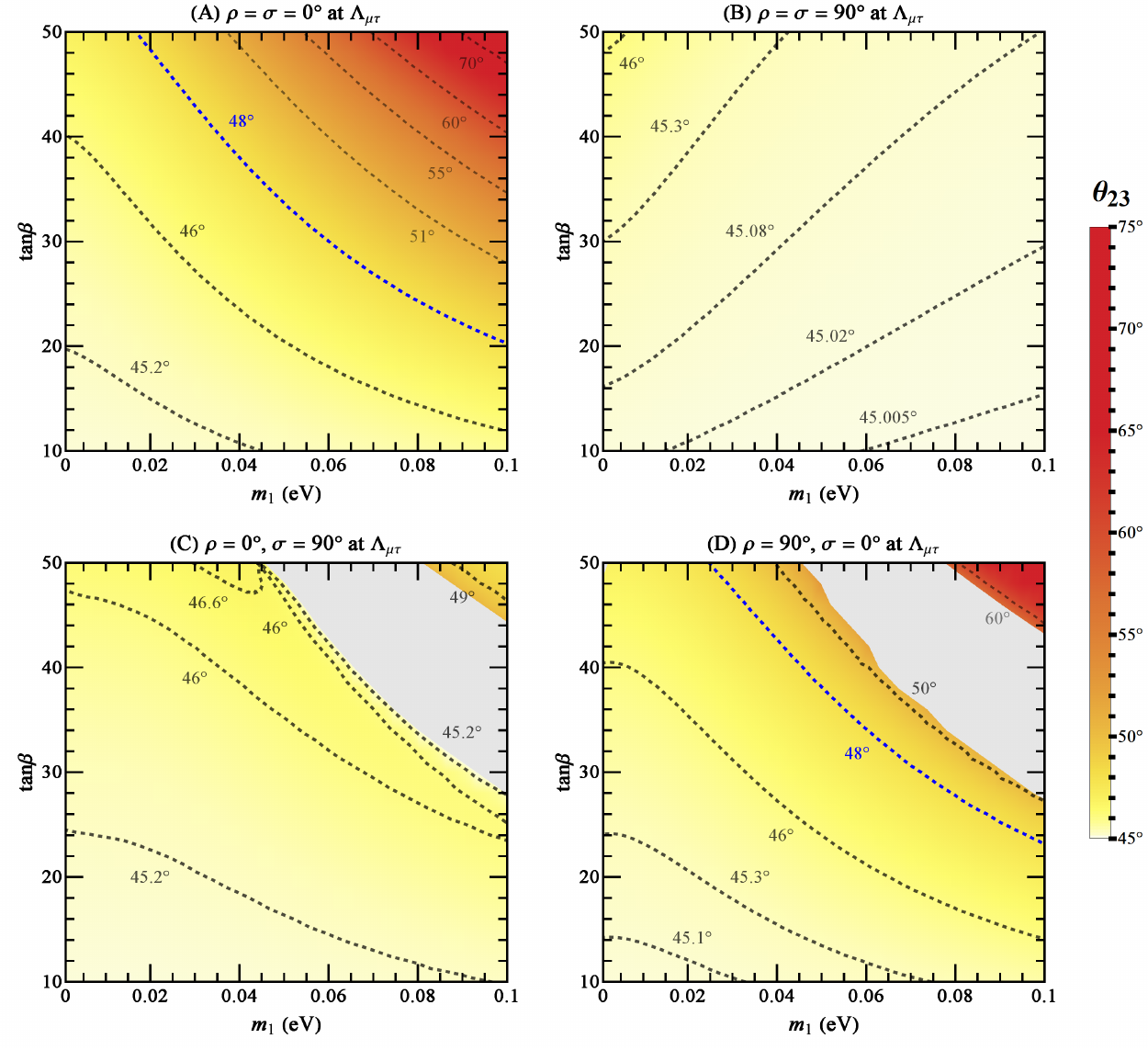}
\vspace{-0.3cm}
\caption{Majorana neutrinos: the allowed region of $\theta^{}_{23}$ at
$\Lambda^{}_{\rm EW}$ as functions of $m^{}_1 \in [0, 0.1]$ eV and
$\tan\beta \in [10, 50]$ in the MSSM with $m^{}_1 < m^{}_2 < m^{}_3$,
arising from the RGE-induced breaking of $\mu$-$\tau$ reflection symmetry at
$\Lambda^{}_{\mu\tau} \sim 10^{14}$ GeV \cite{Huan:2018lzd}. Here the
dashed curves are the contours for some typical values of $\theta^{}_{23}$,
the blue one is compatible with the best-fit result of $\theta^{}_{23}$
obtained in Ref.~\cite{Capozzi:2018ubv}, and the grey-shaded region is
essentially excluded in this analysis.}
\label{Fig:theta23}
\end{center}
\end{figure}
\begin{figure}[t!]
\begin{center}
\includegraphics[width=15.4cm]{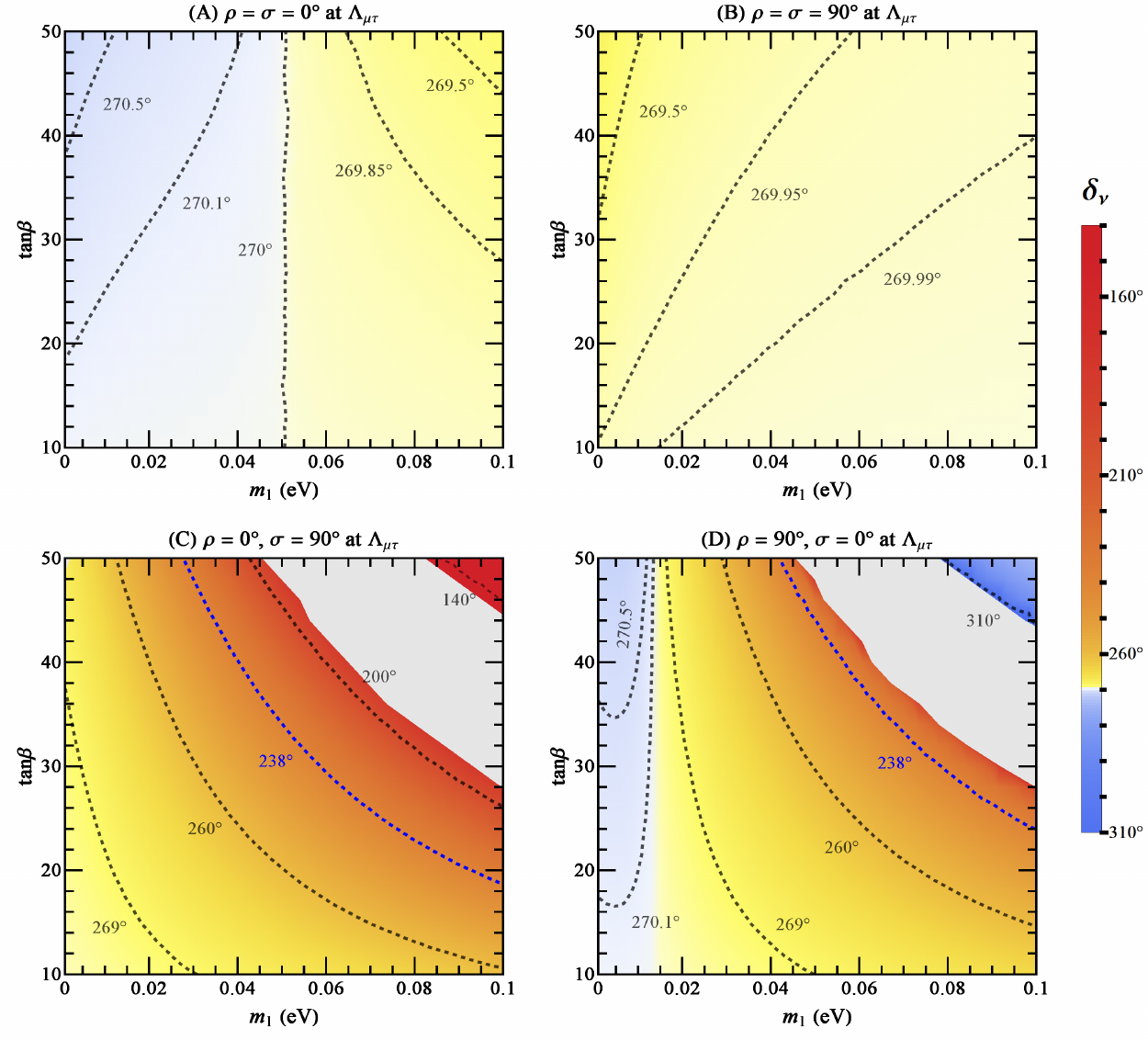}
\vspace{-0.25cm}
\caption{Majorana neutrinos: the allowed region of $\delta^{}_\nu$ at
$\Lambda^{}_{\rm EW}$ as functions of $m^{}_1 \in [0, 0.1]$ eV and
$\tan\beta \in [10, 50]$ in the MSSM with $m^{}_1 < m^{}_2 < m^{}_3$,
arising from the RGE-induced breaking of $\mu$-$\tau$ reflection symmetry
at $\Lambda^{}_{\mu\tau} \sim 10^{14}$ GeV \cite{Huan:2018lzd}.
Here the dashed curves are the contours for some typical values of $\delta^{}_\nu$,
the blue one is compatible with the best-fit result of $\delta^{}_\nu$
obtained in Ref.~\cite{Capozzi:2018ubv}, and the grey-shaded region is
essentially excluded in this analysis.}
\label{Fig:delta}
\end{center}
\end{figure}

\subsection{RGE-induced effects on Dirac neutrinos}
\label{section 5.3}

Although our main interest is in the Majorana nature of three active neutrinos, by which
their small masses can be naturally understood with the help of the seesaw mechanisms,
it also makes sense to consider the possibility that the three known neutrinos may simply
have the Dirac nature and their mass term respects the $\mu$-$\tau$ reflection symmetry
at a superhigh energy scale $\Lambda^{}_{\mu\tau}$. So far a lot of attention has been
paid to building natural models of Dirac neutrinos and exploring their phenomenology
(see, e.g., Refs.~\cite{Luo:2008yc,Mohapatra:1986bd,Arkani-Hamed:2000oup,Abel:2004tt,
Ko:2005sh,Hagedorn:2005kz,Memenga:2013vc,Liu:2012axa,Ding:2013eca,Aranda:2013gga,
Esmaili:2015pna,Zhang:2015wua,Ma:2016mwh,Borah:2017leo,Wang:2016lve}). Before the massive
neutrinos are experimentally verified to be the Majorana particles, it certainly makes
sense to study the Dirac neutrinos and their flavor textures.

Here we assume that the finite masses of three active neutrinos simply arise from the
following standard Yukawa interactions:
\begin{eqnarray}
-{\cal L}^{}_{\rm Dirac} = \overline{\ell^{}_{\rm L}} \hspace{0.05cm}
Y^{}_{\nu} \widetilde{H} N^{}_{\rm R} + {\rm h.c.} \; ,
\label{eq:Dirac}
\end{eqnarray}
which is exactly the same as the first term in Eq.~(\ref{eq:seesaw1}).
In the MSSM framework the charged lepton and neutrino sectors are associated with
the Higgs doublets $H^{}_1$ (with the hypercharge $+1/2$ and the vacuum expectation
value $v\cos\beta/\sqrt{2}$) and $H^{}_2$ (with the hypercharge $-1/2$ and the
vacuum expectation value $v \sin\beta/\sqrt{2}$), respectively~\cite{Mei:2003gn}.
After spontaneous electroweak symmetry breaking, ${\cal L}^{}_{\rm Dirac}$ in
Eq.~(\ref{eq:Dirac}) becomes
\begin{eqnarray}
-{\cal L}^{\prime}_{\rm Dirac} = \overline{\nu^{}_{\rm L}} \hspace{0.05cm}
M^{}_{\rm D} N^{}_{\rm R} + {\rm h.c.} \; ,
\label{eq:Dirac-mass}
\end{eqnarray}
where $M^{}_{\rm D} = Y^{}_{\nu}\langle H\rangle = Y^{}_\nu v/\sqrt{2}$
with $v \simeq 246$ GeV. As pointed out below Eq.~(\ref{eq:seesaw-mass-matrix}),
the Dirac neutrino mass matrix $M^{}_{\rm D}$ is in general neither Hermitian nor
symmetric. To constrain the arbitrary flavor texture of $M^{}_{\rm D}$, the $\mu$-$\tau$
reflection symmetry can be invoked as a good example~\cite{Xing:2017mkx}. The key
point is to require the Dirac mass term ${\cal L}^{\prime}_{\rm Dirac}$ to keep
invariant under the flavor-changing CP transformations
\begin{eqnarray}
\nu^{}_{\rm L} \to {\cal P} (\nu^{}_{\rm L})^c \; , \quad
N^{}_{\rm R} \to {\cal P} (N^{}_{\rm R})^c \; ,
\label{eq:Dirac-reflection}
\end{eqnarray}
in which the orthogonal $(\mu, \tau)$-associated permutation matrix has been given
in Eq.~(\ref{eq:S-tranformation}). Then $M^{}_{\rm D}$ has to satisfy the
relation $M^{}_{\rm D} = {\cal P} M^*_{\rm D} {\cal P}$, and thus its flavor texture
can be constrained to the explicit form shown in Eq.~(\ref{eq:seesaw-constraint2}).
One may diagonalize $M^{}_{\rm D}$ with the help of a bi-unitary transformation of
the form $U^\dagger M^{}_{\rm D} U^\prime = D^{}_\nu$, where $U$ is just the PMNS
matrix in the basis that allows $M^{}_l = D^{}_l$ to hold for the three charged
leptons. As a result, we have
\begin{eqnarray}
H^{}_\nu \equiv M^{}_{\rm D} M^{\dagger}_{\rm D} = U D^2_\nu U^\dagger
= {\cal P} H^*_\nu {\cal P} \; ,
\label{eq:H}
\end{eqnarray}
whose nine matrix elements satisfy both the Hermitian conditions
$\langle H\rangle^{}_{\beta\alpha} = \langle H\rangle^{*}_{\alpha\beta}$ (for
$\alpha, \beta = e, \mu, \tau$) and the nontrivial
relationships $\langle H\rangle^{}_{e\tau} = \langle H\rangle^{*}_{e\mu}$ and
$\langle H\rangle^{}_{\tau\tau} = \langle H\rangle^{}_{\mu\mu}$~\cite{Xing:2017mkx}.
It is therefore straightforward to show that $\theta^{}_{23} = \pi/4$ and
$\delta^{}_\nu = \pm \pi/2$ exactly hold in the standard parametrization of $U$,
as constrained by the $\mu$-$\tau$ reflection symmetry of
${\cal L}^{\prime}_{\rm Dirac}$ at a given energy scale $\Lambda^{}_{\mu\tau}$.

If $\Lambda^{}_{\mu\tau} \gg \Lambda^{}_{\rm EW}$ holds as naturally expected
from the point of view of model building, the RGE-induced radiative corrections
to $Y^{}_\nu$, $M^{}_{\rm D}$ or $H^{}_\nu$ need to be taken into account. The
one-loop RGE of $Y^{}_\nu$ can be expressed
as~\cite{Cheng:1973nv,Machacek:1983fi,Grzadkowski:1987tf,Lindner:2005as}
\begin{eqnarray}
16\pi^{2} \frac{{\rm d} Y^{}_\nu}{{\rm d} t} = \left[ \alpha^{}_\nu
+ C^{}_{\nu} Y^{}_\nu Y^\dagger_\nu + C^{}_{l} Y^{}_{l}Y^{\dagger}_{l}
\right] Y^{}_\nu \; ,
\label{eq:Dirac-RGE}
\end{eqnarray}
where $t \equiv \ln \left(\mu/\Lambda^{}_{\mu\tau}\right)$,
$C^{}_\nu = -C^{}_l = 3/2$ and $\alpha^{}_\nu \simeq -9 g^2_1/20
-9 g^2_2/4 + 3 y^2_t$ in the SM framework, or
$C^{}_\nu = 3 C^{}_l = 3$ and $\alpha^{}_\nu \simeq -3 g^2_1/5
-3 g^2_2 + 3 y^2_t$ in the MSSM framework. Note that in
Eq.~(\ref{eq:Dirac-RGE}) the $Y^{}_\nu Y^\dagger_\nu$ term is completely
negligible because the neutrino masses are extremely small as
compared with $\langle H\rangle \simeq 174~{\rm GeV}$, and
$Y^{}_l Y^\dagger_l = {\rm Diag}\{y^2_e, y^2_\mu, y^2_\tau\}$ holds in
the chosen $M^{}_l = D^{}_l$ basis. In this case the RGE of $H^{}_\nu$
is given by
\begin{eqnarray}
16\pi^{2}_{}\frac{{\rm d} H^{}_\nu}{{\rm d} t} = 2 \alpha^{}_\nu
H^{}_\nu + C^{}_l \left[H^{}_\nu Y^{}_l Y^\dagger_l + Y^{}_l Y^\dagger_l
H^{}_\nu\right] \; ,
\label{eq:H-RGE}
\end{eqnarray}
where we have omitted the tiny $H^2_\nu = \big(Y^{}_\nu Y^\dagger_\nu\big)^2$ term.
The integral solutions to Eqs.~(\ref{eq:Dirac-RGE}) and (\ref{eq:H-RGE}) at
$\Lambda^{}_{\rm EW}$ are found to be~\cite{Xing:2017mkx}
\begin{eqnarray}
M^{}_{\rm D} (\Lambda^{}_{\rm EW}) =
I^{}_\nu (\Lambda^{}_{\rm EW}) \left[T^{}_l (\Lambda^{}_{\rm EW})
\cdot M^{}_{\rm D} (\Lambda^{}_{\mu\tau}) \right] \; ,
\label{eq:Dirac-integral-M}
\end{eqnarray}
and
\begin{eqnarray}
H^{}_\nu (\Lambda^{}_{\rm EW}) =
I^{2}_\nu (\Lambda^{}_{\rm EW}) \left[T^{}_l (\Lambda^{}_{\rm EW})
\cdot H^{}_\nu (\Lambda^{}_{\mu\tau}) \cdot T^{}_l (\Lambda^{}_{\rm EW}) \right] \; ,
\label{eq:Dirac-integral-M}
\end{eqnarray}
where $T^{}_l = {\rm Diag}\{I^{}_e , I^{}_\mu , I^{}_\tau\}$, and the
scale-dependent one-loop RGE evolution functions $I^{}_\nu$ and
$I^{}_\alpha$ (for $\alpha = e, \mu, \tau$) are defined as~\cite{Zhang:2020lsd}
\begin{eqnarray}
I^{}_\nu (\mu) & \equiv & \exp\left[-\frac{1}{16 \pi^2}
\int^{\ln(\Lambda^{}_{\mu\tau}/\Lambda^{}_{\rm EW})}_{\ln(\mu/\Lambda^{}_{\rm EW})}
\alpha^{}_\nu(t) \hspace{0.05cm} {\rm d} t\right] \; ,
\nonumber \\
I^{}_\alpha (\mu) & \equiv & \exp\left[-\frac{C^{}_l}{16 \pi^2}
\int^{\ln(\Lambda^{}_{\mu\tau}/\Lambda^{}_{\rm EW})}_{\ln(\mu/\Lambda^{}_{\rm EW})}
y^2_\alpha(t) \hspace{0.05cm} {\rm d} t\right] \; .
\label{eq:Dirac-evolution-function}
\end{eqnarray}
Here again it is an excellent approximation to take $I^{}_e \simeq I^{}_\mu \simeq 1$
and $I^{}_\tau = 1 + \Delta^{}_\tau$ with
\begin{eqnarray}
\Delta^{}_\tau (\mu) \simeq -\frac{C^{}_l}{16 \pi^2}
\int^{\ln(\Lambda^{}_{\mu\tau}/\Lambda^{}_{\rm EW})}_{\ln(\mu/\Lambda^{}_{\rm EW})}
y^2_\tau(t) \hspace{0.05cm} {\rm d} t \;
\label{eq:Dirac-Delta-tau}
\end{eqnarray}
being a small quantity at $\mu = \Lambda^{}_{\rm EW}$ in the SM or in the MSSM
with $\tan\beta \lesssim 30$ for the Dirac neutrinos. Figure~\ref{Fig:loop-functions}
illustrates how $I^{}_\nu$ and $\Delta^{}_\tau$ evolve with the energy scale $\mu$.

Taking the parametrization of $U$ as given in Eq.~(\ref{eq:U}), let us make use of
Eq.~(\ref{eq:Dirac-integral-M}) to calculate $\Delta\theta^{}_{ij}$
(for $ij = 12, 13, 23$) and $\Delta\delta^{}_\nu$ that have been defined
in Eq.~(\ref{eq:Deviation}) for the Dirac neutrinos running from the
$\mu$-$\tau$ reflection symmetry scale $\Lambda^{}_{\mu\tau}$ down to
the electroweak scale $\Lambda^{}_{\rm EW}$. We obtain the neutrino
masses~\cite{Xing:2017mkx}
\footnote{Note that the sign of $\Delta^{}_\tau$ defined in Ref.~\cite{Xing:2017mkx} is
opposite to the one defined here in Eq.~(\ref{eq:Dirac-Delta-tau}).}
\begin{eqnarray}
m^{}_1 (\Lambda^{}_{\rm EW}) & \simeq & I^{}_\nu
\left[ 1 + \frac{\Delta^{}_\tau}{2} \left( 1 - c^2_{12} c^2_{13}
\right) \right] m^{}_1 (\Lambda^{}_{\mu\tau}) \; ,
\nonumber \\
m^{}_2 (\Lambda^{}_{\rm EW}) & \simeq & I^{}_\nu
\left[ 1 + \frac{\Delta^{}_\tau}{2} \left( 1 - s^2_{12} c^2_{13}
\right) \right] m^{}_2 (\Lambda^{}_{\mu\tau}) \; ,
\nonumber \\
m^{}_3 (\Lambda^{}_{\rm EW}) & \simeq & I^{}_\nu
\left[ 1 + \frac{\Delta^{}_\tau}{2} \hspace{0.05cm} c^2_{13}
\right] m^{}_3 (\Lambda^{}_{\mu\tau}) \; ,
\label{eq:Dirac-mass-evolution}
\end{eqnarray}
in which $\theta^{}_{12}$ and $\theta^{}_{13}$ take their values at
$\Lambda^{}_{\rm EW}$. Moreover, the RGE-induced corrections to $\theta^{}_{ij}$ and
$\delta^{}_\nu$ are found to be
\begin{eqnarray}
\Delta \theta^{}_{12} & \simeq & -\frac{\Delta^{}_\tau}{2} c^{}_{12} s^{}_{12}
\Big[ c^2_{13} \hspace{0.05cm} \xi^{}_{21} + s^2_{13} \left(\xi^{}_{31} - \xi^{}_{32}
\right)\Big] \; ,
\nonumber \\
\Delta \theta^{}_{13} & \simeq & -\frac{\Delta^{}_\tau}{2} c^{}_{13} s^{}_{13}
\Big[ c^2_{12} \hspace{0.05cm} \xi^{}_{31} + s^2_{12} \hspace{0.05cm}
\xi^{}_{32}\Big] \; ,
\nonumber \\
\Delta \theta^{}_{23} & \simeq & -\frac{\Delta^{}_\tau}{2}
\Big[ s^2_{12} \hspace{0.05cm} \xi^{}_{31} + c^2_{12} \hspace{0.05cm}
\xi^{}_{32}\Big] \; ,
\label{eq:Dirac-angle-evolution}
\end{eqnarray}
and
\begin{eqnarray}
\Delta \delta^{}_\nu \simeq \Delta^{}_\tau \frac{c^{}_{12} s^{}_{12}
\eta^{}_\nu}{2 s^{}_{13}} \Big[\xi^{}_{32} - \xi^{}_{31}
- \frac{s^{2}_{13}}{c^{2}_{12} s^{2}_{12}}
\left(c^4_{12} \hspace{0.05cm} \xi^{}_{32} - s^4_{12} \hspace{0.05cm} \xi^{}_{31}
+ \xi^{}_{21}\right)\Big] \; ,
\label{eq:Dirac-phase-evolution}
\end{eqnarray}
where $\xi^{}_{ij} \equiv \left(m^{2}_i + m^{2}_j\right)/\left(m^{2}_i -
m^{2}_j\right) = \left(m^{2}_i + m^{2}_j\right)/\Delta m^2_{ij}$ (for
$i, j = 1, 2, 3$) with the values of three neutrino masses being input at
$\Lambda^{}_{\rm EW}$, and $\eta^{}_\nu \equiv \sin\delta^{}_\nu = \pm 1$
denote two options of $\delta^{}_\nu (\Lambda^{}_{\mu\tau})$ in the
$\mu$-$\tau$ reflection symmetry limit.

Eq.~(\ref{eq:Dirac-mass-evolution}) tells us that the smallness of
$\Delta^{}_\tau$ assures the neutrino masses to evolve from
$\Lambda^{}_{\mu\tau}$ down to $\Lambda^{}_{\rm EW}$ almost to the same extent.
Among the three flavor mixing angles, $\theta^{}_{12}$ is most sensitive to
the RGE-induced running effect because $\Delta\theta^{}_{12}$ is dominated
by the term proportional to $\xi^{}_{21} = \left(m^{}_1 + m^{}_2\right)^2/
\Delta m^2_{21}$ and can be enhanced by the smallness of $\Delta m^2_{21}$.
In contrast, $\theta^{}_{13}$ is most insensitive to the radiative correction
since $\Delta\theta^{}_{13}$ is suppressed not only by the smallness of
$\Delta^{}_\tau$ but also by the smallness of $s^{}_{13}$. As can be seen from
Eq.~(\ref{eq:Dirac-phase-evolution}), the CP-violating phase $\delta^{}_\nu$
may have a relatively appreciable running effect as a result of
$\Delta\delta^{}_\nu \propto s^{-1}_{13}$. To illustrate, we quote the
numerical example given in Ref.~\cite{Huan:2018lzd} --- Figure~\ref{Fig:Dirac}
shows the allowed regions of $\theta^{}_{23}$ and $\delta^{}_\nu$ at
$\Lambda^{}_{\rm EW}$ as the functions of $m^{}_1$ and $\tan\beta$ in the MSSM
framework with $m^{}_1 < m^{}_2 < m^{}_3$. In this case it is possible to
attribute the observed values of $\theta^{}_{23}$ and
$\delta^{}_\nu$ at low energies to the RGE-induced $\mu$-$\tau$ reflection
symmetry breaking at a superhigh energy scale.
\begin{figure}[t!]
\begin{center}
\includegraphics[width=15.4cm]{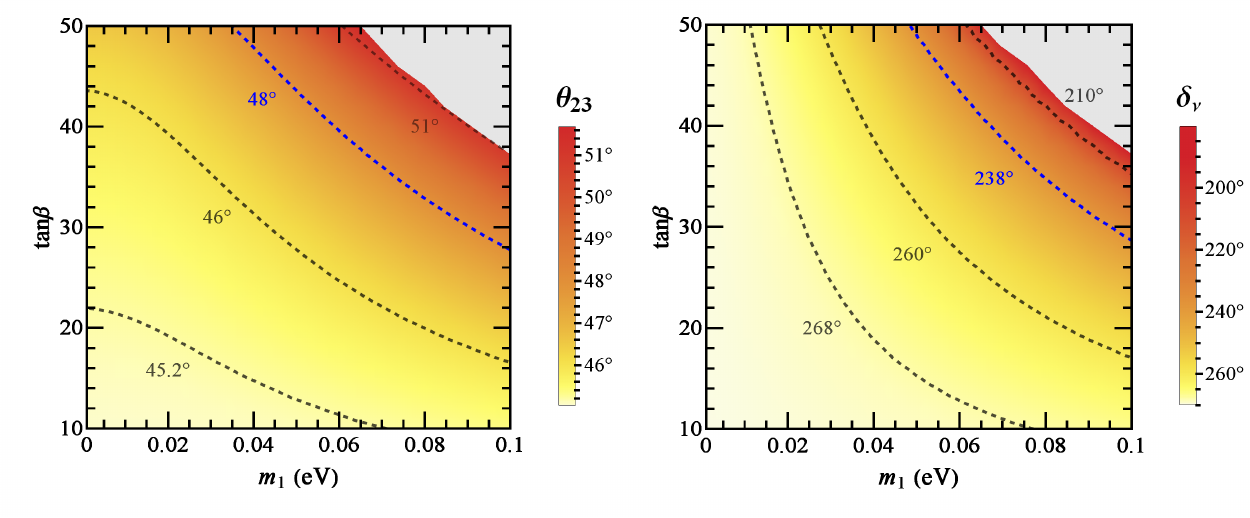}
\vspace{-0.4cm}
\caption{Dirac neutrinos: the allowed regions of $\theta^{}_{23}$ and $\delta^{}_\nu$
at $\Lambda^{}_{\rm EW}$ as functions of $m^{}_1 \in [0, 0.1]$ eV and
$\tan\beta \in [10, 50]$ in the MSSM with $m^{}_1 < m^{}_2 < m^{}_3$,
arising from the RGE-induced breaking of $\mu$-$\tau$ reflection symmetry
at $\Lambda^{}_{\mu\tau} \sim 10^{14}$ GeV \cite{Huan:2018lzd}.
Here the dashed curves are the contours for some typical values of $\theta^{}_{23}$
and $\delta^{}_\nu$, the blue ones are compatible with the best-fit results
of $\theta^{}_{23}$ and $\delta^{}_\nu$ obtained in Ref.~\cite{Capozzi:2018ubv},
and the grey-shaded region is essentially excluded in this analysis.}
\label{Fig:Dirac}
\end{center}
\end{figure}

\subsection{Simplified RGEs when $m^{}_1 = 0$ or $m^{}_3 = 0$}
\label{section 5.4}

Since the absolute neutrino mass scale has not been determined from any of
the available experiments, current neutrino oscillation data {\it do} allow the
possibility of either $m^{}_1 = 0$ or $m^{}_3 = 0$. On the other hand, some
neutrino mass models and specific flavor symmetries may also dictate $m^{}_1$ or
$m^{}_3$ to vanish or vanishingly small at a given energy scale
(see the examples given in section~\ref{section 3}, section~\ref{section 6.4}
and section~\ref{section 7.1}, and a recent review in Ref.~\cite{Xing:2020ald}).
In this case one of the Majorana CP-violating phases can be rotated away, and
therefore the diagonal Majorana phase matrix $P^{}_\nu$ in Eq.~(\ref{eq:U}) is
simplified to $P^{}_\nu = {\rm Diag}\{1, e^{{\rm i}\sigma}, 1\}$ no matter whether
$m^{}_1 = 0$ or $m^{}_3 = 0$. Taking account of the constraints of $\mu$-$\tau$
reflection symmetry on the PMNS matrix $U$ together with $m^{}_1 = 0$ or
$m^{}_3 = 0$ at a superhigh energy scale $\Lambda^{}_{\mu\tau}$, one may
simplify the one-loop RGEs obtained in sections~\ref{section 5.2} and
\ref{section 5.3} to arrive at some much simpler results for the effects of soft
symmetry breaking at $\Lambda^{}_{\rm EW}$. Let us elaborate on the relevant
RGEs below.

As for the Majorana neutrinos with either $\nu^{}_1$ or $\nu^{}_3$ being massless,
the RGE-induced $\mu$-$\tau$ reflection symmetry breaking effects at the one-loop
level can be summarized as
\begin{eqnarray}
m^{}_1 (\Lambda^{}_{\rm EW}) \simeq 0 \; ,
\nonumber \\
m^{}_2 (\Lambda^{}_{\rm EW}) \simeq
I^{}_\kappa \Big[ 1 + \left( 1 - s^2_{12} c^2_{13} \right) \Delta^{}_\tau
\Big] m^{}_2 (\Lambda^{}_{\mu\tau}) \; ,
\nonumber \\
m^{}_3 (\Lambda^{}_{\rm EW}) \simeq
I^{}_\kappa \Big[ 1 + c^2_{13} \Delta^{}_\tau \Big]
m^{}_3 (\Lambda^{}_{\mu\tau}) \; ,
\nonumber \\
\Delta \theta^{}_{12} \simeq -\frac{1}{2} c^{}_{12} s^{}_{12}
\left(1 - s^2_{13} \hspace{0.05cm} \zeta^{\eta^{}_\sigma}_{32} \right)
\Delta^{}_\tau \; ,
\nonumber \\
\Delta \theta^{}_{13} \simeq -\frac{1}{2} c^{}_{13} s^{}_{13}
\left(c^2_{12} + s^2_{12} \hspace{0.05cm} \zeta^{\eta^{}_\sigma}_{32}\right)
\Delta^{}_\tau \; ,
\nonumber \\
\Delta \theta^{}_{23} \simeq -\frac{1}{2} \left(s^2_{12} +
c^2_{12} \hspace{0.05cm} \zeta^{-\eta^{}_\sigma}_{32}\right)
\Delta^{}_\tau \;
\nonumber \\
\Delta \delta^{}_\nu \simeq -\frac{c^{}_{12} s^{}_{12} \eta^{}_\delta}{2 s^{}_{13}}
\left[1 - \zeta^{-\eta^{}_\sigma}_{32} + \frac{s^{2}_{13}}{c^{2}_{12} s^{2}_{12}}
\left(1 - s^4_{12} + c^4_{12} \hspace{0.05cm} \zeta^{-\eta^{}_\sigma}_{32}
\right)\right] \Delta^{}_\tau \; , \hspace{0.5cm}
\nonumber \\
\Delta \sigma \simeq c^{}_{12} s^{}_{12} s^{}_{13} \eta^{}_\delta
\left(1 - \zeta^{-\eta^{}_\sigma}_{32}\right) \Delta^{}_\tau \;
\label{eq:Majorana-m1=0}
\end{eqnarray}
in the $m^{}_1 (\Lambda^{}_{\mu\tau}) = 0$ case; and
\begin{eqnarray}
m^{}_1 (\Lambda^{}_{\rm EW}) \simeq I^{}_\kappa
\Big[ 1 + \left( 1 - c^2_{12} c^2_{13} \right) \Delta^{}_\tau
\Big] m^{}_1 (\Lambda^{}_{\mu\tau}) \; ,
\nonumber \\
m^{}_2 (\Lambda^{}_{\rm EW}) \simeq I^{}_\kappa
\Big[ 1 + \left( 1 - s^2_{12} c^2_{13} \right) \Delta^{}_\tau
\Big] m^{}_2 (\Lambda^{}_{\mu\tau}) \; ,
\nonumber \\
m^{}_3 (\Lambda^{}_{\rm EW}) \simeq 0 \; ,
\nonumber \\
\Delta \theta^{}_{12} \simeq -\frac{1}{2} c^{}_{12} s^{}_{12}
c^2_{13} \hspace{0.05cm} \zeta^{-\eta^{}_\sigma}_{21}
\Delta^{}_\tau \; ,
\nonumber \\
\Delta \theta^{}_{13} \simeq + \frac{1}{2} c^{}_{13} s^{}_{13}
\Delta^{}_\tau \; ,
\nonumber \\
\Delta \theta^{}_{23} \simeq +\frac{1}{2} \Delta^{}_\tau \; ,
\nonumber \\
\Delta \delta^{}_\nu \simeq \frac{s^{}_{13} \eta^{}_\delta}{2 c^{}_{12} s^{}_{12}}
\hspace{0.05cm} \big(1 - 2 s^2_{12} - \zeta^{\eta^{}_\sigma}_{21}\big)
\hspace{0.05cm} \Delta^{}_\tau \; ,
\nonumber \\
\Delta \sigma \simeq \frac{s^{}_{12} s^{}_{13} \eta^{}_\delta}{2 c^{}_{12}}
\hspace{0.05cm}\big(1 + \zeta^{\eta^{}_\sigma}_{21}\big) \hspace{0.05cm}
\Delta^{}_\tau \;
\label{eq:Majorana-m3=0}
\end{eqnarray}
in the $m^{}_3 (\Lambda^{}_{\mu\tau}) = 0$ case, where $I^{}_\kappa$
and $\Delta^{}_\tau$ are defined respectively in
Eqs.~(\ref{eq:evolution-function}) and (\ref{eq:Delta-tau}), and
$\zeta^{}_{ij} \equiv \left(m^{}_i - m^{}_j\right)
/\left(m^{}_i + m^{}_j\right)$ is defined at $\Lambda^{}_{\rm EW}$.
Note that $I^{}_\kappa$, $\Delta^{}_\tau$, $\theta^{}_{12}$ and
$\theta^{}_{13}$ in Eqs.~(\ref{eq:Majorana-m1=0}) and (\ref{eq:Majorana-m3=0})
take their values at $\Lambda^{}_{\rm EW}$. In addition,
$\theta^{}_{23}(\Lambda^{}_{\mu\tau}) = \pi/4$,
$\delta^{}_\nu(\Lambda^{}_{\mu\tau}) = \pm\pi/2$ and
$\sigma(\Lambda^{}_{\mu\tau}) = 0$ or $\pi/2$ have been input, and
thus $\eta^{}_\delta \equiv \sin\delta^{}_\nu = \pm 1$ and
$\eta^{}_\sigma \equiv \cos 2\sigma = \pm 1$ are defined at
$\Lambda^{}_{\mu\tau}$ as in section~\ref{section 5.2}.
Some discussions are in order.
\begin{itemize}
\item     The lightest neutrino mass keeps zero during the one-loop RGE
evolution from $\Lambda^{}_{\mu\tau}$ down to $\Lambda^{}_{\rm EW}$, as well
known (see, e.g., Refs.~\cite{Ohlsson:2013xva,Antusch:2005gp,Mei:2005qp,
Casas:1999tg,Mei:2003gn}). The two nonzero neutrino masses acquire the RGE-induced
corrections which are all proportional to $I^{}_\kappa$ in either the
$m^{}_1 (\Lambda^{}_{\mu\tau}) = 0$ case or the $m^{}_3 (\Lambda^{}_{\mu\tau}) = 0$
case.
\item     Among the three flavor mixing angles, the RGE-induced correction to
$\theta^{}_{13}$ is always suppressed by the smallness of $s^{}_{13}$ no matter
whether the neutrino mass ordering is normal or inverted. The magnitude of
$\Delta \theta^{}_{12}$ can be enhanced by
$\zeta^{}_{21} = \left(m^{}_1 + m^{}_2\right)^2/\Delta m^2_{21}$
in the $m^{}_3 (\Lambda^{}_{\mu\tau}) = 0$ case if $\eta^{}_\sigma = +1$ holds,
but it only has a mild running effect in the $m^{}_1 (\Lambda^{}_{\mu\tau}) = 0$
case. In comparison, the magnitude of $\Delta \theta^{}_{23}$ is roughly
$\Delta^{}_\tau/2$ for either $m^{}_1 (\Lambda^{}_{\mu\tau}) = 0$ or
$m^{}_3 (\Lambda^{}_{\mu\tau}) = 0$, but its signs are opposite in these two
different cases.
\item     The RGE-induced correction to $\delta^{}_\nu$ is enhanced by $1/s^{}_{13}$
in the $m^{}_1 (\Lambda^{}_{\mu\tau}) = 0$ case, but it is suppressed by
$s^{}_{13}$ in the $m^{}_3 (\Lambda^{}_{\mu\tau}) = 0$ case. The running behaviors
of $\sigma$ are suppressed by $s^{}_{13}$ in both neutrino mass ordering cases.
\end{itemize}

As for the Dirac neutrinos with either $\nu^{}_1$ or $\nu^{}_3$ being massless,
one may similarly obtain the RGE-induced $\mu$-$\tau$ reflection symmetry breaking
effects by simplifying the one-loop RGEs obtained in Eqs.~(\ref{eq:Dirac-mass-evolution}),
(\ref{eq:Dirac-angle-evolution}) and (\ref{eq:Dirac-phase-evolution}):
\begin{eqnarray}
m^{}_1 (\Lambda^{}_{\rm EW}) \simeq 0 \; ,
\nonumber \\
m^{}_2 (\Lambda^{}_{\rm EW}) \simeq I^{}_\nu
\left[ 1 + \frac{1}{2} \left( 1 - s^2_{12} c^2_{13}
\right) \Delta^{}_\tau \right] m^{}_2 (\Lambda^{}_{\mu\tau}) \; ,
\nonumber \\
m^{}_3 (\Lambda^{}_{\rm EW}) \simeq I^{}_\nu
\left[ 1 + \frac{1}{2} c^2_{13} \Delta^{}_\tau
\right] m^{}_3 (\Lambda^{}_{\mu\tau}) \; ,
\nonumber \\
\Delta \theta^{}_{12} \simeq -\frac{1}{2} c^{}_{12} s^{}_{12}
\left(1 - s^2_{13} \hspace{0.05cm} \xi^{}_{32}\right) \Delta^{}_\tau \; ,
\nonumber \\
\Delta \theta^{}_{13} \simeq -\frac{1}{2} c^{}_{13} s^{}_{13}
\left( c^2_{12} + s^2_{12} \hspace{0.05cm} \xi^{}_{32}\right) \Delta^{}_\tau \; ,
\nonumber \\
\Delta \theta^{}_{23} \simeq -\frac{1}{2}
\left( s^2_{12} + c^2_{12} \hspace{0.05cm} \xi^{}_{32}\right) \Delta^{}_\tau \; ,
\nonumber \\
\Delta \delta^{}_\nu \simeq \frac{c^{}_{12} s^{}_{12}
\eta^{}_\nu}{2 s^{}_{13}} \Big[\xi^{}_{32} - 1
- \frac{s^{2}_{13}}{c^{2}_{12} s^{2}_{12}}
\left(c^4_{12} \hspace{0.05cm} \xi^{}_{32} - s^4_{12} + 1\right)\Big] \Delta^{}_\tau \;
\label{eq:Dirac-m1=0}
\end{eqnarray}
in the $m^{}_1 (\Lambda^{}_{\mu\tau}) = 0$ case; and
\begin{eqnarray}
m^{}_1 (\Lambda^{}_{\rm EW}) \simeq I^{}_\nu
\left[ 1 + \frac{1}{2} \left( 1 - c^2_{12} c^2_{13}
\right) \Delta^{}_\tau \right] m^{}_1 (\Lambda^{}_{\mu\tau}) \; ,
\nonumber \\
m^{}_2 (\Lambda^{}_{\rm EW}) \simeq I^{}_\nu
\left[ 1 + \frac{1}{2} \left( 1 - s^2_{12} c^2_{13}
\right) \Delta^{}_\tau \right] m^{}_2 (\Lambda^{}_{\mu\tau}) \; ,
\nonumber \\
m^{}_3 (\Lambda^{}_{\rm EW}) \simeq 0 \; ,
\nonumber \\
\Delta \theta^{}_{12} \simeq -\frac{1}{2} c^{}_{12} s^{}_{12}
c^2_{13} \hspace{0.05cm} \xi^{}_{21} \Delta^{}_\tau \; ,
\nonumber \\
\Delta \theta^{}_{13} \simeq +\frac{1}{2} c^{}_{13} s^{}_{13}
\Delta^{}_\tau \; ,
\nonumber \\
\Delta \theta^{}_{23} \simeq +\frac{1}{2} \Delta^{}_\tau \; ,
\nonumber \\
\Delta \delta^{}_\nu \simeq -\frac{s^{}_{12} s^{}_{13}
\eta^{}_\nu}{c^{}_{12}} \Delta^{}_\tau \; ,
\label{eq:Dirac-m3=0}
\end{eqnarray}
in the $m^{}_3 (\Lambda^{}_{\mu\tau}) = 0$ case, where
$\xi^{}_{ij} \equiv \left(m^{2}_i + m^{2}_j\right)/\left(m^{2}_i - m^{2}_j\right)$
with the values of $m^{}_i$ and $m^{}_j$ being input at the
electroweak scale $\Lambda^{}_{\rm EW}$, and $\eta^{}_\nu \equiv
\sin\delta^{}_\nu = \pm 1$ represent two options of $\delta^{}_\nu$ in the
$\mu$-$\tau$ reflection symmetry limit at $\Lambda^{}_{\mu\tau}$.
One can easily see that the RGEs of $m^{}_i$ (for $i = 1, 2, 3$),
$\Delta\theta^{}_{ij}$ (for $ij = 12, 13, 23$) and $\Delta\delta^{}_\nu$ for the
Dirac neutrinos are formally similar to those for the Majorana neutrinos in the
$m^{}_1 (\Lambda^{}_{\mu\tau}) = 0$ case, and thus their running behaviors
including possible enhancement or suppression in magnitude are also similar to
those for the Majorana neutrinos. The same is true in the
$m^{}_3 (\Lambda^{}_{\mu\tau}) = 0$ case. Note, however, that the RGEs in
Eqs.~(\ref{eq:Dirac-m1=0}) and (\ref{eq:Dirac-m3=0}) are more or less simpler
than those in Eqs.~(\ref{eq:Majorana-m1=0}) and (\ref{eq:Majorana-m3=0}),
simply because the Dirac neutrinos do not involve any uncertainties associated
with the Majorana phase $\sigma$.

The simplified RGEs obtained above allow us to see the correlations
among $\Delta\theta^{}_{ij}$, $\Delta\delta^{}_\nu$ and $\Delta \sigma$ in
a more transparent way. In the case of either $m^{}_1 = 0$ or $m^{}_3 = 0$,
however, the RGE-induced $\mu$-$\tau$ reflection symmetry breaking effects
are expected to be less significant as compared with the case of a nearly
degenerate neutrino mass spectrum.
A numerical example of this kind will be given in
section~\ref{section 6.4} for the minimal seesaw model (see also
Ref.~\cite{Zhao:2020bzx}).

\setcounter{equation}{0}
\section{Applications of $\mu$-$\tau$ reflection symmetry}
\label{section 6}

\subsection{Bounds on charged lepton flavor violation}
\label{section 6.1}

Although lepton flavor violation has been verified by quite a number of robust
neutrino oscillation experiments~\cite{Workman:2022ynf}, there has never been any
evidence for the similar effects in the charged lepton
sector~\cite{Cheng:1980tp,Bernstein:2013hba,Lindner:2016bgg}.
Since lepton flavor conservation
associated with the three charged leptons is essentially a consequence of the
vanishing masses of three active neutrinos in the SM, it is naturally expected that
the phenomena of charged lepton flavor violation should take place once we go beyond
the SM by taking into account the observed non-degenerate neutrino masses and
significant lepton flavor mixing effects. This is indeed the case, simply because the
massive neutrinos can mediate the radiative decays $\beta^- \to \alpha^- + \gamma$
(for $\alpha, \beta = e, \mu, \tau$ and $m^{}_\beta > m^{}_\alpha$) at the one-loop
level as illustrated by Figure~\ref{Fig:cLFV}. In comparison, the dominant tree-level
decay modes $\beta^- \to \alpha^- + \overline{\nu}^{}_\alpha + \nu^{}_\beta$ are
insensitive to the tiny masses of three active neutrinos, so the ratios
\begin{eqnarray}
\xi^{}_{\alpha\beta} \equiv \frac{\Gamma\big(\beta^- \to \alpha^- + \gamma\big)}
{\Gamma\big(\beta^- \to \alpha^- + \overline{\nu}^{}_\alpha
+ \nu^{}_\beta\big)} \;
\label{eq:LFV-ratio}
\end{eqnarray}
should automatically vanish when the neutrino masses and lepton flavor mixing effects
are switched off. Focusing only on the contributions from the three active and
light Majorana neutrinos $\nu^{}_i$ (for $i = 1, 2, 3$) and assuming the $3\times 3$
PMNS matrix $U$ to be exactly unitary, one may calculate the ratio of the decay rate of
$\mu^- \to e^- + \gamma$ to that of $\mu^- \to e^- + \overline{\nu}^{}_e + \nu^{}_\mu$
as a typical example and arrive at a formidably suppressed result as follows~\cite{Minkowski:1977sc,Petcov:1976ff,Bilenky:1977du,
Cheng:1976uq,Marciano:1977wx,Lee:1977qz,Lee:1977tib}:
\begin{eqnarray}
\xi^{}_{e\mu} & = & \frac{3 \alpha^{}_{\rm em}}
{32 \pi} \left|\sum^3_{i=1} U^*_{\mu i} U^{}_{e i} \frac{m^2_i}{M^2_W}
\right|^2 = \frac{3 \alpha^{}_{\rm em}}
{32 \pi} \left|\sum^3_{i=2} U^*_{\mu i} U^{}_{e i}
\frac{\Delta m^2_{i1}}{M^2_W}\right|^2
\nonumber \\
& \lesssim & {\cal O}\left(10^{-54}\right) \; ,
\label{eq:mu-e-gamma}
\end{eqnarray}
where $\alpha^{}_{\rm em}$ is the fine structure constant of electromagnetic
interactions, $M^{}_W$ is the $W$-boson mass, and
a numerical estimate has been made by using the experimental data listed
in Table~\ref{table2}. It becomes obvious that this decay mode
will be completely forbidden if the neutrino masses are vanishing as in the SM.
Note that the rate of $\mu^- \to e^- + \gamma$ obtained in Eq.~(\ref{eq:mu-e-gamma})
is roughly forty orders of magnitude smaller than the sensitivity of today's
measurements~\cite{Workman:2022ynf}, so it will be hopeless to observe such
charged-lepton-flavor-violating processes unless their reaction rates can be
significantly enhanced by a kind of new physics beyond the
SM~\cite{Bernstein:2013hba,Lindner:2016bgg}.
\begin{figure}[t!]
\begin{center}
\includegraphics[width=8.5cm]{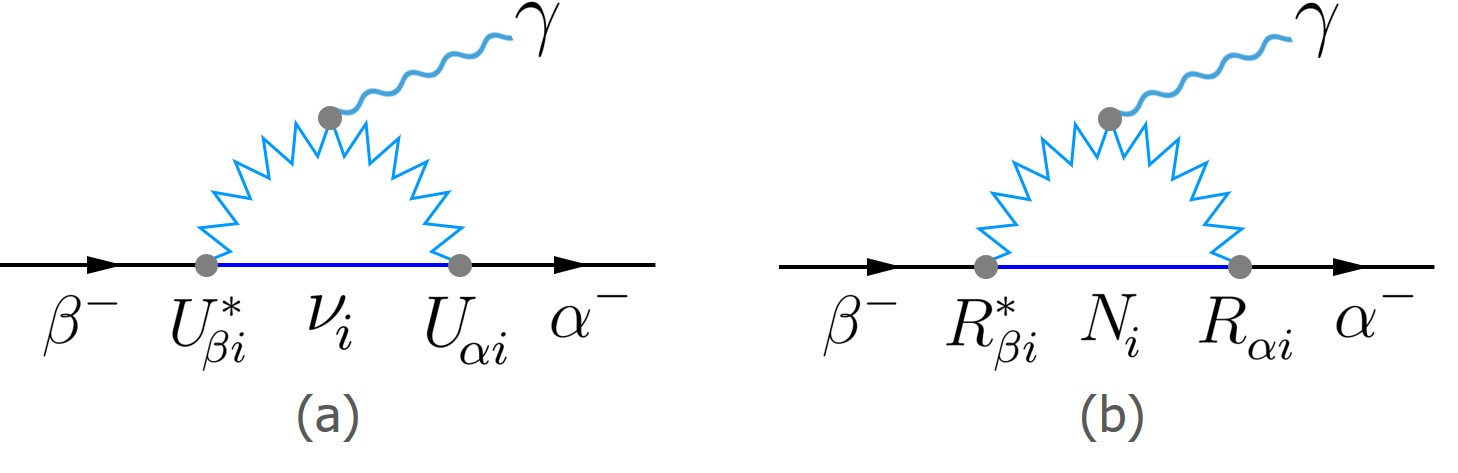}
\vspace{-0.4cm}
\caption{The one-loop Feynman diagrams for the charged-lepton-flavor-violating
$\beta^- \to \alpha^- + \gamma$ decays mediated by (a) the light Majorana neutrinos
$\nu^{}_i$ and (b) the heavy Majorana neutrinos $N^{}_i$ (for $i = 1, 2, 3$),
where $\alpha$ and $\beta$ run over the $e$, $\mu$ and $\tau$ flavors and
$m^{}_\beta > m^{}_\alpha$ holds.}
\label{Fig:cLFV}
\end{center}
\end{figure}

Now that the canonical seesaw mechanism introduced in section~\ref{section 2.4}
is the most popular theoretical framework for neutrino mass generation based on
a most reasonable and economical extension of the SM, let us take a look at the
contributions of those heavy Majorana neutrinos $N^{}_i$ to the radiative decay
modes $\beta^- \to \alpha^- + \gamma$ and discuss the constraints of $\mu$-$\tau$
reflection symmetry on them. As shown in Figure~\ref{Fig:cLFV}, the flavor mixing
factors $U^{}_{\alpha i} U^*_{\beta i}$ and $R^{}_{\alpha i} R^*_{\beta i}$ are
associated with the flavor-changing neutral current processes
$\beta^- \to \alpha^- + \gamma$ mediated respectively by $\nu^{}_i$ and $N^{}_i$
(for $i = 1, 2, 3$). Thanks to the unitarity condition
$U U^\dagger + R R^\dagger = {\cal I}$ given in Eq.~(\ref{eq:seesaw-unitarity})
and the huge mass hierarchies $m^{}_i \ll M^{}_W \ll M^{}_i$, the ratio of the
decay rate of $\beta^- \to \alpha^- + \gamma$ to that of
$\beta^- \to \alpha^- + \overline{\nu}^{}_\alpha + \nu^{}_\beta$
can now be expressed as~\cite{Xing:2020ivm,Zhang:2021tsq,Zhang:2021jdf}
\begin{eqnarray}
\xi^{}_{\alpha\beta} & \simeq &
\frac{3\alpha^{}_{\rm em}}{2\pi} \left|
\sum^{3}_{i=1} U^{}_{\alpha i} U^\ast_{\beta i} \left(-\frac{5}{6}
+ \frac{1}{4}\cdot \frac{m^2_i}{M^2_W}\right) -\frac{1}{3} \sum^{3}_{i=1}
R^{}_{\alpha i} R^\ast_{\beta i} \right|^2
\nonumber \\
& = & \frac{3\alpha^{}_{\rm em}}{8\pi} \left|
\sum^{3}_{i=1} U^{}_{\alpha i} U^\ast_{\beta i} \left(1 - \frac{1}{2}\cdot
\frac{m^2_i}{M^2_W}\right) \right|^2 \; ,
\label{eq:beta-alpha-gamma}
\end{eqnarray}
up to the leading order of $R^{}_{\alpha i} R^*_{\beta i}$. It is very clear that
Eq.~(\ref{eq:mu-e-gamma}) can be reproduced from Eq.~(\ref{eq:beta-alpha-gamma})
if the heavy degrees of freedom are switched off and the unitarity of $U$ is
restored. Since $U$ is non-unitary in the seesaw framework, the sum of
$U^{}_{\alpha i} U^*_{\beta i}$ with respect to the index $i$ does not vanish.
In this case it should be quite safe to neglect the $m^2_i/M^{2}_W$ terms in
Eq.~(\ref{eq:beta-alpha-gamma}). Then we are left with a straightforward but
strong constraint on the {\it effective} vertex of the unitarity hexagon
illustrated in Figure~\ref{Fig:unitarity-hexagon}:
\begin{eqnarray}
\left| \sum^{3}_{i=1} R^{}_{\alpha i} R^\ast_{\beta i} \right| =
\left| \sum^{3}_{i=1} U^{}_{\alpha i} U^\ast_{\beta i} \right| =
\sqrt{\frac{8\pi}{3 \alpha^{}_{\rm em}} \xi^{}_{\alpha\beta}}
\simeq 33.88 \sqrt{\xi^{}_{\alpha\beta}} \; ,
\label{eq:vertex}
\end{eqnarray}
where $\alpha^{}_{\rm em} \simeq 1/137$ has been input.
This result offers a realistic way to probe or constrain the seesaw-induced
unitarity violation in radiative $\beta^- \to \alpha^- + \gamma$
decays~\cite{Fernandez-Martinez:2016lgt,Xing:2020ivm}. Note that the coefficient
of $\sqrt{\xi^{}_{\alpha\beta}}$ in Eq.~(\ref{eq:vertex}) differs from the one
obtained in the {\it minimal unitarity violation} scheme by a factor $5/3$ (i.e.,
$\sqrt{24\pi/\left(25 \alpha^{}_{\rm em}\right)} \simeq 20.33$
\cite{Antusch:2006vwa,Antusch:2014woa,Calibbi:2017uvl}), simply because the
latter does not include the one-loop matching between the seesaw mechanism
and its low-energy effective operators~\cite{Zhang:2021tsq,Zhang:2021jdf}.

As pointed out in section~\ref{section 2.4}, the exact seesaw formula
assures that $R = {\cal P} R^* \zeta^\prime$ is a natural consequence of
$U = {\cal P} U^* \zeta$ in the $\mu$-$\tau$ reflection symmetry limit. To be
more explicit, it is $U^{}_{e i} = \eta^{}_i U^*_{e i}$ and
$U^{}_{\mu i} = \eta^{}_i U^*_{\tau i}$ that lead to
$R^{}_{e i} = \eta^{\prime}_i R^*_{e i}$ and
$R^{}_{\mu i} = \eta^{\prime}_i R^*_{\tau i}$, where $\eta^{}_i = \pm 1$ and
$\eta^{\prime}_i = \pm 1$ (for $i = 1, 2, 3$). Then it is straightforward to
prove
\begin{eqnarray}
\left| \sum^{3}_{i=1} U^{}_{e i} \hspace{0.072cm} U^\ast_{\mu i} \right| & = &
\left| \sum^{3}_{i=1} \eta^2_i \hspace{0.072cm} U^{\ast}_{e i} \hspace{0.072cm}
U^{}_{\tau i} \right| = \left| \sum^{3}_{i=1} U^{}_{e i} \hspace{0.072cm}
U^{\ast}_{\tau i} \right| \; ,
\nonumber \\
\left| \sum^{3}_{i=1} R^{}_{e i} R^\ast_{\mu i} \right| & = &
\left| \sum^{3}_{i=1} \eta^{\prime 2}_i R^{\ast}_{e i} R^{}_{\tau i} \right|
= \left| \sum^{3}_{i=1} R^{}_{e i} R^{\ast}_{\tau i} \right| \; .
\label{eq:vertex-mu-tau}
\end{eqnarray}
In this case we arrive at a simple but striking prediction
$\xi^{}_{e\tau} = \xi^{}_{e\mu}$, which can in principle be tested in a
precision measurement. In practice, however, it is extremely difficult to detect
such lepton-flavor-violating radiative decays of the charged leptons. The
present experimental upper bounds on the branching fractions of
$\beta^- \to \alpha^- + \gamma$ are~\cite{Workman:2022ynf}
\begin{eqnarray}
{\cal B}\big(\mu^- \to e^- + \gamma\big) & < &
4.2 \times 10^{-13} \; ,
\nonumber \\
{\cal B}\big(\tau^- \to e^- + \gamma\big) & < &
3.3 \times 10^{-8} \; ,
\nonumber \\
{\cal B}\big(\tau^- \to \mu^- + \gamma\big) & < &
4.2 \times 10^{-8} \; ,
\label{eq:branching-ratios}
\end{eqnarray}
at the $90\%$ confidence level. In comparison, we have
${\cal B}\big(\mu^- \to e^- + \overline{\nu}^{}_e + \nu^{}_\mu\big)
\simeq 100\%$, ${\cal B}\big(\tau^- \to e^- + \overline{\nu}^{}_e
+ \nu^{}_\tau\big) \simeq 17.82\%$ and ${\cal B}\big(\tau^- \to \mu^-
+ \overline{\nu}^{}_\mu + \nu^{}_\tau\big) \simeq 17.39\%$~\cite{Workman:2022ynf}.
As a result, $\xi^{}_{e\mu} < 4.20 \times 10^{-13}$,
$\xi^{}_{e\tau} < 1.85 \times 10^{-7}$ and
$\xi^{}_{\mu\tau} < 2.42 \times 10^{-7}$.
Combining these upper limits with Eq.~(\ref{eq:vertex}) leads us to the constraints
\begin{eqnarray}
\left| \sum^{3}_{i=1} U^{}_{e i} \hspace{0.038cm} U^\ast_{\mu i} \right| =
\left| \sum^{3}_{i=1} R^{}_{e i} \hspace{0.025cm} R^\ast_{\mu i} \right| & < &
2.20 \times 10^{-5} \; ,
\nonumber \\
\left| \sum^{3}_{i=1} U^{}_{e i} \hspace{0.043cm} U^\ast_{\tau i} \right| =
\left| \sum^{3}_{i=1} R^{}_{e i} \hspace{0.045cm} R^\ast_{\tau i} \right| & < &
1.46 \times 10^{-2} \; ,
\nonumber \\
\left| \sum^{3}_{i=1} U^{}_{\mu i} U^\ast_{\tau i} \right| =
\left| \sum^{3}_{i=1} R^{}_{\mu i} R^\ast_{\tau i} \right| & < &
1.66 \times 10^{-2} \; .
\label{eq:vertex-limit}
\end{eqnarray}
But when the $\mu$-$\tau$ reflection symmetry prediction
$\xi^{}_{e\tau} = \xi^{}_{e\mu}$ below Eq.~(\ref{eq:vertex-mu-tau}) is taken
into account, one will obtain a much more stringent constraint~\cite{Xing:2022cvu}
\begin{eqnarray}
\left| \sum^{3}_{i=1} U^{}_{e i} U^\ast_{\tau i} \right| =
\left| \sum^{3}_{i=1} R^{}_{e i} R^\ast_{\tau i} \right| <
2.20 \times 10^{-5} \; ,
\label{eq:vertex-limit2}
\end{eqnarray}
which is about three orders of magnitude smaller than the corresponding upper
bound obtained in Eq.~(\ref{eq:vertex-limit}). These results tell us that the
unitarity hexagon in Figure~\ref{Fig:unitarity-hexagon} can be treated as
the effective unitarity triangles, as the ``size" of the effective vertex
is expected to be at most at the ${\cal O}(10^{-2})$ level.
This interesting example shows that the $\mu$-$\tau$ reflection
symmetry may help a lot in establishing a kind of correlation between the
active-sterile neutrino mixing parameters.

Of course, we have assumed the $\mu$-$\tau$ reflection symmetry to be
exact at low energies in obtaining Eq.~(\ref{eq:vertex-limit2}). The
obtained stringent constraint can be relaxed to some extent
if the effects of soft $\mu$-$\tau$ reflection symmetry breaking are
taken into consideration, but here we refrain from going into details.

\subsection{Implications for lepton number violation}
\label{section 6.2}

As a minimal but powerful working flavor symmetry, the $\mu$-$\tau$
reflection symmetry may also constrain some lepton-number-violating
processes. Here let us take three categories of examples for the sake
of illustration.
\begin{figure}[t]
\begin{center}
\includegraphics[width=4.4in]{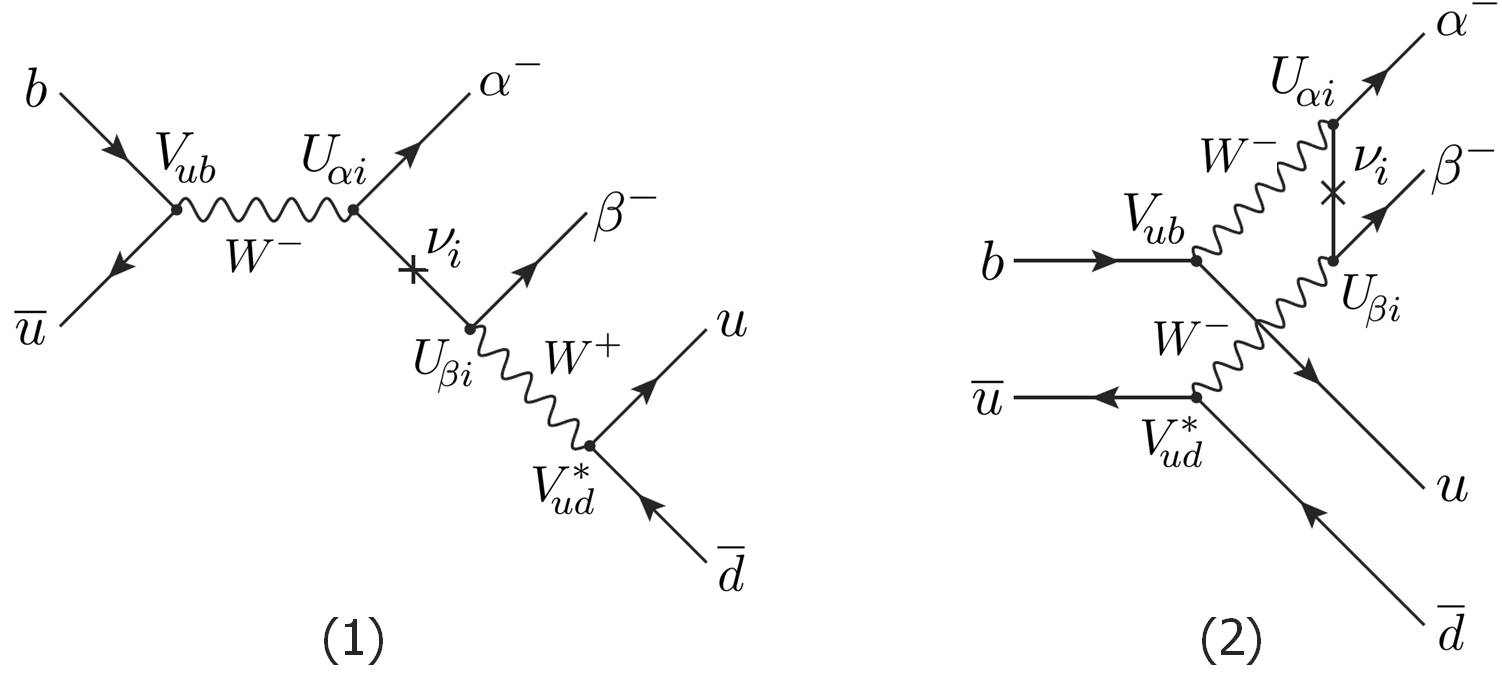}
\vspace{-0.25cm}
\caption{The tree-level Feynman diagrams for the lepton-number-violating decay
modes $B^-_u \to \pi^+ \alpha^- \beta^-$ (for $\alpha, \beta = e, \mu, \tau$),
where $\nu^{}_i$ (for $i = 1, 2, 3$) denote the three Majorana neutrino mass
eigenstates, $U^{}_{\alpha i}$ and $U^{}_{\beta i}$ are the PMNS matrix elements,
while $V^{}_{ud}$ and $V^{}_{ub}$ stand for the CKM matrix elements.}
\label{Fig:Bu-decay}
\end{center}
\end{figure}

(1) The lepton-number-violating decay modes $B^-_u \to \pi^+ \alpha^- \beta^-$
(for $\alpha, \beta = e, \mu, \tau$) mediated by the light Majorana neutrinos
$\nu^{}_i$ (for $i = 1, 2, 3$), as illustrated by Figure~\ref{Fig:Bu-decay},
which involve all the six effective Majorana neutrino mass terms
$\langle m\rangle^{}_{\alpha\beta}$ defined in Eq.~(\ref{eq:M-reconstruction}).
Similar to the $0\nu 2\beta$ decay, whose rate is proportional to
$|\langle m\rangle^{}_{ee}|^2$, the rates of $B^-_u \to \pi^+ \alpha^- \beta^-$
decays are simply proportional to
$|\langle m\rangle^{}_{\alpha\beta}|^2$~\cite{Doi:1985dx}. As a result,
\begin{eqnarray}
\Gamma \big(B^-_u \to \pi^+ e^- \tau^-\big) & = &
\Gamma \big(B^-_u \to \pi^+ e^- \mu^-\big) \; ,
\nonumber \\
\Gamma \big(B^-_u \to \pi^+ \tau^- \tau^-\big) & = &
\Gamma \big(B^-_u \to \pi^+ \mu^- \mu^-\big) \; ,
\label{eq:Bu-decays}
\end{eqnarray}
thanks to $\langle m\rangle^{}_{e\tau} = \langle m\rangle^{*}_{e\mu}$ and
$\langle m\rangle^{}_{\tau\tau} = \langle m\rangle^{*}_{\mu\mu}$ in the
$\mu$-$\tau$ reflection symmetry limit. For the time being, however,
there are only very preliminary upper bounds on the branching ratios
of $B^-_u \to \pi^+ e^- e^-$, $B^-_u \to \pi^+ e^- \mu^-$ and
$B^-_u \to \pi^+ \mu^- \mu^-$ decays,
namely ${\cal B}(B^-_u \to \pi^+ e^- e^-) < 2.3 \times
10^{-8}$ and ${\cal B}(B^-_u \to \pi^+ e^- \mu^-) < 1.5 \times
10^{-7}$ at the $90\%$ confidence level and
${\cal B}(B^-_u \to \pi^+ \mu^- \mu^-) < 4.0 \times 10^{-9}$ at the
$95\%$ confidence level~\cite{Workman:2022ynf}. In comparison, a rough estimate
shows that the branching ratios of such rare decay modes are expected to be of
${\cal O}(10^{-35})$ for $|\langle m\rangle^{}_{\alpha\beta}| \sim 0.1 ~{\rm eV}$
in the standard three-flavor scheme~\cite{Doi:1985dx}, which are too small
to be experimentally accessible. So these decay channels seem hopeless
to provide any meaningful constraints on $|\langle m\rangle^{}_{ee}|$,
$|\langle m\rangle^{}_{e\mu}|$ and $|\langle m\rangle^{}_{\mu\mu}|$, in contrast
with the more realistic and feasible $0\nu 2\beta$-decay experiments. From an
experimental point of view, however, searching for such lepton-number-violating
processes has been an important task in the upcoming precision measurement era, 
which is characterized by the LHCb experiment at the high-luminosity
LHC~\cite{LHCb:2018roe} and the Belle-II experiments at the KEK super-$B$
factory~\cite{Belle-II:2018jsg}.
Of course, one may also search for the similar
lepton-number-violating decay modes $K^- \to \pi^+ \alpha^- \beta^-$ at the
high-intensity frontier of particle physics. But in the latter case $\alpha$ and
$\beta$ can only run over $e$ and $\mu$ flavors for the kinematical reason, and
hence it is unable to fully test the $\mu$-$\tau$ reflection symmetry predictions.

(2) The lepton-number-violating decay modes $H^{--} \to \alpha^- + \beta^-$ and
$H^- \to \alpha^- + \nu^{}_\beta$ (for $\alpha, \beta = e, \mu, \tau$) 
in the type-II seesaw mechanism~\cite{Konetschny:1977bn,Magg:1980ut,
Schechter:1980gr,Cheng:1980qt,Lazarides:1980nt,Mohapatra:1980yp}. 
This seesaw scenario contains an $\rm SU(2)^{}_{\rm L}$ Higgs triplet 
$\Delta$ which interacts with both the SM lepton doublet $\ell^{}_{\rm L}$
and the SM Higgs doublet $H$, leading to lepton number violation. In this case the
gauge- and Lorentz-invariant neutrino mass term can be expressed as
\begin{eqnarray}
-{\cal L}^{}_\Delta = \frac{1}{2} \hspace{0.05cm} \overline{\ell^{}_{\rm L}}
\hspace{0.05cm} Y^{}_\Delta \Delta \hspace{0.05cm} {\rm i} \hspace{0.05cm} \sigma^{}_2
\big(\ell^{}_{\rm L}\big)^c - \lambda^{}_\Delta M^{}_\Delta H^T \hspace{0.05cm}
{\rm i} \hspace{0.05cm} \sigma^{}_2 \Delta H
+ {\rm h.c.} \; ,
\label{eq:type-II}
\end{eqnarray}
where $Y^{}_\Delta$ and $\lambda^{}_\Delta$ stand respectively for the Yukawa
coupling matrix and the scalar coupling coefficient, and the mass scale
$M^{}_\Delta$ is expected to be far above the vacuum expectation value of the
SM Higgs field (i.e., $M^{}_\Delta \gg v \simeq 246 ~{\rm GeV}$).
One may acquire the dimension-five Weinberg operator
by integrating out the heavy degrees of freedom, and achieve the effective
Majorana neutrino mass term after spontaneous electroweak symmetry breaking. The
corresponding seesaw formula is $M^{}_\nu = \lambda^{}_\Delta Y^{}_\Delta v^2/M^{}_\Delta$
for the symmetric mass matrix of three active Majorana neutrinos, where the smallness
of $\lambda^{}_\Delta$ is consistent with 't Hooft's naturalness criterion (namely,
setting $\lambda^{}_\Delta = 0$ will enhance the symmetry of ${\cal L}^{}_\Delta$
--- lepton number conservation~\cite{tHooft:1979rat,Xing:2009in}). A salient feature
of the type-II seesaw model is that there are totally seven physical Higgs bosons:
two doubly-charged Higgs bosons $H^{\pm\pm}$, two singly-charged Higgs bosons $H^\pm$,
the SM-like Higgs boson (CP-even), and two new neutral Higgs bosons (one is CP-even and
the other is CP-odd). The lepton-number-violating decays $H^{--} \to \alpha^- + \beta^-$
and $H^- \to \alpha^- + \nu^{}_\beta$ turn out to be the promising collider signatures
to probe the type-II seesaw mechanism~\cite{FileviezPerez:2008jbu,delAguila:2008cj}.
Their branching ratios are given respectively by
\begin{eqnarray}
{\cal B}\big(H^{--} \to \alpha^- + \beta^-\big) \equiv
\frac{\displaystyle \Gamma\big(H^{--} \to \alpha^- + \beta^-\big)}
{\displaystyle\sum_{\alpha, \beta}
\Gamma\big(H^{--} \to \alpha^- + \beta^-\big)} =
\frac{\displaystyle 2 \left|\langle m \rangle^{}_{\alpha\beta}
\right|^2}{\displaystyle \left(1 + \delta^{}_{\alpha\beta}\right) \sum_i m^2_i} \; ,
\nonumber \\
{\cal B}\big(H^{-} \to \alpha^- + \nu^{}_{\rm invi}\big)
\equiv \frac{\displaystyle \sum^{}_{\beta} \Gamma\big(H^{-} \to \alpha^-
+ \nu^{}_{\beta}\big)}{\displaystyle \sum_{\alpha, \beta}
\Gamma\big(H^{-} \to \alpha^- + \nu^{}_\beta\big)} =
\frac{\displaystyle \langle m \rangle^{2}_\alpha}{\displaystyle \sum_i m^2_i} \; ,
\label{eq:H-decays}
\end{eqnarray}
where $\alpha$ and $\beta$ run over $e$, $\mu$ and $\tau$ flavors, ``invi"
means that the {\it invisible} neutrinos are summed over their three flavors,
the effective neutrino mass terms $\langle m\rangle^{}_{\alpha\beta}$ and
$\langle m\rangle^{}_\alpha$ are expressed in terms of the neutrino masses
$m^{}_i$ and the PMNS matrix elements $U^{}_{\alpha i}$ and
$U^{}_{\beta i}$ (for $i = 1, 2, 3$) in Eqs.~(\ref{eq:M-reconstruction})
and (\ref{eq:new-mass-squared}) respectively. It is straightforward to
check that ${\cal B}\big(H^{--} \to \alpha^- + \beta^-\big) \propto
\left|\langle m \rangle^{}_{\alpha\beta}\right|^2$ are sensitive to two of
the three CP-violating phases of $U$ (or their combinations)~\cite{Hu:2021ziw},
but ${\cal B}\big(H^{-} \to \alpha^- + \nu^{}_{\rm invi}\big) \propto
\langle m\rangle^2_\alpha$ depend only on the CP-violating phase
$\delta^{}_\nu$~\cite{Xing:2013woa}.
Given the lightest neutrino mass to be $0.1 ~{\rm eV}$, for
example, the branching ratios of both $H^{--} \to \alpha^- + \beta^-$ and
$H^{-} \to \alpha^- + \nu^{}_{\rm invi}$ are expected to be of ${\cal O}(10^{-1})$
no matter whether the neutrino mass ordering is normal or
inverted~\cite{Garayoa:2007fw}. So the experimental observability of such
rare processes mainly depends on the production rate of $H^{--}$ and $H^{-}$ at
a high-energy collider. Since the $\mu$-$\tau$ reflection
symmetry assures $\langle m\rangle^{}_{e\tau} = \langle m\rangle^{*}_{e\mu}$,
$\langle m\rangle^{}_{\tau\tau} = \langle m\rangle^{*}_{\mu\mu}$ and
$\langle m\rangle^2_\tau = \langle m\rangle^2_\mu$ to hold, we immediately
arrive at
\begin{eqnarray}
{\cal B}\big(H^{--} \to e^- + \tau^-\big) & = &
{\cal B}\big(H^{--} \to e^- + \mu^-\big) \; ,
\nonumber \\
{\cal B}\big(H^{--} \to \tau^- + \tau^-\big) & = &
{\cal B}\big(H^{--} \to \mu^- + \mu^-\big) \; ,
\nonumber \\
{\cal B}\big(H^{-} \to \tau^- + \nu^{}_{\rm invi}\big) & = &
{\cal B}\big(H^{-} \to \mu^- + \nu^{}_{\rm invi}\big) \; .
\label{eq:H-decays2}
\end{eqnarray}
Whether the type-II seesaw mechanism is true remains an open question. Even if this
mechanism were really verified with the experimental discoveries of those hypothetical
Higgs bosons like $H^{\pm\pm}$ and $H^\pm$ in the foreseeable future, it would still
be a big challenge to probe the lepton-number-violating decay modes of $H^{\pm\pm}$ and
$H^\pm$ under discussion. Nevertheless, the latter may serve as a good example to
illustrate how powerful the $\mu$-$\tau$ reflection symmetry can be in constraining the
lepton flavors.

(3) The lepton-number-violating $0\nu2\beta$ decay modes $(A, Z) \to (A, Z+2) + 2 e^-$,
where the atomic mass number $A$ and the atomic number $Z$ are both
even~\cite{Furry:1939qr}. Such rare transitions depend on the effective
electron-neutrino mass term $\langle m\rangle^{}_{ee}$ given by
Eq.~(\ref{eq:2beta-decay}) in the standard three-flavor scheme. Taking the
parametrization of $U$ in Eq.~(\ref{eq:U}), we explicitly have the
expression of $\big|\langle m\rangle^{}_{ee}\big|$ as follows:
\begin{eqnarray}
\big|\langle m\rangle^{}_{ee}\big| = \left|\left[ m^{}_1 c^2_{12} \hspace{0.02cm}
e^{2\hspace{0.02cm} {\rm i} \hspace{0.02cm} \rho} + m^{}_2 s^2_{12} \hspace{0.02cm}
e^{2 \hspace{0.02cm} {\rm i} \hspace{0.02cm} \sigma} \right] c^2_{13}
\hspace{0.02cm} e^{2\hspace{0.02cm} {\rm i} \hspace{0.02cm} \delta^{}_\nu}
+ m^{}_3 s^2_{13}\right| \; .
\label{eq:mee}
\end{eqnarray}
Although this effective neutrino mass term does not directly involve the
$\mu$ and $\tau$ flavors, it can be simplified in the $\mu$-$\tau$ reflection
symmetry limit (denoted by the ``$\blacktriangle$" symbol):
\begin{eqnarray}
\big|\langle m\rangle^{\blacktriangle}_{ee}\big| = \left| m^{}_3 s^2_{13} -
\left[ m^{}_1 \eta^{}_\rho c^2_{12} + m^{}_2 \eta^{}_\sigma s^2_{12}\right]
c^2_{13}\right| \; ,
\label{eq:mee-mu-tau}
\end{eqnarray}
where $\eta^{}_\rho \equiv \cos 2\rho = \pm 1$ and $\eta^{}_\sigma \equiv
\cos 2\sigma = \pm 1$ are defined, and $\delta^{}_\nu = \pm\pi/2$ have been taken
into account~\cite{Xing:2017cwb,Nath:2018zoi}. In this case the magnitude of
$\langle m\rangle^{\blacktriangle}_{ee}$ is dependent upon the unknown absolute neutrino
mass scale (e.g., the value of $m^{}_1$) and has four-fold ambiguities arising from
$\eta^{}_\rho$ and $\eta^{}_\sigma$. If the neutrino mass ordering is inverted
(i.e., $m^{}_3 < m^{}_1 < m^{}_2$), the smallness of $\theta^{}_{13}$ and the
remarkable deviation of $\theta^{}_{12}$ from $\pi/4$ will assure that a significant
cancellation among the three terms of $\big|\langle m\rangle^{\blacktriangle}_{ee}\big|$
cannot happen. This observation is true no matter what values $\eta^{}_\rho$ and
$\eta^{}_\sigma$ may take. If the neutrino mass ordering is normal (i.e.,
$m^{}_1 < m^{}_2 < m^{}_3$) and $\eta^{}_\rho = \eta^{}_\sigma$ holds, then
the combination $\left(m^{}_1 c^2_{12} + m^{}_2 s^2_{12}\right) c^2_{13}$ is expected
to be much larger than $m^{}_3 s^2_{13}$, in which case a significant cancellation
cannot occur either. But when $\eta^{}_\rho = -\eta^{}_\sigma$ holds in the normal
neutrino mass ordering case, it is possible for the three terms of
$\big|\langle m\rangle^{\blacktriangle}_{ee}\big|$ cancel one another in a significant
way, leading us to a strongly suppressed (and even vanishingly small) value of
$\big|\langle m\rangle^{\blacktriangle}_{ee}\big|$~\cite{Xing:2003jf,Xing:2015zha,
Xing:2016ymd}. Note that such interesting observations are also valid even if the
$\mu$-$\tau$ reflection symmetry is more or less broken,
as numerically illustrated in Refs.~\cite{Cao:2019hli,Xing:2016ymd}
by using the available neutrino oscillation data.

The above three simple examples, together with neutrino-antineutrino oscillations
that have been briefly discussed in section~\ref{section 2.3},
tell us that the $\mu$-$\tau$ reflection symmetry can really be used to constrain
different lepton-number-violating processes and lead to some testable consequences,
although its soft breaking effects may somewhat affect those intriguing
constraints or predictions obtained in the exact symmetry limit. Some more examples
of this kind~\cite{Rodejohann:2011mu} can similarly be discussed.

\subsection{Constraints on cosmic UHE neutrino flavors}
\label{section 6.3}

In the past decade the km$^3$-volume IceCube Neutrino Observatory at the South
Pole has observed a flux of TeV-PeV astrophysical
neutrinos~\cite{IceCube:2013cdw,IceCube:2013low,IceCube:2014stg},
achieved the robust evidence for the blazar TXS 0506+056 as the first known
astronomical source of high-energy neutrinos~\cite{IceCube:2018cha,IceCube:2018dnn},
and discovered the first ultrahigh-energy (UHE) cosmic neutrino event on the
Glashow resonance~\cite{IceCube:2021rpz}. These groundbreaking scientific discoveries
bring us to the exciting frontier of cosmic UHE neutrino astronomy, making it more
promising to search for the point-like neutrino sources so as to study the puzzling
origin of UHE cosmic rays. In this regard a further precision measurement of the flavor
distribution of cosmic UHE neutrinos will help a lot to distinguish between the
$p \gamma$ and $p p$ collisions --- two expected dominant production mechanisms of
UHE neutrinos in a distant cosmic accelerator~\cite{Xing:2011zza,Gaisser:1994yf}.

It is commonly believed that the UHE cosmic-ray protons ought to interact with
the ambient protons or photons in a cosmic accelerator, producing a huge
amount of charged and neutral pions via the energetic proton-proton ($pp$) or
proton-photon ($p\gamma$) collisions. The UHE cosmic neutrinos and antineutrinos
can then be produced via the primary $\pi^\pm$ decay modes
($\pi^+ \to \mu^+ + \nu^{}_\mu$ and $\pi^- \to \mu^- + \overline{\nu}^{}_\mu$)
and the secondary $\mu^\pm$ decay modes ($\mu^+ \to e^+ + \overline{\nu}^{}_\mu
+ \nu^{}_e$ and $\mu^- \to e^- + \nu^{}_\mu + \overline{\nu}^{}_e$), and their
flavor distributions in the dominant $pp$ and $p\gamma$ collisions are given
respectively as
\begin{eqnarray}
pp ~{\rm collision}: \hspace{0.2cm}
f^{\rm S}_{\nu^{}_e} : f^{\rm S}_{\overline{\nu}^{}_e} :
f^{\rm S}_{\nu^{}_\mu} : f^{\rm S}_{\overline{\nu}^{}_\mu} :
f^{\rm S}_{\nu^{}_\tau} : f^{\rm S}_{\overline{\nu}^{}_\tau} & = &
\frac{1}{6} : \frac{1}{6} : \frac{1}{3} : \frac{1}{3} : 0 : 0 \; ,
\nonumber \\
p\gamma ~{\rm collision}: \hspace{0.2cm}
f^{\rm S}_{\nu^{}_e} : f^{\rm S}_{\overline{\nu}^{}_e} :
f^{\rm S}_{\nu^{}_\mu} : f^{\rm S}_{\overline{\nu}^{}_\mu} :
f^{\rm S}_{\nu^{}_\tau} : f^{\rm S}_{\overline{\nu}^{}_\tau} & = &
\frac{1}{3} : 0 : \frac{1}{3} : \frac{1}{3} : 0 : 0 \; ,
\label{eq:UHE-source}
\end{eqnarray}
where the superscript ``S" means the flavor composition at the source, the
$\pi^+$ and $\pi^-$ events are expected to be equally produced in the $pp$
collision as a result of the isospin symmetry, and only the $\pi^+$ events
can be generated in the $p\gamma$ collision thanks to the conservation of
electric charges. The realistic flavor ratios of cosmic UHE neutrinos and
antineutrinos may be more or less affected by the multiple-pion production
channels and the energy dependence~\cite{Zhou:2020oym}, but the precision
measurements of such flavor distributions at a neutrino telescope will be
greatly helpful to probe the true production mechanism of high-energy
astrophysical neutrinos. In this connection the $\mu$-$\tau$ reflection
symmetry breaking is expected to have an impact on some of the observable
quantities at the IceCube detector and other possible neutrino
detectors~\cite{Xing:2006xd,Pakvasa:2007dc,Xing:2012sj}.

Here let us reexamine the effects of $\mu$-$\tau$ reflection symmetry
breaking on the flavor distributions of cosmic UHE neutrinos and
antineutrinos at a detector on the earth~\cite{Zhou:2020oym}.
Now that a cosmic accelerator is most likely to be extragalactic, its distance
from a terrestrial neutrino telescope is expected to be much larger than
the neutrino oscillation lengths characterized respectively by $\Delta m^2_{21}$
and $\Delta m^2_{31}$ even if the cosmic neutrino and antineutrino fluxes
are very energetic. In this case the energy-dependent oscillation terms of
$P (\nu^{}_\beta \to \nu^{}_\alpha)$ and
$P (\overline{\nu}^{}_\beta \to \overline{\nu}^{}_\alpha)$ will be averaged
out, and thus the flavor composition of UHE neutrinos and antineutrinos at
the detector can be expressed as
\begin{eqnarray}
f^{\rm D}_{\nu^{}_\alpha} & = & \sum_\beta f^{\rm S}_{\nu^{}_\beta}
P (\nu^{}_\beta \to \nu^{}_\alpha) = \sum_\beta \sum_i
\big|U^{}_{\alpha i}\big|^2 \big|U^{}_{\beta i}\big|^2
f^{\rm S}_{\nu^{}_\beta} \; ,
\nonumber \\
f^{\rm D}_{\overline{\nu}^{}_\alpha} & = & \sum_\beta
f^{\rm S}_{\overline{\nu}^{}_\beta}
P (\overline{\nu}^{}_\beta \to \overline{\nu}^{}_\alpha) = \sum_\beta \sum_i
\big|U^{}_{\alpha i}\big|^2 \big|U^{}_{\beta i}\big|^2
f^{\rm S}_{\overline{\nu}^{}_\beta} \; ,
\label{eq:detector}
\end{eqnarray}
where $\alpha$ and $\beta$ run over $e$, $\mu$ and $\tau$, and the superscript
``D" means the detector. Given the standard parametrization of $U$ in
Eq.~(\ref{eq:U}) and the $\mu$-$\tau$ interchange asymmetries $\Delta^{}_i$
defined in Eq.~(\ref{eq:mu-tau-U-breaking}), the moduli of nine PMNS matrix
elements in Eq.~(\ref{eq:detector}) read as
\begin{eqnarray}
\pmatrix{ \big|U^{}_{e 1}\big|^2 & \big|U^{}_{e 2}\big|^2
& \big|U^{}_{e 3}\big|^2 \cr
\big|U^{}_{\mu 1}\big|^2 & \big|U^{}_{\mu 2}\big|^2
& \big|U^{}_{\mu 3}\big|^2 \cr
\big|U^{}_{\tau 1}\big|^2 & \big|U^{}_{\tau 2}\big|^2
& \big|U^{}_{\tau 3}\big|^2 \cr}
= \left[ \frac{1}{2} \pmatrix{ 2 c^2_{12} c^2_{13} & 2 s^2_{12} c^2_{13}
& 2 s^2_{13} \cr
s^2_{12} + c^2_{12} s^2_{13} & c^2_{12} + s^2_{12} s^2_{13}
& c^2_{13} \cr
s^2_{12} + c^2_{12} s^2_{13} & c^2_{12} + s^2_{12} s^2_{13}
& c^2_{13} \cr} \right.
\nonumber \\
\hspace{5.35cm}\left. \hspace{-0.07cm} + \frac{1}{2} \pmatrix{0 & 0 & 0 \cr
-\Delta^{}_1 & -\Delta^{}_2 & -\Delta^{}_3 \cr
+\Delta^{}_1 & +\Delta^{}_2 & +\Delta^{}_3 \cr}\right] \; .
\label{eq:U-moduli}
\end{eqnarray}
It is clear that $\big|U^{}_{\alpha i}\big|^2$ will be simplified if
$\Delta^{}_i$ (for $i = 1, 2, 3$) vanish in the $\mu$-$\tau$ reflection symmetry
limit. Substituting Eqs.~(\ref{eq:UHE-source}) and (\ref{eq:U-moduli}) into
Eq.~(\ref{eq:detector}), we obtain the flavor composition of UHE neutrinos and
antineutrinos at the detector as follows.

(1) As for the $pp$ collisions at the source, the flavor distribution at
the telescope is
\begin{eqnarray}
f^{\rm D}_{\nu^{}_e} & = & f^{\rm D}_{\overline{\nu}^{}_e}
= \frac{1}{6} + \frac{1}{6} \sum_i \big|U^{}_{e i}\big|^2 \Delta^{}_i
= \frac{1}{6} \big(1 - 2\Delta\big) \; ,
\nonumber \\
f^{\rm D}_{\nu^{}_\mu} & = & f^{\rm D}_{\overline{\nu}^{}_\mu}
= \frac{1}{6} + \frac{1}{6} \sum_i \big|U^{}_{\mu i}\big|^2 \Delta^{}_i
= \frac{1}{6} \big(1 + \Delta + \overline{\Delta}\big) \; ,
\nonumber \\
f^{\rm D}_{\nu^{}_\tau} & = & f^{\rm D}_{\overline{\nu}^{}_\tau}
= \frac{1}{6} + \frac{1}{6} \sum_i \big|U^{}_{\tau i}\big|^2 \Delta^{}_i
= \frac{1}{6} \big(1 + \Delta - \overline{\Delta}\big) \; ,
\label{eq:detector-pp}
\end{eqnarray}
where $\Delta \equiv -\left(c^2_{12} c^2_{13} \Delta^{}_1 + s^2_{12} c^2_{13}
\Delta^{}_2 + s^2_{13} \Delta^{}_3\right)/2$ and
$\overline{\Delta} \equiv \left(\Delta^2_1 + \Delta^2_2 + \Delta^2_3\right)/2$
are defined. If the detector does not distinguish between the
neutrino and antineutrino events, then only
$f^{\rm D}_\alpha \equiv f^{\rm D}_{\nu^{}_\alpha}
+ f^{\rm D}_{\overline{\nu}^{}_\alpha}$ (for $\alpha = e, \mu, \tau$) can be
measured. In this situation we have
\begin{eqnarray}
f^{\rm D}_{e} : f^{\rm D}_{\mu} : f^{\rm D}_{\tau} =
\frac{1}{3} \big(1 - 2\Delta\big) :
\frac{1}{3} \big(1 + \Delta + \overline{\Delta}\big) :
\frac{1}{3} \big(1 + \Delta - \overline{\Delta}\big) \; ,
\label{eq:detector-pp-ratio}
\end{eqnarray}
which will be reduced to the democratic flavor ratio
$f^{\rm D}_{e} : f^{\rm D}_{\mu} : f^{\rm D}_{\tau} = 1/3 : 1/3 : 1/3$
if the $\mu$-$\tau$ reflection symmetry is exact (i.e.,
$\overline{\Delta} = \Delta = 0$)~\cite{Learned:1994wg}.

(2) As for the $p\gamma$ collisions at the source, the flavor distribution at
the telescope is
\begin{eqnarray}
f^{\rm D}_{\nu^{}_e} & = &
\frac{1}{3} - \frac{1}{3} \sum_i \big|U^{}_{e i}\big|^2
\big|U^{}_{\tau i}\big|^2
= \frac{1}{3} \big(1 - \Xi - \Delta\big) \; ,
\nonumber \\
f^{\rm D}_{\nu^{}_\mu} & = &
\frac{1}{3} - \frac{1}{3} \sum_i \big|U^{}_{\mu i}\big|^2
\big|U^{}_{\tau i}\big|^2
= \frac{1}{6} \big(1 + \Xi + \overline{\Delta}\big) \; ,
\nonumber \\
f^{\rm D}_{\nu^{}_\tau} & = &
\frac{1}{3} - \frac{1}{3} \sum_i \big|U^{}_{\tau i}\big|^2
\big|U^{}_{\tau i}\big|^2
= \frac{1}{6} \big(1 + \Xi + 2\Delta - \overline{\Delta}\big) \;
\label{eq:detector-pgamma}
\end{eqnarray}
for the cosmic UHE neutrinos; and
\begin{eqnarray}
f^{\rm D}_{\overline{\nu}^{}_e} & = &
\frac{1}{3} \sum_i \big|U^{}_{e i}\big|^2 \big|U^{}_{\mu i}\big|^2
= \frac{1}{3} \big(\Xi - \Delta\big) \; ,
\nonumber \\
f^{\rm D}_{\overline{\nu}^{}_\mu} & = &
\frac{1}{3} \sum_i \big|U^{}_{\mu i}\big|^2 \big|U^{}_{\mu i}\big|^2
= \frac{1}{6} \big(1 - \Xi + 2\Delta + \overline{\Delta}\big) \; ,
\nonumber \\
f^{\rm D}_{\overline{\nu}^{}_\tau} & = &
\frac{1}{3} \sum_i \big|U^{}_{\tau i}\big|^2 \big|U^{}_{\mu i}\big|^2
= \frac{1}{6} \big(1 - \Xi - \overline{\Delta}\big) \;
\label{eq:detector-pgamma2}
\end{eqnarray}
for the cosmic UHE antineutrinos,
where $\Xi \equiv c^2_{13} \left(c^2_{12} s^2_{12} c^2_{13} + s^2_{13}\right)$
is defined~\cite{Zhou:2020oym}. In this case it is straightforward to check
that the neutrino-antineutrino-indistinguishable flavor ratio
$f^{\rm D}_{e} : f^{\rm D}_{\mu} : f^{\rm D}_{\tau}$ is actually independent
of $\Xi$ and given also by Eq.~(\ref{eq:detector-pp-ratio}).

Taking account of the best-fit values of $\theta^{}_{12}$, $\theta^{}_{13}$,
$\theta^{}_{23}$ and $\delta^{}_\nu$ listed in Table~\ref{table2}, we have
$\Delta^{}_1 \simeq 0.059$ (or $0.021$),
$\Delta^{}_2 \simeq -0.158$ (or $0.115$) and
$\Delta^{}_3 \simeq 0.099$ (or $-0.136$) as a typical numerical example for the
normal (or inverted) neutrino mass ordering. The corresponding values of
$\Delta$, $\overline{\Delta}$ and $\Xi$ are $\Delta \simeq 2.28 \times 10^{-3}$
(or $-2.27 \times 10^{-2}$), $\overline{\Delta} \simeq 1.91 \times 10^{-2}$
(or $1.61 \times 10^{-2}$) and $\Xi \simeq 0.224$ (or $0.224$) in the normal
(or inverted) mass ordering case. It is obvious that the running IceCube
neutrino detector is insensitive to the small effects of $\mu$-$\tau$ reflection
symmetry breaking, but the future neutrino telescopes may hopefully measure or
constrain $\Delta$ and $\overline{\Delta}$ provided the experimental sensitivities
to those flavor distribution quantities shown
in Eqs.~(\ref{eq:detector-pp})---(\ref{eq:detector-pgamma2}) can finally reach
the challenging $1\%$ or even $0.1\%$ level.
A recent detailed analysis of this issue can be found in Ref.~\cite{Zhou:2020oym},
where an interesting hexagonal diagram description of the flavor distributions of
cosmic UHE neutrinos and antineutrinos at a terrestrial neutrino telescope has
been introduced and discussed.

\subsection{Constraints on the minimal seesaw scenario}
\label{section 6.4}

A remarkable and nontrivial application of the $\mu$-$\tau$ reflection symmetry
to the canonical seesaw mechanism has been discussed in section~\ref{section 2.4}
by starting from the empirical conjecture $U = {\cal P} U^* \zeta$. Although such a
simple working flavor symmetry is very powerful in constraining the texture of the
Dirac neutrino mass matrix $M^{}_{\rm D}$ and the pattern of the active-sterile
neutrino mixing matrix $R$ via the exact seesaw formula in
Eq.~(\ref{eq:exact-seesaw}), it leaves little constraint on the texture of the
Majorana neutrino mass matrix $M^{}_{\rm R}$. Imposing the same symmetry on
$M^{}_{\rm R}$ is certainly a feasible way to enhance the predictive power of
the seesaw mechanism (see, e.g.,
Refs.~\cite{Xing:2015fdg,Mohapatra:2015gwa,Xing:2019edp,Zhao:2021dwc}), and this
enhancement will be further amplified in the minimal canonical seesaw model
which contains only two right-handed neutrino fields~\cite{Xing:2020ald,Kleppe:1995zz,
Ma:1998zg,Frampton:2002qc,Guo:2006qa}.

The two right-handed neutrino fields $N^{}_{\mu \rm R}$ and $N^{}_{\tau \rm R}$
in the minimal seesaw model form a column vector
$N^{\prime}_{\rm R} = (N^{}_{\mu \rm R} , N^{}_{\tau \rm R})^T$,
where the subscripts ``$\mu$" and ``$\tau$" are adopted in correspondence
to the flavors of $\nu^{}_\mu$ and $\nu^{}_\tau$ in the active neutrino sector.
In this case the neutrino mass terms in Eq.~(\ref{eq:seesaw-mass-matrix}) can
be reduced to ${\cal L}^\prime_{\rm MSS}$ in
Eq.~(\ref{eq:minimal-seesaw-mass-matrix}),
where $\widehat{M}^{}_{\rm D}$ and $\widehat{M}^{}_{\rm R}$ are the $3\times 2$
and $2\times 2$ mass matrices, respectively. Now let us assume that
$\nu^{}_{\rm L}$ and $N^{\prime}_{\rm R}$ undergo the $\mu$-$\tau$ reflection
transformations
\begin{eqnarray}
\nu^{}_{\rm L} \to {\cal P} (\nu^{}_{\rm L})^c \; , \quad
N^{\prime}_{\rm R} \to {\cal Z} (N^{\prime}_{\rm R})^c \; ,
\label{eq:minimal-seesaw-reflection}
\end{eqnarray}
where the $3\times 3$ orthogonal $(\mu, \tau)$-associated permutation matrix
$\cal P$ has been given in Eq.~(\ref{eq:S-tranformation}), and its $2\times 2$
counterpart $\cal Z$ is given as follows:
\begin{eqnarray}
{\cal Z} = {\cal Z}^T = {\cal Z}^\dagger =
\pmatrix{ 0 & 1 \cr 1 & 0 \cr} \; .
\label{eq:S2-tranformation}
\end{eqnarray}
Then the invariance of ${\cal L}^\prime_{\rm MSS}$ under the transformations made
in Eq.~(\ref{eq:minimal-seesaw-reflection}) requires
\begin{eqnarray}
\widehat{M}^{}_{\rm D} = {\cal P} \widehat{M}^{*}_{\rm D} {\cal Z} \; , \quad
\widehat{M}^{}_{\rm R} = {\cal Z} \widehat{M}^{*}_{\rm R} {\cal Z} \; .
\label{eq:minimal-seesaw-constraint}
\end{eqnarray}
The flavor textures of $\widehat{M}^{}_{\rm D}$ and $\widehat{M}^{}_{\rm R}$ are
therefore constrained to the forms given in Eq.~(\ref{eq:minimal-seesaw-textures}),
and the corresponding active Majorana neutrino mass matrix
is expressed by the approximate seesaw formula
$\widehat{M}^{}_\nu \simeq -\widehat{M}^{}_{\rm D} \widehat{M}^{-1}_{\rm R}
\widehat{M}^{T}_{\rm D}$ in Eq.~(\ref{eq:minimal-seesaw-formula}). Since
$\widehat{M}^{}_\nu$ is of rank two and respects the $\mu$-$\tau$ reflection
symmetry, both the mass spectrum and flavor mixing pattern of three active
neutrinos can be well constrained.

Once the minimal seesaw mechanism is constrained by the $\mu$-$\tau$ reflection
symmetry at a given energy scale, its predictive power and testability will be
greatly enhanced. In this case, of course, there still exist
some ambiguities associated with the neutrino mass ordering and the CP-violating
phases. To be more explicit, we may either have a special normal neutrino
mass ordering
\begin{eqnarray}
m^{}_1 & = & 0 \; ,
\nonumber \\
m^{}_2 & = & \sqrt{\Delta m^2_{21}} \simeq 8.61 \times 10^{-3}~{\rm eV} \; ,
\nonumber \\
m^{}_3 & = & \sqrt{\Delta m^2_{31}} \simeq 5.01 \times 10^{-2}~{\rm eV} \; ;
\label{eq:MSS-NMO}
\end{eqnarray}
or have a special inverted neutrino mass ordering
\begin{eqnarray}
m^{}_1 & = & \sqrt{\left|\Delta m^2_{31}\right|} \simeq 4.92
\times 10^{-2}~{\rm eV} \; ,
\nonumber \\
m^{}_2 & = & \sqrt{\left|\Delta m^2_{32}\right|} \simeq 4.99
\times 10^{-2}~{\rm eV} \; ,
\nonumber \\
m^{}_3 & = & 0 \; ,
\label{eq:MSS-IMO}
\end{eqnarray}
where the best-fit values of $\Delta m^2_{21}$ and $\Delta m^2_{31}$ listed in
Table~\ref{table2} have been input for a numerical illustration. Current
neutrino oscillation data slightly favor the normal mass
ordering~\cite{Gonzalez-Garcia:2021dve,Capozzi:2021fjo}, and the
next-generation experiments (e.g., JUNO~\cite{JUNO:2015zny,JUNO:2021vlw},
Hyper-Kamiokande~\cite{Hyper-KamiokandeProto-:2015xww} and
DUNE~\cite{DUNE:2015lol})
are expected to determine which neutrino mass ordering is really true. On the
other hand, the next-generation precision measurements of the cosmic microwave
background anisotropies and large scale structures may help to
probe the absolute neutrino mass scale and thus examine whether $m^{}_1$ or
$m^{}_3$ is vanishing or vanishingly small~\cite{Xing:2019vks}. Constrained by
the $\mu$-$\tau$ reflection symmetry, the PMNS matrix in the minimal seesaw
model can be simplified to
\begin{eqnarray}
U = P^{}_l \pmatrix{c^{}_{12} c^{}_{13} & s^{}_{12} c^{}_{13} &
\mp {\rm i} \hspace{0.05cm} s^{}_{13} \cr
\frac{-1}{\sqrt 2} \big(s^{}_{12} \pm {\rm i} \hspace{0.05cm} c^{}_{12} s^{}_{13} \big)
& \frac{1}{\sqrt 2} \big(c^{}_{12} \mp {\rm i} \hspace{0.05cm} s^{}_{12} s^{}_{13} \big)
& \frac{1}{\sqrt 2} c^{}_{13} \cr \vspace{-0.45cm} \cr
\frac{1}{\sqrt 2} \big(s^{}_{12} \mp {\rm i} \hspace{0.05cm} c^{}_{12} s^{}_{13} \big)
& \frac{-1}{\sqrt 2}\big(c^{}_{12} \pm {\rm i} \hspace{0.05cm} s^{}_{12} s^{}_{13} \big)
& \frac{1}{\sqrt 2} c^{}_{13} \cr} P^{}_\nu \; ,
\label{eq:MSS-U}
\end{eqnarray}
where Eq.~(\ref{eq:reflection-phases}) has been used,
$P^{}_\nu = {\cal I}$ for $\sigma = 0$ or
$P^{}_\nu = {\rm Diag}\big\{1, {\rm i}, 1\big\}$ for $\sigma = \pi/2$,
and $P^{}_l = {\rm Diag}\{e^{{\rm i}\phi^{}_e} , e^{{\rm i}\phi^{}_\mu} ,
e^{{\rm i}\phi^{}_\tau}\}$ with $\phi^{}_e = 0$ or $\pi/2$ and
$\phi^{}_\mu + \phi^{}_\tau = \left( 2n + 1\right)\pi - 2\phi^{}_e$
(for $n = 0, \pm 1, \pm 2, \cdots$). Since $m^{}_1 = 0$ or $m^{}_3 = 0$ assures
that one of the Majorana phases associated with either $\nu^{}_1$ or
$\nu^{}_3$ can be eliminated~\cite{Xing:2020ald}, $U$ contains only two
CP-violating phases $\delta^{}_\nu$ and $\sigma$ whose values are fixed
by the $\mu$-$\tau$ reflection symmetry with only two-fold ambiguities.
With the help of Eqs.~(\ref{eq:MSS-NMO}), (\ref{eq:MSS-IMO}) and
(\ref{eq:MSS-U}), one may easily reconstruct the flavor texture of
$\widehat{M}^{}_\nu$ in Eq.~(\ref{eq:minimal-seesaw-formula}). We
obtain~\cite{Xing:2022hst}
\begin{eqnarray}
\left|\langle m\rangle^{}_{ee}\right| =
\left\{\begin{array}{ll}
\hspace{-0.15cm} \displaystyle
m^{}_2 s^2_{12} c^2_{13} \mp m^{}_3 s^2_{13} \; ,
& \quad\quad (m^{}_1 = 0)  \\ \vspace{-0.45cm} \\
\hspace{-0.15cm} c^2_{13} \displaystyle
\left(m^{}_1 c^2_{12} \pm m^{}_2 s^2_{12} \right) \; ,
& \quad\quad (m^{}_3 = 0)  \end{array} \right.
\nonumber \\
\left|\langle m\rangle^{}_{e\mu}\right| =
\left\{\begin{array}{l}
\hspace{-0.15cm} \displaystyle
\frac{c^{}_{13}}{\sqrt 2} \left[m^{2}_2 c^2_{12} s^2_{12} +
\left(m^{}_2 s^2_{12} \pm m^{}_3 \right)^2 s^2_{13}\right]^{1/2} , \\
\hspace{-0.15cm} \displaystyle
\frac{c^{}_{13}}{\sqrt 2} \left[\left( m^{}_1 \mp m^{}_2 \right)^2
c^2_{12} s^2_{12} + \left( m^{}_1 c^2_{12} \pm m^{}_2 s^2_{12}\right)^2
s^2_{13}\right]^{1/2} , \end{array} \right.
\nonumber \\
\left|\langle m\rangle^{}_{\mu\mu}\right| =
\left\{\begin{array}{l}
\hspace{-0.15cm} \displaystyle
\frac{1}{2} \left\{\left[ m^{}_2 \left( c^2_{12} - s^2_{12} s^2_{13}\right)
\pm m^{}_3 c^2_{13}\right]^2 + 4 m^2_2 c^2_{12} s^2_{12} s^2_{13}
\right\}^{1/2} , \\ \vspace{-0.4cm} \\
\hspace{-0.15cm} \displaystyle
\frac{1}{2} \left\{\left[ m^{}_1 \left( s^2_{12} -
c^2_{12} s^2_{13} \right) \pm m^{}_2 \left( c^2_{12} - s^2_{12} s^2_{13}
\right)\right]^2 \right. \\
\hspace{0.2cm} + \left. 4 \left(m^{}_1 \mp m^{}_2\right)^2 c^2_{12} s^2_{12}
s^2_{13}\right\}^{1/2} , \end{array} \right.
\nonumber \\
\left|\langle m\rangle^{}_{\mu\tau}\right| =
\left\{\begin{array}{l}
\hspace{-0.15cm} \displaystyle
\frac{1}{2} \left[ m^{}_3 c^2_{13} \mp m^{}_2 \left( c^2_{12} +
s^{2}_{12} s^2_{13}\right)\right] \; , \\ \vspace{-0.4cm} \\
\hspace{-0.15cm} \displaystyle
\frac{1}{2} \left[ m^{}_2 \left( c^2_{12} + s^2_{12} s^2_{13} \right)
\pm m^{}_1 \left( s^2_{12} + c^2_{12} s^2_{13}\right)\right] \; ,
\end{array} \right.
\label{eq:MSS-matrix-elements}
\end{eqnarray}
in which the relevant sign ambiguities originate from $\sigma = 0$ or
$\pi/2$. Note that all the expressions in Eq.~(\ref{eq:MSS-matrix-elements})
are actually insensitive to the two-fold uncertainties of $\delta^{}_\nu$ (i.e.,
$\delta^{}_\nu = \pm \pi/2$) in the $\mu$-$\tau$ reflection symmetry limit.
Table~\ref{table3} illustrates the magnitudes of these effective Majorana
neutrino mass terms at the electroweak scale, where the best-fit values of
$\theta^{}_{12}$ and $\theta^{}_{13}$ listed in Table~\ref{table2} together
with the numerical results given in Eqs.~(\ref{eq:MSS-NMO}) and
(\ref{eq:MSS-IMO}) have been input. The smallness of
$\big|\langle m\rangle^{}_{\alpha\beta}\big|$ (for $\alpha, \beta = e, \mu, \tau$)
makes it very challenging to probe the lepton-number-violating processes
mediated by the light Majorana neutrinos $\nu^{}_i$
(for $i = 1, 2, 3$)~\cite{Rodejohann:2011mu,Bilenky:2014uka,
Dolinski:2019nrj}, but the next-generation $0\nu 2\beta$ experiments
are quite promising to explore the parameter space of
$\big|\langle m\rangle^{}_{ee}\big|$ for the inverted neutrino mass ordering
including the $m^{}_3 = 0$ case.
\begin{table}[t]
\caption{A numerical illustration of the sizes of $\langle m\rangle^{}_{ee}$,
$\langle m\rangle^{}_{e\mu}$, $\langle m\rangle^{}_{\mu\mu}$ and
$\langle m\rangle^{}_{\mu\tau}$ in the minimal seesaw model constrained by the
$\mu$-$\tau$ reflection symmetry at the electroweak scale, where the best-fit
values of $\Delta m^2_{21}$, $\Delta m^2_{31}$ (or $\Delta m^2_{32}$),
$\theta^{}_{12}$ and $\theta^{}_{13}$ listed in Table~\ref{table2} have already
been input.}
\vspace{0.2cm}
\label{table3}
\begin{indented}
\item[]\begin{tabular}{lllll} \br
& \multicolumn{2}{c}{\hspace{0.5cm} $m^{}_1 = 0$}
& \multicolumn{2}{c}{\hspace{0.5cm} $m^{}_3 = 0$} \\ \mr
& \hspace{0.5cm} $\sigma = 0$
& \hspace{0.1cm} $\sigma = \pi/2$
& \hspace{0.5cm} $\sigma = 0$
& \hspace{0.1cm} $\sigma = \pi/2$ \\ \mr
\vspace{0.1cm}
$\big|\langle m\rangle^{}_{ee}\big|$ (eV)
& \hspace{0.5cm} $1.45 \times 10^{-3}$
& \hspace{0.1cm} $3.67 \times 10^{-3}$
& \hspace{0.5cm} $4.83 \times 10^{-2}$
& \hspace{0.1cm} $1.91 \times 10^{-2}$
\\ \vspace{0.1cm}
$\big|\langle m\rangle^{}_{e\mu}\big|$ (eV)
& \hspace{0.5cm} $6.13 \times 10^{-3}$
& \hspace{0.1cm} $5.65 \times 10^{-3}$
& \hspace{0.5cm} $5.13 \times 10^{-3}$
& \hspace{0.1cm} $3.19 \times 10^{-2}$
\\ \vspace{0.1cm}
$\big|\langle m\rangle^{}_{\mu\mu}\big|$ (eV)
& \hspace{0.5cm} $2.75 \times 10^{-2}$
& \hspace{0.1cm} $2.15 \times 10^{-2}$
& \hspace{0.5cm} $2.43 \times 10^{-2}$
& \hspace{0.1cm} $1.22 \times 10^{-2}$
\\ \vspace{0.1cm}
$\big|\langle m\rangle^{}_{\mu\tau}\big|$ (eV)
& \hspace{0.5cm} $2.15 \times 10^{-2}$
& \hspace{0.1cm} $2.75 \times 10^{-2}$
& \hspace{0.5cm} $2.54 \times 10^{-2}$
& \hspace{0.1cm} $9.68 \times 10^{-3}$
\\ \br
\end{tabular}
\end{indented}
\end{table}

Note that it is certainly more natural to expect the minimal seesaw mechanism
constrained by the $\mu$-$\tau$ reflection symmetry to take effect
at a superhigh energy scale $\Lambda^{}_{\mu\tau}$. In this case one
should consider the radiative corrections to the
relevant neutrino mass spectrum and flavor mixing parameters at the
electroweak scale $\Lambda^{}_{\rm EW}$ by using the RGEs. With the
help of the analytical results obtained in Eqs.~(\ref{eq:Majorana-m1=0})
and (\ref{eq:Majorana-m3=0}), it is straightforward for us to calculate
the RGE-induced $\mu$-$\tau$ reflection symmetry breaking effects.
To illustrate, let us simply consider the possibility
of $\delta^{}_\nu = -\pi/2$ (i.e., $\eta^{}_\delta = -1$)
at $\Lambda^{}_{\mu\tau} \simeq 10^{14}~{\rm GeV}$
and input the best-fit values $\theta^{}_{12} \simeq 33.45^\circ$ and
$\theta^{}_{13} \simeq 8.62^\circ$ at $\Lambda^{}_{\rm EW} \simeq 10^{2}~{\rm GeV}$.
Eqs.~(\ref{eq:MSS-NMO}) and (\ref{eq:MSS-IMO}) lead us to
$\zeta^{}_{32} \simeq 0.707$ in the $m^{}_1 (\Lambda^{}_{\mu\tau}) = 0$ case
and $\zeta^{}_{21} \simeq 7.564 \times 10^{-3}$ in the
$m^{}_3 (\Lambda^{}_{\mu\tau}) = 0$ case at $\Lambda^{}_{\rm EW}$. As for
the RGE-induced radiative corrections, we quote
$I^{}_\kappa (\Lambda^{}_{\rm EW}) \simeq 0.748$ and
$\Delta^{}_\tau (\Lambda^{}_{\rm EW}) \simeq 2.822 \times 10^{-5}$ in the SM;
$I^{}_\kappa (\Lambda^{}_{\rm EW}) \simeq 0.879$ and
$\Delta^{}_\tau (\Lambda^{}_{\rm EW}) \simeq -1.354 \times 10^{-3}$ in the MSSM
with $\tan\beta = 10$; or $I^{}_\kappa (\Lambda^{}_{\rm EW}) \simeq 0.871$ and
$\Delta^{}_\tau (\Lambda^{}_{\rm EW}) \simeq -1.346 \times 10^{-2}$ in the MSSM
with $\tan\beta = 30$~\cite{Zhang:2020lsd}. Then we immediately arrive at
\begin{eqnarray}
\frac{m^{}_2 (\Lambda^{}_{\rm EW})}{m^{}_2 (\Lambda^{}_{\mu\tau})} \simeq
\frac{m^{}_3 (\Lambda^{}_{\rm EW})}{m^{}_3 (\Lambda^{}_{\mu\tau})}
\simeq I^{}_\kappa (\Lambda^{}_{\rm EW}) \simeq
\left\{\begin{array}{l}
\hspace{-0.15cm} 0.75 \quad ({\rm SM}) \; ,
\\ \vspace{-0.6cm} \\
\hspace{-0.15cm} 0.88 \quad (\tan\beta = 10) \; ,
\\ \vspace{-0.6cm} \\
\hspace{-0.15cm} 0.87 \quad (\tan\beta = 30) \; ,
\end{array} \right.
\label{eq:mass-ratios}
\end{eqnarray}
in the $m^{}_1 (\Lambda^{}_{\mu\tau}) = 0$ case; and the same results for the
ratios $m^{}_1 (\Lambda^{}_{\rm EW})/m^{}_1 (\Lambda^{}_{\mu\tau})$ and
$m^{}_2 (\Lambda^{}_{\rm EW})/m^{}_2 (\Lambda^{}_{\mu\tau})$ in the
$m^{}_3 (\Lambda^{}_{\mu\tau}) = 0$ case. The numerical results of
$\Delta\theta^{}_{12}$, $\Delta\theta^{}_{13}$, $\Delta\theta^{}_{23}$,
$\Delta\delta^{}_\nu$ and $\Delta\sigma$ are obtained for either
$\sigma (\Lambda^{}_{\mu\tau}) = 0$ or $\sigma (\Lambda^{}_{\mu\tau}) = \pi/2$,
and they are listed in Table~\ref{table4}. One can see that in the SM framework
the numerical results of $\Delta\theta^{}_{ij}$, $\Delta\delta^{}_\nu$ and
$\Delta\sigma$ are so tiny that they should be experimentally indistinguishable.
In the MSSM framework the RGE-induced effects can be somewhat enhanced by
taking relatively larger values of $\tan\beta$, simply because of
$\Delta{}_\tau (\Lambda^{}_{\rm EW}) \propto \left(1 + \tan^2\beta\right)$.
For example, $\Delta\theta^{}_{12}$ is significantly enhanced for
$m^{}_3 (\Lambda^{}_{\mu\tau}) = 0$ and $\sigma (\Lambda^{}_{\mu\tau}) = 0$
in the MSSM with a sufficiently large $\tan\beta$, since
$\Delta\theta^{}_{12}$ is proportional to both
$\zeta^{-1}_{21} = \left(m^{}_1 + m^{}_2\right)^2/\Delta m^2_{21} \simeq 132.2$
and $\left(1 + \tan^2\beta\right) \gtrsim 10^2$ for $\tan\beta \gtrsim 10$
and thus gets enhanced in this case. In comparison, $\Delta\delta^{}_\nu$
and $\Delta\sigma$ can be remarkably enhanced for the same reasons if
$m^{}_3 (\Lambda^{}_{\mu\tau}) = 0$ and $\sigma (\Lambda^{}_{\mu\tau}) = \pi/2$
are taken in the MSSM, although these two quantities are suppressed
by the smallness of $\theta^{}_{13}$ as shown in Eq.~(\ref{eq:Majorana-m3=0}).
Furthermore, the small magnitude of $\Delta\theta^{}_{23}$ in
the MSSM with $\tan\beta = 30$ implies that the RGE-induced
correction to $\theta^{}_{23} (\Lambda^{}_{\mu\tau}) = \pi/4$ at
$\Lambda^{}_{\rm EW}$ is too mild to resolve the octant issue of
$\theta^{}_{23}$ in the minimal seesaw scenario combined with the
$\mu$-$\tau$ reflection symmetry~\cite{Xing:2022hst}.
\begin{table}[t]
\begin{center}
\caption{A numerical illustration of the RGE-induced radiative corrections
to $\theta^{}_{12}$, $\theta^{}_{13}$, $\theta^{}_{23}$, $\delta^{}_\nu$
and $\sigma$ in the minimal seesaw model constrained by the $\mu$-$\tau$
reflection symmetry, where $\Lambda^{}_{\mu\tau} \simeq 10^{14}~{\rm GeV}$
and $\delta (\Lambda^{}_{\mu\tau}) = -\pi/2$ have been taken, and the results
of ${\cal O}(10^{-5})$ degrees or smaller have been denoted as
$\sim 0^\circ$~\cite{Xing:2022hst}.}
\vspace{0.3cm}
\label{table4}
\begin{tabular}{c|cc|cc|cc} \hline\hline
$m^{}_1 (\Lambda^{}_{\mu\tau}) = 0$
& \multicolumn{2}{c}{SM}
& \multicolumn{2}{c}{MSSM ($\tan\beta = 10$)}
& \multicolumn{2}{c}{MSSM ($\tan\beta = 30$)}
\\ \hline
$\sigma (\Lambda^{}_{\mu\tau})$
& $0$
& $\pi/2$
& $0$
& $\pi/2$
& $0$
& $\pi/2$
\\ \hline
$\Delta\theta^{}_{12}$
& $\sim 0^\circ$
& $\sim 0^\circ$
& $0.018^\circ$
& $0.017^\circ$
& $0.175^\circ$
& $0.172^\circ$
\\
$\Delta\theta^{}_{13}$
& $\sim 0^\circ$
& $\sim 0^\circ$
& $0.005^\circ$
& $0.006^\circ$
& $0.051^\circ$
& $0.064^\circ$
\\
$\Delta\theta^{}_{23}$
& $-0.001^\circ$
& $-0.0006^\circ$
& $0.050^\circ$
& $0.031^\circ$
& $0.497^\circ$
& $0.307^\circ$
\\
$\Delta\delta^{}_\nu$
& $-0.0004^\circ$
& $0.001^\circ$
& $0.021^\circ$
& $-0.056^\circ$
& $0.207^\circ$
& $-0.552^\circ$
\\
$\Delta\sigma$
& $\sim 0^\circ$
& $\sim 0^\circ$
& $-0.002^\circ$
& $0.0015^\circ$
& $-0.022^\circ$
& $0.015^\circ$
\\ \hline\hline
$m^{}_3 (\Lambda^{}_{\mu\tau}) = 0$
& \multicolumn{2}{c}{SM}
& \multicolumn{2}{c}{MSSM ($\tan\beta = 10$)}
& \multicolumn{2}{c}{MSSM ($\tan\beta = 30$)}
\\ \hline
$\sigma (\Lambda^{}_{\mu\tau})$
& $0$
& $\pi/2$
& $0$
& $\pi/2$
& $0$
& $\pi/2$
\\ \hline
$\Delta\theta^{}_{12}$
& $-0.048^\circ$
& $\sim 0^\circ$
& $2.306^\circ$
& $0.0001^\circ$
& $22.925^\circ$
& $0.001^\circ$
\\
$\Delta\theta^{}_{13}$
& $0.0001^\circ$
& $0.0001^\circ$
& $-0.006^\circ$
& $-0.006^\circ$
& $-0.056^\circ$
& $-0.056^\circ$
\\
$\Delta\theta^{}_{23}$
& $0.0008^\circ$
& $0.0008^\circ$
& $-0.039^\circ$
& $-0.039^\circ$
& $-0.386^\circ$
& $-0.386^\circ$
\\
$\Delta\delta^{}_\nu$
& $-0.0001^\circ$
& $0.034^\circ$
& $0.005^\circ$
& $-1.646^\circ$
& $0.048^\circ$
& $-16.360^\circ$
\\
$\Delta\sigma$
& $-0.00008^\circ$
& $-0.011^\circ$
& $0.004^\circ$
& $0.505^\circ$
& $0.038^\circ$
& $5.019^\circ$
\\ \hline\hline
\end{tabular}
\end{center}
\end{table}

\subsection{Constraints on the leptogenesis mechanism}
\label{section 6.5}

It is well known that the canonical seesaw mechanism~\cite{Fritzsch:1975sr,
Minkowski:1977sc,Yanagida:1979as,GellMann:1980vs,Glashow:1979nm,Mohapatra:1979ia}
can not only offer a natural explanation of the smallness of three active
neutrino masses but also provide a big bonus --- the thermal leptogenesis
mechanism~\cite{Fukugita:1986hr,DiBari:2021fhs} which allows us to naturally
interpret the cosmological matter-antimatter asymmetry signified by the observed
baryon-to-photon ratio $\eta \equiv n^{}_{\rm b}/n^{}_\gamma =
\left(6.14 \pm 0.19\right) \times 10^{-10}$~\cite{Workman:2022ynf}. The
essential point of this mechanism is that the lepton-number-violating and
out-of-equilibrium decays of a heavy Majorana neutrino $N^{}_i$ through
the Yukawa interactions are also CP-violating and may thus produce a net
lepton-antilepton asymmetry in the early Universe,
and the latter can subsequently be converted to a
baryon-antibaryon asymmetry of the Universe with the help of the $B-L$
conserving sphaleron process~\cite{Klinkhamer:1984di,Kuzmin:1985mm}.
Here let us consider the CP-violating asymmetries between the decay
modes $N^{}_i \to \ell^{}_\alpha + H$ and their CP-conjugate counterparts
$N^{}_i \to \overline{\ell^{}_\alpha} + \overline{H}$ (for $\alpha = e, \mu,
\tau$ and $i = 1, 2, 3$)~\cite{Xing:2011zza,Endoh:2003mz}:
\begin{eqnarray}
\varepsilon^{}_{i \alpha} & \equiv &
\frac{\Gamma\big(N^{}_i \to \ell^{}_\alpha + H\big) -
\Gamma\big(N^{}_i \to \overline{\ell^{}_\alpha} +
\overline{H}\big)}{\displaystyle \sum_\alpha \left[\Gamma\big(N^{}_i \to
\ell^{}_\alpha + H\big) + \Gamma\big(N^{}_i \to \overline{\ell^{}_\alpha} +
\overline{H}\big)\right]}
\nonumber \\
& = & \frac{1}{8\pi \big(Y^\dagger_\nu Y^{}_\nu\big)^{}_{ii}} \sum_{j \neq i}
\left\{ {\rm Im}\left[\big(Y^*_\nu\big)^{}_{\alpha i}
\big(Y^{}_\nu\big)^{}_{\alpha j} \big(Y^\dagger_\nu Y^{}_\nu\big)^{}_{ij}
\right] {\cal F}\big(x^{}_{ji}\big) \right.
\nonumber \\
&& + \left. {\rm Im}\left[\big(Y^*_\nu\big)^{}_{\alpha
i} \big(Y^{}_\nu\big)^{}_{\alpha j} \big(Y^\dagger_\nu Y^{}_\nu\big)^*_{ij}
\right] {\cal G}\big(x^{}_{ji}\big) \right\} \; ,
\label{eq:CP-asymmetry}
\end{eqnarray}
in which the asymmetry has been normalized to the total decay rate of $N^{}_i$
so as to make the relevant Boltzmann equations linear in flavor space
\cite{Davidson:2008bu}, and the two loop functions are given by
${\cal F}\big(x^{}_{ji}\big) = \sqrt{x^{}_{ji}}
\left\{1 + 1/\big(1-x^{}_{ji}\big) + \big(1+x^{}_{ji}\big)
\ln \left[x^{}_{ji}/\big(1+x^{}_{ji}\big)\right]\right\}$ and
${\cal G}\big(x^{}_{ji}\big) = 1/\big(1-x^{}_{ji}\big)$ in the SM framework
with $x^{}_{ji} \equiv M^2_j/M^2_i$. If all the interactions in the era of
thermal leptogenesis are blind to the lepton flavors, one may simply focus on
the total flavor-independent CP-violating asymmetries
\begin{eqnarray}
\varepsilon^{}_i = \sum_\alpha \varepsilon^{}_{i\alpha} =
\frac{1}{8\pi \big(Y^\dagger_\nu Y^{}_\nu\big)^{}_{ii}} \sum_{j \neq i} {\rm
Im} \left[\big(Y^\dagger_\nu Y^{}_\nu\big)^2_{ij}\right]
{\cal F}\big(x^{}_{ji}\big) \; .
\label{eq:unflavored}
\end{eqnarray}
In the $M^{}_1 \ll M^{}_2 \ll M^{}_3$ case, for example, it is found that
$\eta \simeq -9.6 \times 10^{-3} \varepsilon^{}_1 \kappa^{}_{\rm f}$ holds as
a good approximation~\cite{Buchmuller:2002rq,Buchmuller:2003gz}, where
$\kappa^{}_{\rm f}$ denotes the efficiency factor determined by solving of the
Boltzmann equations of the heavy Majorana neutrino $N^{}_1$ and the lepton
number densities. So it is possible to successfully account for the
observational value of $\eta$ if a specific seesaw model allows for
$\varepsilon^{}_1 \kappa^{}_{\rm f} \sim -6.4 \times 10^{-8}$.

Instead of going into details of any specific seesaw models and their
consequences on thermal leptogenesis, we are more
interested in possible nontrivial constraints of the $\mu$-$\tau$ reflection
symmetry on both the flavored CP-violating asymmetries $\varepsilon^{}_{i \alpha}$
and the unflavored CP-violating asymmetries $\varepsilon^{}_{i}$. As
shown in Eq.~(\ref{eq:seesaw-constraint}), the $\mu$-$\tau$ reflection
symmetry requires $M^{}_{\rm D} = Y^{}_\nu \langle H\rangle$ and $M^{}_{\rm R}$
to satisfy the conditions $M^{}_{\rm D} = {\cal P} M^{*}_{\rm D} {\cal T}$ and
$M^{}_{\rm R} = {\cal T}^T M^{*}_{\rm R} {\cal T}$ in the canonical seesaw
framework, where $\cal P$ is given in Eq.~(\ref{eq:S-tranformation}) and $\cal T$
is an arbitrary unitary matrix. So $Y^{}_\nu = {\cal P} Y^*_\nu {\cal T}$ holds.
Note that the rates of $N^{}_i$ decays are usually calculated in the basis of
$M^{}_{\rm R} = D^{}_N$ by neglecting tiny non-unitary effects of the PMNS
matrix $U$. In this safe approximation we are left with ${\cal T} = {\cal I}$
from $D^{}_N = {\cal T}^T D^{}_N {\cal T}$ as the simplest choice,
together with $Y^{}_\nu \simeq R D^{}_N/\langle H\rangle$ and
\begin{eqnarray}
U & \simeq & P^{}_l \pmatrix{ c^{}_{12} c^{}_{13} & \hat{s}^*_{12}
c^{}_{13} & \hat{s}^*_{13} \cr
-\hat{s}^{}_{12} c^{}_{23} -
c^{}_{12} \hat{s}^{}_{13} \hat{s}^*_{23} & c^{}_{12} c^{}_{23} -
\hat{s}^*_{12} \hat{s}^{}_{13} \hat{s}^*_{23} & c^{}_{13}
\hat{s}^*_{23} \cr
\hat{s}^{}_{12} \hat{s}^{}_{23} - c^{}_{12}
\hat{s}^{}_{13} c^{}_{23} & -c^{}_{12} \hat{s}^{}_{23} -
\hat{s}^*_{12} \hat{s}^{}_{13} c^{}_{23} & c^{}_{13} c^{}_{23} \cr} \; ,
\nonumber \\
R & \simeq & P^{}_l \pmatrix{ \hat{s}^*_{14} &
\hat{s}^*_{15} & \hat{s}^*_{16} \cr
\hat{s}^*_{24} & \hat{s}^*_{25} &
\hat{s}^*_{26} \cr
\hat{s}^*_{34} & \hat{s}^*_{35} &
\hat{s}^*_{36} \cr} + {\cal O}\left(s^3_{ij}\right) \; ,
\label{eq:U+R}
\end{eqnarray}
where $c^{}_{ij} \equiv \cos\theta^{}_{ij}$, $s^{}_{ij} \equiv \sin\theta^{}_{ij}$,
and $\hat{s}^{}_{ij} \equiv s^{}_{ij} e^{{\rm i} \delta^{}_{ij}}$
with all the flavor mixing angles $\theta^{}_{ij}$ lying in the first quadrant,
and all the $s^{}_{ij}$ terms of $R$ (for $i = 1, 2, 3$ and $j = 4, 5, 6$) are
expected to be below or far below ${\cal O}\big(0.1\big)$~\cite{Antusch:2006vwa}.
Given the $\mu$-$\tau$
reflection symmetry conditions $U = {\cal P} U^*$ and $R = {\cal P} R^*$,
for example, some of the flavor mixing angles and CP-violating phases in
Eq.~(\ref{eq:U+R}) can be well constrained by Eqs.~(\ref{A10}) and
(\ref{A11}) in~\ref{Appendix A}: $\theta^{}_{23} \simeq \pi/4$,
$\delta^{}_\nu \simeq \pm\pi/2$,
$\delta^{}_{12} \simeq 0$ or $\pi$,
$\delta^{}_{13} \simeq 0$ or $\pi$, and
$\delta^{}_{23} \simeq \pm\pi/2$ for the PMNS matrix $U$; and
$\theta^{}_{2j} \simeq \theta^{}_{3j}$,
$\delta^{}_{1j} \simeq 0$ or $\pi$, and
$\delta^{}_{2j} + \delta^{}_{3j} \simeq \pm\pi/2$ (for
$j = 4, 5, 6$) associated with the active-sterile flavor mixing matrix $R$.

We proceed to examine possible impacts of the above constraint conditions on
the CP-violating asymmetries $\varepsilon^{}_{i \alpha}$ and
$\varepsilon^{}_{i}$ (for $i = 1, 2, 3$ and $\alpha = e, \mu, \tau$).
First of all, we find
\begin{eqnarray}
\big(Y^\dagger_\nu Y^{}_\nu\big)^{}_{11} & \simeq &
\frac{M^{2}_1}{\langle H\rangle^2} \left( s^2_{14} +
s^2_{24} + s^2_{34}\right) \simeq
\frac{M^{2}_1}{\langle H\rangle^2} \left( s^2_{14} +
2 s^2_{24}\right) \; ,
\nonumber \\
\big(Y^\dagger_\nu Y^{}_\nu\big)^{}_{22} & \simeq &
\frac{M^{2}_2}{\langle H\rangle^2} \left( s^2_{15} +
s^2_{25} + s^2_{35}\right) \simeq
\frac{M^{2}_2}{\langle H\rangle^2} \left( s^2_{15} +
2 s^2_{25}\right) \; ,
\nonumber \\
\big(Y^\dagger_\nu Y^{}_\nu\big)^{}_{33} & \simeq &
\frac{M^{2}_3}{\langle H\rangle^2} \left( s^2_{16} +
s^2_{26} + s^2_{36}\right) \simeq
\frac{M^{2}_3}{\langle H\rangle^2} \left( s^2_{16} +
2 s^2_{26}\right) \; ,
\nonumber \\
\big(Y^\dagger_\nu Y^{}_\nu\big)^{}_{12} & \simeq &
\frac{M^{}_1 M^{}_2}{\langle H\rangle^2} \left( \hat{s}^{}_{14} \hat{s}^*_{15}
+ \hat{s}^{}_{24} \hat{s}^*_{25} + \hat{s}^{}_{34} \hat{s}^*_{35}\right)
\nonumber \\
& \simeq & \frac{M^{}_1 M^{}_2}{\langle H\rangle^2}
\left[ s^{}_{14} s^{}_{15} + 2 s^{}_{24} s^{}_{25} \cos\left(\delta^{}_{24}
- \delta^{}_{25}\right)\right] \; ,
\nonumber \\
\big(Y^\dagger_\nu Y^{}_\nu\big)^{}_{13} & \simeq &
\frac{M^{}_1 M^{}_3}{\langle H\rangle^2} \left( \hat{s}^{}_{14} \hat{s}^*_{16}
+ \hat{s}^{}_{24} \hat{s}^*_{26} + \hat{s}^{}_{34} \hat{s}^*_{36}\right)
\nonumber \\
& \simeq & \frac{M^{}_1 M^{}_3}{\langle H\rangle^2}
\left[ s^{}_{14} s^{}_{16} + 2 s^{}_{24} s^{}_{26} \cos\left(\delta^{}_{24}
- \delta^{}_{26}\right)\right] \; ,
\nonumber \\
\big(Y^\dagger_\nu Y^{}_\nu\big)^{}_{23} & \simeq &
\frac{M^{}_2 M^{}_3}{\langle H\rangle^2} \left( \hat{s}^{}_{15} \hat{s}^*_{16}
+ \hat{s}^{}_{25} \hat{s}^*_{26} + \hat{s}^{}_{35} \hat{s}^*_{36}\right)
\nonumber \\
& \simeq & \frac{M^{}_2 M^{}_3}{\langle H\rangle^2}
\left[ s^{}_{15} s^{}_{16} + 2 s^{}_{25} s^{}_{26} \cos\left(\delta^{}_{25}
- \delta^{}_{26}\right)\right] \; ,
\label{eq:unflavored-mu-tau}
\end{eqnarray}
which are all real as constrained by the $\mu$-$\tau$ reflection symmetry. As a
direct consequence, we are left with $\varepsilon^{}_{i} \simeq 0$.
In other words, the unflavored leptogenesis mechanism does not work
in the $\mu$-$\tau$ reflection symmetry limit. In the flavor-dependent
case we have
\begin{eqnarray}
{\rm Im}\left[\big(Y^*_\nu\big)^{}_{\alpha i} \big(Y^{}_\nu\big)^{}_{\alpha j}
\big(Y^\dagger_\nu Y^{}_\nu\big)^{}_{ij} \right] & \simeq &
\frac{M^2_i M^2_j}{\langle H\rangle^4} {\rm Im} \left[ R^*_{\alpha i}
R^{}_{\alpha j} \left(R^\dagger R\right)^{}_{ij}\right] \; ,
\nonumber \\
{\rm Im}\left[\big(Y^*_\nu\big)^{}_{\alpha i} \big(Y^{}_\nu\big)^{}_{\alpha j}
\big(Y^\dagger_\nu Y^{}_\nu\big)^*_{ij} \right] & \simeq &
\frac{M^2_i M^2_j}{\langle H\rangle^4} {\rm Im} \left[ R^*_{\alpha i}
R^{}_{\alpha j} \left(R^\dagger R\right)^{*}_{ij}\right] \; ,
\label{eq:flavored-mu-tau}
\end{eqnarray}
where the elements $\left(R^\dagger R\right)^{}_{ij} \simeq
\langle H\rangle^2 \left[D^{-1}_N \left(Y^\dagger_\nu Y^{}_\nu\right)
D^{-1}_N\right]^{}_{ij}$ (for $j \neq i$)
are real as we have shown in Eq.~(\ref{eq:unflavored-mu-tau}). Moreover,
$R^*_{e 1} R^{}_{e 2} \simeq \hat{s}^{}_{14} \hat{s}^*_{15} \simeq
s^{}_{14} s^{}_{15}$, $R^*_{e 1} R^{}_{e 3} \simeq \hat{s}^{}_{14}
\hat{s}^*_{16} \simeq s^{}_{14} s^{}_{16}$ and $R^*_{e 2} R^{}_{e 3}
\simeq \hat{s}^{}_{15} \hat{s}^*_{16} \simeq s^{}_{15} s^{}_{16}$ are
also real. But it is easy to check that
\begin{eqnarray}
R^*_{\mu 1} R^{}_{\mu 2} & \simeq & \hat{s}^{}_{24} \hat{s}^*_{25}
\simeq \hat{s}^{*}_{34} \hat{s}^{}_{35}
\simeq R^{}_{\tau 1} R^{*}_{\tau 2} \; ,
\nonumber \\
R^*_{\mu 1} R^{}_{\mu 3} & \simeq & \hat{s}^{}_{24} \hat{s}^*_{26}
\simeq \hat{s}^{*}_{34} \hat{s}^{}_{36}
\simeq R^{}_{\tau 1} R^{*}_{\tau 3} \; ,
\nonumber \\
R^*_{\mu 2} R^{}_{\mu 3} & \simeq & \hat{s}^{}_{25} \hat{s}^*_{26}
\simeq \hat{s}^{*}_{35} \hat{s}^{}_{36}
\simeq R^{}_{\tau 2} R^{*}_{\tau 3} \;
\label{eq:flavored-mu-tau2}
\end{eqnarray}
are in general complex. So we arrive at
\begin{eqnarray}
{\rm Im}\left[\big(Y^*_\nu\big)^{}_{\mu i} \big(Y^{}_\nu\big)^{}_{\mu j}
\big(Y^\dagger_\nu Y^{}_\nu\big)^{}_{ij} \right] \simeq
- {\rm Im}\left[\big(Y^*_\nu\big)^{}_{\tau i} \big(Y^{}_\nu\big)^{}_{\tau j}
\big(Y^\dagger_\nu Y^{}_\nu\big)^{}_{ij} \right] \; ,
\nonumber \\
{\rm Im}\left[\big(Y^*_\nu\big)^{}_{\mu i} \big(Y^{}_\nu\big)^{}_{\mu j}
\big(Y^\dagger_\nu Y^{}_\nu\big)^{*}_{ij} \right] \simeq
- {\rm Im}\left[\big(Y^*_\nu\big)^{}_{\tau i} \big(Y^{}_\nu\big)^{}_{\tau j}
\big(Y^\dagger_\nu Y^{}_\nu\big)^{*}_{ij} \right] \; .
\label{eq:flavored-mu-tau3}
\end{eqnarray}
As a result, we simply obtain $\varepsilon^{}_{i e} \simeq 0$ and
$\varepsilon^{}_{i\tau} \simeq - \varepsilon^{}_{i\mu}$, which are certainly
consistent with $\varepsilon^{}_i = \varepsilon^{}_{i e} +
\varepsilon^{}_{i\mu} + \varepsilon^{}_{i\tau} \simeq 0$. It is therefore
possible to realize the $\mu$-flavored or $\tau$-flavored thermal leptogenesis
in the $\mu$-$\tau$ reflection symmetry limit~\cite{Xing:2019edp}.
Of course, a realistic and viable model-building exercise of this kind
will depend on some further assumptions for those unknown flavor parameters,
including the heavy Majorana neutrino masses, the active-sterile flavor mixing
angles and the corresponding CP-violating phases (see, e.g., Refs.~\cite{Xing:2015fdg,Zhao:2020bzx,Zhao:2021dwc,Ahn:2008hy,He:2011kn,
Liu:2017frs,Samanta:2018efa,Samanta:2019yeg,Nishi:2020oas,Zhao:2021tgi}).

\setcounter{equation}{0}
\section{More specific examples and extensions}
\label{section 7}

\subsection{The littlest seesaw and partial $\mu$-$\tau$ symmetry}
\label{section 7.1}

The littlest seesaw model is a further simplified version of the minimal seesaw
scenario built in the basis where both the $3\times 3$ charged lepton mass
matrix $M^{}_l$ and the $2\times 2$ right-handed neutrino mass matrix
$\widehat{M}^{}_{\rm R}$ are diagonal and
positive~\cite{King:2015dvf,King:2016yvg,deMedeirosVarzielas:2022fbw}.
Its most salient feature is that the $3\times 2$ Dirac mass matrix
\begin{eqnarray}
\widehat{M}^{}_{\rm D} = \pmatrix{
A^{}_{e\mu} & A^{}_{e\tau} \cr
A^{}_{\mu\mu} & A^{}_{\mu\tau} \cr
A^{}_{\tau\mu} & A^{}_{\tau\tau}} \;
\label{eq:littlest}
\end{eqnarray}
satisfy the conditions
\begin{eqnarray}
A^{}_{e\mu} = 0 \; , \quad A^{}_{\mu\mu} = A^{}_{\tau\mu} \; , \quad
A^{}_{\tau\tau} \propto A^{}_{\mu\tau} \propto A^{}_{e\tau} \; .
\label{eq:littlest-conditions}
\end{eqnarray}
Comparing Eq.~(\ref{eq:littlest-conditions})
with the flavor texture of $\widehat{M}^{}_{\rm D}$
constrained by the $\mu$-$\tau$ reflection symmetry
in Eq.~(\ref{eq:minimal-seesaw-textures}), which requires
$A^{}_{e \tau} = A^*_{e \mu}$, $A^{}_{\tau\tau} = A^*_{\mu\mu}$ and
$A^{}_{\tau\mu} = A^*_{\mu\tau}$ to hold, one can see that the littlest
seesaw model does not really respect the $\mu$-$\tau$ reflection symmetry.
However, this model may possess a {\it partial} $\mu$-$\tau$ symmetry and
thus be able to successfully produce $\theta^{}_{23} = \pi/4$ and
$\delta^{}_\nu = \pm\pi/2$~\cite{King:2018kka,King:2019tbt}.
Note that the predictions of the littlest seesaw
model are not only a partial $\mu$-$\tau$ symmetry but also a
specific correlation between the two nonzero neutrino masses.
To see these points clearly, let us consider two viable textures of
$M^{}_\nu \simeq -\widehat{M}^{}_{\rm D} \widehat{M}^{-1}_{\rm R}
\widehat{M}^{T}_{\rm D}$ obtained in two special littlest seesaw
scenarios~\cite{King:2015dvf,King:2016yvg,King:2018kka,King:2019tbt}:
\begin{eqnarray}
{\rm Scenario~A}: && M^{}_\nu = m^{}_* \left[\pmatrix{
1 & 3 & 1 \cr 3 & 9 & 3 \cr 1 & 3 & 1} + r \hspace{0.03cm}
e^{+{\rm i}\varphi} \pmatrix{
0 & 0 & 0 \cr 0 & 1 & 1 \cr 0 & 1 & 1}\right] \; ,
\nonumber \\
{\rm Scenario~B}: && M^{}_\nu = m^{}_* \left[\pmatrix{
1 & 1 & 3 \cr 1 & 1 & 3 \cr 3 & 3 & 9} + r \hspace{0.03cm}
e^{-{\rm i}\varphi} \pmatrix{
0 & 0 & 0 \cr 0 & 1 & 1 \cr 0 & 1 & 1}\right] \; ,
\label{eq:patterns}
\end{eqnarray}
where $m^{}_*$ denotes the mass scale of three light Majorana neutrinos,
$\varphi = 2\pi/3$, and $r$ is a real and positive free parameter. Although
$M^{}_\nu$ seems to have little to do with the $\mu$-$\tau$ flavor symmetry,
one find that $H^{}_\nu \equiv M^{}_\nu M^\dagger_\nu$ has the texture
\begin{eqnarray}
H^{}_\nu = 11 m^{2}_* \pmatrix{
1 & \pm\left(1 - {\rm i} \hspace{0.02cm} 2\sqrt{3}\right)
& \mp\left(1 + {\rm i} \hspace{0.02cm} 2\sqrt{3}\right) \cr
\pm\left(1 + {\rm i} \hspace{0.02cm} 2\sqrt{3}\right) & 19
& 17 - {\rm i} \hspace{0.02cm} 4\sqrt{3} \cr
\mp\left(1 - {\rm i} \hspace{0.02cm} 2\sqrt{3}\right) &
17 + {\rm i} \hspace{0.02cm} 4\sqrt{3} & 19} \;
\label{eq:Hs}
\end{eqnarray}
for $ r = 11$, where two possible sign options just correspond to scenarios
A and B of $M^{}_\nu$ in Eq.~({\ref{eq:patterns}).
This special but interesting case can be obtained in
a supersymmetric model with the $\rm S^{}_{4 L} \times S^{}_{\rm 4 R}
\times U(1) \times U(1)^\prime$ flavor symmetry~\cite{King:2019tbt}.
It is easy to verify
\begin{eqnarray}
H^{}_\nu = \pmatrix{ -1 & 0 & 0 \cr 0 & 1 & 0 \cr 0 & 0 & 1}
{\cal P} H^*_\nu {\cal P} \pmatrix{ -1 & 0 & 0 \cr 0 & 1 & 0 \cr
0 & 0 & 1} \; ,
\label{eq:H-mu-tau-like}
\end{eqnarray}
and therefore $H^{}_\nu$ essentially respects the $\mu$-$\tau$ reflection
symmetry~\cite{Xing:2017mkx,King:2018kka,King:2019tbt}. In this sense
one may simply argue that the two special littlest seesaw scenarios under
discussion possess a {\it partial} $\mu$-$\tau$ reflection symmetry.

Note that the above two scenarios are phenomenologically equivalent to
each other, as their difference reflected in the textures of $H^{}_\nu$
can actually be eliminated by redefining the phases of three charged
lepton fields~\cite{King:2019tbt}. This point can also become transparent
when making the transformation $U^\dagger H^{}_\nu U = D^2_\nu$, where
$U$ is just the PMNS lepton flavor mixing matrix and its main part
can be uniquely determined:
\begin{eqnarray}
U = P^{}_l \pmatrix{ \frac{2}{\sqrt 6} & \frac{1}{\sqrt 6} a^{}_+
& \frac{1}{\sqrt 6} a^{}_- \cr
\vspace{-0.45cm} \cr
\frac{-1}{\sqrt 6} & \frac{1}{\sqrt 6} a^{}_+ + \frac{\rm i}{2}
a^{}_- & \frac{1}{\sqrt 6} a^{}_- - \frac{\rm i}{2} a^{}_+ \cr
\vspace{-0.45cm} \cr
\frac{-1}{\sqrt 6} & \frac{1}{\sqrt 6} a^{}_+ - \frac{\rm i}{2} a^{}_-
& \frac{1}{\sqrt 6} a^{}_- + \frac{\rm i}{2} a^{}_+} P^{}_\nu \; ,
\label{eq:U-like}
\end{eqnarray}
where $a^{2}_{\pm} = 1 \pm 11/\big(3\sqrt{17}\big)$, namely
$a^{}_+ \simeq 1.3745$ and $a^{}_- \simeq 0.3327$. As a result, we have
\begin{eqnarray}
\theta^{}_{12} & = & \arctan\left(\frac{a^{}_+}{2}\right)
\simeq 34.50^\circ \; ,
\nonumber \\
\theta^{}_{13} & = & \arcsin\left(\frac{a^{}_-}{\sqrt 6}\right)
\simeq 7.81^\circ \; ,
\nonumber \\
\theta^{}_{23} & = & 45^\circ \; , \quad \delta^{}_\nu = \pm 90^\circ \; ,
\label{eq:U-like-parameters}
\end{eqnarray}
together with the normal neutrino mass spectrum
\begin{eqnarray}
m^{}_1 & = & 0 \; ,
\nonumber \\
m^{}_2 & = & \sqrt{\frac{33}{2} \left(13 - 3\sqrt{17}\right)}
\hspace{0.1cm} m^{}_* \simeq 3.23 \hspace{0.05cm} m^{}_* \; ,
\nonumber \\
m^{}_3 & = & \sqrt{\frac{33}{2} \left(13 + 3\sqrt{17}\right)}
\hspace{0.1cm} m^{}_* \simeq 20.46 \hspace{0.05cm} m^{}_* \; .
\label{eq:U-like-masses}
\end{eqnarray}
Therefore, $\Delta m^2_{21}/\Delta m^2_{31} =
\left(m^{}_2/m^{}_3\right)^2 \simeq 0.025$.
Note that the normal structural hierarchy of
$H^{}_\nu$ in Eq.~(\ref{eq:Hs}), namely $\left|(H^{}_\nu)^{}_{33}\right|
= \left|(H^{}_\nu)^{}_{22}\right| > \left|(H^{}_\nu)^{}_{23}\right|
\gg \left|(H^{}_\nu)^{}_{13}\right| = \left|(H^{}_\nu)^{}_{12}\right|
> \left|(H^{}_\nu)^{}_{11}\right|$, implies that this model definitely
disfavors the inverted neutrino mass ordering.
Comparing these interesting results with the experimental data listed
in Table~\ref{table2}, one can see that the model predictions for
$\theta^{}_{13}$ and $\Delta m^2_{21}/\Delta m^2_{31}$ are both smaller
than their $3\sigma$ global-fit intervals. A natural and straightforward
way out of such slight disagreements is to take into account the RGE-induced
radiative corrections to the relevant flavor parameters in the MSSM
framework by setting $\Lambda^{}_{\mu\tau} \sim 10^{14} ~{\rm GeV}$ and
inputting sufficiently large values of $\tan\beta$, as done in Refs.~\cite{King:2019tbt,King:2016yef,Geib:2017bsw}.

The first benchmark model to realize the littlest seesaw idea is based
on the flavor symmetry group
$\rm S^{}_4 \times Z^{}_3 \times Z^\prime_3$~\cite{King:2015dvf},
where the $\rm S^{}_4$ flavor group helps produce the proper vacuum
alignments so as to achieve the flavor textures of $M^{}_\nu$ in
Eq.~(\ref{eq:patterns}), the $\rm Z^{}_3$ group is chosen to fix
$\varphi = 2\pi/3$, and the $\rm Z^\prime_3$ group is introduced to generate
the strong mass hierarchy of three charged leptons. Another possible way of
building the littlest seesaw model is to make use of the
$\rm S^{}_4 \times U(1)$ flavor group in a semi-direct supersymmetric
framework~\cite{King:2016yvg}, and the phase parameter $\varphi$ of
$M^{}_\nu$ can be fixed to be associated with one of the cube roots of unity
(i.e., $\omega = \exp\left({\rm i}\varphi\right)$ and $\omega^3 =1$) by
extending the $\rm S^{}_4 \times U(1)$ symmetry to
$\rm S^{}_4 \times U(1) \times \left(Z^{}_3\right)^5$. To accommodate the
partial $\mu$-$\tau$ reflection symmetry for scenario A of $M^{}_\nu$ in
Eq.~(\ref{eq:patterns}), one may invoke an
$\rm S^{}_{4 L} \times S^{}_{4 R} \times U(1) \times U(1)^\prime$ flavor
symmetry in a supersymmetric littlest seesaw scenario~\cite{King:2019tbt},
where the two right-handed neutrino fields transform as a doublet in
$\rm S^{}_{4 R}$. Although this seesaw model is quite complicated at
a given ultrahigh energy scale, its low-energy consequences are rather
simple and instructive as have been discussed above.

\subsection{Zero textures with a $\rm Z^{}_2$ reflection symmetry}
\label{section 7.2}

If some elements of the quark or lepton mass matrices are vanishing or vanishingly
small, as dictated by a kind of underlying flavor symmetry, it will be possible
to establish one or more testable correlations between the observable flavor mixing
quantities and the quark or lepton mass ratios~\cite{Xing:2019vks,Fritzsch:1999ee}.
A combination of this popular texture-zero approach with the $\mu$-$\tau$ reflection
symmetry may help determine or constrain the absolute neutrino mass scale. For
example, confronting the $\mu$-$\tau$ reflection symmetry constraints with current
experimental data on neutrino oscillations and the cosmological bound on neutrino
masses indicates that the texture zero $\langle m\rangle^{}_{ee} = 0$ (or
$\langle m\rangle^{}_{\mu\tau} = 0$) is essentially allowed for the normal (or
inverted) neutrino mass ordering with $\rho = 0$ and $\sigma = \pi/2$ or
$\rho = \pi/2$ and $\sigma = 0$ (or $\rho = \pi/2$ and
$\sigma = 0$)~\cite{Xing:2017cwb}. Along this line of thought, some interesting
attempts have made to build a more predictive neutrino mass model (see, for
instance, Refs.~\cite{Nishi:2016wki,Nishi:2018vlz,Liu:2018hwg,Liu:2018cka,
Yang:2020qsa,Yang:2021xob}).

Here let us briefly introduce a typical example of this kind~\cite{Yang:2020qsa},
in which the Hermitian charged lepton mass matrix $M^{}_l$ and the symmetric
Majorana neutrino mass matrix $M^{}_\nu$ not only respect the $\mu$-$\tau$
reflection symmetry but also have the texture zeros
$\big(M^{}_l\big)^{}_{11} = \big(M^{}_\nu\big)^{}_{ee} = 0$. To be more explicit,
\begin{eqnarray}
M^{}_l = \pmatrix{
0 & A^{}_l & A^{*}_l \cr
A^{*}_l & B^{}_l & C^{}_l \cr
A^{}_l & C^*_l & B^{}_l} \; , \quad
M^{}_\nu = \pmatrix{
0 & A^{}_\nu & A^{*}_\nu \cr
A^{}_\nu & B^{}_\nu & C^{}_\nu \cr
A^{*}_\nu & C^{}_\nu & B^{*}_\nu} \; ,
\label{eq:lepton-M}
\end{eqnarray}
where $B^{}_l$ and $C^{}_\nu$ are real. It is straightforward to
check that such textures of $M^{}_l$ and $M^{}_\nu$ satisfy the
$\mu$-$\tau$ reflection symmetry:
\begin{eqnarray}
M^{}_l = {\cal P} M^{*}_l {\cal P} \; , \quad
M^{}_\nu = {\cal P} M^{*}_\nu {\cal P} \; .
\label{eq:lepton-mu-tau}
\end{eqnarray}
After making the unitary transformations
$O^\dagger_l M^{}_l M^\dagger_l O^{}_l = D^2_l$
and $O^\dagger_\nu M^{}_\nu O^*_\nu = D^{}_\nu$, where
$D^{}_l = {\rm Diag}\big\{m^{}_e , m^{}_\mu , m^{}_\tau\big\}$ and
$D^{}_\nu = {\rm Diag}\big\{m^{}_1 , m^{}_2 , m^{}_3\big\}$, one may immediately
obtain the PMNS lepton flavor mixing matrix $U = O^{\dagger}_l O^{}_\nu$. Note that
the $\mu$-$\tau$ reflection symmetry condition $U = O^{}_\nu = {\cal P} U^* \zeta$
in the basis of $M^{}_l = D^{}_l$ is essentially supported by current
neutrino oscillation data, and hence we have to require that the contribution
of $O^{}_l$ to $U$ be insignificant so as to assure $U = O^{\dagger}_l O^{}_\nu$
to be compatible with the experimental data in the present situation. To see
that such a requirement is actually satisfiable, we make a common {\it bimaximal}
transformation for $M^{}_l$ and $M^{}_\nu$ in Eq.~(\ref{eq:lepton-M}):
\begin{eqnarray}
O^\dagger_{\rm BM} M^{}_l O^{}_{\rm BM} = M^\prime_l \; , \quad
O^\dagger_{\rm BM} M^{}_\nu O^{*}_{\rm BM} = M^\prime_\nu \; ,
\label{eq:lepton-M2}
\end{eqnarray}
where
\begin{eqnarray}
O^{}_{\rm BM} =
\pmatrix{ 1 & 0 & 0 \cr
0 & \frac{1}{\sqrt 2} & \frac{1}{\sqrt 2} \cr \vspace{-0.4cm} \cr
0 & \frac{-1}{\sqrt 2} & \frac{1}{\sqrt 2} \cr} \; ,
\label{eq:BM}
\end{eqnarray}
and
\begin{eqnarray}
M^{\prime}_l = \pmatrix{
0 & {\rm i} \sqrt{2} \hspace{0.04cm} {\rm Im} \left(A^{}_l\right)
& \sqrt{2} \hspace{0.04cm} {\rm Re} \left(A^{}_l\right) \cr
-{\rm i} \sqrt{2} \hspace{0.04cm} {\rm Im} \left(A^{}_l\right)
& \hspace{0.035cm} B^{}_l - {\rm Re} \left(C^{}_l\right) \hspace{0.035cm}
& {\rm i} \hspace{0.04cm} {\rm Im} \left(C^{}_l\right) \cr
\sqrt{2} \hspace{0.04cm} {\rm Re} \left(A^{}_l\right)
& -{\rm i} \hspace{0.04cm} {\rm Im} \left(C^{}_l\right)
& B^{}_l + {\rm Re} \left(C^{}_l\right) \cr} \; ,
\nonumber \\
M^{\prime}_\nu = \pmatrix{
0 & {\rm i} \sqrt{2} \hspace{0.04cm} {\rm Im} \left(A^{}_\nu\right)
& \sqrt{2} \hspace{0.04cm} {\rm Re} \left(A^{}_\nu\right) \cr
{\rm i} \sqrt{2} \hspace{0.04cm} {\rm Im} \left(A^{}_\nu\right)
& {\rm Re} \left(B^{}_\nu\right) - C^{}_\nu
& {\rm i} \hspace{0.04cm} {\rm Im} \left(B^{}_\nu\right) \cr
\sqrt{2} \hspace{0.04cm} {\rm Re} \left(A^{}_\nu\right)
& {\rm i} \hspace{0.04cm} {\rm Im} \left(B^{}_\nu\right)
& {\rm Re} \left(B^{}_\nu\right) + C^{}_\nu \cr} \; .
\label{eq:lepton-M3}
\end{eqnarray}
Of course, $O^{}_{\rm BM}$ does not affect the PMNS matrix $U$ because it
will be cancelled out. In this new basis $M^\prime_l$ and $M^\prime_\nu$
still share a parallel one-zero texture. But it is necessary to break such
an elegant structural parallelism in order to avoid a significant
cancellation between $O^\dagger_l$ and $O^{}_\nu$ in calculating $U$.
To do so, one may simply assume ${\rm Re}\left(A^{}_l\right) = 0$ so as to
achieve two more zero elements in $M^\prime_l$ --- a Fritzsch-like texture
which has been well studied in the literature (see, e.g.,
Refs.~\cite{Xing:2003zd,Matsuda:2006xa,Gupta:2012fsl,Barranco:2012ci}).
But here one should keep in mind that such an assumption
is purely phenomenological.
The strong mass hierarchy $m^{}_e \ll m^{}_\mu \ll m^{}_\tau$ implies that
the nonzero elements of $M^\prime_l$ should be strongly hierarchical too,
such that the phenomenological
conditions $\big|{\rm Im}\left(A^{}_l\right)\big| \ll \big|B^{}_l
- {\rm Re}\left(C^{}_l\right)\big| \ll \big|B^{}_l +
{\rm Re}\left(C^{}_l\right)\big|$ and $\big|{\rm Im}\left(C^{}_l\right)\big|
\sim \big|B^{}_l - {\rm Re}\left(C^{}_l\right)\big|$ hold. Then
the leading $(1,2)$ and $(2,3)$ rotation angles of $O^{}_l$ are expected
to be $\sqrt{m^{}_e/m^{}_\mu} \simeq 0.069$ and
$m^{}_\mu/m^{}_\tau \simeq 0.059$, respectively~\cite{Xing:2003zd}. In this
case the PMNS matrix $U$ is dominated by $O^{}_\nu$, and the small
contribution of $O^{}_l$ provides a natural source of slight $\mu$-$\tau$
reflection symmetry breaking. Moreover, the vanishing entry of $M^{}_\nu$
allows for a correlation among the three neutrino masses and thus makes
the neutrino mass spectrum determinable with the help of current neutrino
oscillation data. A detailed analysis of the phenomenological
consequences of this interesting lepton mass scenario has been done in
Ref.~\cite{Yang:2020qsa}.

It is worth pointing out that the Hermitian up- and down-type quark mass
matrices may also take the $\rm Z^{}_2$ reflection-invariant zero textures
like that of $M^{}_l$ in Eq.~(\ref{eq:lepton-M}):
\begin{eqnarray}
M^{}_{\rm u} = \pmatrix{
0 & A^{}_{\rm u} & A^{*}_{\rm u} \cr
A^{*}_{\rm u} & B^{}_{\rm u} & C^{}_{\rm u} \cr
A^{}_{\rm u} & C^*_{\rm u} & B^{}_{\rm u}} \; , \quad
M^{}_{\rm d} = \pmatrix{
0 & A^{}_{\rm d}  & A^{*}_{\rm d} \cr
A^{*}_{\rm d} & B^{}_{\rm d} & C^{}_{\rm d} \cr
A^{}_{\rm d} & C^{*}_{\rm d} & B^{}_{\rm d}} \; ,
\label{eq:quark-M}
\end{eqnarray}
which satisfy a similar $\rm Z^{}_2$ reflection symmetry
\begin{eqnarray}
M^{}_{\rm u} = {\cal P} M^{*}_{\rm u} {\cal P} \; , \quad
M^{}_{\rm d} = {\cal P} M^{*}_{\rm d} {\cal P} \; .
\label{eq:quark-Z2}
\end{eqnarray}
A pursuit of such a possible {\it universal} flavor textures of fermion mass
matrices is based on the phenomenological conjecture that the charged fermions
and massive neutrinos may have a common or parallel mass generation and flavor
mixing mechanism. At least the charged leptons and quarks should share
some similarities in their flavor dynamics. For the sake of illustration, let
us make a slightly different bimaximal transformations for Hermitian
$M^{}_{\rm u}$ and $M^{}_{\rm d}$ as follows~\cite{Yang:2020qsa}:
\begin{eqnarray}
O^{\prime\dagger}_{\rm BM} M^{}_{\rm u} O^{\prime}_{\rm BM}
= M^\prime_{\rm u} \; , \quad
O^{\prime\dagger}_{\rm BM} M^{}_{\rm d} O^{\prime}_{\rm BM}
= M^\prime_{\rm d} \; ,
\label{eq:quark-M2}
\end{eqnarray}
where
\begin{eqnarray}
O^{\prime}_{\rm BM} =
\pmatrix{ 1 & 0 & 0 \cr
0 & \frac{\rm i}{\sqrt 2} & \frac{\rm i}{\sqrt 2} \cr \vspace{-0.4cm} \cr
0 & \frac{-1}{\sqrt 2} & \frac{1}{\sqrt 2} \cr} \; .
\label{eq:BM2}
\end{eqnarray}
The texture of $M^\prime_{\rm q}$ (for $\rm q = u$ or $\rm d$) turns out to be
\begin{eqnarray}
M^\prime_{\rm q} =
\pmatrix{ 0 & -\frac{1 - {\rm i}}{\sqrt 2} X^{}_{\rm q} &
\frac{1 + {\rm i}}{\sqrt 2} X^{\prime}_{\rm q} \cr \vspace{-0.4cm} \cr
-\frac{1 + {\rm i}}{\sqrt 2} X^{}_{\rm q} &
B^{}_{\rm q} - {\rm Im}\left(C^{}_{\rm q}\right) &
-{\rm i} \hspace{0.04cm} {\rm Re}\left(C^{}_{\rm q}\right) \cr \vspace{-0.4cm} \cr
\frac{1 - {\rm i}}{\sqrt 2} X^{\prime}_{\rm q} &
{\rm i} \hspace{0.04cm} {\rm Re}\left(C^{}_{\rm q}\right) &
B^{}_{\rm q} + {\rm Im}\left(C^{}_{\rm q}\right)} \; ,
\label{eq:quark-M3}
\end{eqnarray}
where
\begin{eqnarray}
X^{}_{\rm q} & = & {\rm Re}\left(A^{}_{\rm q}\right)
+ {\rm Im}\left(A^{}_{\rm q}\right) \; ,
\nonumber \\
X^{\prime}_{\rm q} & = & {\rm Re}\left(A^{}_{\rm q}\right)
- {\rm Im}\left(A^{}_{\rm q}\right) \; .
\label{eq:quark-M4}
\end{eqnarray}
So $X^{\prime}_{\rm q} = 0$, or equivalently ${\rm Re}\left(A^{}_{\rm q}\right)
= {\rm Im}\left(A^{}_{\rm q}\right)$, allows us to obtain an extra independent
texture zero for $M^\prime_{\rm q}$. It is well known that the resulting parallel
textures of $M^\prime_{\rm u}$ and $M^\prime_{\rm d}$ can fit current experimental
data on the quark mass spectrum and CKM flavor mixing parameters~\cite{Xing:2019vks,
Fritzsch:1999ee,Gupta:2012fsl,Xing:2015sva,Fritzsch:2021ipb}, and thus the
original $\rm Z^{}_2$ reflection-invariant textures of $M^{}_{\rm u}$ and
$M^{}_{\rm d}$ in Eq.~(\ref{eq:quark-M}) should also be viable.

\subsection{The $\mu$-$\tau$ reflection symmetry of (3+1) flavors}
\label{section 7.3}

The conjecture that there might exist a light sterile neutrino species (of mass
$\sim 1~{\rm eV}$) in addition to the three active neutrino flavors, which was
primarily motivated by the LSND~\cite{LSND:1996ubh,LSND:2001aii} and
MiniBooNE~\cite{MiniBooNE:2007uho,MiniBooNE:2018esg} accelerator neutrino
(or antineutrino) oscillation experiments and the reactor antineutrino
anomaly~\cite{Mention:2011rk,Mueller:2011nm}, has been a subject of intense scholarly
debate~\cite{DayaBay:2014fct,DayaBay:2016qvc,MINOS:2020iqj}. From the phenomenological
point of view, the $(3+1)$ active-sterile flavor mixing scheme deserves further
attention if it can simultaneously interpret all or most the available neutrino
oscillation data~\cite{Kopp:2013vaa,Gariazzo:2015rra,Gariazzo:2017fdh,Dentler:2018sju}.
Since the active-sterile flavor mixing angles are experimentally constrained to be
very small, the dominant $3\times 3$ flavor mixing submatrix $U$ for the active
neutrinos in the overall $4\times 4$ flavor mixing matrix $\cal U$ should still
have the approximate $\mu$-$\tau$ symmetry relations
$\big|U^{}_{\mu i}\big| \simeq \big|U^{}_{\tau i}\big|$ (for $i = 1, 2, 3$). In
fact, the small active-sterile flavor mixing effects are expected to slightly break
the exact $\mu$-$\tau$ permutation symmetry of three active neutrinos in the $(3+1)$
neutrino mixing scheme, because $\nu^{}_\mu$ and $\nu^{}_\tau$ may mix with
the sterile flavor in different ways such that the equality between
$\big|U^{}_{\mu i}\big|$ and $\big|U^{}_{\tau i}\big|$ in the $4\times 4$
unitary matrix $\cal U$ is in general not guaranteed~\cite{Xing:2015fdg,Barry:2011wb,
Barry:2011fp,Merle:2014eja,Rivera-Agudelo:2015vza,Borah:2016fqj,Sarma:2018bgf}.
So it is also of
great interest to consider the $\mu$-$\tau$ reflection symmetry in the same scheme,
from the perspective of either model building or pure phenomenology. In particular,
the origin of a light sterile neutrino needs an acceptable theoretical reason.

In the basis where the flavor eigenstates of three charged leptons are identical
with their corresponding mass eigenstates (i.e., $M^{}_l = D^{}_l$), let us
simply assume that the $4\times 4$ active-sterile neutrino mixing matrix $\cal U$
to be unitary. To be specific,
\begin{eqnarray}
\pmatrix{ \nu^{}_e \cr \nu^{}_\mu \cr \nu^{}_\tau \cr \nu^{}_s}
= \pmatrix{
{\cal U}^{}_{e 1} & {\cal U}^{}_{e 2} & {\cal U}^{}_{e 3}
& {\cal U}^{}_{e 4} \cr
{\cal U}^{}_{\mu 1} & {\cal U}^{}_{\mu 2} & {\cal U}^{}_{\mu 3}
& {\cal U}^{}_{\mu 4} \cr
{\cal U}^{}_{\tau 1} & {\cal U}^{}_{\tau 2} & {\cal U}^{}_{\tau 3}
& {\cal U}^{}_{\tau 4} \cr
{\cal U}^{}_{s 1} & {\cal U}^{}_{s 2} & {\cal U}^{}_{s 3} & {\cal U}^{}_{s 4}}
\pmatrix{ \nu^{}_1 \cr \nu^{}_2 \cr \nu^{}_3 \cr \nu^{}_4} \; ,
\label{eq:(3+1)-mixing}
\end{eqnarray}
where $\nu^{}_s$ and $\nu^{}_4$ stand respectively for the flavor and mass
eigenstates of a light sterile neutrino. Note, however, that by definition
the sterile neutrino flavor $\nu^{}_s$ does not directly participate in the
standard weak interactions. So it is actually the $3\times 4$ submatrix of
$\cal U$ that appears in the standard charged-current weak interactions:
\begin{eqnarray}
\pmatrix{ \nu^{}_e \cr \nu^{}_\mu \cr \nu^{}_\tau}^{}_{\hspace{-0.15cm} \rm L}
= \pmatrix{
{\cal U}^{}_{e 1} & {\cal U}^{}_{e 2} & {\cal U}^{}_{e 3}
& {\cal U}^{}_{e 4} \cr
{\cal U}^{}_{\mu 1} & {\cal U}^{}_{\mu 2} & {\cal U}^{}_{\mu 3}
& {\cal U}^{}_{\mu 4} \cr
{\cal U}^{}_{\tau 1} & {\cal U}^{}_{\tau 2} & {\cal U}^{}_{\tau 3}
& {\cal U}^{}_{\tau 4}}
\pmatrix{ \nu^{}_1 \cr \nu^{}_2 \cr \nu^{}_3 \cr
\nu^{}_4}^{}_{\hspace{-0.15cm} \rm L} \; .
\label{eq:(3+1)-mixing2}
\end{eqnarray}
One may trivially extend the $\mu$-$\tau$ reflection symmetry relationship
$U = {\cal P} U^* \zeta$, which has been extracted from the experimentally favored
relations $\big|U^{}_{\mu i}\big| = \big|U^{}_{\tau i}\big|$ (for $i = 1, 2, 3$),
to the $(3+1)$ active-sterile flavor mixing case:
${\cal U} = {\cal S} {\cal U}^* \xi$, where
\begin{eqnarray}
{\cal S} = {\cal S}^T = {\cal S}^\dagger = \pmatrix{
{\cal P} & {\bf 0} \cr
{\bf 0}^T & 1} \;
\label{eq:(3+1)-transformation}
\end{eqnarray}
with ${\bf 0}$ denoting the $3\times 1$ zero matrix,
and $\xi = {\rm Diag}\{\eta^{}_1, \eta^{}_2, \eta^{}_3, \eta^{}_4\}$ with
$\eta^{}_i = \pm 1$.
Then $\big|{\cal U}^{}_{\mu i}\big| = \big|{\cal U}^{}_{\tau i}\big|$
holds (for $i = 1, 2, 3, 4$)~\cite{Chakraborty:2019rjc}, as in the standard
three-flavor mixing case.

But it is nontrivial to identify that behind ${\cal U} = {\cal S} {\cal U}^* \xi$
is a $\mu$-$\tau$ reflection symmetry for the overall neutrino mass term, because the
latter is strongly model-dependent. Here we follow the spirit of the canonical seesaw
mechanism but assume one of the three right-handed neutrino fields to be subject to
the mass scale $\mu^{}_s \sim {\cal O}(1) ~{\rm eV}$~\cite{Barry:2011wb,Zhao:2022trb}.
In this contrived case the effective neutrino mass terms can be expressed as
\begin{eqnarray}
-{\cal L}^{}_{\rm m} & = & \overline{\nu^{}_{\rm L}} \hspace{0.05cm}
M^{}_{\rm D} N^{\prime}_{\rm R} + \overline{\nu^{}_{\rm L}} \hspace{0.05cm}
M^{}_{\rm S} \nu^{\prime}_s + \frac{1}{2} \hspace{0.05cm}
\overline{(N^{\prime}_{\rm R})^c} \hspace{0.05cm} M^{}_{\rm R} N^{\prime}_{\rm R}
+ \frac{1}{2} \hspace{0.05cm}
\overline{(\nu^{\prime}_s)^c} \hspace{0.05cm} \mu^{}_s \nu^{\prime}_s + {\rm h.c.}
\nonumber \\
& = & \frac{1}{2} \hspace{0.05cm} \overline{\pmatrix{
\nu^{}_{\rm L} \hspace{-0.25cm} & (\nu^{\prime}_s)^c \hspace{-0.25cm}
& (N^{\prime}_{\rm R})^c}}
\pmatrix{ {\bf 0} & M^{}_{\rm S} & M^{}_{\rm D} \cr
M^{T}_{\rm S} & \mu^{}_s & {\bf 0}^\prime \cr
M^T_{\rm D} & {\bf 0}^{\prime T} & M^{}_{\rm R} \cr}
\pmatrix{ (\nu^{}_{\rm L})^c \cr \nu^{\prime}_s \cr N^{\prime}_{\rm R} \cr}
+ {\rm h.c.} \; ,
\label{eq:sterile-seesaw}
\end{eqnarray}
where $N^{\prime}_{\rm R} = (N^{\prime}_{\mu \rm R} , N^{\prime}_{\tau \rm R})^T$
denotes the column vector of two heavy right-handed neutrino fields like that
given in Eq.~(\ref{eq:minimal-seesaw-mass-matrix}) for the minimal seesaw
scenario, the mass matrices $M^{}_{\rm D}$, $M^{}_{\rm S}$ and $M^{}_{\rm R}$ are
respectively of $3\times 2$, $3\times 1$ and $2\times 2$, and the zero matrices
$\bf 0$ and ${\bf 0}^\prime$ are accordingly of $3\times 3$ and
$1\times 2$. As the mass scale of $M^{}_{\rm R}$ is expected to be far above the
electroweak scale and hence considerably higher
than the mass scale of $M^{}_{\rm D}$, one may integrate out the heavy degrees
of freedom and arrive at an effective $4\times 4$ Majorana mass matrix for
the three active neutrinos and one light sterile neutrino:
\begin{eqnarray}
{\cal M} \simeq \pmatrix{
-M^{}_{\rm D} M^{-1}_{\rm R} M^T_{\rm D} & M^{}_{\rm S} \cr
M^T_{\rm S} & \mu^{}_s} \; .
\label{eq:sterile-seesaw-matrix}
\end{eqnarray}
Switching off $M^{}_{\rm S}$ will make the sterile neutrino completely isolated
from the active neutrinos. So it should be natural to assume that the
mass scale of $M^{}_{\rm S}$ is far below $\mu^{}_s$. Such an assumption not only
allows us to estimate $m^{}_4 \simeq \mu^{}_s$ for the sterile neutrino mass and
${\cal U}^{}_{\alpha 4} \simeq (M^{}_{\rm S})^{}_{\alpha 1}/\mu^{}_s$ for the
active-sterile flavor mixing matrix elements, where $\alpha$ runs over $e$, $\mu$
and $\tau$, but also helps us to achieve an approximate seesaw-like formula for
the effective mass matrix of three active Majorana neutrinos~\cite{Barry:2011wb}:
\begin{eqnarray}
M^{}_\nu \simeq -M^{}_{\rm D} M^{-1}_{\rm R} M^T_{\rm D}
- M^{}_{\rm S} \mu^{-1}_s M^T_{\rm S} \; .
\label{eq:sterile-seesaw-formula}
\end{eqnarray}
Within this active-sterile flavor mixing scenario, the flavor eigenstate $\nu^{}_s$
appearing in Eq.~(\ref{eq:(3+1)-mixing}) can indirectly participate in the weak
charged-current interactions if its left-handed state is identified as
$(\nu^\prime_s)^c$ that has been introduced in Eq.~(\ref{eq:sterile-seesaw}).

Like the $\rm Z^{}_2$ reflection transformations made in
Eq.~(\ref{eq:minimal-seesaw-reflection}), here we require that ${\cal L}^{}_{\rm m}$
in Eq.~(\ref{eq:sterile-seesaw}) keep invariant under the transformations
\begin{eqnarray}
\nu^{}_{\rm L} \to {\cal P} (\nu^{}_{\rm L})^c \; , \quad
N^{\prime}_{\rm R} \to {\cal Z} (N^{\prime}_{\rm R})^c \; , \quad
\nu^\prime_s \to (\nu^\prime_s)^c \; ,
\label{eq:sterile-seesaw-transformation}
\end{eqnarray}
where $\cal P$ and $\cal Z$ have been given in Eqs.~(\ref{eq:S-tranformation}) and
(\ref{eq:S2-tranformation}), respectively. Then we are left with the constraint
conditions
\begin{eqnarray}
M^{}_{\rm D} = {\cal P} M^{*}_{\rm D} {\cal Z} \; , \quad
M^{}_{\rm R} = {\cal Z} M^{*}_{\rm R} {\cal Z} \; , \quad
M^{}_{\rm S} = {\cal P} M^{*}_{\rm S} \; , \quad
\mu^{}_s = \mu^*_s \; .
\label{eq:sterile-seesaw-constraint}
\end{eqnarray}
Note that $\nu^\prime_s$ and $N^\prime_{\rm R}$ transform
separately and in different ways in Eq.~(\ref{eq:sterile-seesaw-transformation}),
because the former is just a single light sterile neutrino field while the
latter is a vector column containing two heavy right-handed neutrino fields
corresponding to the $\mu$ and $\tau$ flavors.
Substituting Eq.~(\ref{eq:sterile-seesaw-constraint}) into
Eq.~(\ref{eq:sterile-seesaw-formula}), we immediately prove that
$M^{}_\nu = {\cal P} M^*_\nu {\cal P}$ holds for the $3\times 3$ active Majorana
neutrino mass matrix. Provided Eq.~(\ref{eq:sterile-seesaw-constraint}) is inserted
into Eq.~(\ref{eq:sterile-seesaw-matrix}), we find that the $4\times 4$
effective Majorana neutrino mass matrix satisfies
\begin{eqnarray}
{\cal M} = {\cal S} {\cal M}^{*} {\cal S} \; ,
\label{eq:sterile-seesaw-mu-tau}
\end{eqnarray}
where $\cal S$ has been defined in Eq.~(\ref{eq:(3+1)-transformation}). In this
way we have realized the $\mu$-$\tau$ reflection symmetry for the $(3+1)$
active-sterile flavor mixing scenario. As an interesting by-product of the
contrived seesaw models of this kind, it has already been shown that the
lepton-number-violating and
CP-violating decays of those two heavy right-handed neutrinos may successfully
produce a net baryon-antibaryon asymmetry of the Universe
via the leptogenesis mechanism~\cite{Zhao:2022trb}.

Of course, an extra flavor symmetry (e.g., the $L^{}_e - L^{}_\mu - L^{}_\tau$
symmetry~\cite{Petcov:2004rk,Shaposhnikov:2006nn,Lindner:2010wr})
should be needed to naturally suppress the mass scale of $\nu^\prime_s$ while
keeping the other two right-handed neutrinos unaffected in such a seesaw scenario.
Another interesting seesaw model of this kind is to extend the canonical seesaw
mechanism by including an additional gauge singlet fermion~\cite{Zhang:2011vh}, so
as to accommodate an eV-scale sterile neutrino in a somewhat more natural way.
The model building strategies in this regard certainly deserve further
investigation, and a systematic study of the soft $\mu$-$\tau$ reflection
symmetry breaking effects is also desirable.

\subsection{Matter effects and nonstandard interactions}
\label{section 7.4}

Regarding neutrino (or antineutrino) oscillations in a normal (electrically
neutral and unpolarized) medium, it has been noticed
that the matter effect~\cite{Wolfenstein:1977ue,Mikheyev:1985zog,Mikheev:1986wj}
{\it does} respect the $\mu$-$\tau$ reflection symmetry under
discussion~\cite{Xing:2015fdg,Harrison:2002et,Xing:2010ez}.
Here the essential point is that the tree-level matter potential
term in the effective Hamiltonian ${\cal H}^{}_{\rm eff}$, which is responsible for
the evolution of three active neutrinos in matter, satisfies the relationship
${\cal H}^{}_{\rm eff} = {\cal P} {\cal H}^*_{\rm eff} {\cal P}$ as a consequence
of $U = {\cal P} U^* \zeta$ with ${\cal P}$ and $\zeta$ being given in and below
Eq.~(\ref{eq:S-tranformation}). This interesting observation will be invalid,
however, if some nonstandard interactions (NSIs) of the active neutrinos with
matter are involved~\cite{Wolfenstein:1977ue}, or if the one-loop electroweak
radiative corrections to the matter potential are taken into
account~\cite{Botella:1986wy,Mirizzi:2009td}. Here let us take a
close look at these two issues.

In the $\mu$-$\tau$ reflection symmetry limit (i.e., $U = {\cal P} U^* \zeta$),
one may prove ${\cal H}^{}_{\rm eff} = {\cal P} {\cal H}^*_{\rm eff} {\cal P}$
at the tree level. The tree-level form of ${\cal H}^{}_{\rm eff}$
can be expressed as~\cite{Kuo:1989qe,Xing:2003ez}:
\begin{eqnarray}
{\cal H}^{}_{\rm eff} = \frac{1}{2 E}
U D^2_\nu U^\dagger + {\cal V}^{}_{\rm tree} \equiv
\frac{1}{2 E} \widetilde{U} \widetilde{D}^{2}_\nu \widetilde{U}^\dagger \; ,
\label{eq:Heff}
\end{eqnarray}
where $E$ denotes the energy of a neutrino beam travelling in the medium,
$D^{}_\nu$ has been given in Eq.~(\ref{eq:M-mass-diagonalization}),
${\cal V}^{}_{\rm tree} = {\rm Diag}\{{\cal V}^{}_{\rm cc} + {\cal V}^{}_{\rm nc} ,
{\cal V}^{}_{\rm nc} , {\cal V}^{}_{\rm nc}\}$ is the matter potential arising
from the coherent forward neutrino scattering with electrons via the weak
charged-current (cc) interactions and with electrons, protons and neutrons
via the weak neutral-current (nc) interactions, $\widetilde{U}$ is the effective
PMNS matrix in matter, and $\widetilde{D}^{}_\nu = {\rm Diag}\{\widetilde{m}^{}_1 ,
\widetilde{m}^{}_2 , \widetilde{m}^{}_3\}$ with $\widetilde{m}^{}_i$ (for
$i = 1, 2, 3$) being the effective neutrino masses in matter.
At the tree level ${\cal V}^{}_{\rm cc} = \sqrt{2}
\hspace{0.04cm} G^{}_{\rm F} N^{}_e$
and ${\cal V}^{}_{\rm nc} = -G^{}_{\rm F}/\sqrt{2}
\left[\left(1 - 4\sin^2\theta^{}_{\rm w} \right) \left(N^{}_e - N^{}_p\right)
+ N^{}_n\right]$ hold, where $G^{}_{\rm F}$ is the Fermi coupling constant,
$\theta^{}_{\rm w}$ denotes the Weinberg angle of weak interactions, and
$N^{}_e$, $N^{}_p$ and $N^{}_n$ stand respectively for the number densities of
electrons, protons and neutrons in matter. Given the fact that $N^{}_e = N^{}_p$
holds for a normal medium, one is simply
left with ${\cal V}^{}_{\rm nc} = - G^{}_{\rm F} N^{}_n/\sqrt{2}$. Substituting
$U = {\cal P} U^* \zeta$ into Eq.~(\ref{eq:Heff}) and taking into account
${\cal P} {\cal V}^{}_{\rm tree} = {\cal V}^{}_{\rm tree} {\cal P}$, we obtain
\begin{eqnarray}
{\cal H}^{}_{\rm eff} = {\cal P} \left[\frac{1}{2 E}
U^* D^2_\nu U^T + {\cal V}^{}_{\rm tree} \right] {\cal P}
= {\cal P} {\cal H}^*_{\rm eff} {\cal P} \; .
\label{eq:Heff2}
\end{eqnarray}
Now that ${\cal H}^{}_{\rm eff}$ can also be expressed in terms of $\widetilde{U}$
and $\widetilde{D}^{}_\nu$, Eq.~(\ref{eq:Heff2}) leads us to
\begin{eqnarray}
\widetilde{U} \widetilde{D}^{2}_\nu \widetilde{U}^\dagger =
{\cal P} \widetilde{U}^* \widetilde{D}^{2}_\nu \widetilde{U}^T {\cal P} \; ,
\label{eq:Heff3}
\end{eqnarray}
from which we immediately arrive at
$\widetilde{U} = {\cal P} \widetilde{U}^* \tilde{\zeta}$, where $\tilde{\zeta}$
denotes an arbitrary diagonal phase matrix. This $\mu$-$\tau$ reflection
symmetry relationship for the matter-corrected PMNS matrix $\widetilde{U}$ implies
that $\big|\widetilde{U}^{}_{\mu i}\big| = \big|\widetilde{U}^{}_{\tau i}\big|$
is a natural result of $\big|U^{}_{\mu i}\big| = \big|U^{}_{\tau i}\big|$
(for $i = 1, 2, 3$). In other words, the tree-level matter potential
${\cal V}^{}_{\rm tree}$ in ${\cal H}^{}_{\rm eff}$ respects the $\mu$-$\tau$
reflection symmetry for neutrino oscillations in a
medium~\cite{Harrison:2002et,Xing:2010ez,Xing:2022efm}.

If the active neutrinos interact with normal matter in a way beyond the SM, their
oscillation behaviors can be modified even at the tree level. Here we focus only
on the neutral-current  NSIs described by the dimension-six four-fermion
operators~\cite{Wolfenstein:1977ue,Ohlsson:2012kf}
\begin{eqnarray}
-{\cal L}^{}_{\rm NSI} = 2\sqrt{2} \hspace{0.05cm} G^{}_{\rm F}
\Big(\overline{\nu^{}_\alpha} \hspace{0.03cm}
\gamma^\rho P^{}_{\rm L} \nu^{}_\beta\Big)
\Big(\overline{f} \hspace{0.03cm} \gamma^{}_\rho P^{}_{\rm C} f\Big)
\epsilon^{f{\rm C}}_{\alpha\beta} + {\rm h.c.} \; ,
\label{eq:NSI}
\end{eqnarray}
where $\rm C = L$ or $\rm R$, $P^{}_{\rm L, R} = \left(1 \mp \gamma^{}_5\right)/2$
are the left- and right-handed chiral projection operators, $f$ runs over
$e$, $u$ and $d$ in matter, and the dimensionless parameters
$\epsilon^{f {\rm C}}_{\alpha\beta}$ characterize the strengths of NSIs (for
$\alpha, \beta = e, \mu, \tau$). In this case the tree-level matter potential
${\cal V}^{}_{\rm tree}$ can be expressed as
\begin{eqnarray}
{\cal V}^{}_{\rm tree} = \sqrt{2} \hspace{0.05cm} G^{}_{\rm F} N^{}_e
\pmatrix{1 + \epsilon^{}_{ee} & \epsilon^{}_{e\mu} & \epsilon^{}_{e\tau} \cr
\epsilon^{*}_{e\mu} & \epsilon^{}_{\mu\mu} & \epsilon^{}_{\mu\tau} \cr
\epsilon^{*}_{e\tau} & \epsilon^{*}_{\mu\tau} & \epsilon^{}_{\tau\tau}}
-\frac{1}{\sqrt 2} G^{}_{\rm F} N^{}_n {\cal I} \; ,
\label{eq:NSI-potential}
\end{eqnarray}
where
\begin{eqnarray}
\epsilon^{}_{\alpha\beta} \equiv \sum_f \left(\epsilon^{f {\rm L}}_{\alpha\beta} +
\epsilon^{f {\rm R}}_{\alpha\beta}\right) \frac{N^{}_f}{N^{}_e} \;
\label{eq:NSI-epsilon}
\end{eqnarray}
with $N^{}_f$ (for $f = e, u, d$) being the number density of electrons, $u$ quarks
or $d$ quarks in matter. It is obvious that ${\cal V}^{}_{\rm tree}$ will not
respect the $\mu$-$\tau$ reflection symmetry unless its NSI parameters satisfy
the conditions
\begin{eqnarray}
\epsilon^{}_{ee} = \epsilon^{*}_{ee} \; , \quad
\epsilon^{}_{e\mu} = \epsilon^{*}_{e\tau} \; , \quad
\epsilon^{}_{\mu\mu} = \epsilon^{}_{\tau\tau} =
\epsilon^{*}_{\tau\tau} = \epsilon^{*}_{\mu\mu} \; .
\label{eq:NSI-mu-tau}
\end{eqnarray}
Then the questions become whether such a possibility can be realized in a
specific flavor symmetry model and whether it is interesting in neutrino
phenomenology.

In this connection an instructive flavor symmetry model, which allows both the
Majorana neutrino mass term and the neutral-current NSI part to respect the
$\mu$-$\tau$ reflection symmetry, has recently been proposed~\cite{Liao:2019qbb}.
The key ideas of this model are to generate the neutrino mass term with the help
of an $\rm S^{}_4 \times Z^{}_4$ flavor symmetry, and to produce sizable NSIs
with the introduction of new electroweak doublets and a charged
gauge-singlet scalar together with a global $\rm Z^{}_2$
symmetry (see, e.g., Ref.~\cite{Babu:2019mfe}).
It turns out that both the relationship $M^{}_\nu = {\cal P} M^*_\nu {\cal P}$
and the conditions in Eq.~(\ref{eq:NSI-mu-tau}) can be satisfied, and therefore
${\cal H}^{}_{\rm eff} = {\cal P} {\cal H}^*_{\rm eff} {\cal P}$ holds even
in the presence of NSIs.
At this point it is worth mentioning that the interesting relation
$\big|\epsilon^{}_{e\mu}\big| = \tan\theta^{}_{23} \big|\epsilon^{}_{e\tau}\big|$,
which comes from a purely phenomenological observation~\cite{Liao:2016orc}
but happens to be satisfied in the $\mu$-$\tau$ reflection symmetry limit
(namely, $\epsilon^{}_{e \mu} = \epsilon^{*}_{e\tau}$
and $\theta^{}_{23} = \pi/4$), will efficiently reduce the parameter redundancy
associated with neutrino oscillations~\cite{Liao:2016orc} and thus enhance the
experimental testability of possible NSIs in the upcoming long-baseline
experiments~\cite{SajjadAthar:2021prg}.

Next let us consider the one-loop electroweak radiative corrections to
${\cal V}^{}_{\rm tree}$ and examine the resulting effect of $\mu$-$\tau$
reflection symmetry breaking in the absence of any NSIs. The point
is that the tree-level universality of weak neutral-current contributions in
${\cal V}^{}_{\rm tree}$ will be slightly violated at the one-loop
level~\cite{Botella:1986wy,Mirizzi:2009td}. Because such quantum corrections
are suppressed by the factors $G^{}_{\rm F} m^2_\alpha$ (for $\alpha = e,
\mu, \tau$) as compared with the corresponding tree-level potential terms
${\cal V}^{}_{e} = {\cal V}^{}_{\rm cc} + {\cal V}^{}_{\rm nc}$ and
${\cal V}^{}_\mu = {\cal V}^{}_\tau = {\cal V}^{}_{\rm nc}$, they are negligibly
small in both ${\cal V}^{}_e$ and ${\cal V}^{}_\mu$. In view of the strong
charged-lepton mass hierarchy $m^{}_e \ll m^{}_\mu \ll m^{}_\tau$ as
illustrated in Figure~\ref{Fig:fermion-masses}, one finds
\begin{eqnarray}
{\cal V}^{}_e - {\cal V}^{}_\mu & = &
\sqrt{2} \hspace{0.04cm} G^{}_{\rm F} N^{}_e \; ,
\nonumber \\
{\cal V}^{}_\tau - {\cal V}^{}_\mu & = &
\frac{3 G^2_{\rm F} m^2_\tau}{2 \pi^2} \left[\left(N^{}_p + N^{}_n \right)
\ln \frac{M^2_W}{m^2_\tau} - N^{}_p - \frac{2}{3} N^{}_n\right] \; ,
\label{eq:one-loop}
\end{eqnarray}
where $M^{}_W$ is the $W$-boson mass. Taking $N^{}_p = N^{}_e$ for a neutral
medium and inputting $G^{}_{\rm F} \simeq 1.1664 \times 10^{-5}~{\rm GeV}^{-2}$,
$M^{}_W \simeq 80.377 ~{\rm GeV}$ and
$m^{}_\tau \simeq 1.777~{\rm GeV}$~\cite{Workman:2022ynf}, we have
\begin{eqnarray}
r \equiv \frac{{\cal V}^{}_\tau - {\cal V}^{}_\mu}
{{\cal V}^{}_e - {\cal V}^{}_\mu} =
\frac{3 G^{}_{\rm F} m^2_\tau}{2\sqrt{2} \hspace{0.05cm} \pi^2}
\left[\frac{N^{}_p + N^{}_n}{N^{}_p} \ln \frac{M^2_W}{m^2_\tau}
- \frac{3 N^{}_p + 2 N^{}_n}{3 N^{}_p} \right] \; .
\label{eq:r}
\end{eqnarray}
So we have $r \simeq 5.4 \times 10^{-5}$ for $N^{}_n = N^{}_p$ and
$r \simeq 1.1 \times 10^{-4}$ for $N^{}_n = 3 N^{}_p$ as two typical numerical
examples. After such one-loop corrections to the tree-level matter potential are
included, the effective Hamiltonian in Eq.~(\ref{eq:Heff}) becomes
\begin{eqnarray}
{\cal H}^{}_{\rm eff} = \frac{1}{2 E} U D^2_\nu U^\dagger
+ {\cal V}^{}_{\rm tree} + {\cal V}^{}_{\rm loop}
\equiv \frac{1}{2 E} \widetilde{U} \widetilde{D}^2_\nu \widetilde{U}^\dagger \; ,
\label{eq:Heff-one-loop}
\end{eqnarray}
where ${\cal V}^{}_{\rm loop} = {\rm Diag}\{0, 0, r {\cal V}^{}_{\rm cc}\}$.
In this case $U = {\cal P} U^* \zeta$ does not lead to
${\cal H}^{}_{\rm eff} = {\cal P} {\cal H}^{*}_{\rm eff} {\cal P}$ anymore,
simply because ${\cal P}$ and ${\cal V}^{}_{\rm loop}$ do not commute with
each other. The tree-level $\mu$-$\tau$ reflection symmetry in
matter, as characterized by $\widetilde{U} = {\cal P} \widetilde{U}^* \xi$,
is therefore broken to some extent by $r \neq 0$ at the one-loop level.

To illustrate the nontrivial effect of ${\cal V}^{}_{\rm loop}$, we proceed
to consider the intriguing tree-level Toshev relation in the standard
parametrizations of $U$ and $\widetilde{U}$
~\cite{Toshev:1991ku,Freund:2001pn,Zhou:2011xm}:
\begin{eqnarray}
\sin 2 \tilde{\theta}^{}_{23}
\sin\tilde{\delta}^{}_\nu = \sin 2\theta^{}_{23} \sin\delta^{}_\nu \; ,
\label{eq:Toshev}
\end{eqnarray}
which provides a straightforward connection between
$\big(\theta^{}_{23}, \delta^{}_\nu\big) = \big(\pi/4, \pm \pi/2\big)$
in vacuum and $\big(\tilde{\theta}^{}_{23}, \tilde{\delta}^{}_\nu\big)
= \big(\pi/4, \pm \pi/2\big)$ in matter as a consequence of the
$\mu$-$\tau$ reflection symmetry. At the one-loop level, however, we
find~\cite{Xing:2022efm}
\begin{eqnarray}
\frac{{\rm d}}{{\rm d} a} \ln\left( \sin 2\tilde{\theta}^{}_{23}
\sin \tilde{\delta}^{}_\nu \right) = - r \kappa \; ,
\label{eq:dToshev}
\end{eqnarray}
where $a = 2\sqrt{2}\hspace{0.07cm} G^{}_{\rm F} N^{}_e E$ denotes the matter
parameter, and $\kappa$ is a complicated nonlinear function of the effective
neutrino oscillation parameters $\tilde{\theta}^{}_{12}$, $\tilde{\theta}^{}_{13}$,
$\tilde{\theta}^{}_{23}$, $\tilde{\delta}^{}_\nu$, $\Delta \widetilde{m}^{2}_{21}$,
$\Delta \widetilde{m}^{2}_{31}$ and $\Delta \widetilde{m}^{2}_{32}$ with
$\Delta \widetilde{m}^{2}_{ij} \equiv \widetilde{m}^2_i -
\widetilde{m}^2_j$ in matter. A formal solution to Eq.~(\ref{eq:dToshev}) is
\begin{eqnarray}
R \equiv \frac{\sin 2\tilde{\theta}^{}_{23} \sin\tilde{\delta}^{}_\nu}
{\sin 2\theta^{}_{23} \sin\delta^{}_\nu} =
\exp\left[-r \int^a_0 \kappa {\rm d} a\right] \; .
\label{eq:Toshev-one-loop}
\end{eqnarray}
One can see that $R = 1$ exactly holds in the $r = 0$ case (i.e., at the tree level
of coherent forward neutrino scattering with matter), and an appreciable deviation of
$R$ from one is possible if $a$ is large enough.

A numerical illustration of $R$ evolving with the matter parameter $a$ can be found
in Ref.~\cite{Xing:2022efm} for neutrino and
antineutrino oscillations in dense matter. Figure~\ref{Fig:R-plot} is an
example of this kind in the case of a neutrino beam travelling in matter and
a normal neutrino mass ordering. It is obvious that $R \simeq 1$ appears as an
excellent approximation for $a \lesssim 0.1~{\rm eV}^2$, a parameter space
suitable for all the long-baseline neutrino oscillation experiments on the
earth, but the quantum correction
becomes significant as $a$ goes up. Note that $R$ will approach zero if
$a \gtrsim 10^3~{\rm eV}^2$ holds, as a straightforward consequence of either
$\tilde{\delta}^{}_\nu \to 0$ (or $\pi$) or $\tilde{\theta}^{}_{23} \to \pi/2$. As
shown in Figure~\ref{Fig:R-plot}, there is a local peak in the curve of $R$ when $a$ is
near a few ${\rm eV}^2$ and up to about $10~{\rm eV}^2$. This phenomenon arises from
the fact that $\tilde{\theta}^{}_{23}$ is almost unchanged in this region of $a$
but $\tilde{\delta}^{}_\nu$ crosses its threshold value $\tilde{\delta}^{}_\nu = -\pi/2$,
and therefore $R \propto \sin\tilde{\delta}^{}_\nu/\sin\delta^{}_\nu = 1$ undergoes
a local peak~\cite{Xing:2022efm}.
But a similar peak cannot appear in the inverted neutrino mass
ordering case, although $R$ may appreciably deviate from one in the
$a \gtrsim 1~{\rm eV}^2$ region~\cite{Xing:2022efm}.
\begin{figure}[t!]
\begin{center}
\includegraphics[width=7cm]{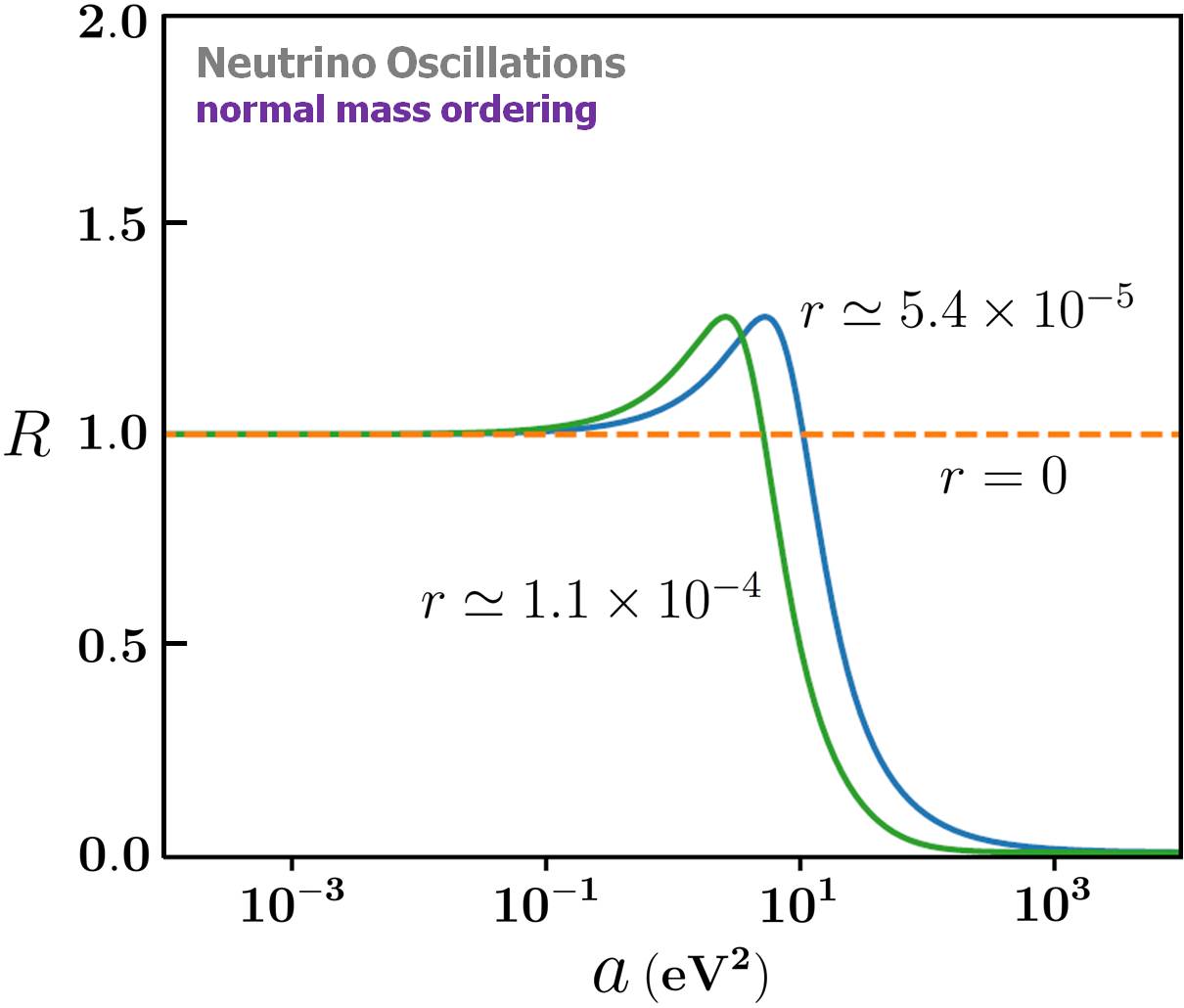}
\vspace{-0.3cm}
\caption{An illustration of the one-loop radiative correction to the tree-level
Toshev relation for neutrino oscillations in dense matter, where the normal
neutrino mass ordering has been assumed~\cite{Xing:2022efm}.}
\label{Fig:R-plot}
\end{center}
\end{figure}

Note that it is extremely difficult to reach
$a \gtrsim 1 ~{\rm eV}^2$ in any realistic circumstances of dense matter.
For example, $a \sim 1.5 \times 10^{-4} ~{\rm eV}^2$ can be obtained for
solar neutrinos with $E \sim 10 ~{\rm MeV}$~\cite{Xing:2018lob}; and
$a \sim 1.5 ~{\rm eV}^2$ becomes possible for a neutrino beam of the
same energy passing through a typical white dwarf~\cite{Luo:2019efb}.
But of course there seems no way to really observe the effect of $R \neq 1$
by making use of a white dwarf or other dense celestial bodies as a medium of
neutrino oscillations. So the above discussions are only conceptually
interesting for the time being.

Finally, it is worth emphasizing that $R$ itself is a rephasing-invariant
quantity and its evolution with $a$ makes sense no matter what asymptotic
behaviors $\tilde{\theta}^{}_{23}$ and $\tilde{\delta}^{}_\nu$ may have in the
chosen parametrization of $\widetilde{U}$. This point can be easily seen from
\begin{eqnarray}
R = \frac{\widetilde{\cal J}^{}_\nu}{{\cal J}^{}_\nu} \cdot
\frac{\big|U^{}_{e 1}\big| \big|U^{}_{e 2}\big| \big|U^{}_{e 3}\big|}
{\big|\widetilde{U}^{}_{e 1}\big| \big|\widetilde{U}^{}_{e 2}\big|
\big|\widetilde{U}^{}_{e 3}\big|} \; ,
\label{eq:R-invariant}
\end{eqnarray}
where $\widetilde{\cal J}^{}_\nu$ is the matter-corrected counterpart of the
Jarlskog invariant ${\cal J}^{}_\nu$ in vacuum.
Given the tree-level matter potential ${\cal V}^{}_{\rm tree}$
without any NSIs, $R = 1$ has been proved in
Refs.~\cite{Xing:2018lob,Chiu:2010da} no matter whether $U$ and $\widetilde{U}$
respect the $\mu$-$\tau$ reflection symmetry or not.

\setcounter{equation}{0}
\section{Summary and outlook}
\label{section 8}

The past twenty-five years marked a new and exciting golden epoch in neutrino
physics. Today we are surely aware, more than ever, of where we are and
where we are going in the experimental, phenomenological and theoretical
aspects.
\begin{itemize}
\item     On the experimental side, the very fact of flavor oscillations of
massive neutrinos has been firmly established. Among the six independent neutrino
oscillation parameters in the standard three-flavor scheme, $\Delta m^2_{21}$,
$\big|\Delta m^2_{31}\big|$ (or $\big|\Delta m^2_{32}\big|$), $\theta^{}_{12}$,
$\theta^{}_{13}$ and $\theta^{}_{23}$ have been measured to a good degree of
accuracy~\cite{Workman:2022ynf}, and a preliminary but encouraging evidence for
$\delta^{}_\nu \neq 0$ or $\pi$ has also been achieved~\cite{T2K:2019bcf}.
The primary goals of the next-generation neutrino oscillation experiments
are to pin down the neutrino mass ordering (i.e., the sign of $\Delta m^2_{31}$
or $\Delta m^2_{32}$) and the Dirac CP-violating phase $\delta^{}_\nu$. An
unambiguous determination of the octant of $\theta^{}_{23}$ and more stringent
tests of the unitarity of the $3\times 3$ PMNS matrix $U$ are our important
targets too. Meanwhile, a number of non-oscillation experiments are underway
towards probing the Majorana nature of massive neutrinos and determining the
absolute neutrino mass scale~\cite{SajjadAthar:2021prg}.

\item     On the phenomenological side, some salient features of the neutrino
mass spectrum and lepton flavor mixing pattern have been found. The
three-flavor global analysis of all the available experimental data has
proved to be a successful method to achieve the parameter space of neutrino
oscillations~\cite{Gonzalez-Garcia:2021dve,Capozzi:2021fjo},
which has provided not only very helpful hints for model building
but also rather useful guidelines for estimating relevant lepton-flavor-violating
or lepton-number-violating processes and for designing new neutrino experiment
facilities. The possibility of a $\mu$-$\tau$ reflection symmetry in the
neutrino sector~\cite{Xing:2015fdg}, characterized by
$\big|U^{}_{\mu i}\big| \simeq \big|U^{}_{\tau i}\big|$ (for $i = 1, 2, 3$)
as indicated by a global fit of current neutrino oscillation data, is just
a good example of this kind and the main subject of this paper.

\item     On the theoretical side, a lot of important progress has been made
towards a much deeper understanding of the origin of tiny neutrino masses and
the dynamics of lepton flavor mixing and CP violation. The canonical seesaw
mechanism and its simple variations, which naturally attribute the smallness of
three active neutrino masses to the possible existence of a few heavy degrees of
freedom at a superhigh energy scale, have been extensively and comprehensively
studied~\cite{Xing:2020ald,King:2003jb}. A very recent but remarkable development
in this regard is the complete one-loop matching of the canonical seesaw model
onto the SM effective field theory~\cite{Zhang:2021tsq,Zhang:2021jdf,Li:2022ipc,
Zhang:2022osj}, making it possible to consistently formulate the low-energy
{\it seesaw} effective field theories beyond the dimension-five Weinberg operator.
To interpret the observed large lepton flavor mixing effects, various instructive
flavor symmetry groups (e.g., $\rm S^{}_4$, $\rm A^{}_4$, $\rm A^{}_5$, $\Delta (27)$,
$\rm T^\prime$, and modular symmetries~\cite{Feruglio:2017spp}) have been tried
and many interesting models have been built~\cite{Xing:2019vks,Altarelli:2010gt,
King:2013eh,Feruglio:2019ybq,Ishimori:2010au}. Encouragingly, these larger
symmetry groups can easily accommodate the $\mu$-$\tau$ reflection symmetry
when they are combined with the generalized CP symmetries~\cite{Xing:2015fdg}.
\end{itemize}
Though many praiseworthy theoretical attempts have been made,
``there has never been a {\it convincing quantitative} model
of the neutrino masses" as once argued by Witten~\cite{Witten:2000dt}. It was
also Weinberg's regret for a failure in building more realistic models to
explain the observed patterns of quark and lepton masses, including the tiny
neutrino masses~\cite{Weinberg:2020zba}. In this situation a less ambitious
approach, motivated by the principle of {\it Occam's razor}, is to pursue
the {\it minimal} seesaw models~\cite{Xing:2020ald} and the {\it minimal}
flavor symmetries~\cite{Xing:2015fdg} so as to account for the available
experimental data with fewer free parameters and make some testable predictions
for the upcoming precision measurements. It is expected that such simplified
scenarios may hopefully shed light on a true fundamental (ultraviolet complete)
theory of neutrino mass generation and flavor mixing, if they are located in
the {\it neutrino landscape} instead of the {\it neutrino swampland} at low
energies~\cite{Vafa:2005ui,Palti:2019pca} --- the former is consistent with
all or most the experimental and observational facts.

It is in this spirit that we have summarized the latest progress made in
exploring the $\mu$-$\tau$ reflection symmetry as the minimal flavor symmetry
associated closely with large lepton flavor mixing
angles and significant CP-violating effects. A brief introduction to the
flavor-changing CP transformation for the three active neutrino fields
has been given, and a detailed description of how to identify the $\mu$-$\tau$
reflection symmetry for either the effective Majorana neutrino mass
term or the canonical seesaw mechanism has been presented. We have elucidated
the {\it translational} and {\it rotational} $\mu$-$\tau$ reflection symmetries,
which can give rise to the phenomenologically favored $\rm TM^{}_1$ flavor
mixing pattern and can also be combined with the canonical seesaw mechanism.
We have paid particular attention to the soft breaking of $\mu$-$\tau$ reflection
symmetry induced by the RGE running from a superhigh energy scale to the
electroweak scale, including the special but interesting case with either
$m^{}_1 = 0$ or $m^{}_3 = 0$. Some intriguing consequences of the $\mu$-$\tau$
reflection symmetry in neutrino physics, such as its implications for lepton
flavor violation and lepton number violation, its constraint on the flavor
distribution of ultrahigh-energy cosmic neutrinos at a neutrino telescope, its
application to the minimal seesaw model and its impact on leptogenesis, have
been discussed. We have finally given a few more specific examples for possible
extensions or applications of the $\mu$-$\tau$ reflection symmetry, including
the littlest seesaw model, the zero textures of lepton and quark mass matrices
with a $\rm Z^{}_2$ reflection symmetry, the $\mu$-$\tau$ reflection symmetry
in a seesaw-induced $(3+1)$ active-sterile flavor mixing scenario, and matter
effects and nonstandard interactions related to neutrino oscillations.

At this point it is worthwhile to emphasize that massive neutrinos definitely
play a unique role in both ``cosmic flavor physics" --- the studies of relevant
flavor problems in cosmology and astrophysics and ``dark flavor physics" ---
the studies of those flavor issues associated with dark matter and even dark
sector of the Universe~\cite{Xing:2019vks}. Being the electrically neutral
fermions which are only sensitive to the standard weak interactions (and
gravitational interactions), the active neutrinos are likely to mix with
the sterile neutrinos and dark fermions in a variety of new physics models.
Such a conjecture seems quite natural in the sense that the active neutrinos
are not {\it that} active and those relic active neutrinos in the early Universe
(i.e., the cosmic neutrino background) {\it do} serve as warm dark matter.
In this sense one may visit the islands of sterile neutrinos and dark fermions,
which are essentially isolated from the SM paradise as illustrated in
Figure~\ref{Fig:flavor-islands}, through the portals of active-sterile flavor
mixing and active-dark flavor mixing. The canonical seesaw mechanism has
already provided a very good example for the active-sterile flavor mixing, as
discussed in the present paper.
One may certainly go beyond the canonical seesaw framework
by introducing some more sterile degrees of freedom in the fermion sector,
if such new hypothetical particles are well motivated in particle physics,
astrophysics and cosmology.
\begin{figure}[t!]
\begin{center}
\includegraphics[width=13cm]{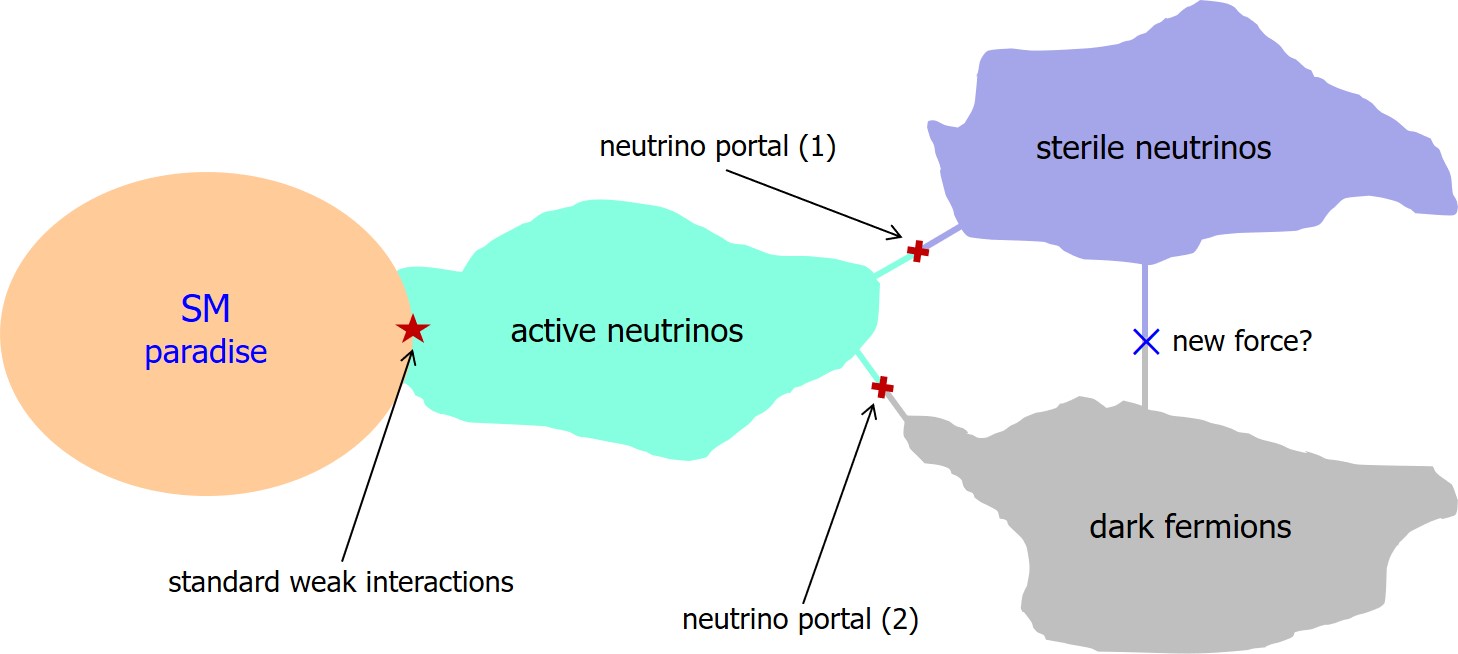}
\vspace{-0.18cm}
\caption{An illustration of the hypothetical islands of sterile neutrinos and
dark fermions, which may be connected to the island of active neutrinos
through the portals of active-sterile flavor mixing and active-dark flavor
mixing.}
\label{Fig:flavor-islands}
\end{center}
\end{figure}

Let us end with a naive conjecture that the three sterile neutrino species
mix separately with the three active neutrino species and the three dark
fermion species, while the active and dark flavors do not directly interact
with each other. In this case the full parameter space of flavor mixing
among the three categories of neutral fermions can be described by a unitary
$9\times 9$ matrix $\cal U$ which contains 36 Euler-like two-dimensional
unitary matrices $O^{}_{ij}$~\cite{Han:2021qum}:
\begin{eqnarray}
{\cal U} = \pmatrix{ {\cal I} & {\bf 0} & {\bf 0} \cr
{\bf 0} & {\cal I} & {\bf 0} \cr
{\bf 0} & {\bf 0} & U^{\prime\prime}_0} L^{}_3
\pmatrix{ {\cal I} & {\bf 0} & {\bf 0} \cr
{\bf 0} & U^{\prime}_0 & {\bf 0} \cr
{\bf 0} & {\bf 0} & {\cal I}} L^{}_2 L^{}_1
\pmatrix{ U^{}_0 & {\bf 0} & {\bf 0} \cr
{\bf 0} & {\cal I} & {\bf 0} \cr
{\bf 0} & {\bf 0} & {\cal I}} \; ,
\label{eq:U9}
\end{eqnarray}
where $\cal I$ denotes the $3\times 3$ identity matrix, $\bf 0$ stands
for the $3\times 3$ zero matrix, and the three $3\times 3$ unitary matrices
read as $U^{}_0 = O^{}_{23} O^{}_{13} O^{}_{12}$,
$U^\prime_0 = O^{}_{56} O^{}_{46} O^{}_{45}$ and
$U^{\prime\prime}_0 = O^{}_{89} O^{}_{79} O^{}_{78}$,
which characterize the primary flavor mixing sectors of three active neutrinos, three
sterile neutrinos and three dark fermions, respectively; and the three $9\times 9$
matrices are
\begin{eqnarray}
L^{}_1 =
O^{}_{36} O^{}_{26} O^{}_{16} O^{}_{35} O^{}_{25}
O^{}_{15} O^{}_{34} O^{}_{24} O^{}_{14} \; ,
\nonumber \\
L^{}_2 =
O^{}_{39} O^{}_{29} O^{}_{19} O^{}_{38} O^{}_{28}
O^{}_{18} O^{}_{37} O^{}_{27} O^{}_{17} \; ,
\nonumber \\
L^{}_3 =
O^{}_{69} O^{}_{59} O^{}_{49} O^{}_{68} O^{}_{58}
O^{}_{48} O^{}_{67} O^{}_{57} O^{}_{47} \; ,
\label{eq:U92}
\end{eqnarray}
which measure the interplay between any two of the three sectors~\cite{Han:2021qum}.
It is obvious that the flavor mixing angles of $L^{}_1$ and $L^{}_2$ are naturally
suppressed, and the three flavor sectors will automatically become decoupled if the
off-diagonal elements of $L^{}_1$, $L^{}_2$ and $L^{}_3$ are all switched off. Does
the flavors of those real and yet-to-be-discovered neutral fermions really have
such a clustered mixing pattern?

\section*{Acknowledgements}

This paper is dedicated to the memory of Prof. Harald Fritzsch (1943 --- 2022), 
the most important collaborator and friend of mine in my career. 
I am greatly indebted to Di Zhang, Zhen-hua Zhao, Shun Zhou and Ye-Ling Zhou
for many useful discussions, constructive comments and friendly helps during
my writing of this article. My research work is supported in part by the National
Natural Science Foundation of China under grants No. 12075254 and No. 11835013.

\appendix

\section{}
\label{Appendix A}

In the canonical seesaw mechanism, a complete Euler-like parametrization of $U$, $R$,
$S$ and $Q$ in Eq.~(\ref{eq:seesaw-diagonalization}) in terms of fifteen flavor mixing
angles and fifteen CP-violating phases can be expressed as
follows~\cite{Xing:2007zj,Xing:2011ur}:
\begin{eqnarray}
\pmatrix{ U & R \cr S & Q \cr} =
\pmatrix{ P^{}_l & {\bf 0} \cr {\bf 0} & U^\prime_0 \cr}
\pmatrix{ A & \tilde{R} \cr R^\prime & B \cr}
\pmatrix{ U^{}_0 & {\bf 0} \cr {\bf 0} & {\cal I} \cr} \; ,
\label{A1}
\end{eqnarray}
where $P^{}_l = {\rm Diag}\{e^{{\rm i} \phi^{}_e} , e^{{\rm i} \phi^{}_\mu} ,
e^{{\rm i} \phi^{}_\tau}\}$, $U^{}_0$ and $U^\prime_0$ are the $3\times 3$ unitary
matrices describing the respective primary flavor mixing effects of active and sterile
neutrinos, and $A$ (or $B$) characterizes a slight deviation of $U = P^{}_l A U^{}_0$
(or $Q = U^\prime_0 B$) from $U^{}_0$ (or $U^\prime_0$). Here we focus only on
$U$ and $R$ that appear in the weak charged-current interactions of massive
Majorana neutrinos $\nu^{}_i$ and $N^{}_i$ (for $i = 1, 2, 3$)
as shown in Eq.~(\ref{eq:cc-seesaw}). To be explicit,
\begin{eqnarray}
U^{}_0 = \pmatrix{ c^{}_{12} c^{}_{13} & \hat{s}^*_{12}
c^{}_{13} & \hat{s}^*_{13} \cr
-\hat{s}^{}_{12} c^{}_{23} -
c^{}_{12} \hat{s}^{}_{13} \hat{s}^*_{23} & c^{}_{12} c^{}_{23} -
\hat{s}^*_{12} \hat{s}^{}_{13} \hat{s}^*_{23} & c^{}_{13}
\hat{s}^*_{23} \cr
\hat{s}^{}_{12} \hat{s}^{}_{23} - c^{}_{12}
\hat{s}^{}_{13} c^{}_{23} & -c^{}_{12} \hat{s}^{}_{23} -
\hat{s}^*_{12} \hat{s}^{}_{13} c^{}_{23} & c^{}_{13} c^{}_{23} \cr} \; ,
\label{A2}
\end{eqnarray}
and
\begin{eqnarray}
A & = & \pmatrix{ c^{}_{14} c^{}_{15} c^{}_{16} & 0 & 0 \cr
A^{}_{21} & c^{}_{24} c^{}_{25} c^{}_{26} & 0 \cr
A^{}_{31} & A^{}_{32} & c^{}_{34} c^{}_{35} c^{}_{36} \cr} \; ,
\nonumber \\
\tilde{R} & = & \pmatrix{ \hat{s}^*_{14} c^{}_{15} c^{}_{16} &
\hspace{0.24cm} \hat{s}^*_{15} c^{}_{16} \hspace{0.24cm} & \hat{s}^*_{16} \cr
\tilde{R}^{}_{21} & \tilde{R}^{}_{22} & c^{}_{16} \hat{s}^*_{26} \cr
\tilde{R}^{}_{31} & \tilde{R}^{}_{32} &
c^{}_{16} c^{}_{26} \hat{s}^*_{36} \cr} \; ,
\label{A3}
\end{eqnarray}
where
\begin{eqnarray}
A^{}_{21} & = & - c^{}_{14} c^{}_{15} \hat{s}^{}_{16} \hat{s}^*_{26}
- c^{}_{14} \hat{s}^{}_{15} \hat{s}^*_{25} c^{}_{26}
- \hat{s}^{}_{14} \hat{s}^*_{24} c^{}_{25} c^{}_{26} \; ,
\nonumber \\
A^{}_{31} & = & - c^{}_{14} c^{}_{15} \hat{s}^{}_{16} c^{}_{26} \hat{s}^*_{36}
+ c^{}_{14} \hat{s}^{}_{15} \hat{s}^*_{25} \hat{s}^{}_{26} \hat{s}^*_{36}
- c^{}_{14} \hat{s}^{}_{15} c^{}_{25} \hat{s}^*_{35} c^{}_{36}
\nonumber \\
&& + \hat{s}^{}_{14} \hat{s}^*_{24} c^{}_{25} \hat{s}^{}_{26} \hat{s}^*_{36}
+ \hat{s}^{}_{14} \hat{s}^*_{24} \hat{s}^{}_{25} \hat{s}^*_{35} c^{}_{36}
- \hat{s}^{}_{14} c^{}_{24} \hat{s}^*_{34} c^{}_{35} c^{}_{36} \; ,
\nonumber \\
A^{}_{32} & = & - c^{}_{24} c^{}_{25} \hat{s}^{}_{26} \hat{s}^*_{36}
- c^{}_{24} \hat{s}^{}_{25} \hat{s}^*_{35} c^{}_{36}
- \hat{s}^{}_{24} \hat{s}^*_{34} c^{}_{35} c^{}_{36} \; ;
\nonumber \\
\tilde{R}^{}_{21} & = & - \hat{s}^*_{14} c^{}_{15} \hat{s}^{}_{16} \hat{s}^*_{26}
- \hat{s}^*_{14} \hat{s}^{}_{15} \hat{s}^*_{25} c^{}_{26}
+ c^{}_{14} \hat{s}^*_{24} c^{}_{25} c^{}_{26} \; ,
\nonumber \\
\tilde{R}^{}_{22} & = & - \hat{s}^*_{15} \hat{s}^{}_{16} \hat{s}^*_{26}
+ c^{}_{15} \hat{s}^*_{25} c^{}_{26} \; ,
\nonumber \\
\tilde{R}^{}_{31} & = &
- \hat{s}^*_{14} c^{}_{15} \hat{s}^{}_{16} c^{}_{26} \hat{s}^*_{36}
+ \hat{s}^*_{14} \hat{s}^{}_{15} \hat{s}^*_{25} \hat{s}^{}_{26} \hat{s}^*_{36}
- \hat{s}^*_{14} \hat{s}^{}_{15} c^{}_{25} \hat{s}^*_{35} c^{}_{36}
\nonumber \\
&& - c^{}_{14} \hat{s}^*_{24} c^{}_{25} \hat{s}^{}_{26} \hat{s}^*_{36}
- c^{}_{14} \hat{s}^*_{24} \hat{s}^{}_{25} \hat{s}^*_{35} c^{}_{36}
+ c^{}_{14} c^{}_{24} \hat{s}^*_{34} c^{}_{35} c^{}_{36} \; ,
\nonumber \\
\tilde{R}^{}_{32} & = & - \hat{s}^*_{15} \hat{s}^{}_{16} c^{}_{26} \hat{s}^*_{36}
- c^{}_{15} \hat{s}^*_{25} \hat{s}^{}_{26} \hat{s}^*_{36}
+ c^{}_{15} c^{}_{25} \hat{s}^*_{35} c^{}_{36} \; .
\label{A4}
\end{eqnarray}
In these equations, $c^{}_{ij} \equiv \cos\theta^{}_{ij}$, $s^{}_{ij} \equiv
\sin\theta^{}_{ij}$ and $\hat{s}^{}_{ij} \equiv s^{}_{ij} \hspace{0.05cm}
e^{{\rm i}\delta^{}_{ij}}$ with $\theta^{}_{ij}$ lying in the first quadrant
(for $ij = 12$, $13$, $\cdots$). It is obvious that
the triangular matrix $A$ signifies slight deviation of the PMNS matrix $U$
from $U^{}_0$. Since the active-sterile flavor mixing effects are expected
to be strongly suppressed~\cite{Antusch:2006vwa,Fernandez-Martinez:2016lgt,
Blennow:2016jkn,Hu:2020oba,Wang:2021rsi}, one may make the following
reasonable approximations~\cite{Xing:2011ur}:
\begin{eqnarray}
A \simeq {\cal I} - \frac{1}{2} \pmatrix{
\displaystyle\sum^6_{i = 4} s^2_{1i}
& 0 & 0 \cr
2 \displaystyle\sum^6_{i = 4} \hat{s}^{}_{1i} \hat{s}^*_{2i}
& \displaystyle\sum^6_{i = 4} s^2_{2i} & 0 \cr
2 \displaystyle\sum^6_{i = 4} \hat{s}^{}_{1i} \hat{s}^*_{3i}
& 2 \displaystyle\sum^6_{i = 4} \hat{s}^{}_{2i} \hat{s}^*_{3i}
& \displaystyle\sum^6_{i = 4} s^2_{3i} \cr} + {\cal O}\big(s^4_{ij}\big) \; ,
\nonumber \\
\tilde{R} \simeq \pmatrix{
\hat{s}^*_{14} & \hat{s}^*_{15} & \hat{s}^*_{16} \cr
\hat{s}^*_{24} & \hat{s}^*_{25} & \hat{s}^*_{26} \cr
\hat{s}^*_{34} & \hat{s}^*_{35} & \hat{s}^*_{36} \cr}
+ {\cal O}\big(s^3_{ij}\big) \; ,
\label{A5}
\end{eqnarray}
where all the $s^{}_{ij}$ terms (for $i = 1, 2, 3$ and $j = 4, 5, 6$) have
been constrained to be below or far below ${\cal O}\big(0.1\big)$. So
$A \simeq {\cal I}$ and $U \simeq P^{}_l U^{}_0$ hold up to small corrections
of ${\cal O}\big(s^2_{ij}\big)$.

Now let us consider the conditions $U = {\cal P} U^*$ and $R = {\cal P} R^*$
in the $\mu$-$\tau$ reflection symmetry limit, as discussed in
section~\ref{section 2.4}. The parameters of $U = P^{}_l A U^{}_0$ and
$R = P^{}_l \tilde{R}$ can then be constrained, but the
exact constraint relations are too complicated to be useful. So it is more
instructive to impose the $\mu$-$\tau$ reflection symmetry on
$U \simeq P^{}_l U^{}_0$ and $R \simeq P^{}_l \tilde{R}$ by neglecting the
small non-unitary corrections to $U^{}_0$ and keeping only the leading terms
of $R$. In this case we obtain the approximate constraint
conditions~\cite{Xing:2022oob}
\begin{eqnarray}
e^{{\rm i} 2\phi^{}_e} \simeq e^{{\rm i} 2\left(\phi^{}_e
- \delta^{}_{12}\right)} \simeq e^{{\rm i} 2\left(\phi^{}_e
- \delta^{}_{13}\right)} \simeq 1 \;
\label{A6}
\end{eqnarray}
from $U^{}_{e i} \simeq U^*_{e i}$ (for $i = 1, 2, 3$), and
\begin{eqnarray}
\left(s^{}_{12} c^{}_{23} + c^{}_{12} s^{}_{13} s^{}_{23}
e^{{\rm i} \delta^{}_\nu}\right) e^{{\rm i}\left(\phi^{}_\mu
+ \phi^{}_\tau + 2\delta^{}_{12} + \delta^{}_{23}\right)}
\simeq - \left(s^{}_{12} s^{}_{23} - c^{}_{12} s^{}_{13} c^{}_{23}
e^{-{\rm i}\delta^{}_\nu}\right) \; ,
\nonumber \\
\left(c^{}_{12} c^{}_{23} - s^{}_{12} s^{}_{13} s^{}_{23}
e^{{\rm i} \delta^{}_\nu}\right) e^{{\rm i}\left(\phi^{}_\mu
+ \phi^{}_\tau + \delta^{}_{23}\right)}
\simeq - \left(c^{}_{12} s^{}_{23} + s^{}_{12} s^{}_{13} c^{}_{23}
e^{-{\rm i}\delta^{}_\nu}\right) \; ,
\nonumber \\
s^{}_{23} e^{{\rm i}\left(\phi^{}_\mu + \phi^{}_\tau
- \delta^{}_{23}\right)} \simeq c^{}_{23} \;
\label{A7}
\end{eqnarray}
from $U^{}_{\mu i } \simeq U^*_{\tau i}$ (for $i = 1, 2, 3$), where
$\delta^{}_\nu \equiv \delta^{}_{13} - \delta^{}_{12} - \delta^{}_{23}$
has been defined; as well as
\begin{eqnarray}
e^{{\rm i} 2\left(\phi^{}_e - \delta^{}_{14}\right)} \simeq
e^{{\rm i} 2\left(\phi^{}_e - \delta^{}_{15}\right)}
\simeq e^{{\rm i} 2\left(\phi^{}_e - \delta^{}_{16}\right)} \simeq 1 \;
\label{A8}
\end{eqnarray}
from $R^{}_{e i} \simeq R^*_{e i}$ (for $i = 1, 2, 3$), and
\begin{eqnarray}
s^{}_{24} e^{{\rm i}\left(\phi^{}_\mu + \phi^{}_\tau - \delta^{}_{24}
- \delta^{}_{34}\right)} \simeq s^{}_{34} \; ,
\nonumber \\
s^{}_{25} e^{{\rm i}\left(\phi^{}_\mu + \phi^{}_\tau - \delta^{}_{25}
- \delta^{}_{35}\right)} \simeq s^{}_{35} \; ,
\nonumber \\
s^{}_{26} e^{{\rm i}\left(\phi^{}_\mu + \phi^{}_\tau - \delta^{}_{26}
- \delta^{}_{36}\right)} \simeq s^{}_{36} \;
\label{A9}
\end{eqnarray}
from $R^{}_{\mu i } \simeq R^*_{\tau i}$ (for $i = 1, 2, 3$). As a consequence,
we arrive at
\begin{eqnarray}
\theta^{}_{23} \simeq \frac{\pi}{4} \; , \quad
\delta^{}_\nu \simeq \pm\frac{\pi}{2} \; , \quad
\delta^{}_{12} \simeq 0 ~{\rm or}~ \pi \; , \quad
\delta^{}_{13} \simeq 0 ~{\rm or}~ \pi \; , \quad
\delta^{}_{23} \simeq \pm \frac{\pi}{2} \; ,
\nonumber \\
\phi^{}_e \simeq 0 ~{\rm or}~ \pi \; , \quad
\phi^{}_\mu + \phi^{}_\tau \simeq \delta^{}_{23} \simeq \pm \pi/2 \;
\label{A10}
\end{eqnarray}
for the PMNS matrix $U$; and
\begin{eqnarray}
\theta^{}_{2j} \simeq \theta^{}_{3j} \; , \quad
\delta^{}_{1j} \simeq 0 ~{\rm or}~ \pi \; , \quad
\delta^{}_{2j} + \delta^{}_{3j} \simeq
\phi^{}_\mu + \phi^{}_\tau \simeq \pm\frac{\pi}{2} \;
\label{A11}
\end{eqnarray}
for the active-sterile flavor mixing matrix $R$, where
$j = 4, 5, 6$. Note that $\phi^{}_e$, $\phi^{}_\mu$ and
$\phi^{}_\tau$ are unobservable in any realistic experiments,
but they should not be ignored when using the $\mu$-$\tau$ reflection
symmetry to constrain those observable flavor mixing angles and
CP-violating phases for the sake of consistency.

\section*{References}

\bibliographystyle{iopart-num}
\bibliography{CCPreferences}

\end{document}